%
\input lanlmac
%
\input epsf
%

\noblackbox
\def\inbar{\,\vrule height1.5ex width.4pt depth0pt}
\def\IC{\relax\hbox{$\inbar\kern-.3em{\rm C}$}}
\def\IR{\relax{\rm I\kern-.18em R}}
\font\cmss=cmss10 \font\cmsss=cmss10 at 7pt
\def\IZ{\relax\ifmmode\mathchoice
{\hbox{\cmss Z\kern-.4em Z}}{\hbox{\cmss Z\kern-.4em Z}}
{\lower.9pt\hbox{\cmsss Z\kern-.4em Z}}
{\lower1.2pt\hbox{\cmsss Z\kern-.4em Z}}\else{\cmss Z\kern-.4em Z}\fi}
 \def\zb{{\bar{\vphantom\i z}}}
\def\np{Nucl. Phys. } \def\pl{Phys. Lett. }
\def\pr{Phys. Rev. } \def\prl{Phys. Rev. Lett. }

\def\CS{{\cal S}}\def\CP{{\cal P}}\def\CJ{{\cal J}}
\def\CM{{\cal M}}\def\CV{{\cal V}}\def\CZ{{\cal Z}}
\def\p{\partial}

\def\DM{{\p\over \p \mu}}
\def\pb{\overline{\partial}}
\def\psib{\overline{\psi}}
\def\poin{Poincar\'e}
\def\d{{\rm d}}
\def\Sgh{S_{\rm gh}}
\def\lapl{\,\raise.5pt\hbox{$\mbox{.09}{.09}$}\,}
\def\ee{{\rm e}^}
\def\gc{g\dup_c}
\def\mod{{\rm mod}}
\def\t{\theta}
\def\hath{\hat{\theta}}
\font\manual=manfnt \def\dbend{\lower3.5pt\hbox{\manual\char127}}
\def\danger#1{\smallskip\noindent\rlap\dbend%
\indent{\bf #1}\par\nobreak\vskip-1.5pt\nobreak}

\def\exercise#1{\bgroup\narrower\footnotefont\baselineskip\footskip\bigbreak
\hrule\medskip\nobreak\noindent {\bf Exercise}. {\it #1\/}\par\nobreak}
\def\endexercise{\medskip\nobreak\hrule\bigbreak\egroup}

\def\figin{\epsfcheck\figin}\def\figins{\epsfcheck\figins}
\def\epsfcheck{\ifx\epsfbox\UnDeFiNeD
\message{(NO epsf.tex, FIGURES WILL BE IGNORED)}
\gdef\figin##1{\vskip2in}\gdef\figins##1{\hskip.5in}
\else\message{(FIGURES WILL BE INCLUDED)}%
\gdef\figin##1{##1}\gdef\figins##1{##1}\fi}
\def\DefWarn#1{}
\def\figinsert{\goodbreak\midinsert}
\def\ifig#1#2#3{\DefWarn#1\xdef#1{fig.~\the\figno}
\writedef{#1\leftbracket fig.\noexpand~\the\figno}%
\figinsert\figin{\centerline{#3}}\medskip\centerline{\vbox{\baselineskip12pt
\advance\hsize by -1truein\noindent\footnotefont{\bf Fig.~\the\figno:} #2}}
\bigskip\endinsert\global\advance\figno by1}


\def\nl{\par\nobreak}
\def\slcu{3.2}
\def\slcfac{3.9}
\def\slqu{3.10}
\def\sqliouv{2}
\def\stdeqg{4}
\def\spia{2.2}
\def\sskpztd{4.3}
\def\soln{(5.14)}
\def\stdcst{5}
\def\sdsmmcl{6.1}
\def\stdstes{5.3}
\def\stdstms{5.4}
\def\sbsap{5.6}
\def\sMMTI{7}
\def\sstagpf{7.3}
\def\sKdV{7.4}
\def\ssIM{7.5}
\def\ssmmm{7.6}
\def\sscsotmc{7.7}
\def\sdotcl{8.2}
\def\splsdi{(8.17)}
\def\smmttff{9}
\def\sltlo{10.2}
\def\swapmm{10.3}
\def\wvfnht{10.19}
\def\slascomm{11}
\def\ssmlaco{11.4}
\def\ssmlft{11.6}
\def\sfsdcft{12}
\def\ssposme{13.6}
\def\sscatt{13}
\def\srabsm{14.2}
\def\swvbs{14.3}
\def\ssnpsm{13.5}
\def\sTrr{13.9}
\def\srsva{14.1}
\def\svoccm{14}
\def\ssastdcc{14.4}
\def\ssascsc{14.5}
\def\sadfp{15}
\def\wvfnspl{(\hbox {A.}1)}
\def\wvfnsmi{(\hbox {A.}2)}
\def\asyi{(\hbox {A.}6)}

\lref\rDavidetal{F. David, Nucl. Phys. B257[FS14] (1985) 45, 543\semi
J. Ambj{\o}rn, B. Durhuus and J. Fr\"ohlich, Nucl. Phys. B257[FS14]
(1985) 433; J. Fr\"ohlich, in: Lecture Notes in Physics, Vol. 216,
ed. L. Garrido (Springer, Berlin, 1985)\semi
V. A. Kazakov, I. K. Kostov and A. A. Migdal, Phys. Lett. 157B (1985) 295;
D. Boulatov, V. A. Kazakov, I. K. Kostov and A. A. Migdal,
Phys. Lett. B174 (1986) 87; Nucl. Phys. B275[FS17] (1986) 641.}
\lref\rBIPZ{E. Br\'ezin, C. Itzykson, G. Parisi and J.-B. Zuber,
Comm. Math. Phys. 59 (1978) 35.}
\lref\rBIZ{D. Bessis, C. Itzykson, and J.-B. Zuber,
Adv. Appl. Math. 1 (1980) 109.}
\lref\rKPZ{V. G. Knizhnik, A. M. Polyakov, and A. B. Zamolodchikov,
Mod. Phys. Lett. A3 (1988) 819.}
\lref\rDDK{F. David, Mod. Phys. Lett. A3 (1988) 1651\semi
J. Distler and H. Kawai, Nucl. Phys. B321 (1989) 509.}
\lref\rDS{M. Douglas and S. Shenker, Nucl. Phys. B335 (1990) 635.}%
\lref\rBK{E. Br\'ezin and V. Kazakov, Phys. Lett. B236 (1990) 144.}%
\lref\rGM{D. Gross and A. Migdal, Phys. Rev. Lett. 64 (1990) 127;
Nucl. Phys. B340 (1990) 333.}%
\lref\rD{M. R. Douglas, Phys. Lett. B238 (1990) 176.}
\lref\rGGPZ{P. Ginsparg, M. Goulian, M. R. Plesser, and J. Zinn-Justin,
``$(p,q)$ string actions,'' Nucl. Phys. B342 (1990) 539.}
\lref\rGZaplob{P. Ginsparg and J. Zinn-Justin,
``Action principle and large order behavior of non-perturbative gravity'',
LA-UR-90-3687 / SPhT/90-140 (1990), published in proceedings of 1990
Carg\`ese workshop\semi
P. Ginsparg and J. Zinn-Justin, ``Large order behavior of nonperturbative
gravity,'' Phys. Lett. B255 (1991) 189.}
\lref\rkade{I. K. Kostov,  Nucl. Phys. B326 (1989) 583.}
\lref\rAGnotes{L. Alvarez-Gaum\'e, ``Random surfaces, statistical mechanics,
and string theory'', Lausanne lectures, winter 1990.}
\lref\rbook{P. Ginsparg and G. Moore, ``{\it Lectures on 2d gravity
and 2d string theory\/}, {\bf the book},'' Cambridge University Press, to appear later in 1993.}
\lref\rNotes{E. Br\'ezin, ``Large $N$ limit and discretized two-dimensional
quantum gravity'', in {\it Two dimensional quantum gravity and random
surfaces\/}, proceedings of Jerusalem winter school (90/91), edited by D.
Gross, T. Piran, and S. Weinberg\semi
D. Gross, ``The $c=1$ matrix models'',
in proceedings of Jerusalem winter school (90/91)\semi
J. Ma\~nes and Y. Lozano, ``Introduction to Nonperturbative 2d quantum
gravity'', Barcelona preprint UB-ECM-PF3/91.}
\lref\rDGZ{P. Di Francesco, P. Ginsparg, and J. Zinn-Justin,
``2D Gravity and Random Matrices,'' Physics Reports, to appear (1993).}
\lref\rAW{O. Alvarez and P. Windey, Nucl. Phys. B348 (1991) 490.}
\lref\rGD{I. M. Gel'fand and L. A. Dikii, Russian Math. Surveys 30:5 (1975)
77\semi
I. M. Gel'fand and L. A. Dikii, Funct. Anal. Appl. 10 (1976) 259.}
\lref\rdOneGK{D.J. Gross and I. Klebanov, Nucl. Phys. B344 (1990) 475;
Nucl. Phys. B354 (1991) 459.}
\lref\rCMM{M. L. Mehta, Comm. Math. Phys. 79 (1981) 327\semi
S. Chadha, G. Mahoux and M. L. Mehta, J. Phys. A14 (1981) 579\semi
C. Itzykson and J.B. Zuber, J. Math. Phys. 21 (1980) 411.}
\lref\rhar{Itzykson Zuber, Mehta, Harishandra, Duistermaat-Eckmann}
\lref\rWtp{E. Witten, Nucl. Phys. B340 (1990) 281.}
\lref\rKBK{V. Kazakov, Phys. Lett. 119A (1986) 140\semi
D. Boulatov and V. Kazakov, Phys. Lett. 186B (1987) 379.}
\lref\rising{E. Br\'ezin, M. Douglas, V. Kazakov, and S. Shenker, Phys. Lett.
B237 (1990) 43\semi
D. Gross and A. Migdal, Phys. Rev. Lett. 64 (1990) 717.}
\lref\rIYL{C. Crnkovi\'c, P. Ginsparg, and G. Moore,
Phys. Lett. B237 (1990) 196.}
\lref\rZJ{J. Zinn-Justin, {\it Quantum Field Theory and Critical Phenomena},
Oxford Univ. Press (1989).}
\lref\rBMP{E. Br\'ezin, E. Marinari, and G. Parisi, Phys. Lett. B242 (1990)
35.}
\lref\rDSS{M. Douglas, N. Seiberg, and S. Shenker, Phys. Lett. B244 (1990)
381.}
\lref\rpgtr{P. Ginsparg, ``Some statistical mechanical models and conformal
field theories,'' lectures given at Trieste spring school, 1989, published in
M.  Green and A. Strominger, eds., {\it Superstrings '89}, World Scientific
1990.}
\lref\rMoore{G. Moore,
``Geometry of the string equations,'' Comm. Math. Phys. 133 (1990) 261.}
\lref\rNeu{H. Neuberger, ``Regularized string and flow equations,''
Nucl. Phys. B352 (1991) 689.}
\lref\rDIFK{P. Di Francesco and D. Kutasov, Nucl. Phys. B342 (1990) 589;
and Princeton preprint PUPT-1206 (1990) published in proceedings of Carg\`ese
workshop (1990).}
\lref\rGTW{ A. Gupta, S. Trivedi and M. Wise, Nucl. Phys. B340 (1990) 475.}
\lref\rBerKl{M. Bershadsky and I. Klebanov, Phys. Rev. Lett. 65 (1990) 3088.}
\lref\rGKn{D. Gross and I. Klebanov, Nucl. Phys. B359 (1991) 3\semi
D. Gross, I. Klebanov, and M. Newman, Nucl. Phys. B350 (1991) 621.}
\lref\rColl{G. Mandal, A. Sengupta, and S. Wadia,  Mod. Phys. Lett. A6 (1991)
1465\semi
K. Demeterfi, A. Jevicki, and J.P. Rodrigues, Nucl. Phys. B362 (1991) 173.}
\lref\jevscatii{K. Demeterfi, A. Jevicki, and J. Rodrigues, ``Scattering
Amplitudes and Loop Corrections in Collective String Field Theory (II),''
Nucl. Phys. B365 (1991)499.}
\lref\rGLi{M. Goulian and M. Li, Phys. Rev. Lett. 66 (1991) 2051.}
\lref\rthooft{G. 't Hooft, Nucl. Phys. B72 (1974) 461.}
\lref\rpoly{A. M. Polyakov, Phys. Lett. 103B (1981) 207, 211.}
\lref\rO{O. Alvarez, Nucl. Phys. B216 (1983) 125.}
\lref\rMMHK{N. E. Mavromatos and J. L. Miramontes, Mod.Phys.Lett. A4 (1989)
1847\semi
E. d'Hoker and  P. S. Kurzepa, Mod.Phys.Lett. A5 (1990) 1411.}
\lref\rSliouv{E. D'Hoker, ``Continuum approaches to 2-D gravity'',
UCLA/91/TEP/41, review talk at Stonybrook Strings and
Symmetries conference, May 1991.}
\lref\rFrlh{D. Friedan, Les Houches lectures summer 1982, in {\it Recent
Advances in Field Theory and Statistical Physics\/}, J.-B. Zuber and R. Stora
eds, (North Holland, 1984).}
\lref\rOA{O. Alvarez, in {\it Unified String Theories},
M. Green and D. Gross, eds., (World Scientific, Singapore, 1986).}
\lref\rGS{M. B. Green and J. Schwarz, Phys. Lett. 149B (1984)
117.}
\lref\rkco{I. Kostov, ``Strings embedded in Dynkin Diagrams'',
SACLAY-SPHT-90-133 (1990), published in proceedings of Carg\`ese Workshop
(1990); Phys. Lett. B266 (1991) 42.}
\lref\rtmr{S. Kharchev, A. Marshakov, A. Mironov, A. Morozov, and A. Zabrodin,
``Unification of All String Models with $c<1$'' (hep-th/9111037),
Phys. Lett. B275 (1992) 311.}
\lref\rtadt{T. Tada, Phys. Lett. B259 (1991) 442.}
\lref\rmrdt{M. Douglas, ``The two-matrix model'', published in proceedings of
1990 Carg\`ese workshop.}
\lref\rdvv{R. Dijkgraaf, H. Verlinde, and E. Verlinde, ``Notes on topological
string theory and 2D quantum gravity'', Princeton preprint PUPT-1217, published
in proceedings of Carg\`ese workshop (1990)\semi
R. Dijkgraaf, ``Topological field theory and 2d quantum gravity'',
in proceedings of Jerusalem winter school (90/91).}
\lref\rfrdc{F. David, ``Nonperturbative effects in 2D gravity and matrix
models,'' Saclay-SPHT-90-178, published in proceedings of Carg\`ese workshop
(1990).}
\lref\witzwie{E. Witten and B. Zwiebach, ``Algebraic Structures and
Differential Geometry in 2D String Theory'' (hep-th/9201056),
Nucl. Phys. B377 (1992) 55.}
\lref\kazmig{V. A. Kazakov and A. A. Migdal, Nucl. Phys. B311 (1988) 171.}
\lref\pgrev{P. Ginsparg, ``Matrix models of 2d gravity,'' Trieste Lectures
(July, 1991), LA-UR-91-4101 (hep-th/9112013).}
\lref\bilal{A. Bilal, ``2d gravity from matrix models,'' Johns Hopkins
Lectures, CERN TH5867/90.}
\lref\kazrv{For a recent review see V. Kazakov, ``Bosonic strings and string
field theories in one-dimensional target space,'' LPTENS 90/30,  published
in proceedings of 1990 Carg\`ese workshop.}
\lref\klebrev{I. Klebanov, ``String theory in two dimensions'',
Trieste lectures, spring 1991, Princeton preprint PUPT--1271
(hep-th/9108019).}
\lref\nati{N. Seiberg, ``Notes on Quantum Liouville Theory and Quantum
Gravity,'' in {\it Common Trends in Mathematics and Quantum Field Theory\/},
Proc. of the 1990 Yukawa International Seminar, Prog. Theor. Phys. Suppl 102,
and in {\it Random surfaces and quantum gravity\/},
proceedings of 1990 Carg\`ese workshop, edited by
O. Alvarez, E. Marinari, and P. Windey, Plenum (1991).}
\lref\mart{E. Martinec, ``An Introduction to 2d Gravity and Solvable String
Models'' (hep-th/9112019), lectures at 1991 Trieste spring school, Rutgers
preprint RU-91-51.}
\lref\kdf{P. Di Francesco and D. Kutasov, Phys. Lett. 261B (1991) 385\semi
P. Di Francesco and D. Kutasov, ``World sheet and space time physics in two
dimensional (super) string theory'' (hep-th/9109005),
Nucl. Phys. B375 (1992) 119.}
\lref\kutrev{D. Kutasov, ``Some properties of (non) critical Strings'',
Trieste lectures, spring 1991, Princeton preprint PUPT--1277
(hep-th/9110041).}
\lref\bkstach{T. Banks, ``The tachyon potential in string theory,''
Nucl. Phys. B361 (1991) 166.}
\lref\rBPZ{A. A. Belavin, A. M. Polyakov and A. B. Zamolodchikov,
Nucl. Phys. B241 (1984) 333.}
\lref\FMS{D. Friedan, E. Martinec, and S. Shenker, Nucl. Phys. B271 (1986) 93.}
\lref\opform{E. Witten, Comm. Math. Phys. 113 (1988) 529\semi
L. Alvarez-Gaume, C. Gomez, G. Moore, and C. Vafa,
Nucl. Phys. B303 (1988) 455\semi
C. Vafa, Phys. Lett. 206B (1988) 421\semi
G. Segal, ``The definition of conformal field theory,'' unpublished.}
\lref\joetalk{J. Polchinski, ``Remarks on the Liouville Field Theory,''
UTTG-19-90,
published in Strings '90, Texas A\&M, Coll. Station Wkshp (1990) 62.}
\lref\mss{G. Moore, N. Seiberg, and M. Staudacher,
Nucl. Phys. {\bf 362} (1991) 665.}
\lref\volodya{V. Kazakov, Mod. Phys. Lett {\bf A4} (1989) 2125.}
\lref\fdavid{F. David, Loop equations and non-perturbative
effects in two dimensional quantum gravity,''
Mod. Phys. Lett. A5 (1990) 1019.}
\lref\fdavidi{F. David, ``Phases of the large $N$ matrix model and
non-perturbative effects in 2d gravity,'' Nucl. Phys. B348 (1991) 507.}
\lref\bdss{T. Banks, M. Douglas, N. Seiberg and S. Shenker, Phys. Lett.
{\bf 238B} (1990) 279.}
\lref\jakiew{E. D'Hoker and R. Jackiw, \prl {\bf 50} (1983) 1719; \pr {\bf D26}
(1982) 3517\semi
E. D'Hoker, D. Freedman and R. Jackiw, \pr {\bf D28} (1983) 2583.}
\lref\rstaudyl{M. Staudacher, ``The Yang--Lee edge singularity on a dynamical
planar random surface,'' Nucl. Phys. B336 (1990) 349.}
\lref\takht{L. Takhtadjan and P.G. Zograf,
Funct. Anal. Appl. {\bf 19} (1985) 67\semi
L. Takhtadjan, 
Proc. Symp. Pure Math. {\bf 49} (1989) 581.}
\lref\takhtii{F. Smirnoff and L. Takhtadjan, ``Towards a quantum
Liouville theory with $c>1$'' Univ. of Boulder preprint}
\lref\gervais{J.-L. Gervais and A. Neveu, \np
{\bf 199} (1982) 59; {\bf B209} (1982) 125; {\bf B224} (1983) 329;
{\bf 238} (1984) 125, 396; \pl {\bf 151B} (1985) 271;
J.-L. Gervais, LPTENS 89/14; 90/4.}
\lref\gervcorr{J.-L. Gervais, ``Gravity-Matter Couplings from
Liouville Theory,'' LPTENS-91/22 (hep-th/9205034).}
\lref\crtthrn{T.L. Curtright and C.B. Thorn, \prl {\bf 48} (1982) 1309;
E. Braaten, T. Curtright and C. Thorn, \pl {\bf 118B}
(1982) 115; Ann. Phys. {\bf 147} (1983) 365;
E. Braaten, T. Curtright, G. Ghandour and C. Thorn, \prl {\bf 51}
(1983) 19; Ann. Phys. {\bf 153} (1984) 147.}
\lref\reedsim{M. Reed and B. Simon, {\it Methods of Modern Mathematical
Physics\/}, Academic Press (1972) vol 1.}
\lref\mms{E. Martinec, G. Moore, and N. Seiberg,
``Boundary operators in 2-d gravity'' (hep-th/9109055),
Phys. Lett. {\bf 263B} (1991) 190.}
\lref\MTW{C. W. Misner, K. S. Thorne, and J. Wheeler, {\it Gravitation\/},
W.H. Freeman and Co. (1973).}
\lref\Wald{R. Wald, {\it General Relativity\/}, Univ. of Chicago Press (1984).}
\lref\wheeler{J. A. Wheeler, ``Geometrodynamics and the issue of the final
state''  in {\it Relativity, Groups, and Topology\/}, C. M. DeWitt and B. S.
DeWitt, eds., Gordon and Breach, N.Y. (1964).}
\lref\dewitt{B.S. DeWitt, ``Quantum Theory of Gravity, I: Canonical Theory,''
Phys. Rev. {\bf 160} (1967) 1113.}
\lref\laflamme{See R. Laflamme, ``Introduction and
Applications of Quantum Cosmology,'' 1991 Gift lectures
for a recent review of quantum cosmology.}
\lref\bouwknegt{P. Bouwknegt, J. McCarthy and K. Pilch,
``Fock space resolutions of the virasoro highest weight modules with $c\le1$,''
Lett. Math. Phys. 23 (1991) 193;
``BRST analysis of physical states for 2-d gravity coupled to $c\le1$
matter,'' Comm. Math. Phys. 145 (1992) 541; and reviews
CERN-TH-6646-92 (hep-th/9209034), CERN-TH-6645-92,
CERN-TH-6279-91 (hep-th/9110031).}
\lref\kacbook{V. Kac,
{\it Infinite Dimensional Lie Algebras\/}, Cambridge (1985).}
\lref\grndrng{E. Witten,
Ground ring of two-dimensional string theory'' (hep-th/9108004),
Nucl. Phys. B373 (1992) 187.}
\lref\joei{J. Polchinski, ``Critical Behavior of Random Surfaces in One
Dimension,'' Nucl. Phys. {\bf B346} (1990) 253.}
\lref\dj{S.R. Das and A. Jevicki, ``String Field Theory and Physical
Interpretation of D=1 Strings,'' Mod. Phys. Lett. {\bf A5} (1990) 1639.}
\lref\wadia{A.M. Sengupta and S.R. Wadia,
``Excitations and interactions in $d=1$ string theory,''
Int. Jour. Mod. Phys. {\bf A6} (1991) 1961.}
\lref\ambj{J. Ambjorn, J. Jurkiewicz, Yu.M. Makeenko,
``Multiloop correlators for two-dimensional quantum gravity,''
Phys. Lett. B251 (1990) 517.}
\lref\bkz{E. Br\'ezin, V. A. Kazakov, and Al. B. Zamolodchikov,
Nucl. Phys. B338 (1990) 673.}
\lref\GMil{D. J. Gross and N. Miljkovic, Phys. Lett.
{\bf 238B} (1990) 217.}
\lref\gzj{P. Ginsparg and J. Zinn-Justin, ``2D gravity + 1d matter,''
Phys. Lett. {\bf 240B} (1990) 333.}
\lref\parisi{G. Parisi, Phys. Lett. {\bf 238B} (1990) 209,213;
Europhys. Lett. 11 (1990) 595.}
\lref\danieli{U. Danielsson, ``Symmetries and special states in two-dimensional
string theory'' (hep-th/9112061), Nucl. Phys. B380 (1992) 83.}
\lref\joesea{J. Polchinski, ``Classical limit of 1+1 Dimensional
String Theory,'' Nucl. Phys. {\bf B362} (1991) 125.}
\lref\gki{D. Gross and I. Klebanov,
``Fermionic String Field Theory of $c=1$ 2D Quantum Gravity,''
Nucl. Phys. {\bf B352} (1991) 671.}
\lref\bz{B. Zwiebach, ``Closed string field theory:
quantum action and the B-V master equation''
(hep-th/9206084), Nucl. Phys. B390 (1993) 33.}
\lref\danielii{U. Danielssohn, ``A Study of Two Dimensional String Theory''
(hep-th/9205063) PhD Thesis, Princeton Univ (1992).}
\lref\mrpl{Moore and Plesser, ``Classical
Scattering in $1+1$ Dimensional String Theory'' (hep-th/9203060),
to appear in Phys. Rev. D.}
\lref\gradsh{I.S. Gradshteyn and I.M. Ryzhik, {\it Tables of Integrals,
Series, and Products\/}, Academic Press (1980).}
\lref\abram{M. Abramowitz and I. Stegun, {\it Handbook of Mathematical
Functions\/}, Dover (1968).}
\lref\rcftrev{P. Ginsparg, ``Applied conformal field theory,'' Les Houches
Session XLIV, 1988, in {\it Fields, Strings, and Critical Phenomena\/}, ed.\ by
E. Br\'ezin and J. Zinn-Justin, North Holland (1989), and references therein.}
\lref\moore{G. Moore, ``Double-scaled field theory at $c=1$,''
Nucl. Phys. {\bf B368} (1992) 557.}
\lref\msi{G. Moore and N. Seiberg, ``From loops to fields in
2d gravity,'' Int. Jour. Mod. Phys. {\bf A7} (1992) 2601.}
\lref\mpr{G. Moore, R. Plesser, and S. Ramgoolam,
``Exact S-Matrix for 2D String Theory'' (hep-th/9111035) Nucl. Phys.
{\bf B377}(1992)143.}
\lref\dmrpl{R. Dijkgraaf, G. Moore, and R. Plesser,
``The Partition Function of 2D String Theory,'' hep-th/9208031,
submitted to Nucl. Phys. B}
\lref\mrsg{G. Moore, ``Gravitational Phase Transitions and
the Sine-Gordon Model,'' Yale preprint YCTP-P1-92, hep-th/9203061.}
\lref\seibshen{N. Seiberg and S. Shenker, ``A note on background independence''
(hep-th/9201017), Phys. Rev. D45 (1992) 4581.}
\lref\farkra{H. M. Farkas and I. Kra, {\it Riemann Surfaces\/}, Springer
(1980).}
\lref\fuchsprs{H. Poincar\'e, {\it Papers on Fuchsian Functions},
J. Stillwell, transl., Springer Verlag 1985.}
\lref\abz{A.B. Zamolodchikov, ``On the entropy of random surfaces,''
Phys. Lett. {\bf 117B} (1982) p. 87.}
\lref\wvii{D. Minic, J. Polchinski, and Z. Yang, ``Translation-invariant
backgrounds in 1+1 dimensional string theory,''
Nucl. Phys. {\bf B362} (1991) 125.}
\lref\wviii{J. Avan and A. Jevicki,
``Classical integrability and higher symmetries of collective field
theory,'' Phys. Lett. {\bf 266B} (1991) 35;
``Quantum integrability and exact eigenstates of the collective string
field theory,'' Phys. Lett. {\bf 272B} (1991) 17;
``Algebraic Structures and Eigenstates for Integrable Collective Field
Theories'' (hep-th/9202065);
``Interacting Theory of Collective and Topological Fields in
2 Dimensions'' (hep-th/9209036).}
\lref\avjvwinft{J. Avan and A. Jevicki,``String field actions from
W-infinity'' (hep-th/9111028), Mod. Phys. Lett. A7 (1992) 357.}
\lref\wix{S.R. Das, A. Dhar, G. Mandal, S. R. Wadia,
``Gauge theory formulation of the $c{=}1$ matrix model:
symmetries and discrete states'' (hep-th/9110021)
Int. J. Mod. Phys. A7 (1992) 5165;
``Bosonization of nonrelativistic fermions and $W_\infty$ algebra,''
Mod. Phys. Lett. {\bf A7} (1992) 71\semi
A. Dhar, G. Mandal, S. R. Wadia,
``Classical Fermi fluid and geometric action for $c=1$'' (hep-th/9204028)
IASSNS-HEP-91-89;
``Non-relativistic fermions, coadjoint orbits of $w_\infty$ and string field
theory at $c=1$'' (hep-th/9207011),  TIFR-TH-92-40.}
\lref\nehari{Z. Nehari, {\it Conformal Mapping}. McGraw-Hill 1952.}
\lref\gunning{R. Gunning, {\it Lectures on Riemann Surfaces},
Princeton University Press (1966).}
\lref\cardy{J. Cardy, ``Conformal invariance and surface critical behavior,''
Nucl. Phys. {\bf B240} (1984) 514\semi
J. Cardy, ``Boundary conditions, fusion rules and the Verlinde formula,''
Nucl. Phys. B324 (1989) 581.}
\lref\hempel{J.A. Hempel, Bull. Lond. Math. Soc. {\bf 20} (1988) 97.}
\lref\polgrav{J. Polchinski, ``A two-dimensional model for quantum gravity,''
Nucl. Phys. {\bf B324} (1989) 123.}
\lref\kutseib{D. Kutasov and N. Seiberg, ``Number of degrees of freedom,
density of states, and tachyons in string theory and CFT,''
Nucl. Phys. B358 (1991) 600.}
\lref\kutseibss{D. Kutasov and N. Seiberg, Phys. Lett. B251 (1990) 67.}
\lref\faddeev{L. Faddeev, in {\it Methods in Field Theory\/},
Les Houches Summer 1975, ed.\ by R. Balian and J. Zinn-Justin,
North Holland/World Scientific (1976/1981).}
\lref\vermster{E. Verlinde, Nucl. Phys. B381 (1992) 141.}
\lref\sidney{S. Coleman, {\it Aspects of Symmetry\/}, Cambridge (1985).}
\lref\tomunpb{T. Banks, unpublished. In this work Banks
showed how to derive the linear equation for the cosmological
constant wavefunction in pure gravity from the nonlinear
equation.}
\lref\zamosq{Al. B. Zamolodchikov and A.B. Zamolodchikov,
``Factorized S matrices in two-dimensions as the exact solutions of
certain relativistic quantum field models,''
Ann. Phys. 120 (1979) 253.}
\lref\polss{A. M. Polyakov, ``Self-tuning fields and resonant correlations in
2-d gravity,'' Mod. Phys. Lett. A6 (1991) 635.}
\lref\klebpasq{I. Klebanov and A. Pasquinucci,
``Correlation functions from two-dimensional string Ward identities''
(hep-th/9204052) PUPT-1313.}
\lref\minicyang{G. Minic and Z. Yang, ``Is $S = 1$ for $c = 1$?''
Phys. Lett. B274 (1992) 27.}
\lref\lowe{D. Lowe, ``Unitarity Relations in $c=1$ Liouville Theory''
(hep-th/9204084), Mod. Phys. Lett. A7 (1992) 2647.}
\lref\bakas{I. Bakas, Phys. Lett. {\bf B228} (1989) 57.}
\lref\hull{C.M. Hull, ``The Geometry of $W$-Gravity,'' QMW/PH/91/6.}
\lref\pope{C.N. Pope, L.J. Romans, and X. Shen, ``A brief history of
$W_\infty$,'' CTP TAMU-89/90, published in
Coll. Station Wkshp (1990) 287.}
\lref\kawaii{M. Fukuma, H. Kawai, and R. Nakayama, ``Infinite
Dimensional Grassmannian Structure of Two-Dimensional Quantum
Gravity,'' Comm. Math. Phys. {\bf 143} (1992) 371.}
\lref\rfknm{M. Fukuma, H. Kawai, R. Nakayama,
``Continuum schwinger-dyson equations and universal structures in
two-dimensional quantum gravity,'' Int. J. Mod. Phys. A6 (1991) 1385\semi
M. Bowick, A. Morozov, Danny Shevitz, ``Reduced unitary matrix models and the
hierarchy of tau functions,'' Nucl. Phys. B354 (1991) 496\semi
By Yu. Makeenko, A. Marshakov, A. Mironov, A. Morozov, ``Continuum versus
discrete virasoro in one matrix models,'' Nucl. Phys. B356 (1991) 574.}
\lref\kutmarsei{D. Kutasov, E. Martinec, and N. Seiberg, ``Ground Rings and
Their Modules in 2D Gravity with $c\leq 1$ Matter'' (hep-th/9111048),
Phys. Lett. B276 (1992) 437.}
\lref\wittbh{E. Witten, ``On String theory and black holes,''
Phys. Rev. D44 (1991) 314.}
\lref\zamii{A.B. Zamolodchikov and Al.B. Zamolodchikov,
``Massless factorized scattering and sigma models with topological terms,''
Nucl. Phys. B379 (1992) 602.}
\lref\witosft{E. Witten, Phys. Rev. D46 (1992) 5467; and IASSNS-HEP-92-63
(hep-th/9210065).}
\lref\kleblow{I. Klebanov and D. Lowe, Nucl. Phys. B363 (1991) 543.}
\lref\berkut{M. Bershadsky and D. Kutasov,
Phys. Lett. B274 (1992) 331-337; Nucl. Phys. B382 (1992) 213.}
\lref\klebpol{I.R. Klebanov and A.M. Polyakov, ``Interactions of Discrete
States in Two-Dimensional String Theory,''
Mod. Phys. Lett. A6 (1991) 3273.}
\lref\klebward{I.R. Klebanov, ``Ward Identities in Two-Dimensional
String Theory,'' Mod. Phys. Lett. A7 (1992) 723.}
\lref\flm{I. Frenkel, J. Lepowsky, A. Meurman,
{\it Vertex Operator Algebras and the Monster Group\/},
Academic Press (1988).}
\lref\cthrm{A. Zamolodchikov, ``Irreversibility of the flux of the
renormalization group in a 2D field theory,'' JETP Lett. {\bf 43} (1986) 730.}
\lref\amit{D.J. Amit, Y.Y. Goldschmidt, and G. Goldin, ``Renormalisation group
analysis of the phase transition in the 2D Coulomb gas, Sine-Gordon theory and
XY model,'' J. Phys. {\bf A13} (1980) 585.}
\lref\benlee{B. Lee in ``Methods in Field Theory,'' Les Houches, 1975,
section 6.3.}
\lref\rDrSo{V. G. Drinfel'd and V. V. Sokolov, Jour. Sov. Math. (1985)
1975\semi
G. Segal and G. Wilson, Pub. Math. I.H.E.S. 61 (1985) 5.}
\lref\grschwtt{M. Green, J. Schwarz, and E. Witten, {\it Superstring theory\/},
Cambridge Univ. Press (1987).}
\lref\kostaudon{I. Kostov and M. Staudacher,
``Multicritical phases of the $O(N)$ model on a random lattice''
(hep-th/9203030), Nucl. Phys. B384 (1992) 459.}
\lref\koststau{I. Kostov and M. Staudacher,
``Strings in discrete and continuous target spaces: a comparison''
(hep-th/9208042),  RU-92-21, Submitted to Phys. Lett. B.}
\lref\martshat{E. Martinec and S. Shatashvili, ``Black hole physics and
liouville theory,'' Nucl. Phys. B368 (1992) 338.}
\lref\xyzmdl{N. Andrei and J. Loewenstein, Phys. Lett. 91B (1980) 401\semi
V. Korepin, Theor. and Math. Physics 41 (1979) 953.}
\lref\msrcft{G. Moore and N. Seiberg, ``Lectures on RCFT,''
RU-89-32-mc, lectures at Trieste Spring School 1989,
published in Trieste Superstrings, edited by A. Strominger and M. Green,
World Scientific (1989),
also in Banff NATO ASI (1989) 236.}
\lref\bouwlect{P. Bouwknegt and K. Schoutens, ``$W$-Symmetry in
Conformal Field Theory'' (hep-th/9210010), CERN-TH-6583/92.}
\lref\lzi{B. Lian and G. Zuckerman,
Phys. Lett. B254 (1991) 417;  Phys. Lett. B266 (1991) 21;
Comm. Math. Phys. 135 (1991) 547; Comm. Math. Phys. 145 (1992) 561.}
\lref\rbong{B. Lian, ``Semi-infinite homology and 2d quantum gravity,''
PhD thesis, Yale Univ. (1991).}
\lref\lziii{B. Lian and G. Zuckerman,
``New perspectives on the brst algebraic structure of string theory,''
(hep-th/9211072) TORONTO-9211072.}
\lref\sarmadi{M.H. Sarmadi, ``The ring structure of
chiral operators for minimal models coupled to 2D  gravity,'' IC/92/301}
\lref\catoptric{J.J. Atick,  G. Moore, and A. Sen, ``Catoptric
Tadpoles,'' Nucl. Phys. {\bf B307} (1988).}
\lref\eguchi{T. Eguchi, H. Kanno, S.-K. Yang, ``$W_{\infty}$-Algebra
in Two-Dimensional Black Hole,'' NIP-NI92004}
\lref\dgseminar{D. Gross, Seminar at Yale, Oct. 27, 1992.}
\lref\cllnwlck{Callan and Wilczek, Nucl. Phys. {\bf B340} (1990) 366.}
\lref\msisc{See, e.g., \msi\ section 4}
\lref\unpbshd{G. Moore and R. Plesser, ``Dynamical determination
of the string coupling constant,'' unpublished.}
\lref\kutflow{D. Kutasov, ``Irreversibility of the
Renormalization Group Flow in Two Dimensional Quantum Gravity,''
Mod. Phys. Lett. A7 (1992) 2943.}
\lref\rnonor{ G. Harris and E. Martinec, Phys. Lett. B245 (1990) 384\semi
E. Brezin and H. Neuberger,
Phys. Rev. Lett. 65 (1990) 2098; Nucl.Phys. B350 (1991) 513.}
\lref\tanii{N. Sakai and Y. Tanii, ``Operator product expansion and topological
states in $c=1$ Matter Coupled to 2D Gravity'' (hep-th/9111049),
Prog. Theor. Phys. Supp. 110 (1992) 117;
``Factorization and topological states in $c=1$ matter coupled to 2D gravity''
(hep-th/9108027), Phys. Lett. B276 (1992) 41.}
\lref\rfftech{Y. Kitazawa, Phys. Lett. B265 (1991) 262\semi
N. Sakai and Y. Tanii, Prog. Theor. Phys. 86 (1991) 547\semi
V. Dotsenko, Mod. Phys. Lett. A6 (1991) 3601.}
\lref\rmabetal{R. Myers, ``New dimensions for old strings,''
Phys. Lett. 199B (1987) 371\semi
I. Antoniadis, C. Bachas, John Ellis, and D.V. Nanopoulos,
``An expanding universe in string theory,'' Nucl. Phys. B328 (1989) 117.}
\lref\rsmth{M. E. Agishtein and A. A. Migdal,
Int. J. Mod. Phys. {\bf C1} (1990) 165; Nucl. Phys. {\bf B350} (1991) 690\semi
F. David, ``What is the intrinsic geometry of two-dimensional quantum
gravity?'' Nucl. Phys. B368 (1992) 671\semi
S. Jain and S. Mathur, ``World sheet geometry and baby universes
in 2-d quantum gravity,'' Phys. Lett. B286 (1992) 239\semi
H. Kawai, N. Kawamoto, T. Mogami and Y. Watabiki,
``Transfer Matrix Formalism for Two-Dimensional Quantum Gravity and Fractal
Structures of Space-time,'' INS-969 (hep-th/9302133).}
\lref\rdijklect{R. Dijkgraaf, lectures in this volume.}
\lref\rtatastuff{S. Das, ``Matrix models and black holes'' (hep-th/9210107),
Mod. Phys. Lett. A8 (1993) 69\semi
A. Dhar, G. Mandal, S. Wadia, ``Stringy quantum effects
in two-dimensional black hole'' (hep-th/9210120),
Mod. Phys. Lett. A7 (1992) 3703\semi
T. Yoneya, ``Matrix models and 2-d critical string theory:
2-D black hole by $c=1$ matrix model'' (hep-th/9211079), UT-KOMABA-92-13.}
\lref\rCGHS{C. Callan, S. Giddings, J. Harvey, and A. Strominger,
``Evanescent Black Holes'' (hep-th/9111056), Phys. Rev. D45 (1992) 1005.}
\lref\rcecvaf{S. Cecotti and C. Vafa, Nucl. Phys. B367 (1991) 359.}
\lref\rwoso{E. Witten, ``Some remarks about string field theory,''
Princeton preprint 86-1188 (1986), Physica Scripta T15 (1987) 70,
and in Marstrand Nobel Sympos. (1986) 70.}
\lref\rChTh{A. Chodos and C. Thorn, ``Making the massless string massive,''
Nucl. Phys. B72 (1974) 509.}
\lref\rDotFat{Vl. S. Dotsenko and V. A. Fateev, ``Conformal algebra and
multipoint correlation functions in 2d statistical models,'' Nucl. Phys.
B240[FS12] (1984) 312\semi Vl. S. Dotsenko and V. A. Fateev, ``Four-point
correlation functions and the operator algebra in 2d conformal invariant
theories with central charge $c\le1$,'' Nucl. Phys. B251[FS13] (1985) 691.}
\lref\rbolo{T. Banks and M. O'Loughlin, ``Nonsingular lagrangians for
two-dimensional black holes'' (hep-th/9212136), RU-92-61.} 
\lref\rkbm{Ivan Kostov, ``Two point correlator for the $d=1$ closed bosonic
string,'' Phys. Lett. (1988) B215\semi
S. Ben-Menahem, ``Two and three point functions in the $d=1$ matrix model,''
Nucl. Phys. B364 (1991) 681.}
\lref\rcmpf{C.G. Callan, E.J. Martinec, M.J. Perry, D. Friedan,
``Strings in background fields,'' Nucl. Phys. B262 (1985) 593.}
\lref\getzler{E. Getzler, ``Batalin-Vilkovisky Algebras and
Two-Dimensional Toplogical Field Theories'' (hep-th/9212043)\semi
P. Horava, ``Spacetime  Diffeomorphisms and Topological $w_\infty$
Symmetry in Two Dimensional Topological String Theory'' (hep-th/9302020);
``Two Dimensional String Theory and the Topological Torus'' 
(hep-th/9202008), Nucl. Phys. B386 (1992) 383.}
\lref\graeme{G. Segal, Lectures at the Isaac Newton Institute,
August 1992, and lectures at Yale University, March 1993.}
\lref\brshkutbh{ M. Bershadsky and D. Kutasov, ``Comment on gauged WZW
theory,'' Phys. Lett. B266 (1991) 345.}
\lref\russo{Jorge G. Russo,
``Black hole formation in $c = 1$ string field theory'' (hep-th/9211057),
Phys. Lett. B300 (1993) 336.}
\lref\wittconf{Edward Witten, ``Two-dimensional string theory and black holes''
(hep-th/9206069), Lecture given at Conf.\ on Topics in Quantum Gravity,
Cincinnati, OH, Apr 3--4, 1992.}
\lref\dvvbh{R. Dijkgraaf, H. Verlinde, and E. Verlinde,
``String propagation in a black hole geometry,''
Nucl. Phys. B371 (1992) 269.}
\lref\mukhivafa{S. Mukhi and C. Vafa, ``Two dimensional black hole as a
topological coset model of $c=1$ string theory'' (hep-th/9301083).}
\lref\wittnmatrix{Edward Witten, ``The $n$ matrix model and gauged WZW
models,'' Nucl. Phys. B371 (1992) 191.}
\lref\distlervafa{J. Distler and C. Vafa, in proceedings of Carg\`ese 1990
workshop;
``A critical matrix model at $c = 1$,'' Mod. Phys. Lett. A6 (1991) 259.}
\lref\mpdoug{J. Minahan and A. Polychronakos, ``Equivalence of Two
Dimensional QCD and the $c=1$ Matrix Model'' (hep-th/9303153)\semi
M. Douglas, ``Conformal Field Theory Techniques for
Large $N$ Group Theory'' (hep-th/9303159).}
\lref\rrrecent{F. David, ``Simplicial quantum gravity and random lattices''
(hep-th/9303127), Lectures given at Les Houches Summer School, July 1992,
Saclay T93/028\semi
A. Morozov, ``Integrability and Matrix Models'' (hep-th/9303139), ITEP-M2/93.}

\Title{\vbox{\baselineskip12pt\hbox{YCTP-P23-92}
\hbox{LA-UR-92-3479}\hbox{hep-th/9304011}}}
{\vbox{\centerline{Lectures on 2D Gravity}
\centerline{and}\vskip2pt
\centerline{2D String Theory}}}

\centerline{\vbox{\hsize3in\centerline{P. Ginsparg}
\bigskip\centerline{ginsparg@xxx.lanl.gov}
\smallskip\centerline{MS-B285}
\centerline{Los Alamos National Laboratory}
\centerline{Los Alamos, NM \ 87545}}
\vbox{\hbox{\quad and\quad}\vskip.75in}
\vbox{\hsize3in\centerline{Gregory Moore}
\bigskip\centerline{moore@castalia.physics.yale.edu}
\smallskip\centerline{Dept.\ of Physics}
\centerline{Yale University}
\centerline{New Haven, CT \ 06511}}}

\vskip .5in

\noindent
These notes are based on lectures delivered at the 1992 Tasi summer school.
They constitute the preliminary version
of a book which will include many corrections and much more useful information.
Constructive comments are welcome.

\vskip.5in\centerline
{\it Lectures given June 11--19, 1992 at TASI summer school, Boulder, CO}

\Date{1992/1993}

\centerline{\bf Contents}\nobreak\medskip{\baselineskip=12pt
 \footnotefont\parskip=0pt\catcode`\@=11 
\def\leaderfill#1#2{\leaders\hbox to 1em{\hss.\hss}\hfill%
\ifx\answ\bigans#1\else#2\fi}
\noindent {0.} {Introduction, Overview, and Purpose} \leaderfill{3}{3} \par
\noindent \quad{0.1.} {Philosophy and Diatribe} \leaderfill{3}{3} \par
\noindent \quad{0.2.} {2D Gravity and 2D String theory} \leaderfill{5}{6} \par
\noindent \quad{0.3.} {Review of reviews} \leaderfill{7}{8} \par
\noindent {1.} {Loops and States in Conformal Field Theory} \leaderfill{8}{9}
\par
\noindent \quad{1.1.} {Lagrangian formalism} \leaderfill{8}{9} \par
\noindent \quad{1.2.} {Hamiltonian formalism} \leaderfill{9}{10} \par
\noindent \quad{1.3.} {Equivalence of states and operators} \leaderfill{10}{11}
\par
\noindent \quad{1.4.} {Gaussian Field with a Background Charge}
\leaderfill{12}{13} \par
\noindent {2.} {2D Euclidean Quantum Gravity I: Path Integral Approach}
\leaderfill{13}{14} \par
\noindent \quad{2.1.} {2D Gravity and Liouville Theory} \leaderfill{13}{15}
\par
\noindent \quad{2.2.} {Path integral approach to 2D Euclidean Quantum Gravity}
\leaderfill{14}{16} \par
\noindent {3.} {Brief Review of the Liouville Theory} \leaderfill{22}{25} \par
\noindent \quad{3.1.} {Classical Liouville Theory} \leaderfill{22}{26} \par
\noindent \quad{3.2.} {Classical Uniformization} \leaderfill{24}{28} \par
\noindent \quad{3.3.} {Quantum Liouville Theory} \leaderfill{26}{29} \par
\noindent \quad{3.4.} {Spectrum of Liouville Theory} \leaderfill{28}{32} \par
\noindent \quad{3.5.} {Semiclassical States} \leaderfill{31}{35} \par
\noindent \quad{3.6.} {Seiberg bound} \leaderfill{33}{37} \par
\noindent \quad{3.7.} {Semiclassical Amplitudes} \leaderfill{35}{39} \par
\noindent \quad{3.8.} {Operator Products in Liouville Theory}
\leaderfill{39}{44} \par
\noindent \quad{3.9.} {Liouville Correlators from Analytic Continuation}
\leaderfill{40}{45} \par
\noindent \quad{3.10.} {Quantum Uniformization} \leaderfill{41}{47} \par
\noindent \quad{3.11.} {Surfaces with boundaries} \leaderfill{48}{54} \par
\noindent {4.} {2D Euclidean Quantum Gravity II: Canonical Approach}
\leaderfill{50}{57} \par
\noindent \quad{4.1.} {Canonical Quantization of Gravitational Theories}
\leaderfill{50}{57} \par
\noindent \quad{4.2.} {Canonical Quantization of 2D Euclidean Quantum Gravity}
\leaderfill{51}{58} \par
\noindent \quad{4.3.} {KPZ states in 2D Quantum Gravity} \leaderfill{52}{60}
\par
\noindent \quad{4.4.} {LZ states in 2D Quantum Gravity} \leaderfill{53}{61}
\par
\noindent \quad{4.5.} {States in 2D Gravity Coupled to a Gaussian Field: more
BRST} \leaderfill{54}{62} \par
\noindent {5.} {2D Critical String Theory} \leaderfill{61}{70} \par
\noindent \quad{5.1.} {Particles in $D$ Dimensions: QFT as 1D Euclidean Quantum
Gravity.} \leaderfill{62}{71} \par
\noindent \quad{5.2.} {Strings in $D$ Dimensions: String Theory as 2D Euclidean
Quantum Gravity} \leaderfill{64}{73} \par
\noindent \quad{5.3.} {2D String Theory: Euclidean Signature}
\leaderfill{66}{76} \par
\noindent \quad{5.4.} {2D String Theory: Minkowskian Signature}
\leaderfill{68}{78} \par
\noindent \quad{5.5.} {Heterodox remarks regarding the ``special states''}
\leaderfill{69}{80} \par
\noindent \quad{5.6.} {Bosonic String Amplitudes and the ``$c>1$ problem''}
\leaderfill{72}{82} \par
\noindent {6.} {Discretized surfaces, matrix models, and the continuum limit}
\leaderfill{75}{86} \par
\noindent \quad{6.1.} {Discretized surfaces} \leaderfill{75}{86} \par
\noindent \quad{6.2.} {Matrix models} \leaderfill{77}{88} \par
\noindent \quad{6.3.} {The continuum limit} \leaderfill{81}{93} \par
\noindent \quad{6.4.} {A first look at the double scaling limit}
\leaderfill{83}{95} \par
\noindent {7.} {Matrix Model Technology I: Method of Orthogonal Polynomials}
\leaderfill{84}{96} \par
\noindent \quad{7.1.} {Orthogonal polynomials} \leaderfill{84}{96} \par
\noindent \quad{7.2.} {The genus zero partition function} \leaderfill{86}{99}
\par
\noindent \quad{7.3.} {The all genus partition function} \leaderfill{88}{101}
\par
\noindent \quad{7.4.} {The Douglas Equations and the KdV hierarchy}
\leaderfill{90}{104} \par
\noindent \quad{7.5.} {Ising Model} \leaderfill{92}{106} \par
\noindent \quad{7.6.} {Multi-Matrix Models} \leaderfill{94}{108} \par
\noindent \quad{7.7.} {Continuum Solution of the Matrix Chains}
\leaderfill{95}{109} \par
\noindent {8.} {Matrix Model Technology II: Loops on the Lattice}
\leaderfill{99}{114} \par
\noindent \quad{8.1.} {Lattice Loop Operators} \leaderfill{99}{114} \par
\noindent \quad{8.2.} {Precise definition of the continuum limit}
\leaderfill{101}{116} \par
\noindent \quad{8.3.} {The Loop Equations} \leaderfill{103}{118} \par
\noindent {9.} {Matrix Model Technology III: Free Fermions from the Lattice}
\leaderfill{105}{121} \par
\noindent \quad{9.1.} {Lattice Fermi Field Theory} \leaderfill{105}{121} \par
\noindent \quad{9.2.} {Eigenvalue distributions} \leaderfill{106}{123} \par
\noindent \quad{9.3.} {Double--Scaled Fermi Theory} \leaderfill{109}{125} \par
\noindent {10.} {Loops and States in Matrix Model Quantum Gravity}
\leaderfill{113}{130} \par
\noindent \quad{10.1.} {Computation of Macroscopic Loops} \leaderfill{113}{130}
\par
\noindent \quad{10.2.} {Loops to Local Operators} \leaderfill{116}{133} \par
\noindent \quad{10.3.} {Wavefunctions and Propagators from the Matrix Model}
\leaderfill{117}{135} \par
\noindent \quad{10.4.} {Redundant operators, singular geometries and contact
terms} \leaderfill{120}{138} \par
\noindent {11.} {Loops and States in the $c=1$ Matrix Model}
\leaderfill{120}{139} \par
\noindent \quad{11.1.} {Definition of the $c=1$ Matrix Model }
\leaderfill{120}{139} \par
\noindent \quad{11.2.} {Matrix Quantum Mechanics} \leaderfill{122}{141} \par
\noindent \quad{11.3.} {Double-Scaled Fermi Field Theory} \leaderfill{127}{147}
\par
\noindent \quad{11.4.} {Macroscopic Loops at $c=1$} \leaderfill{129}{149} \par
\noindent \quad{11.5.} {Wavefunctions and Wheeler--DeWitt Equations}
\leaderfill{134}{154} \par
\noindent \quad{11.6.} {Macroscopic Loop Field Theory and $c=1$ scaling}
\leaderfill{134}{155} \par
\noindent \quad{11.7.} {Correlation functions of Vertex Operators}
\leaderfill{136}{158} \par
\noindent {12.} {Fermi Sea Dynamics and Collective Field Theory}
\leaderfill{139}{160} \par
\noindent \quad{12.1.} {Time dependent Fermi Sea} \leaderfill{139}{160} \par
\noindent \quad{12.2.} {Collective Field Theory} \leaderfill{140}{161} \par
\noindent \quad{12.3.} {Relation to 1+1 dimensional relativistic field theory}
\leaderfill{142}{163} \par
\noindent \quad{12.4.} {$\tau $-space and $\phi $-space} \leaderfill{143}{164}
\par
\noindent \quad{12.5.} {The $w_{\infty }$ Symmetry of the Harmonic Oscillator}
\leaderfill{146}{168} \par
\noindent \quad{12.6.} {The $w_{\infty }$ Symmetry of Free Field Theory}
\leaderfill{148}{171} \par
\noindent \quad{12.7.} {$w_{\infty }$ symmetry of Classical Collective Field
Theory} \leaderfill{149}{172} \par
\noindent {13.} {String scattering in two spacetime dimensions}
\leaderfill{151}{174} \par
\noindent \quad{13.1.} {Definitions of the $S$-Matrix} \leaderfill{151}{174}
\par
\noindent \quad{13.2.} {On the Violation of Folklore} \leaderfill{155}{178}
\par
\noindent \quad{13.3.} {Classical scattering in collective field theory}
\leaderfill{156}{180} \par
\noindent \quad{13.4.} {Tree-Level Collective Field Theory $S$-Matrix}
\leaderfill{158}{182} \par
\noindent \quad{13.5.} {Nonperturbative $S$-matrices} \leaderfill{159}{183}
\par
\noindent \quad{13.6.} {Properties of $S$-Matrix Elements}
\leaderfill{163}{188} \par
\noindent \quad{13.7.} {Unitarity of the $S$-Matrix} \leaderfill{165}{190} \par
\noindent \quad{13.8.} {Generating functional for $S$-matrix elements}
\leaderfill{167}{192} \par
\noindent \quad{13.9.} {Tachyon recursion relations} \leaderfill{169}{195} \par
\noindent \quad{13.10.} {The many faces of $c=1$} \leaderfill{171}{197} \par
\noindent {14.} {Vertex Operator Calculations and Continuum Methods}
\leaderfill{172}{199} \par
\noindent \quad{14.1.} {Review of the Shapiro-Virasoro Amplitude}
\leaderfill{172}{199} \par
\noindent \quad{14.2.} {Resonant Amplitudes and the ``Bulk $S$-Matrix''}
\leaderfill{174}{201} \par
\noindent \quad{14.3.} {Wall vs.\ Bulk Scattering} \leaderfill{177}{204} \par
\noindent \quad{14.4.} {Algebraic Structures of the 2D String: Chiral
Cohomology} \leaderfill{179}{207} \par
\noindent \quad{14.5.} {Algebraic Structures of the 2D String: Closed String
Cohomology} \leaderfill{183}{212} \par
\noindent {15.} {Achievements, Disappointments, Future Prospects}
\leaderfill{184}{213} \par
\noindent \quad{15.1.} {Lessons} \leaderfill{185}{213} \par
\noindent \quad{15.2.} {Disappointments} \leaderfill{186}{214} \par
\noindent \quad{15.3.} {Future prospects and Open Problems}
\leaderfill{187}{215} \par
\noindent Appendix {A.} {Special functions} \leaderfill{188}{217} \par
\noindent \quad{\hbox {A.}1.} {Parabolic cylinder functions}
\leaderfill{188}{217} \par
\noindent \quad{\hbox {A.}2.} {Asymptotics} \leaderfill{189}{218} \par
\catcode`\@=12\bigbreak\bigskip}

\secno-1
\newsec{Introduction, Overview, and Purpose} 

\subsec{Philosophy and Diatribe}

Following the discovery of spacetime anomaly cancellation in 1984 \rGS,
string theory has undergone rapid development in several directions. The
early hope of making direct contact with conventional particle physics
phenomenology has however long since dissipated, and there is as yet
no experimental program for finding even indirect manifestations of
underlying string degrees of freedom in nature.
The question of whether string theory is
``correct'' in the physical sense thus remains impossible to answer for
the foreseeable future. Particle/string theorists nonetheless continue to be
tantalized by the richness of the theory and by its natural ability to provide
a consistent microscopic underpinning for both gauge theory and gravity.

A prime obstacle to our understanding of string theory has been
an inability to penetrate beyond its perturbative expansion. Our
understanding of gauge theory is enormously enhanced by having a
fundamental formulation based on the principle of local gauge invariance
from which the perturbative expansion can be derived. Symmetry breaking and
nonperturbative effects such as instantons admit a clean and intuitive
presentation. In string theory, our lack of
a fundamental formulation is compounded by our
ignorance of the true ground state of the theory. Beginning in 1989, there was
some progress towards extracting
such nonperturbative information from string theory, at least in some simple
contexts. The aim of these lectures is to provide the conceptual
background for this work, and to describe some of its immediate consequences.

In string theory we wish to perform an integral over two dimensional geometries
and a sum over two dimensional topologies,
\eqn\ezsim{Z\sim\sum_{\rm topologies} \int\CD g\,\CD X\ \e{-S}\ ,}
where the spacetime physics (in the case of the bosonic string) resides in the
conformally invariant action
\eqn\esprop{S\propto\int \d^2\xi\,\sqrt g\, g^{ab}\,\del_a X^\mu\,\del_b X^\nu
\,G_{\mu \nu}(X)\ .}
Here $\mu,\nu$ run from $1,\ldots,D$ where $D$ is the number of spacetime
dimensions, $G_{\mu \nu}(X)$ is the spacetime metric,
and the integral $\CD g$ is over worldsheet metrics. Typically we ``gauge-fix''
the worldsheet metric to $g\dup_{ab}=\ee{\ph}\delta_{ab}$, where $\ph$ is known
as the Liouville field.
Following the formulation of string theory in this form (and in particular
following the appearance of the work of Polyakov \rpoly),
there was much work to develop the quantum
Liouville theory (some of which is reviewed in chapt.~\sqliouv\ here), and
conformal field theory itself has been characterized as ``an unsuccessful
attempt to solve the Liouville theory'' \ref\rPolne{A. M. Polyakov, lecture at
Northeastern Univ., spring 1990.}. It has been recognized that
evaluation of the partition function $Z$ in \ezsim\
without taking into account the integral over geometry does not
solve the problem of interest, and moreover does not provide a systematic basis
for a perturbation series in any known parameter.

The program initiated in \refs{\rDS\rBK{--}\rGM}\ relies
on a discretization of the string
worldsheet to provide a method of taking the continuum limit which incorporates
simultaneously the contribution of 2d surfaces with any number of handles.
In one seemingly giant step,
it is thus possible not only to integrate over all possible
deformations of a given genus surface (the analog of the integral over Feynman
parameters for a given loop diagram), but also to sum over all genus
(the analog of the sum over all loop diagrams).
This would in principle free us from the mathematically fascinating but
physically irrelevant problems of calculating conformal field theory
correlation functions on surfaces of fixed genus with fixed moduli (objects
which we never knew how to integrate over moduli or sum over genus anyway).
The progress, however, is limited in the sense that these methods only apply
currently for non-critical strings embedded in dimensions $D\le1$ (or critical
strings embedded in $D\le2$), and the nonperturbative information even in this
restricted context has proven incomplete. Due to familiar problems with lattice
realizations of supersymmetry and chiral fermions, these methods have also
resisted extension to the supersymmetric case.

The developments we shall describe here nonetheless provide at least a
half-step in the correct direction, if only to organize the perturbative
expansion in a most concise way. They have also prompted much useful evolution
of related continuum methods. Our point of view here is that string theories
embedded in $D\le1$ dimensions provide a simple context for testing ideas and
methods of calculation.
Just as we would encounter much difficulty calculating infinite dimensional
functional integrals without some prior experience with their finite
dimensional analogs, 
progress in string theory should be aided by experimentation with systems
possessing a restricted number of degrees of freedom.

While it is occasionally stated that
exactly solvable models are too special to provide useful lessons for
physics, at least one striking
historical example suggests the opposite: Onsager's
exact solution of the Ising model led to many fundamental
ideas in quantum field theory. In particular, ideas associated
with the renormalization group, phase transitions,
non-mean field exponents, and the operator product expansion
all had their origin in the solution of the Ising model. We hope that
this historical example will serve as well as the paradigm for
applications of exactly soluble spacetime solutions in string theory.

\subsec{2D Gravity and 2D String theory}
\subseclab\sstdgatds

%
%

We begin with a quick tour and overview of
2D gravity and 2D string theory, emphasizing the main physical ideas and
lessons we have learned from recent progress in the subject, so that they
are not lost in the lengthy discussion that follows. See also chapt.~\sadfp\
for another appraisal when we are done.

We have learned distinct lessons for 2D gravity and 2D string theory
due to the two-fold  interpretation of the models we discuss:

\ifig\fhhpiv{Correlation functions $\bigl<W(\ell)\bigr>$,
$\bigl<W(\ell)W(\ell)\bigr>$, and $\bigl<W(\ell)W(\ell)W(\ell)\bigr>$.}
{\epsfxsize4.5in\epsfbox{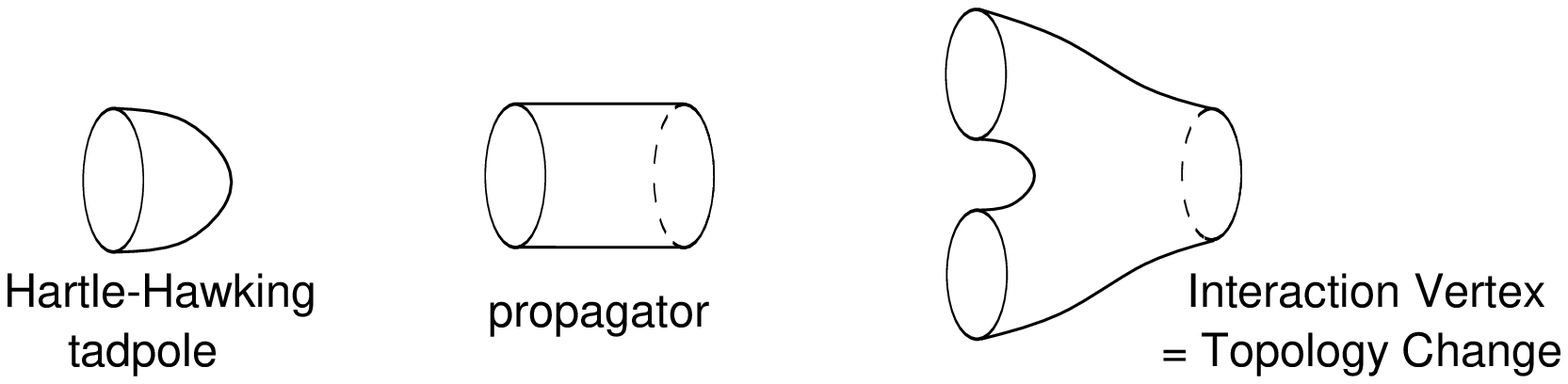}}

\smallskip\noindent{\it 1) As Quantum Gravity\/}\par\nobreak
In this guise, we study a ``field theory of universes.'' We will
introduce an operator (the macroscopic loop operator) $W(\ell,\dots)$ that
creates universes of size $\ell$ (1D and circular, in the case of a 2D target
space). The matrix model allows us to compute correlation functions
$\bigl<W(\ell)\bigr>$, $\bigl<W(\ell)\,W(\ell)\bigr>$, $\bigl<
W(\ell)\,W(\ell)\,W(\ell)\bigr>,\ \dots$, associated respectively with the
Hartle--Hawking wavefunction, with universe propagation, and with topology
change (processes depicted in \fhhpiv). The matrix model even allows both
calculation of the effect of more drastic topology changes and
summation over topology--changing amplitudes.

\smallskip\noindent{\it 2) As Critical String Theory\/}\par\nobreak

We will consider in detail the case for which \esprop\
defines a flat Euclidean two dimensional target spacetime
with coordinates $\phi,X$. Physical interpretations of
the spacetime theory require an analytic continuation to a spacetime of
Minkowskian signature. There are two choices for this continuation, but we
shall argue that the clearest lessons for string theory come from the $c=1$
model that we will construct,
with $\phi$ taken as the Liouville coordinate, and $t=-i X$ as the
Minkowskian time coordinate.

\ifig\fwall{Free, strong, and wall regions of the tachyon potential.
At left, particles are produced as tachyons scatter.}
{\epsfxsize4.5in\epsfbox{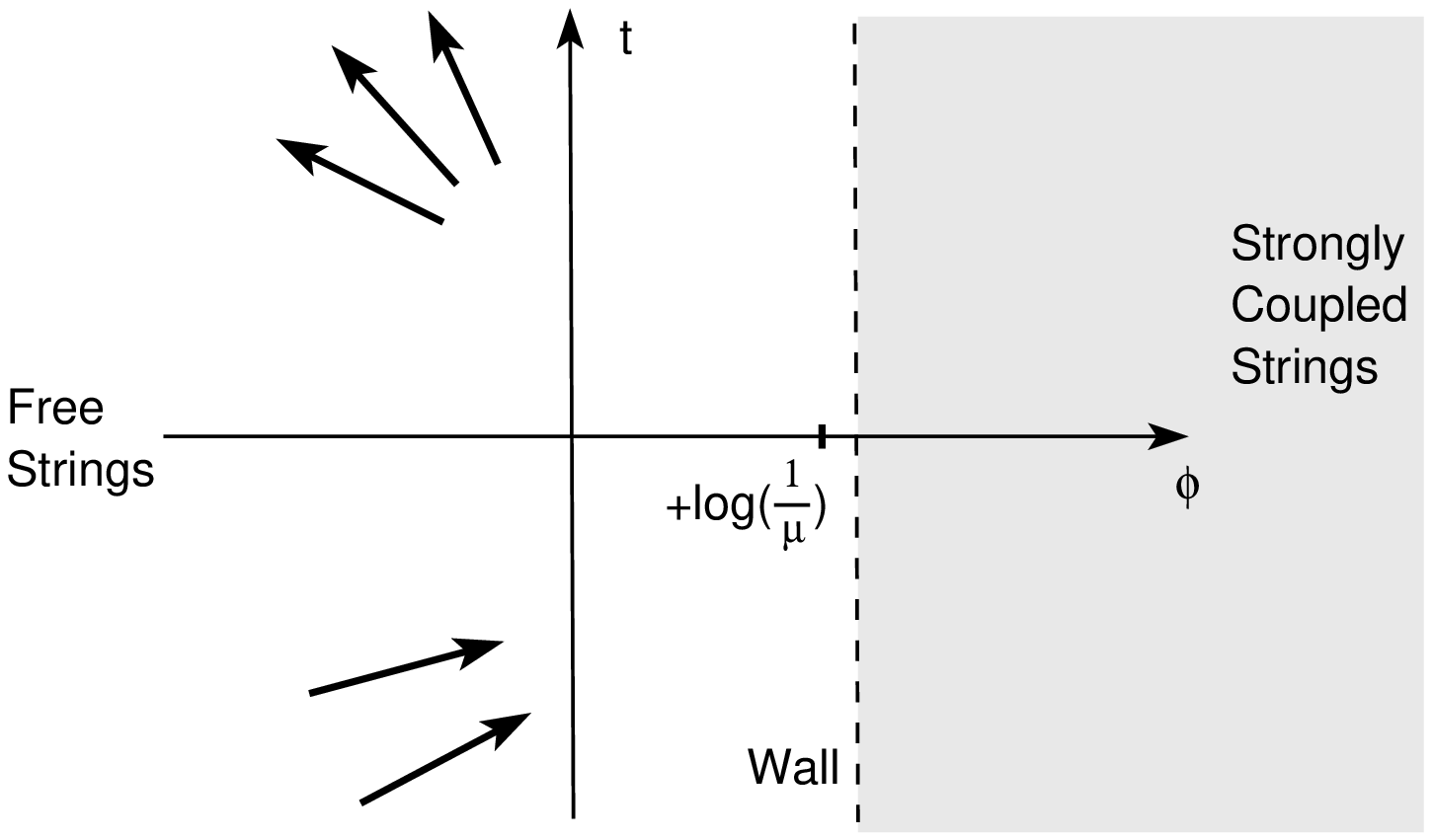}}

This exactly solvable spacetime background is a strange world. Since there are
only two spacetime dimensions, there are no transverse degrees of freedom. The
only on-shell\foot{The graviton and dilaton in 2 spacetime target dimensions
have no on-shell degrees of freedom, i.e.\ are not physical propagating
particles.} field theoretic degree of freedom in the theory is that of a
massless bosonic field $T(t,\phi)$, called the ``massless tachyon'' (because
the field $T$ is mathematically analogous to a field which is tachyonic for any
string propagating in {\it more\/} than 2 spacetime dimensions,
in particular for the $26$-dimensional bosonic string).
Moreover, the world is spatially
inhomogeneous. We shall find (see eq.~\soln) that the spacetime
dilaton field has expectation value $\langle D\rangle=Q\phi/2$, so
that the string coupling varies with the spatial coordinate $\phi$ as
$\kappa\dup_{\rm eff}= \kappa\dup_0\,\ee{\ha Q \phi}$.
At $\phi=-\infty$, strings are
free while at $\phi=+\infty$ strings are infinitely strongly coupled. Finally,
there is a cosmological constant term $\sim \mu \ee\phi$
in the Liouville path integral formulation of this theory that strongly
suppresses contributions to path integrals from large positive values of $\phi$
(we are assuming throughout these notes that the cosmological constant
satisfies $\mu>0$ , unless specified otherwise). Since the interaction turns on
exponentially there is effectively a wall located at $\phi=\log 1/\mu$, often
called ``the Liouville wall'', and the world looks as depicted in \fwall. The
most obvious physical experiment we can perform in this world is to bounce our
massless bosons off the wall. In chapt.~\sscatt, we will show how to compute
exactly the $S$-matrix for such scattering processes. We are also able
to compute the ``flows'' that relate physics in different (time-dependent)
backgrounds.

\subsec{Review of reviews}

Several reviews overlap different portions of the subject matter
covered here.
The Liouville theory is reviewed in \refs{\nati,\joetalk}\ and
references therein. 
The matrix model technology is reviewed in
\refs{\kazmig\rAGnotes\pgrev\bilal\rNotes{--}\rDGZ}
(note that some sections here are adapted directly from \pgrev),
the $c=1$ matrix model is reviewed in \refs{\kazrv,\klebrev}, and
spacetime properties are emphasized in \mart. Other recent reviews are
\rrrecent.

After treating the necessary preliminaries, our viewpoint here is largely
complementary to the other treatments, placing an emphasis on the properties of
macroscopic loops, the Wheeler--DeWitt equation, and the scattering theory in
$D=2$ target space dimensions.
In chapters 1--5, we give an overview of the continuum ideas
and formalism we shall use for treating loops and states in
conformal field theory, for understanding the
classical, semiclassical, and quantum Liouville theories,
and for implementing the path integral
and canonical treatments of 2D Euclidean quantum gravity.
In chapters 6--9, we focus on the discretized approach to 2d quantum gravity
and 2d string theory, and explain various features of the matrix model
approach.  In chapters 10--14, we employ the techniques provided by the
discretized approach to calculate many of the continuum quantities introduced
earlier. In chapter 15, we assess our prospects for the future.
Appendix A contains some useful definitions
and facts about some of the special functions used in these lectures.

Due to lack of spacetime, we will {\it not\/} be reviewing many other important
works on the subject of 2D gravity.
These notes are a preliminary version of a book \rbook,\foot{the modest
expenditure for which will be amply rewarded by
the substantial further enlightenment contained therein.} that will contain
much interesting material omitted from these lecture notes. We welcome
constructive comments concerning typos, inconsistent
conventions, inconsistent references,
sign errors and conceptual errors in these notes.

\newsec{Loops and States in Conformal Field Theory}
\seclab\slascft

We begin here with a review of how a loop
in the context of conformal field theory
can be replaced by a sum of local operators. For simplicity
we will restrict attention to the Gaussian model.
The intuition from this example will be essential to our later extraction of
correlation functions from macroscopic loop amplitudes.

\subsec{Lagrangian formalism}

For simplicity, consider the standard $c=1$ Gaussian model,
\eqn\gauss{S=\int_\Sigma \p X\,\pb X\ ,}
where $\Sigma$ is a surface perhaps with handles and boundaries.
The objects of interest are the path integrals
\eqn\pathint{\int \CD X(z,\zb)\ \ee{-S}\ \prod_i \CO_i(p_i)\ ,}
where we integrate over maps $X:\Sigma \to \IR$.

The space of local operators is spanned by expressions of the form
\eqn\locopers{\CO\sim
\CP(\p X,\,\p^2 X,\,\dots;\ \pb X,\,\pb^2 X,\,\dots)\, \ee{i k X(z,\zb)\ ,}
}
where $\CP$ is a polynomial and the expression is suitably normal-ordered.
When $X$ is a periodic variable, i.e.\ a map $X:\Sigma \to S^1$,
then additional considerations apply: $k$ is quantized and there
are winding modes which allow the definition of a more subtle theory
with an extra zero mode, leading to ``magnetic'' and ``electric''
quantum numbers. See \rcftrev\ for more details.

\subsec{Hamiltonian formalism}

In the radial Hamiltonian formalism (for a review, see e.g.~\rcftrev) we
decompose the fields into modes,
\eqn\modes{\eqalign{i\p X
&=\sum \alpha_n\, z^{-n-1}\ , \qquad [\alpha_n,\alpha_m]=n\,\delta_{n+m}\cr
i\pb X&=\sum \bar\alpha_n\, \zb^{-n-1}\ , \qquad
[\bar\alpha_n,\bar\alpha_m]=n\,\delta_{n+m}\ .\cr}
}
The stress-energy tensor $T=-\half (\p X)^2$ defines a
Virasoro algebra and radial propagation is generated by
the Hamiltonian $L_0 +\bar L_0$.

In terms of the Fock space
\eqn\focksp{\eqalign{\CF_p&={\rm Span}\Bigl\{\prod_n(\alpha_{-n})^{m_n}|p
\rangle\quad n > 0,\ \ m_n \geq 0\Bigr\}\cr
&\alpha_0 |p\rangle = p\, |p\rangle\qquad
\alpha_n |p\rangle =0\quad (n>0)\ ,\cr}}
we see that the states of the theory lie in a Hilbert Space of the form
\eqn\gausshil{\CH=\oplus_{p,\bar p}\ N_{p,\bar p}\,\CF_p\otimes\CF_{\bar p}\ .}
Alternatively, we can think of the Hilbert space as the space of
wavefunctionals $\Psi[X(\sigma)]$. (More precisely, we may introduce
the loop space $L\IR$ and, with an appropriate measure, the
Hilbert space is simply the space $\CH=L^2(L\IR)$ of square integrable maps
of loops into $\IR$.)

\exercise{Coherent states}

If we decompose
\eqn\bcondii{X(\sigma)=x_0 +\sum_{n>0} \ee{i n \sigma} x_n +
\sum_{n<0} \ee{i n \sigma} \bar x_n\ ,}
we may regard $\Psi$ as a function of infinitely
many variables $x_0,x_n,\bar{x}_n$. Using the coherent state representation
for the harmonic oscillators,
$\alpha_n\leftrightarrow n x_n$, for $n>0$;
$\alpha_n\leftrightarrow \del/\del x_n$, for $n<0$; and similarly for
$\bar\alpha$ and $\bar x$, translate the Fock space states
\focksp\ into wavefunctionals $\Psi[X(\sigma)]$.

\endexercise

The Lagrangian and Hamiltonian formalisms are related by the
so-called ``operator formalism''. The basic idea is that
the neighborhood of any
point on the surface $\Sigma$ locally looks like the complex plane and
information about the rest of the surface may be summarized in
a state at infinity.

\subsec{Equivalence of states and operators} 

One of the basic properties of conformal field theory is the one-to-one
correspondence between operators $\CO$ and states $|\CO\rangle$.

To map {\it operators $\to$ states\/}, we associate the state
$|\CO\rangle$ to the operator $\CO(z,\zb)$ according to
\eqn\opst{
\CO\mapsto |\CO\rangle\equiv \lim_{z,\zb\to 0}\CO(z,\zb)|0\rangle\ .
}
Equivalently, we can create a state by performing a path integral on a
hemisphere $D$. To evaluate such a path integral,
the boundary conditions for the field $X(z,\zb)$ must be specified
on the equator, parametrized here by $\sigma$, and the
value of the path integral defines the wavefunction $\Psi[X(\sigma)]$
(i.e.\ the wavefunction for the identity operator).
Insertion of an operator $\CO$ on the hemisphere $D$ gives
the wavefunction $\Psi_\CO$ for the operator $\CO$,
\eqn\wvfn{\Psi_{\CO}\bigl[X(\sigma)\bigr]=
\figins{\vcenter{\epsfxsize30pt\epsfbox{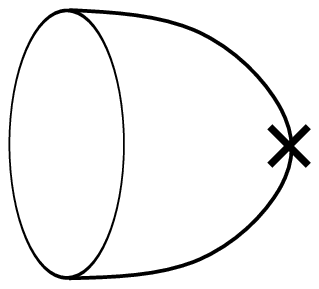}}}\lower8pt\hbox{$\CO$}
=\int_{X|_{\p D}=X(\sigma)} [\d X]\ \ee{-S}\,\CO\ .}
Using the equivalence between the Fock space and wavefunctional
descriptions of the states, the two descriptions
\opst\ and \wvfn\ of the
operator $\mapsto$ state maps are easily seen to be equivalent:
namely $\bigl<X(\sigma)\big|\CO\bigr> =\Psi_{\CO}\bigl[X(\sigma)\bigr]$
(where $\bigl<X(\sigma)\big|$ is a basis of eigenstates of the operator
$X(\sigma)$ in the Fock space representation).

\exercise{Wavefunctions from path integrals}

Consider a disk of radius $r$ in the complex plane, centered
at $z=0$, and consider the path integral \wvfn\ with boundary condition
\eqn\bcond{X(\sigma)=x_0 +\sum_{n>0} \ee{i n \sigma} x_n +
\sum_{n<0} \ee{i n \sigma} \bar x_n}

a) Solve for the classical field configuration $X_{\rm cl}$ satisfying
these boundary conditions. Then shift the
field $X\to X_{\rm cl}+\delta X$ in the path integral, where
$\delta X$ satisfies Dirichlet boundary conditions, and
show by substituting into \wvfn\ that the wavefunction is
\eqn\gndwfgn{
\Psi\bigl[X(\sigma)\bigr]
= C\prod_{n=1}^\infty (2 \pi n) \ee{-2 \pi n^2 |x_n|^2}\ ,
}
where the constant $C$ is determined by imposing some normalization
condition.

b) Why can this be identified with the state $|0\rangle$ ?

c) Repeat this exercise to calculate the wavefunction for
some other simple operators.

d) Describe the wavefunction associated to states created by local operators of
the form \locopers.

\endexercise

\noindent{\it Expansion of loops in terms of local operators\/}
\par\nobreak
Having described the operator $\mapsto$ state mapping, now we wish to consider
the inverse state $\mapsto$ operator mapping.
To ``insert a state'' into the path integral, we cut a hole of radius $r$ out
of the surface $\Sigma$, and insert a state with some wavefunction
$\Psi\bigl[X(\sigma)\bigr]$ on the boundary of the hole
(i.e.\ use $\Psi\bigl[X(\sigma)\bigr]$
as the weight factor in the functional integral over $X(\sigma)$).
We shall see that the new path integral is equivalent to that
obtained by inserting a local operator into \pathint, thus providing a {\it
states $\to$ operators\/} mapping.  The mapping
$\Psi_{\CO}\mapsto\CO$ is clearly linear and one-to-one,
but to show that there is moreover an isomorphism between states and operators,
we need to show that {\it every\/} state is equivalently represented by an
operator insertion.
The basic idea, physically, is to take a state inserted on a hole and by
conformal invariance shrink the hole to arbitrarily small size. Since an
infinitesimally small hole has the same effect as a local operator, the
insertion of a state on the boundary of a hole in the path integral is
equivalent to the insertion of some operator.

In formulae, the {\it states $\mapsto$ operators\/} mapping
equates the insertion of a little hole of radius $r$
on any surface $\Sigma$ in state $|\Psi\rangle$ with
the insertion of a sum of operators at the center of the hole,
\eqn\critloop{W_{\Psi}(r)\equiv\sum_i \langle \CO_i|\Psi\rangle
\,r^{\Delta_i+\bar\Delta_i}\,\CO_i\ .}
Here $\CO_i(z,\zb)$ is a basis of local operators diagonalizing
$L_0+\overline L_0$, and $W_{\Psi}(r)$ is an {\it operator\/}
that inserts a
macroscopic loop of size $r$ with wavefunction $\Psi\bigl[X(\sigma)\bigr]$.

\noindent{\bf Example}: Annular path integral.\par\nobreak
Consider the path integral on a sphere with two holes, or,
after a conformal transformation, on an annulus.
This is given in the operator formalism by
\eqn\annpath{\eqalign{\vcenter{\hbox{$r{=}1$}\vskip5pt\hbox{ \ $\Psi_2$}}
\figins{\vcenter{\epsfxsize30pt\epsfbox{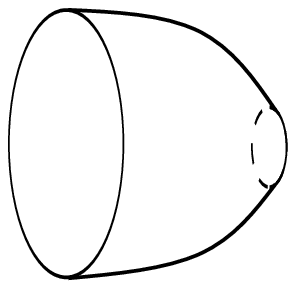}}}
\vcenter{\hbox{$r$}\vskip-3pt\hbox{$\Psi_1$}}
\quad &=\langle \Psi_2|r^{L_0+\overline L_0}|\Psi_1\rangle\cr
&=\sum_i \langle \Psi_2|\CO_i\rangle\ r^{\Delta_i+\bar\Delta_i}\
\langle\CO_i|\Psi_1\rangle\cr
&=\sum_i\ \lower10pt\hbox{$\Psi_2$}
\figins{\vcenter{\epsfxsize30pt\epsfbox{punctan.ps}}}\lower8pt\hbox{$\CO_i$}
\ \ r^{\Delta_i+\bar\Delta_i}\,\langle\CO_i|\Psi_1\rangle\cr
&=\bigl< \Psi_2 | W_{\Psi_1}(r)\bigr>\ ,\cr}}
where $\bigl|W_{\Psi_1}(r)\bigr>$ is the state associated to
the local operator $W_{\Psi_1}(r)$.
For more details on the operator formalism, see e.g.\
\refs{\rBPZ\FMS{--}\opform}.

Our goal is to calculate macroscopic loop amplitudes in 2D gravity. Current
matrix model technology will allow this only for some very specific
states $|\Psi\rangle$, but these states will have overlaps with
sufficiently many interesting operators
that much useful information can be extracted.
To provide a physical framework for interpreting the matrix model results,
we shall first investigate in the next two chapters the spectrum of Liouville
theory and of 2D gravity.

\subsec{Gaussian Field with a Background Charge}
\subseclab\ssctff

For later comparison with the Liouville results, we give here
a brief overview of
gaussian conformal field theory in the presence of a background charge,
also known as Chodos--Thorn/Feigin--Fuks (CTFF) theory.
We consider the action
\eqn\bckchgi{S_{\rm CTFF}=\int \d^2 z\,  \sqrt{\hat g}
\Bigl({1\over 8 \pi}(\hat\nabla\phi)^2
+ {i\alpha_0\over 4 \pi}\phi R(\hat g)\Bigr)\ .}
The additional term leads to the modified stress-energy
\eqn\eetwbc{T=-\ha\del\phi\del\phi+i\alpha\dup_0\,\del^2 \phi\ ,}
which generates a Virasoro algebra with central charge
$$c=1-12\alpha^2_0\ .$$
We see that the effect of the extra term in \eetwbc\ is to shift $c<1$
for $\alpha\dup_0$ real.  Since the stress-energy tensor in \eetwbc\
has an imaginary part, the theory it defines is not unitary for arbitrary
$\alpha\dup_0$.  For particular values of $\alpha_0$, it turns out to
contain a consistent unitary subspace.

Taking the operator product with the modified $T$ of \eetwbc, we find
that the conformal weight of $\ee{\beta\phi}$ is shifted to
$\Delta=\ha\beta(\beta-2\alpha\dup_0)$. The same conformal weights would be
inferred from two-point functions calculated in the presence of a
`background charge' $-\sqrt{2}\alpha\dup_0$ at infinity, so
the modification of $T(z)$ in \eetwbc\ is interpreted as the presence of such
a background charge.
This formalism was originally used by Chodos and Thorn in
\rChTh, and was more recently revived by Feigen and Fuks in a form
used in \rDotFat\
to calculate correlation functions of the $c<1$ conformal field theories.

\exercise{Momentum Shift from Background Charge}

Consider a Gaussian field with background
charge $Q=2i\alpha\dup_0$.  Derive the relation
\eqn\vrtxstp{i p_\phi=\alpha-{Q\over 2}\ ,}
for the momentum $p_\phi$ of the state created by the
vertex operator $\exp(\alpha\phi)$
by considering the state created by the path integral on the disk,
analogous to the example of the Gaussian model in \wvfn.

\endexercise
\newsec{2D Euclidean Quantum Gravity I: Path Integral Approach} 
\seclab\sqliouv

In this chapter, we shall consider the implementation of
quantum gravity as a theory which makes
precise and well-defined sense out of a path integral over
metrics on some spacetime.

\subsec{2D Gravity and Liouville Theory}
\subseclab\sstdglt

Using the principle of general covariance, {\it any\/} quantum field theory
$S_{\rm matter}[X^i]$ in any number of dimensions may be coupled to gravity,
resulting in an action $S[g,X^i]$, where $X^i$ refer
to ``matter'' fields in the theory and $g$ is the metric.

Classically, the theory $S[g,X^i]$ with gravity coupling in two dimensions
is always a conformal field theory. To see this, recall that the
stress energy  tensor of the theory is
${\delta\over\delta g^{\alpha \beta}}S=T_{\alpha \beta}$.
Defining the  Liouville mode as the overall
scale of the metric, $g=\ee{\gamma \phi} \hat g$,
we see that the Liouville equation of motion is $T^{\alpha}_{\alpha}=0$.
This defines a classical conformal field  theory.
In two dimensions, we may pass to local complex
coordinates and write this equation as $T_{z\zb}=0$. Conservation
of energy-momentum then shows that the theory has
a holomorphic energy momentum tensor $T_{zz}=T(z)$.

Quantum mechanically, we try to understand the
(Euclidean) quantum gravitational path integrals
\eqn\grvpth{\langle \CO_1\dots \CO_n\rangle
={1\over Z}\int {\CD g\,\CD X\over {\rm vol}({\it Diff\/})}
\,\ee{- \kappa \int R - \mu A - S[X]}\,\CO_1\dots \CO_n\ ,}
where $\CO_i$ are generally covariant operators.
By fixing a conformal gauge $g\dup_{ab}=\ee{\phi}\delta_{ab}$,
Polyakov \rpoly\ observed that the matter/gravity
theory could be written as a coupled tensor product of
Liouville theory and the ``matter'' theory $S_{\rm matter}$.
The passage to conformal gauge necessarily introduces the
Faddeev--Popov reparametrization ghosts.
At a formal level,  the gauge invariance of the
theory, expressed as the independence of the integral
\grvpth\ to Weyl transformations of the gauge slice,
implies that the coupling of Liouville and matter theories is
itself a conformal field theory.
If the original matter theory was not conformal, however,
then the resulting theory will not be a simple tensor product.

{\bf Example 1}: Coupling the massive Ising model
\eqn\msvis{S= \int \bar{\psi} \p \psi + m \bar{\psi}\psi}
to gravity results in the lagrangian
\eqn\isgrav{S= \int \sqrt g\,
\bigl(\bar{\psi}D\psi + m \bar{\psi}\psi\bigr)
+\int \d^2 z\,\sqrt{\hat g}\Bigl({1\over 8 \pi} (\hat\nabla \phi)^2
+{\mu\over 8 \pi \gamma^2}\,\ee{\gamma \phi}+{Q\over
8 \pi}\phi R(\hat g)\Bigr)\ ,}
where $D$ is the covariant derivative (i.e.\ includes the spin connection).

{\bf Example 2}: Coupling the massive sine-Gordon model
\eqn\singrd{\int \d^2 z\, \sqrt{\hat g}\Bigl(
{1\over 8 \pi} (\hat\nabla X)^2+ m \cos (p X/\sqrt{2})\Bigr)}
to gravity results in the lagrangian
\eqn\pertlag{\eqalign{
S=&\int \d^2 z\, \sqrt{\hat g}\Bigl({1\over 8 \pi} (\hat\nabla X)^2
+ m \ee{\xi \phi} \cos (p X/\sqrt{2})\Bigr)\cr
&+\int \d^2 z\,\sqrt{\hat g}\Bigl({1\over 8 \pi} (\hat\nabla \phi)^2
+{\mu\over 8 \pi \gamma^2}\,\ee{\gamma \phi}+{Q\over
8 \pi}\phi R(\hat g)\Bigr)\ .\cr}}

In both examples, we will see that $Q$ and $\gamma$ are fixed by general
covariance.
The remarkable property of Liouville theory that allows it to associate
an arbitrary quantum field theory with some conformal field theory
deserves to be better understood.

\subsec{Path integral approach to 2D Euclidean Quantum Gravity} 
\subseclab\spia

The first success of the discretized (matrix model) approach,
the focus of our later chapters here, was to reproduce the
critical exponents predicted from continuum (Liouville) methods.
(In fact the coincidence of results served to give post-facto verification of
both methods.)  In this section we review the latter continuum methods from a
fairly formal point of view. A more physical point of view will appear
in the next chapter.

\noindent{\it String susceptibility $\Gamma_{\rm str}$}
\par\nobreak
We consider the continuum partition function
\eqn\epf{Z=\int {\CD g\,\CD X\over {\rm vol}({\it Diff\/})}\
\e{-S_M(X;g)-{\mu_0\over8\pi} \int \d^2\xi\,\sqrt g}\ ,}
where $S_M$ is some conformally invariant action for matter fields coupled to a
two dimensional surface $\Sigma$ with metric $g$, $\mu_0$ is a bare
cosmological constant, and we have symbolically divided the measure by the
``volume'' of the diffeomorphism group (which acts as a local symmetry) of
$\Sigma$ . For the free bosonic string, we take
$S_M={1\over8\pi}\int \d^2\xi\,\sqrt g\, g^{ab}\,
\del_a \vec X\cdot\del_b\vec X$ where the $\vec X(\xi)$ specify the embedding
of $\Sigma$ into flat $D$-dimensional spacetime.

To define \epf, we need to specify the measures for the integrations over
$X$ and $g$ (see, e.g.\ \rFrlh). The measure $\CD X$ is determined by requiring
that $\int \CD_g \delta X\ \ee{-\|\delta X\|^2_g}=1$, where the norm in the
gaussian functional integral is given by
$\|\delta X\|^2_g=\int \d^2\xi\,\sqrt g\,\delta\vec X\cdot\delta\vec X$.
Similarly, the measure $\CD g$ is determined by normalizing
$\int\CD_g \delta g\ \ee{-\half\|\delta g\|^2_g}=1$, where
$\|\delta g\|^2_g=\int \d^2\xi\,\sqrt g\,
(g^{ac}g^{bd}+2g^{ab}g^{cd})\,\delta g\dup_{ab}\,\delta g\dup_{cd}$, and
$\delta g$ represents a metric fluctuation at some point
$g\dup_{ij}$ in the space of metrics on a genus $h$ surface.

The measures $\CD X$ and $\CD g$ are invariant under the group of
diffeomorphisms of the surface, but not necessarily under conformal
transformations $g\dup_{ab}\to\ee\sigma\,g\dup_{ab}$.
Indeed due to the metric dependence in the norm $\|\delta X\|^2_g$, it turns
out that
\eqn\edx{\CD_{\ee\sigma\! g}X = \e{{D\over 48\pi}\, S_L(\sigma)} \CD_g X\ ,}
where
\eqn\eL{S_L( \sigma)=\int \d^2\xi\,\sqrt g\,\Bigl(\half g^{ab}\, \del_a\sigma
\del_b\sigma + R\sigma + \mu\ee\sigma\Bigr)}
is known as the Liouville action.
(This result may be derived diagrammatically, via the Fujikawa method, or via
an index theorem; for a review see \rOA.)

The metric measure $\CD g$ as well has an anomalous variation under conformal
transformations. To express it in a form analogous to \edx, we first need to
recall some basic facts about the domain of integration.
The space of metrics on a compact topological surface $\Sigma$ modulo
diffeomorphisms and Weyl transformations is a finite dimensional compact space
$\CM_h$, known as moduli space. (It is 0-dimensional for genus $h=0$;
2-dimensional for $h=1$; and ($6h-6$)-dimensional for $h\ge2$). If for each
point $\tau\in \CM_h$, we choose a representative metric $\hat g\dup_{ij}$,
then the orbits generated by the diffeomorphism and Weyl groups acting on $\hat
g\dup_{ij}$
generate the full space of metrics on $\Sigma$. Thus given the slice
$\hat g(\tau)$, any metric can be represented in the form
$$f^*g=\ee{\ph}\, \hat g(\tau)\ ,$$
where $f^*$ represents the action of the diffeomorphism $f: \Sigma\to\Sigma$.

Since the integrand of \epf\ is diffeomorphism invariant, the functional
integral would be infinite unless we formally divide out by the volume of orbit
of the diffeomorphism group. This is accomplished by gauge fixing to the slice
$\hat g(\tau)$; the Jacobian that enters can be represented in terms of
Fadeev--Popov ghosts, as familiar from the analogous procedure in gauge theory.
We parametrize an infinitesimal change in the metric as
$$\delta g\dup_{zz}=\grad z \xi_z\ ,
\qquad\delta g\dup_{\zb \zb}=\grad {\zb}\xi_{\zb}$$
(where for convenience we employ complex coordinates, and recall that the
components $g\dup_{z\zb}=g^{\zb z}$ are parametrized by $\ee{\ph}$). The
measure $\CD g$ at $\hat g(\tau)$ splits into an integration $[\d\tau]$ over
moduli, an integration $\CD\ph$ over the conformal factor, and an integration
$\CD\xi\,\CD\bar\xi$ over diffeomorphisms. The change of integration variables
$\CD\delta g\dup_{zz}\,\CD\delta g\dup_{\zb\zb}=(\det\!\grad z
\det\!\grad{\zb})\,\CD\xi\,\CD\bar\xi$ introduces the Jacobian $\det\!\grad z
\det\!\grad{\zb}$ for the change from $\delta g$ to $\xi$. The determinants
in turn can be represented as
\eqn\eFP{\eqalign{\det\!\grad z\,\det\!\grad{\zb}
 &= \int\CD b\,\CD c\,\CD \bar b\,\CD\bar c\
 \e{-\int\d^2\xi\,\sqrt g\, b_{zz} \grad {\zb} c^z
    -\int\d^2\xi\,\sqrt g\, b_{\zb \zb}\grad z c^\zb}\cr
&\qquad\equiv\int\CD({\rm gh})\ \e{-\Sgh(b,c,\bar b,\bar c)}\ ,}}
where $\CD({\rm gh})\equiv \CD b\,\CD c\,\CD \bar b\,\CD\bar c$ is an
abbreviation for the measures associated to the ghosts $b,c,\bar b,\bar c$;
$b_{zz}$ is a holomorphic quadratic differential, and $c^z$ ($c^\zb$)
is a holomorphic (anti-holomorphic) vector.

Finally, the ghost measure $\CD({\rm gh})$ is
not invariant under the conformal transformation $g\to\ee\sigma g$,
instead we have \refs{\rpoly,\rFrlh,\rOA}\
\eqn\eagh{\CD_{\ee\sigma\! g}({\rm gh})
=\e{-{26\over 48 \pi}\, S_L(\sigma,g)}\CD_{g}({\rm gh})\ ,}
where $S_L$ is again the Liouville action \eL.
(In units in which the contribution of a single scalar field to the
conformal anomaly is $c=1$, and hence $c=1/2$ for a single Majorana--Weyl
fermion, the conformal anomaly due to a spin $j$ reparametrization ghost
is given by $c=(-1)^F 2(1+6j(j-1))$. The contribution from a spin
$j=2$ reparametrization ghost is thus $c=-26$.)

We have thus far succeeded to reexpress the partition function \epf\ as
$$Z=\int [\d \tau]\ \CD_{g}\ph\ \CD_g({\rm gh})\ \CD_g X\
\e{-S_M-\Sgh-{\mu_0\over2\pi}\int \d^2\xi\,\sqrt g}\ .$$
Choosing a metric slice $g=\ee{\ph}\hat g$ gives
$$\CD_{\ee{\ph}\hat g}\ph\,\CD_{\ee{\ph}\hat g}({\rm gh})\,\CD_{\ee{\ph} \hat
g}X
= J(\ph,\hat g)\ \CD_{\hat g}\ph\,\CD_{\hat g}({\rm gh})\,\CD_{\hat g}X\ ,$$
where the Jacobian $J(\ph,\hat g)$ is easily calculated for the matter and
ghost sectors $\bigl($\edx\ and \eagh$\bigr)$ but not for the Liouville mode
$\ph$.
The functional integral over $\ph$ is complicated by the implicit metric
dependence in the norm
$$\|\delta \ph\|^2_g=\int \d^2\xi\,\sqrt g\,(\delta \ph)^2
=\int \d^2\xi\,\sqrt {\hat g}\, \ee{\ph}\, (\delta \ph)^2\ ,$$
since only if the $\ee{\ph}$ factor were absent above would the
$\CD_{\hat g}\ph$ measure reduce to that of a free field.

In \rDDK, it is simply {\it assumed\/}\foot{Some attempts to
justify this assumption may be found in \rMMHK.
In the next two chapters, we shall see why this result should be expected from
the canonical Hamiltonian point of view of \crtthrn.}
that the overall Jacobian
$J(\ph,\hat g)$ takes the form of an exponential of a local Liouville-like
action
$\int \d^2\xi\,\sqrt {\hat g}\,(\tilde a\,\hat g^{ab} \del_a \ph \del_b \ph
+\tilde b \hat R \ph + \mu \ee{\tilde c \ph})$, where $\tilde a$, $\tilde b$,
and $\tilde c$ are constants that will be determined by requiring overall
conformal invariance ($\tilde c$ is inserted in anticipation of rescaling of
$\ph$). With this assumption, the partition function \epf\ takes the form
\eqn\epddk{\eqalign{Z=\int [\d \tau]\,\CD_{\hat g}\ph\,
\CD_{\hat g}({\rm gh})\,\CD_{\hat g}X\ &\e{-S_M(X,\hat g)
-\Sgh(b,c,\bar b,\bar c;\,\hat g)}\cr
&\qquad\cdot
\e{-\int \d^2\xi\,\sqrt {\hat g}\,
(\tilde a\,\hat g^{ab} \del_a \ph \del_b \ph
+\tilde b \hat R \ph + \mu \ee{\tilde c \ph})}\cr}}
where the $\ph$ measure is now that of a free field.

The path integral \epddk\ was defined to be reparametrization invariant, and
should depend only on $\ee{\ph}\hat g=g$ (up to diffeomorphism), not on the
specific slice $\hat g$.  Due to diffeomorphism invariance, \epddk\
should thus be invariant under the infinitesimal transformation
\eqn\eir{\delta \hat g=\varepsilon(\xi)\hat g\ ,\quad
\delta\ph=-\varepsilon(\xi)\ ,}
and we can use the known conformal anomalies \edx\ and \eagh\ for
$\ph$, $X$, and the ghosts to determine the constants
$\tilde a,\tilde b,\tilde c$.
Substituting the variations \eir\ in \epddk, we find terms of the form
$$\Bigl({D-26+1\over 48 \pi}+\tilde b\Bigr)\int \d^2\xi\,\sqrt {\hat g}\
\hat R\,\varepsilon\qquad{\rm and}\qquad
(2\tilde a-\tilde b) \int \d^2\xi\,\sqrt {\hat g}\ \varepsilon \lapl\ph\ ,$$
where the $D-26$ on the left is the contribution from the matter and ghost
measures $\CD_{\hat g}X$ and $\CD_{\hat g}({\rm gh})$,
and the additional 1 comes from the $\CD_{\hat g}\ph$ measure.
Invariance under \eir\ thus determines
\eqn\eab{\tilde b={25-D\over48 \pi}\ ,\quad \tilde a=\half \tilde b\ .}
(In general we would substitute here $D\to c_{\rm matter}$, where
$c\dup_{\rm matter}$ is the contribution to the central charge from the
``matter'' sector of the theory.)

Substituting the values of $\tilde a,\tilde b$ into the Liouville action in
\epddk\ gives
\eqn\eabs{{1\over 8 \pi}\int \d^2\xi\,\sqrt {\hat g}\,
\Bigl({25-D\over12}\hat g^{ab}\,\del_a\ph\,\del_b\ph
+{25-D\over6}\hat R\,\ph\Bigr)\ .}
To obtain a conventionally normalized kinetic term
${1\over8 \pi}\int (\del\ph)^2$, we rescale
$\ph\to \sqrt{{12\over25-D}}\,\ph$. (This normalization leads to the leading
short distance expansion $\ph(z)\,\ph(w)\sim -\log(z-w)$.)
In terms of the rescaled $\ph$, we write the Liouville action as
\eqn\ersL{{1\over 8 \pi}\int \d^2\xi\,\sqrt {\hat g}\,
\Bigl(\hat g^{ab}\,\del_a\ph\,\del_b\ph+Q\hat R\,\ph\Bigr)\ ,}
where
\eqn\eQ{Q\equiv\sqrt{{25-D\over3}}\ .}
The energy-momentum tensor
$T=-\half\del\ph\del\ph+{Q\over2}\del^2\ph$ derived from
\ersL\ has leading short distance expansion
$T(z)T(w)\sim {\ha c\dup_{\rm Liouville}/(z-w)^4}+\ldots$,
where $c\dup_{\rm Liouville}=1+3Q^2$. Note that if we substitute \eQ\
into $c\dup_{\rm Liouville}$ and add an additional $c=D-26$ from the matter and
ghost sectors, we find that the
total conformal anomaly vanishes,
$$c\dup_{\rm matter}+c\dup_{\rm Liouv}+ c\dup_{\rm ghost}= D+(26-D)-26=0$$
(consistent with the required overall conformal invariance).

It remains to determine the coefficient $\tilde c$ in \epddk. We have since
rescaled $\ph$, so we write instead $\ee{\gamma\ph}$ and determine
$\gamma$ by the requirement that the
physical metric be $g={\hat g}\,\ee{\gamma\ph}$. Geometrically, this means that
the area of the surface is represented by
$\int \d^2\xi\,\sqrt {\hat g}\,\ee{\gamma\ph}$.
$\gamma$ is thereby determined by the requirement that $\ee{\gamma\ph}$ behave
as a (1,1) conformal field (so
that the combination $\d^2\xi\,\ee{\gamma\ph}$ is conformally invariant). For
the energy-momentum tensor
mentioned after \eQ, the conformal weight\foot{Recall that $\Delta$
is given by the
leading term in the operator product expansion $T(z)\,\ee{\gamma\ph(w)}\sim
{\Delta\,\ee{\gamma\ph}/(z-w)^2}+\ldots\ $. Recall also that for a conventional
energy-momentum tensor $T=-\half\del\ph\del\ph$, the conformal weight of
$\ee{ip\ph}$ is $\Delta=\overline \Delta=p^2/2$.} of $\ee{\gamma\ph}$ is
\eqn\ecw{\Delta(\ee{\gamma\ph})
=\overline\Delta(\ee{\gamma\ph})=-\half\gamma(\gamma-Q)\ .}
Requiring that $\Delta(\ee{\gamma\ph})=\overline \Delta(\ee{\gamma\ph})=1$
determines that
\eqn\eQgam{Q={2/\gamma}+\gamma\ .}
Using \eQ\ and solving for $\gamma$
then gives\foot{One method for choosing the root for $\gamma$ is
to make contact with the classical limit of the Liouville action. Note that the
effective coupling in \eabs\ goes as $(25-D)\inv$ so the classical limit
is given by $D\to-\infty$. In this limit the above choice of root has the
classical $\gamma\to0$ behavior. We shall discuss this issue further in the
next chapter.}
\eqn\ealph{\gamma={1\over\sqrt{12}}\bigl(\sqrt{25-D}-\sqrt{1-D}\bigr)=
{Q\over 2}-\ha\sqrt{Q^2-8} \ .}
%

For spacetime embedding dimension $d\le1$, we find from \eQ\ and \ealph\
that $Q$ and $\gamma$ are both real (with $\gamma\le Q/2$). The $D\le1$
domain is thus where the Liouville theory is well-defined and most easily
interpreted. For $D\ge25$, on the other hand, both $\gamma$ and $Q$ are
imaginary. To define a real physical metric
$g=\ee{\gamma\ph}{\hat g}$, we need to Wick rotate $\ph\to-i\ph$. (This changes
the sign of the kinetic term for $\ph$. Precisely at $D=25$ we can interpret
$X^0=-i\ph$ as a free time coordinate. In other words, for a string naively
embedded in 25 flat euclidean dimensions, the Liouville mode turns out to
provide automatically a single timelike dimension,
dynamically realizing a string embedded in 26 dimensional Minkowski spacetime.
In general for $D\geq 25$,
we must have $c\dup_{\rm Liouville}<1$ and the kinetic
term of the Liouville field changes sign. The conformal mode of the metric in
Euclidean space is a ``wrong sign'' scalar field analogous to the conformal
mode in four-dimensional Euclidean quantum gravity \polgrav, so it would be
useful to make sense of this case (if possible).
Finally, in the regime $1<D<25$, $\gamma$ is complex, and $Q$ is imaginary.
As we shall see later in lurid detail, this problem is equivalent to
the cosmological constant becoming a macroscopic state operator.
Sadly, it is not yet known how to make sense of the Liouville approach for the
regime of most physical interest.

A useful critical exponent that can be calculated in this formalism
is the string susceptibility $\Gamma_{\rm str}$.
We write the partition function for fixed area $A$ as
\eqn\eZA{Z(A)=\int\CD\ph\,\CD X\ \e{-S}\
\delta\Bigl({\textstyle\int} \d^2\xi\,\sqrt {\hat g}
\,\ee{\gamma\ph} - A\Bigr)\ ,}
where for convenience we now group the ghost determinant and integration over
moduli into $\CD X$. We define a string susceptibility $\Gamma_{\rm str}$ by
\eqn\eZainf{Z(A)\sim A^{(\Gamma_{\rm str}-2){\chi/2}-1}\ ,\quad A\to\infty\ ,}
and determine $\Gamma_{\rm str}$ by a simple scaling argument. (Note that for
genus zero,
we have $Z(A)\sim A^{\Gamma_{\rm str}-3}$.)
Under the shift $\ph\to\ph+\rho/\gamma$ for $\rho$ constant, the measure in
\eZA\ does not change. The change in the action \ersL\ comes from the term
$${Q\over 8 \pi}\int \d^2\xi\,\sqrt {\hat g}\, \hat R\,\ph\to
{Q\over 8 \pi}\int \d^2\xi\,\sqrt {\hat g}\, \hat R\,\ph +
{Q\over 8 \pi}{\rho\over\gamma}\int \d^2\xi\,\sqrt {\hat g} \hat R\ .$$
Substituting in \eZA\ and using the Gauss-Bonnet formula
${1\over4 \pi}\int \d^2\xi\,\sqrt {\hat g} \hat R=\chi$ together with the
identity $\delta(\lambda x)=\delta(x)/|\lambda|$ gives
$Z(A)=\ee{-{Q \rho \chi/ 2\gamma}-\rho}\,Z(\ee{-\rho}A)$.
We may now choose $\ee\rho=A$, which results in
$$Z(A)=A^{-Q\chi/2\gamma-1}\,Z(1)= 
A^{(\Gamma_{\rm str}-2){\chi/2}-1}\,Z(1)\ ,$$
and we confirm from \eQ\ and \ealph\ that
\eqn\egam{\Gamma_{\rm str}=2-{Q\over \gamma}
={1\over12}\bigl(D-1-\sqrt{(D-25)(D-1)}\,\bigr)\ .}

In the nomenclature of \rBPZ, so-called ``minimal conformal field theories''
(those with a finite number of primary fields) are specified by a pair of
relatively prime integers $(p,q)$ and have central charge
$D=c_{p,q}=1-6(p-q)^2/pq$.
The unitary discrete series, for example,
is the subset specified by $(p,q)=(m+1,m)$.
After coupling to gravity, the general $(p,q)$ model has critical exponent
$\Gamma_{\rm str}=-2/(p+q-1)$. Notice that $\Gamma_{\rm str}=-1/m$ for the
values $D=1-6/m(m+1)\,)$
of central charge in the unitary discrete series. (In general, the $m^{\rm th}$
order multicritical point of the one-matrix model will turn out to describe the
$(2m-1,2)$ model (in general non-unitary) coupled to gravity, so its critical
exponent $\Gamma_{\rm str}=-1/m$ happens to coincide with that of the $m^{\rm
th}$ member
of the unitary discrete series coupled to gravity.) Notice also that \egam\
ceases to be sensible for $D>1$, an indication of a ``barrier'' at $D=1$ that
has already appeared and will reappear in various guises in what
follows.\foot{The ``barrier'' occurs when coupling gravity to $D=1$ matter in
the language of non-critical string theory, or equivalently in the case of
$d=2$ target space dimensions in the language of critical string theory.
So-called non-critical strings (i.e.\ whose conformal anomaly is compensated by
a Liouville mode) in $D$ dimensions can always be reinterpreted as critical
strings in $d=D+1$ dimensions, where the Liouville mode provides the additional
(interacting) dimension. (The converse, however, is not true since it is not
always possible to gauge-fix a critical string and artificially disentangle the
Liouville mode (see e.g.\ \wittbh).)}

\smallskip
\noindent{\it Dressed operators / dimensions of fields}
\par\nobreak
Now we wish to determine the effective dimension of fields after coupling
to gravity. Suppose that $\Phi_0$ is some spinless primary field in a
conformal field theory with conformal weight $\Delta_0=\Delta(\Phi_0)
=\overline\Delta(\Phi_0)$ before coupling to gravity.
The gravitational ``dressing'' can
be viewed as a form of wave function renormalization that allows $\Phi_0$
to couple to gravity. The dressed operator $\Phi=\ee{\alpha\ph}\Phi_0$ is
required to have dimension (1,1)
so that it can be integrated over the surface $\Sigma$ without breaking
conformal invariance. (This is the same argument used prior to \ealph\ to
determine $\gamma$). Recalling the formula \ecw\ for the conformal weight of
$\ee{\alpha\ph}$, we find that $\alpha$ is determined by the condition
\eqn\ehdr{\Delta_0-\half \alpha(\alpha-Q)=1\ .}

We may now associate a critical exponent $\Delta$ to the behavior of the
one-point function of $\Phi$ at fixed area $A$,
\eqn\eopt{F_\Phi(A)\equiv {1\over Z(A)}\int\CD\ph\,\CD X\ \e{-S}\
\delta\Bigl({\textstyle\int} \d^2\xi\,\sqrt {\hat g}\,\ee{\gamma\ph} - A\Bigr)
\ {\textstyle\int} \d^2\xi\,\sqrt{\hat g}\, \ee{\alpha\ph}\,\Phi_0
\sim A^{1-\Delta}\ .}
This definition conforms to the standard convention that $\Delta<1$ corresponds
to a relevant operator, $\Delta=1$ to a marginal operator, and $\Delta>1$ to an
irrelevant operator (and in particular that relevant operators tend to dominate
in the infrared, i.e.\ large area, limit).

To determine $\Delta$, we employ the same scaling argument that led to \egam.
We shift $\ph\to\ph+\rho/\gamma$ with $\ee\rho=A$ on the right hand side of
\eopt, to find
$$F_\Phi(A)
={A^{-Q\chi/2\gamma-1+\alpha/\gamma}\over A^{-Q\chi/2\gamma-1}}\,F_\Phi(1)
=A^{\alpha/\gamma}\,F_\Phi(1)\ ,$$
where the additional factor of $\ee{\rho\alpha/\gamma}=A^{\alpha/\gamma}$ comes
from the $\ee{\alpha \phi}$ gravitational dressing of $\Phi_0$.
The gravitational scaling dimension $\Delta$ defined in \eopt\ thus satisfies
\eqn\ehba{\Delta=1-\alpha/\gamma\ .}
Solving \ehdr\ for $\alpha$ with the same branch used in \ealph,
\eqn\embeta{\alpha=\half Q-\sqrt{{\textstyle {1\over4}}Q^2-2+2\Delta_0}
={1\over\sqrt{12}}\bigl(\sqrt{25-D}-\sqrt{1-D+24 \Delta_0}\bigr)}
(for which $\alpha\le Q/2$, and $\alpha\to0$ as $D\to-\infty$).
Finally we substitute the above result for $\alpha$ and the value \ealph\ for
$\gamma$ into \ehba, and find\foot{We can also substitute
$\alpha=\gamma(1-\Delta)$ from \ehba\ into \ehdr\ and use
$-\half\gamma(\gamma-Q)=1$ (from before \ealph) to rederive the result
$\Delta-\Delta_0=\Delta(1-\Delta){\gamma^2/2}$ for the difference between the
``dressed weight'' $\Delta$ and the bare weight $\Delta_0$ \rKPZ.}
\eqn\ehf{\Delta
={\sqrt{1-D+24\Delta_0}-\sqrt{1-D}\over\sqrt{25-D}-\sqrt{1-D}}\ .}

We can apply these results to the $(p,q)$
minimal models \rBPZ\ mentioned after \egam.
These have a set of operators labelled by two integers
$r,s$ (satisfying $1\le r\le q-1,\ 1\le s\le p-1$) with bare conformal weights
$\Delta_0=\bigl((pr-qs)^2-(p-q)^2\bigr)/4pq$ (we take $p>q$).
Coupled to gravity, these operators have dressed Liouville exponents
\eqn\edrs{1-\Delta_{r,s}=
{\alpha_{r,s}\over\gamma}={p+q-|pr-qs|\over 2q}
\qquad 1\le r\le q-1,\ 1\le s\le p-1}
$$\Bigl(\hbox{note also that } c=1-6{(p-q)^2\over pq}\quad\Longrightarrow
\quad \gamma=\sqrt{2q\over p}\,,\ \
Q=\sqrt{2p\over q}+\sqrt{2q\over p}\ \Bigr)\ ,$$
in agreement with the weights determined from the $(p,q)$ formalism
(to be discussed in sections {\it\sKdV\/} and {\it\sscsotmc\/})
for the generalized KdV hierarchy (see e.g.\ \refs{\rD,\rGGPZ}).

\newsec{Brief Review of the Liouville Theory} 
\seclab\sbrlt

In this chapter, we touch briefly on some of the highlights of Liouville
theory from the viewpoint advocated in \refs{\nati,\mart,\joetalk,\mss}.
For other points of view on the Liouville theory, see \refs{\jakiew,\takht},
and the sequence of works \gervais. The classical Liouville theory was
extensively studied at the end of the nineteenth century in connection with the
uniformization problem for Riemann surfaces. We will sketch some of this in
sections {\it\slcu, \slqu\/}.

\subsec{Classical Liouville Theory}

We choose some reference metric $\hat g$ on a surface $\Sigma$. The
Liouville theory is the theory of metrics $g$ on $\Sigma$, and the Liouville
field $\phi$ is defined by
\eqn\elidef{g=\ee{\gamma \phi}\hat g\ .}
The action is
\eqn\liouvs{S_{\rm Liouville}=\int \d^2 z\,  \sqrt{\hat g}
\Bigl({1\over 8 \pi} (\hat\nabla \phi)^2
+{Q\over 8 \pi}\phi R(\hat g)\Bigr)
+{\mu\over 8 \pi \gamma^2}\int \d^2 z\,\sqrt{\hat g}\,\ee{\gamma \phi}\ ,}
very similar to the background--charge theory
\bckchgi\ with a pure imaginary background charge $Q=2i\alpha_0$.
The interaction given by $\mu$ (the ``cosmological constant'' term), while
soft, will be seen to have profound effects on the theory.
For the particular choice
\eqn\classqu{Q=2/\gamma\ ,}
the action \liouvs\ defines a classical
conformal field theory, invariant  under the Weyl transformations
\eqn\weyl{
\hat g\to \ee{2\rho} \hat g\qquad \qquad \gamma\phi\to \gamma\phi-2 \rho\ .}
\noindent{\bf Remark}: The linear shift in $\phi$ under a conformal
transformation shows that $\phi$ can be interpreted as a Goldstone
boson for broken Weyl invariance (broken by the choice of $\hat g$).

\exercise{Classical Liouville theory}

a) Using the transformation properties of the Ricci
scalar in two dimensions,
\eqn\curvtrm{R[\ee{2\rho} \hat g]=\ee{-2\rho}\bigl(R[\hat g]
- \hat\nabla^2 2\rho\bigr)\ ,}
compute the change in the action
\liouvs\ under \weyl, and show that
for $Q=2/\gamma$ the change is independent of $\phi$
(so doesn't affect the classical equations of motion).

b) Show that the classical equations of motion for \liouvs\ may be expressed as
\eqn\clsseq{R[g]=-\half \mu\ ,}
i.e.\ they describe a surface with constant negative curvature.
(We take $\mu$ positive.)
Using again \curvtrm\ (in the form $R[g]=\ee{-\gamma \phi}\bigl(R[\hat g]
- \hat\nabla^2 \gamma \phi\bigr)$, with $\phi$ as in \elidef), note that
\clsseq\ is explicitly invariant under \weyl.

\endexercise

The stress-energy tensor following from \liouvs,
$T_{\mu \nu}=-2 \pi {\delta S\over \delta g^{\mu \nu}}$,
takes the familiar Feigin--Fuks form
\eqn\stressenergy{\eqalign{T_{z\zb}&= 0\cr
T_{zz}&=-\half (\p \phi)^2+\half Q \p^2 \phi\cr
T_{\zb\zb}&=-\half (\pb \phi)^2+\half Q \pb^2 \phi\ ,\cr}
}
where we have used the equations of motion in the first line.

Since $\phi$ is a component of a metric, its transformation
law under conformal transformations $z\to w=f(z)$,
\eqn\liouvtmn{\phi\rightarrow \phi
+ {1\over \gamma}\log\Bigl|{\d w\over \d z}\Bigr|^2\ ,}
is more complicated than that of an ordinary scalar field.
In particular, the $U(1)$  current $\del_z\phi$ measuring the Liouville
momentum transforms as
\eqn\crrnt{\p_z \phi\to {\d w\over \d z}\, \p_w \phi +
{\d\over \d z}{1\over \gamma}\log\Bigl|{\d w\over \d z}\Bigr|\ ,}
and the stress tensor $T_{zz}$ transforms as
\eqn\strsstns{T_{zz}\to \Bigl({\d w\over \d z}\Bigr)^2 T_{ww}
+ {1\over\gamma^2}\, S[w;z]\ .}
The object $S[w;z]$ is called the ``Schwartzian derivative'' and has many
equivalent definitions.\foot{It may be considered, for example, as the
integrated version of the conformal anomaly, or it may be defined in terms of
``projective connections.'' For a discussion of the latter concept see
\gunning.} Combining \stressenergy\ and \crrnt, we see that
\eqn\dfschw{\eqalign{
S[w;z]&=-\ha\Bigl(\p_z \log\Bigl|{\d w\over \d z}\Bigr|\Bigr)^2+
{\d^2\over \d z^2}\log\Bigl|{\d w\over \d z}\Bigr|\cr
&=T_{zz}\Bigl(\phi=\log\Bigl|{\d w\over \d z}\Bigr|\Bigr)\cr
&={w'''\over w'}-{3\over 2} \Bigl({w''\over w'}\Bigr)^2\ .\cr}}
The unusual transformation laws \crrnt\ and \strsstns\
have counterparts in the quantum theory, where they result in shifted formulae
for conformal charges and weights in Liouville theory.

\subsec{Classical Uniformization} 
\subseclab\slcu

The central theorem in classical Liouville theory is the
uniformization theorem that characterizes Riemann surfaces.\hfill\break
{\bf Uniformization Theorem}: Every Riemann surface $\Sigma$
is conformally equivalent to
\nobreak
\item{1)} $CP^1$, the Riemann sphere, or
\item{2)} $H$, the Poincar\'e upper half plane, or
\item{3)} A quotient of $H$ by a discrete subgroup $\Gamma\subset SL(2,\IR)$
acting as M\"obius transformations.\par\nobreak\noindent
The first proofs of the uniformization theorem were based on the
existence of solutions to the classical field equations \clsseq;
the standard proofs use potential theory and are nonconstructive (see \farkra).

The upper half plane supports the standard solution of
the Liouville equation, namely the \poin\ metric:
\eqn\pomet{\d s^2 =\ee{\gamma \phi}\,|\d z|^2
= {4\over \mu}{1\over ({\rm Im}\, z)^2}\,|\d z|^2\ ,}
a constant negative curvature solution to \clsseq.
This metric is invariant under the M\"obius transformations
\eqn\emobius{z\to {az+b\over cz+d}\ ,\qquad a,b,c,d\in\IR,\ ad-bc=1\ }
(i.e.\ the group $PSL(2,\IR)=SL(2,\IR)/\IZ_2$), and thus descends to a metric
\eqn\poincare{\d s^2=\ee{\gamma \phi}\,|\d z|^2
= {4\over \mu}{\p A\,\pb B\over\bigl(A(z)-B(\zb)\bigr)^2}\,|\d z|^2}
on the Riemann surface $X=H/\Gamma$
for some locally defined (anti-)holomorphic functions
$A(z)$ $\bigl(B(\zb)\bigr)$. In general,
when we quotient $H$ by the action of a discrete
group $\Gamma$ to define a space $X=H/\Gamma$,
there is a natural projection $\pi:H\to X$ and an ``inverse'' map
\eqn\invmap{f:X\to H\ ,}
known as the ``uniformizing map''.

\exercise{Classical energy-momentum}

a) Evaluate the energy-momentum tensor for the \poin\
metric and show that $T_{zz}=0$.

b) Let $f:X\to H$ be the uniformizing map as in \invmap.
Show that the solution \poincare\
to the field equations has energy-momentum
\eqn\enrgmo{
T_{zz}={1\over \gamma^2} S[f;z]\ ,}
with $S[f;z]$ as in \dfschw.
\endexercise

Note that the uniformizing
map \invmap\ is not well-defined. If we continue
the values of $f$ from some coordinate patch around a nontrivial
cycle, then $f$ will change by the action of $T\in \Gamma$. The nature
of the surface near such a nontrivial curve depends on the
nature of the conjugacy class of $T$, in turn
classified  by the value of the trace. There are three types of
conjugacy classes in $SL(2,\IR)$:
\item{1)} elliptic: $|\Tr\,T|<2$\quad
($T$ conjugate to ${\ \ \cos\lambda\ \sin \lambda\choose
-\sin\lambda\ \cos\lambda}$)\ ,
\item{2)} parabolic: $|\Tr\,T|=2$\quad
($T$ conjugate to ${1\ \lambda\choose 0\ 1}$)\ ,\qquad and
\item{3)} hyperbolic: $|\Tr\,T|>2$\quad
($T$ conjugate to ${\cosh \lambda\ \sinh \lambda\choose
\sinh \lambda\ \cosh \lambda}$)\ .
\par\nobreak\noindent
In cases 1,2), the nontrivial curve surrounds a
puncture on the surface. In case 3), the curve surrounds a
handle. See \mart\ for further discussion.

\exercise{}

a) Show that the Schwartzian derivative is invariant under
independent M\"obius transformation of either $z$ or $f$.

b) Show that although the uniformizing map is not globally
defined, the energy-momentum \enrgmo\ is nonetheless well-defined.

\endexercise

\subsec{Quantum Liouville Theory}

It is more subtle that the theory \liouvs\
is also a {\it quantum\/} conformal field theory.

If $\mu$ were zero and $\phi$ were a free field, we would immediately
conclude that $T$ defines a Virasoro algebra with central
charge $c=1+3 Q^2$, that exponentials $\ee{\alpha \phi}$ have
conformal weight $-\half \alpha(\alpha-Q)$, and that these
operators create states $|\alpha\rangle$ on which we
could construct Feigin--Fuks modules.\foot{A Feigin--Fuks module is a Fock
space
in which the Virasoro algebra is represented by an energy-momentum tensor such
as \stressenergy.} Since $\phi$ is {\it not\/} a free field,\foot{We will see
that the operator product of two exponentials is not given by the free
field expression.} we must be more careful.

We proceed via canonical quantization, first passing from the complex
$z$-plane to cylindrical coordinates $(t,\sigma)$ via
$z=\ee{t+i \sigma}$, and expanding
\eqn\canexp{\eqalign{\phi(\sigma,t)&=\phi\dup_0(t)+\sum_{n\not=0}
{i\over n}\Bigl(a_n(t)\,\ee{-i n \sigma}+ b_n(t)\,\ee{i n \sigma}\Bigr)\cr
\Pi(\sigma,t)&=p_0(t)+\sum_{n\not=0} {1\over
4 \pi}\Bigl(a_n(t)\,\ee{-i n \sigma}+ b_n(t)\,\ee{i n \sigma}\Bigr)\ .\cr}}
Here $a_n^\dagger = a_{-n}$, $b_n^\dagger = b_{-n}$, and we
have the {\it equal-time\/} canonical commutation relations
\eqn\cancomm{\bigl[a_n(t),b_m(t)\bigr]=n\,\delta_{n,m}\ .}
The energy-momentum tensor in canonical variables takes the form
(again using the equations of motion)
\eqn\quanvir{\eqalign{T_{+-}&=0\cr
T_{\pm \pm}&={1\over 8}(4 \pi \Pi\pm \phi')^2-{Q\over
4}(4 \pi \Pi\pm \phi')'+{\mu\over 8 \gamma^2}
\ee{\gamma \phi} + {Q^2\over 8}\ .\cr}
}
The additive factor of $Q^2/8$ in the above arises from the Schwartzian
derivative in the transformation properties \strsstns\ of $T$ when mapping
from the plane to the cylinder $z\to(t,\sigma)$.

In \crtthrn, it was shown that the operators \quanvir\ satisfy a
Virasoro algebra if
\eqn\qanqu{Q={2\over\gamma} + \gamma\ }
(as we derived from another point of view in \eQgam).
Calculating the $[T,T]$ commutator, one finds indeed a central charge
\eqn\quancee{c=1+3 Q^2\ ,}
and calculating commutators with $T$ shows that
exponentials $\ee{\alpha \phi}$ have conformal weight
\eqn\qandel{\Delta(\ee{\alpha \phi})=-\half(\alpha-Q/2 )^2 + Q^2/8\ .}
Note that if we impose the condition $\Delta(\ee{\alpha \phi})+
\Delta_0=1$, to dress an operator with bare weight $\Delta_0$, we rederive the
KPZ equation \embeta.

It is useful to have an intuitive understanding of
eqns.~\eqns{\qanqu{--}\qandel}. Note that
by rescaling $\phi\to {1\over \gamma} \phi$ in \liouvs, we identify $\gamma$
with the coupling constant of the theory. The semiclassical theory is thus
defined by asymptotic expansion as $\gamma\to 0$. We see that \qanqu\ is the
quantum version of the classical condition \classqu\ for conformal invariance.
Similarly, \quancee\ consists of a classical part
($\propto Q^2\propto 1/\gamma^2$) already visible in the
classical transformation law \strsstns, plus a quantum conformal anomaly
($c=1$), familiar for a single scalar field.
Finally, to understand \qandel\ we note that the analog of \crrnt\ in the full
quantum theory is
\eqn\qancrrnt{\p_z \phi\to {\d w\over \d z}\, \p_w \phi +
{\d\over \d z}{Q\over 2}\log\Bigl|{\d w\over \d z}\Bigr|\ .}
In particular, passing from the plane to the cylinder
via the conformal transformation $w=\log z$ we have
$\p_z \phi\to (\p_w \phi - Q/2)/z$,
so the momentum ``shifts'' by $Q/2$. The vertex
operators $\ee{\alpha \phi}$, inserted on the $z$-plane,
create states  with Liouville momentum $p\dup_\phi$ given by
\eqn\liouvmom{i p_\phi=\alpha-{Q\over 2}\ .}
(Note that ``states'' refer to quantization on the cylinder,
and ``Liouville momentum'' refers to the zero mode $p\dup_0$
of $\Pi$ in \canexp.)
We see that the first term in \qandel\ is simply $\half p_\phi^2$,
and the second term is the shift in the energy (relative to the
Gaussian case) due to the ``extra'' central charge.

The above formulae are valid for cosmological constant $\mu\ge0$.
It may seem curious that the quantum formulae for $\mu > 0$ are identical to
those obtained for a free field, i.e.\
 with cosmological constant $\mu=0$ in \liouvs.
(N.B.: a translation of $\phi$ {\it cannot\/} transform one case into the
other.) Heuristically, we can understand this by noting that in the worldsheet
ultraviolet, $\phi\to -\infty$, the interaction term disappears: Quantities
such as $c$ and $\Delta$, determined by the singular terms in operator product
expansion, depend only on the ultraviolet behavior of the theory.

\subsec{Spectrum of Liouville Theory}

We now proceed to study the Hilbert space of the theory, all of whose
subtleties lie in the zero modes $\phi\dup_0(t)$, $p\dup_0(t)$. We can
understand the
physics of these degrees of freedom by studying the action \liouvs\ for field
configurations independent of $\sigma$, i.e., we study the Liouville quantum
mechanics
\eqn\liuvqm{S
=\int \d t\,(\dot \phi^2 + {\mu \over \gamma^2}\,\ee{\gamma \phi})\ .}
The  Hamiltonian $H_0=\half\oint (T_{++}+T_{--})$,
after substituting \quanvir,  takes the form
\eqn\lvqmh{H_0=-\ha\Bigl({\p\over \p \phi\dup_0}\Bigr)^2
+{\mu\over8\gamma^2}\,\ee{\gamma \phi\dup_0} + {1\over 8}Q^2\ .}

\ifig\foscbess{Particle wavefunctions in the exponential potential.}
{\epsfxsize2.75in\epsfbox{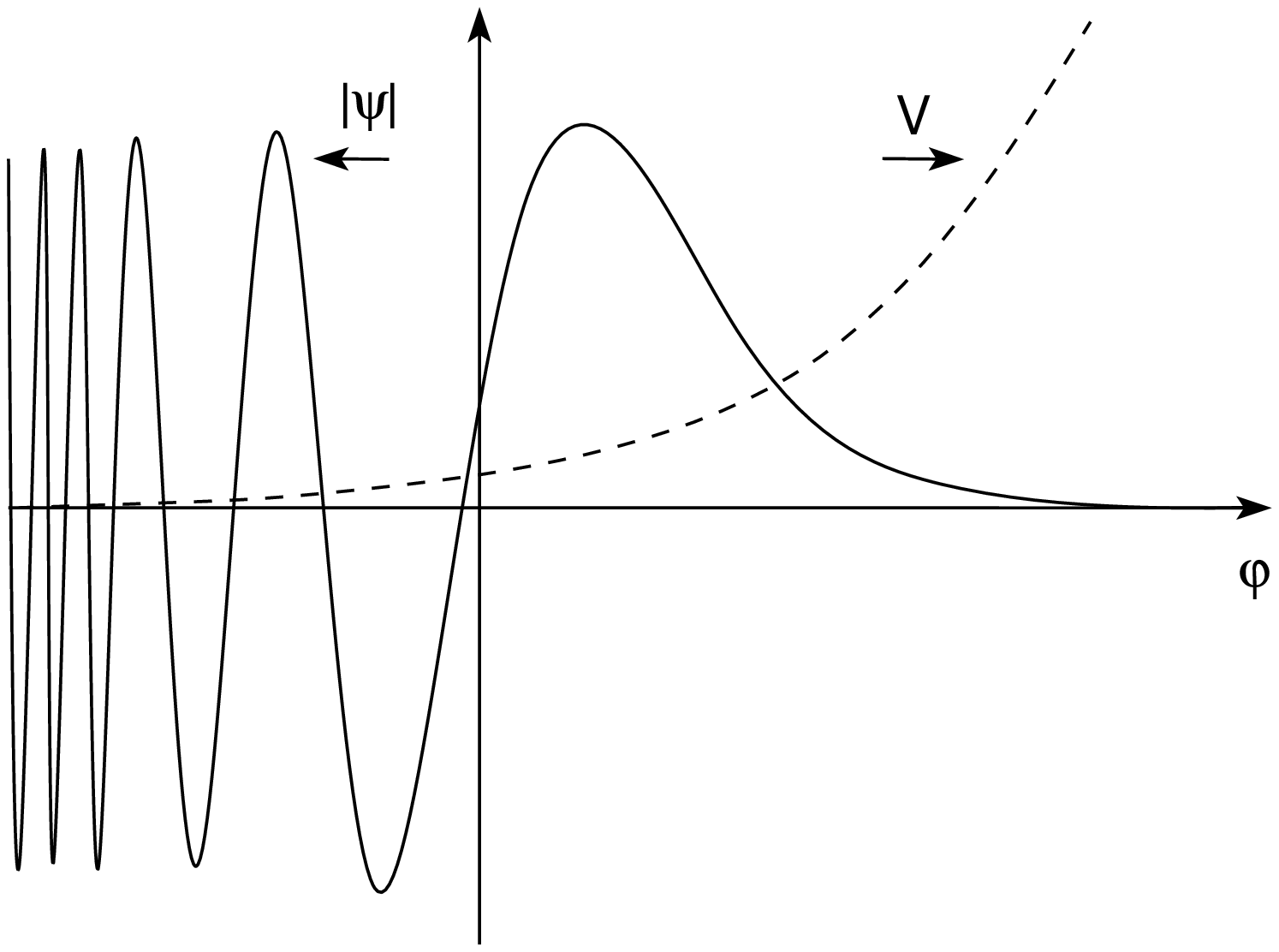}}

The spectrum of $H_0$ is easily understood.
In the worldsheet ultraviolet, $\phi\dup_0\sim-\infty$ (i.e.\ at {\it short\/}
physical distances), the potential disappears so that
normalizable states behave like plane waves,
$\psi\dup_E\sim \sin E \phi\dup_0$,
with energy $\half E^2+{1\over 8}Q^2$.
The exponential growth of the potential prevents
the ``particle'' at $\phi\dup_0$ from penetrating too far to the
right, and hence gives total reflections of any incoming wave.
Because of the total reflection property there is no distinction
between states with $+E$ and $-E$ and we can therefore take
$E>0$. The wavefunctions look as in \foscbess.

The circumference of the 1D ``universe'' in physical units is measured by
the quantity
\eqn\length{\ell=\e{\ha\gamma \phi\dup_0}\ .}
Using $\ell$, it is moreover possible to give an
exact description of the eigenstates of $H_0$ in
terms of Bessel functions. Changing variables to $\ell$,
the eigenvalue equation
$H_0 \psi\dup_E=\Bigl(\half E^2+{1\over8}Q^2\Bigr)\psi\dup_E$ becomes
\eqn\eveqtn{\Bigl(-(\ell {\p\over \p \ell})^2+4 \hat\mu \ell^2 +
{Q^2\over \gamma^2}\Bigr)\psi\dup_E(\ell)
=\Bigl({Q^2\over\gamma^2}+{\hat E^2\over 2}\Bigr)\psi\dup_E\ ,}
where $\hat \mu=\mu/(4 \gamma^4)$ and $\hat E=\sqrt{8}E/\gamma$.
We recognize \eveqtn\ as the Bessel differential equation.
Imposing the boundary condition that $\psi$ decays
for large universes, $\ell\to \infty$, gives the eigenfunction
\eqn\diagsts{\psi\dup_E={1\over \pi}\sqrt{\hat E \sinh\pi \hat E}\,
\,K_{i\hat E}(2\sqrt{\hat \mu}\ell)\ ,}
where $K$ is the modified Bessel function. We have chosen
a $\delta$-function normalization for the states,
\eqn\delnorm{\int_0^\infty {\d \ell\over \ell}\,
\psi\dup_{E}(\ell)\, \psi\dup_{E'}(\ell)=\delta(E-E')\ .}

\exercise{The wall analogy}

Show that the solutions $\psi\dup_E$ behave asymptotically
for $\phi\dup_0\to-\infty$
as $\sin\bigl(\half \gamma E\phi\dup_0+\alpha(E)\bigr)$, where
$$\alpha(E)={\rm arg}\,\Gamma(1+i E)
=\ha\log {\Gamma(1+iE)\over \Gamma(1-iE)} .$$
For intuitive purposes, it is useful to replace the exponentially growing
Liouville potential by an infinite hard wall. Where should this wall be located
for energies of order $E$?

\endexercise

Now we proceed from the zero mode structure of the theory to construct
the full field theory.
Combining the above discussion on zero modes with canonical
quantization, one expects \crtthrn\
the Hilbert space (as a ${\rm Vir}\oplus\overline{\rm Vir}$ representation
space) of the Liouville conformal field theory to take the form
\eqn\curtthr{\CH=\rlap{\hbox{\hskip3pt$\oplus$}}\int_0^\infty
\d E\ \CF_{\Delta(E)}\otimes \overline\CF_{\Delta(E)} .}
Here $\CF$ is a Feigin--Fuks module with weight $\Delta(E)
=\half E^2+{1\over 8}Q^2$ and central charge $c=1+3 Q^2$.
Our notation on the r.h.s.\ of \curtthr\
is meant to indicate a direct integral of Hilbert spaces \reedsim. Each
Feigin--Fuks module is generated by adding oscillator excitations to some
primary state. As usual, this structure may be understood heuristically since
in the worldsheet ultraviolet ($\phi\to - \infty$) the theory becomes free.

In a conventional conformal field theory we are able to associate each
state in the Hilbert space to an operator, and then determine the operator
algebra of the theory.
How do we construct the vertex operators that create states in the space
\curtthr?
According to \liouvmom,
we might expect the
primary fields of Liouville momentum $p\dup_\phi=E$
to have quantum vertex operators
\eqn\prpvrtx{V_E(z,\zb)=\ee{\alpha\phi}
=\ee{i E \phi}\, \ee{\ha Q \phi}\ .}
At this point, however, we begin to encounter some of the
confusing subtleties of Liouville theory {\it cum\/} quantum gravity,
as first emphasized in \refs{\nati,\joetalk}: the operators \prpvrtx\
by no means encompass all of the quantities of interest in the theory.
For example, a natural quantity
in quantum gravity is the ``volume of the universe,''
\eqn\area{A=\int_\Sigma \ee{\gamma \phi}\sqrt{\hat g}\ ,}
given by integrating the area operator $\ee{\gamma \phi}$.
But comparing with \prpvrtx, we see that this operator has imaginary
momentum and cannot correspond to a normalizable state (in the sense of
\delnorm). To obtain some insight into this puzzling observation, we
return to examine the semiclassical theory.


\subsec{Semiclassical States}
\subseclab\ssclst

The semiclassical approximation is an important source of intuition for
understanding Liouville theory.
Classical Liouville theory describes the geometry of
negatively curved surfaces (eq.~\clsseq).
Earlier, we identified $\gamma$ with the coupling constant of the theory (by
rescaling $\phi\to \phi/\gamma$), so
the semiclassical limit is defined by $\gamma\to 0$ asymptotics.
In this limit, we expect the above quantum states to
correspond to specific constant curvature surfaces, and this has been verified
in detail \refs{\nati,\mart}. Briefly, the $\sigma$-independent metric
\eqna\macmet
$$\ee{\gamma \phi(t)}(\d t^2+ \d\sigma^2)
={4\over \mu} {\varepsilon^2\over \sin^2(\varepsilon t)}
(\d t^2+ \d\sigma^2)\qquad
{n \pi\over \varepsilon}<t<{(n+1) \pi\over \varepsilon}\eqno\macmet a$$
(where  $z=\ee{t+i \sigma}$ and $\varepsilon$ is a real number),
is a solution to the Liouville equation \clsseq\ that looks something like
$$\vcenter{\hbox{hyperbolic:\qquad}}
\vcenter{\figin{\epsfysize1in\epsfbox{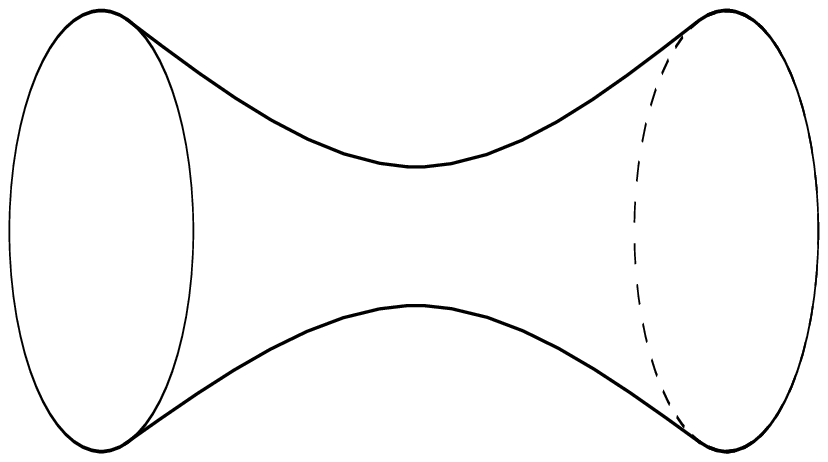}}}\quad.\eqno\macmet b$$

\exercise{Classical field energy}

Use the energy momentum tensor \stressenergy\
to show that the Liouville field configuration \macmet{}\ has
 classical energy $\half (\varepsilon^2/\gamma^2)+ {1\over8}Q^2$.

Recalling that the quantum state $\psi\dup_E$ of \eveqtn\ has energy
$\ha E^2 + {1\over8}Q^2$, this classical field energy allows us to identify
\macmet{}\ as the semiclassical picture of the quantum state
$\psi\dup_E$ for $E=\varepsilon/\gamma$.

\endexercise

As noted in sec.~{\it\slcu\/}, there are three kinds of
local behavior of a solution to the classical Liouville equation,
classified by the monodromy properties of $A,B$ in \poincare.
In the solution \macmet a\ above, we have $A=z^{i \varepsilon}=\overline B$,
thus giving an example of a hyperbolic class.
{}From the semiclassical point of view, we are naturally led to
ask what quantum states correspond to the other two classes, namely the
elliptic and parabolic solutions.
These are given respectively by
\eqna\micmet
$$\eqalign{
\ee{\gamma \phi}(\d t^2+ \d\sigma^2)&={4\over \mu} {\nu^2
\over \sinh^2(\nu t)}(\d t^2+ \d\sigma^2)\qquad t<0\cr
\ee{\gamma \phi}(\d t^2+ \d\sigma^2)&={4\over
\mu} {1\over t^2}(\d t^2+ \d\sigma^2)\qquad t<0\ .\cr}\eqno\micmet a$$
Here $\nu$ is a real number.
An important feature of the solutions \micmet a\ is
their equivalence for both $\pm \nu$.
These solutions look something like
$$\vcenter{\hbox{elliptic, parabolic:\qquad}}
\vcenter{\figin{\epsfysize1in\epsfbox{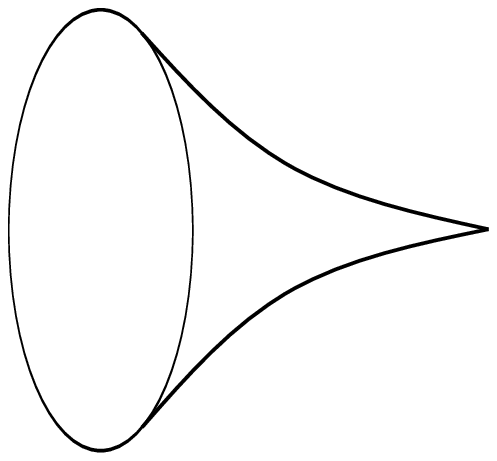}}}\quad,\eqno\micmet b$$
and have energy $-\half (\nu^2/\gamma^2)+ {1\over8}Q^2$.
Quantum mechanically, according to \eqns{\eveqtn{,\ }\diagsts}\ they therefore
correspond to states with imaginary momentum $E=i \nu/\gamma$. The
corresponding wavefunctions are of type $K_\nu(\ell)$, $\nu$ real. These
wavefunctions blow up (as $\ell^{-| \nu|}$) at short distances $\ell\to 0$ and
consequently might appear unphysical. On the contrary, as we saw at the end of
the previous section, operators corresponding to imaginary momentum states,
including for example the volume of the universe, appear quite naturally in 2D
quantum gravity and play an important role.

The distinction between normalizable and non-normalizable states in Liouville
theory, and the necessity to include the non-normalizable states in the theory,
was first emphasized by Seiberg in \nati. Motivated by the geometries
illustrated above, the normalizable states were labelled ``macroscopic
states,'' and the non-normalizable states were labelled ``microscopic states''.
Semiclassically, the macroscopic states do not have a well-defined insertion
point in the intrinsic geometry of the surface. The microscopic states, on the
other hand, correspond semiclassically to the elliptic geometry
pictured in \micmet b, and thus to local operators ---
the operator insertion in this case
is localized at the tip of the ``funnel.''
We discuss these issues further in sec.~{\it\sskpztd\/} below.

\exercise{Curvature Sources}

a) Use the exponential map $z=\ee{t+i \sigma}$ to transform the
solutions \micmet{}\ to the $z$-plane:
\eqn\micmeti{ \ee{\gamma \phi}\,\d z\,\d\zb= {16\over \mu}
{\nu^2 (z\zb)^{\nu}\over (1-(z \zb)^\nu)^2}{\d z\,\d\zb\over z \zb}\ .}

b) Show that the field $\phi$ solves the Liouville equation with source,
\eqn\srce{
{1\over 4 \pi} \Delta \phi - {\mu\over 8 \pi \gamma} \ee{\gamma \phi}
+{1-\nu\over \gamma} \delta^{(2)}(z)=0\ ,}
i.e., such solutions have a source of curvature at $z=0$.

c) Show that the solution corresponds to the choice of
functions (in \poincare):
\eqn\asbs{
A(z)=i {z^\nu+ 1\over z^\nu-1} \qquad\qquad
B(\zb)=-i {\zb^\nu+ 1\over \zb^\nu-1}\ .}

d) Show that the monodromy when $z$ circles around the
puncture is the real elliptic M\"obius transformation
\eqn\monaa{A(z)\to {\cos \pi \nu A(z) - \sin \pi \nu\over
 \sin \pi \nu A(z) + \cos \pi \nu}\ .}

e) Show that a straight line through $z=0$ is conformally
mapped into an angle $ \pi \nu$, and hence we must have
$\nu\geq 0$ on geometric grounds.

f) Repeat the above for the hyperbolic and parabolic cases.

\endexercise

\subsec{Seiberg bound}
\subseclab\sssb

Classically, the metrics \eqns{{\macmet{}}{,\ }{\micmet{}}}\
are invariant under $\nu\to -\nu, E\to -E$. Quantum mechanically, the
wavefunctions $K_{\nu},K_{i E}$ share this invariance, due to
the total reflection property of the Liouville ``wall.''
Turning on the wall by setting $\mu>0$ effectively halves the states that
exist in the ($\mu=0$) free spectrum.

In the DDK/KPZ formalism described in chapt.~\sqliouv, on the other hand,
the choice of root in the KPZ formula \ehf\
affects the scaling properties of the
operator $\ee{\alpha \phi}\,\Phi_0$. Since the
Liouville interaction truncates the spectrum
by half,  we must choose a root. In \refs{\nati,\joetalk}, it is argued that
only those operators with
\eqn\seibound{\alpha\le \half Q}
can exist. This choice of root has many distinguishing properties, some
of which will be noted in later sections:
\item{1)} This root gives a smooth semiclassical limit in
quantum gravity, as we saw in sec.~{\it\spia\/}.
\item{2)} The area element is integrable only for sources satisfying
\seibound. A related fact in terms of deficit angles has appeared in
part (e) in the exercise above (following \monaa).
\item{3)} As we will see in the penultimate paragraph of sec.~{\it\sskpztd\/},
the Wheeler--DeWitt wavefunction for a local operator in quantum
gravity, related to the vertex operator $V(\phi)$
by $\psi(\phi)=\ee{-\ha Q\phi}\,V(\phi)$, must be concentrated
on $\ell\to 0$, that is, where $\phi\to -\infty$.
\item{4)} A closely related question is the nature of gravitational
dressing (see sec.~{\it\spia\/}). Only with the choice of
root \seibound\ do gravitationally dressed
relevant/irrelevant operators grow/decay in the worldsheet
infrared.
\item{5)} As will be seen in sec.~{\it\stdstms\/} below,
the bound \seibound\
has the following spacetime interpretation: When scattering in
a left half-space, incomers must be rightmovers and outgoers
must be leftmovers.

\smallskip\noindent
In addition, there is circumstantial evidence that \seibound\ is correct:
\item{1)} In the matrix model, we will see that
only scaling operators with scaling corresponding to $\alpha\le Q/2$ appear.
\item{2)} In the semiclassical calculations of \crtthrn, there are difficulties
constructing correlation functions of ``wrong branch'' operators.
\item{3)} In the $SL(2,\IR)$ quantum group approach
pursued by Gervais and others, inverse powers of the
metric $\ee{-j \gamma \phi}$, $2j\in \IZ_+$, are easily constructed,  while
the positive powers have thus far eluded construction.

\smallskip\noindent
The above reasoning is qualitative. While the Seiberg bound
\seibound\ is undoubtedly correct, a precise mathematical understanding of the
statement would be useful.

\danger{A common confusion}
There are {\it two\/} distinct novelties in the
description of the Hilbert space of Liouville theory:
(1) Vertex operators with $\alpha>\half Q$
do not exist. (2) States with real momentum $p\dup_\phi=E$,
which formally correspond to vertex operators with
$\alpha=\half Q+iE$, do not have a correspondence
with local operators. These two distinct points are
often confused in the literature.
There is no (obvious) connection between the Seiberg bound, which specifies
the operators that exist, and the difficulty of localizing operators that
correspond to states $|E\rangle$ of energy $\half E^2 + {1\over8}Q^2$.

\danger{An unresolved confusion}
There is some confusion in the literature as to whether
the states $|E\rangle$ have a correspondence with
vertex operators $V_E=\ee{i E \phi}\,\ee{\ha Q\phi}$.
Although the semiclassical pictures of states (e.g.\ \macmet b) makes any
correspondence with {\it local\/} operators seem unlikely \nati,
we shall find these operators necessary
to formulate our scattering theory in two Minkowskian dimensions.
This problem is closely linked to the problem of time in string theory.

\exercise{Seiberg bound and semiclassical limits}

Consider the minimal conformal field theories labelled by
$(2,2m-1)$ in the BPZ classification, as mentioned after \egam.
Show that $\gamma=2/\sqrt{2m-1}$ so that as $m\to \infty$ the semiclassical
approximation becomes valid.  Note that the root chosen in
sec.~{\it\spia\/} on the basis of the semiclassical limit coincides with that
dictated by the bound \seibound.

\endexercise

\subsec{Semiclassical Amplitudes}
\subseclab\sssa

In the semiclassical approximation, we evaluate the amplitudes
\eqn\liuccorri{
\Bigl< \prod_i \ee{\alpha_i \phi(z_i)}\Bigr>\equiv \int [\d\phi]\,
\ee{-S_{\rm Liouville}[\phi]}\,\prod_i\ee{\alpha_i \phi}
}
via the saddle point approximation by first solving the classical equation
\eqn\lqsrc{
{1\over 4 \pi} \Delta \phi - {\mu\over 8 \pi \gamma}\, \ee{\gamma \phi}
+\sum_i \alpha_i\, \delta^{(2)}(z-z_i)=0\ .
}
Integrating \lqsrc\ over the surface $\Sigma$ we find,
since $\mu>0$, that a necessary condition for the existence of a solution is
\eqn\xbz{{1\over \gamma}\Bigl({1\over \gamma}(2-2h)-\sum \alpha_i\Bigr) <0\ .}

\exercise{}

Derive \xbz\ using the Gauss-Bonnet theorem,
$\int {1\over 4 \pi} R\sqrt{g}= \chi=2-2h$.

\endexercise

The particular combination
\eqn\kpzexp{s\equiv   {Q\over 2 \gamma} \chi - \sum {\alpha_i\over \gamma}\ ,}
known as the KPZ exponent \rKPZ, plays an important
role in the theory. Equation \xbz\ says that the semiclassical
KPZ exponent ($Q\to2/\gamma$) is negative.

When $s<0$, there is indeed a solution to \lqsrc, and we can expand
around it to evaluate \liuccorri.  Near $z=z_i$, it follows from \lqsrc\ that
\eqn\appxbh{
\phi\dup_{\rm cl}\sim - \alpha_i \log |z-z_i|^2
}
for $\alpha_i<1/\gamma$. For $\alpha=1/\gamma$ we have instead
\eqn\loglog{
\phi\dup_{\rm cl}\sim - {1\over \gamma}\Bigl( \log |z-z_i|^2 +2 \log\log{1\over
|z-z_i|}\Bigr)\ .
}
In either case, to write the semiclassical amplitudes we must excise disks
$B(r,z_i)$ around $z_i$ of radius $r$ (in the $\hat g$ metric), and define the
regularized action\foot{The need for these subtractions reflects the need for
renormalization of the vertex operators \nati.}
\eqn\regact{\overline S[\phi\dup_{\rm cl}]
\equiv\lim_{r\to 0}\Bigl(\int_{\Sigma-\cup_i B(r,z_i)} \CL
-\sum_i (\Delta_i+\overline\Delta_i)\log r\Bigr)
}
where $\Delta_i$ is the conformal weight of $\ee{\alpha_i \phi}$.
In terms of $\overline S$, the leading
$\gamma\to 0$ asymptotics of the correlator are given by
\eqn\leadact{\Bigl<\prod \ee{\alpha_i \phi(z_i)}\Bigr>_{\rm s.c.}
\sim  \ee{-\overline S[\phi\dup_{\rm cl}]}\ .}

There are also cases of interest with $s\geq 0$, notably, for genus zero
correlators of vertex operators with small total ``Liouville charge''
$\sum \alpha_i$.
In these cases, as described in \nati, we can still perform semiclassical
calculations by fixing the total area of the surface: we insert
$\delta(\int \ee{\gamma \phi} - A)$ into the path integral
to ensure that \lqsrc\ has a solution.
We obtain the $A$ dependence of fixed
area correlators by a scaling argument (similar to that used
before \egam\ and \ehba): shifting
$\phi\to \phi+{1\over \gamma} \log A$ gives
\eqn\fixai{\Bigl< \prod_i \ee{\alpha_i \phi(z_i)}\Bigr>_A=A^{-1-s}
\Bigl< \prod \ee{\alpha_i \phi(z_i)}\Bigr>_{A=1}\ ,
}
with $s$ as in \kpzexp.  To Laplace transform to
fixed cosmological constant, we integrate
\eqn\lplctrmn{
\int_0^\infty {\d A\over A} A^{-s} \ee{-\mu A} = \mu^s\, \Gamma(-s)\ .
}
The UV divergence as $A\to 0$ for $s\geq 0$ (reflecting the
absence of a classical solution without the area constraint)
plays an important role in speculations on the relation
of free field theory to the Liouville theory described in
sections {\it \slcfac, \srabsm, \swvbs\/} below.

Using \fixai, the problem of calculating correlators is reduced to
the case $A=1$.
The semiclassical formulae for genus zero correlators are obtained by averaging
over the space of classical solutions.  These solutions
are obtained for $s\geq 0$ by applying complex M\"obius transformations
from the standard round-sphere metric
\eqn\rndsphr{\d s^2=\ee{\gamma \bar \phi}\,|\d z|^2
={16\over\mu}\,{|\d z|^2\over \bigl(1+|z|^2\bigr)^2}}
(which, unlike \clsseq, has positive curvature $R=+\ha\mu$).
The result is
\eqn\avsl{\eqalign{&\Bigl< \prod \ee{\alpha_i \phi(z_i)}\Bigr>^{\rm
s.c.}_{A=1}=
\int \d^2 a\,\d^2 b\,\d^2 c\,\d^2 d \ \delta^{(2)}(ad-bc-1)
\prod_i \ee{\alpha_i \bar\phi(z_i)}\cr
&\qquad=\prod_i\Bigl({16\over\mu}\Bigr)^{\alpha_i/\gamma}
\int_0^\infty \d\lambda\, \lambda \int_{\IC} \d^2 w
\prod_i {1\over
\bigl(|\lambda z_i+ w|^2+\lambda^{-2}\bigr)^{2 \alpha_i/\gamma}}\ ,\cr}
}
where in the second line we have parametrized $SL(2,\IC)$ elements
by a unitary matrix times an upper triangular matrix, and we have
dropped the volume of $SU(2)$.
See \nati\ for more details.

\smallskip\noindent{\it Semiclassical Seiberg Bound} \refs{\nati,\joetalk}
\par\nobreak
The semiclassical approach provides a key insight into the
Seiberg bound \seibound. Consider the classical equation
\lqsrc\ in the neighborhood of a vertex operator
insertion. {\it If\/} we neglect the cosmological
constant term, the solution must behave as in \appxbh.
To check if this is self-consistent, we insert
\appxbh\ back into \lqsrc\ and note that the neglected term behaves as
\eqn\ngltrm{\ee{\gamma \phi}\sim {1\over |z-z_i|^{2 \alpha_i\gamma} }\ .}
If $\alpha_i\gamma> 1$, the cosmological constant
operator is not integrable at $z=0$ and we expect
trouble. Indeed, the careful considerations leading
to the classification of solutions in sec.~{\it\slcu\/} (following
eq.~\enrgmo)
show that there is no solution for $\alpha_i>1/\gamma \sim Q/2$. The essential
point is that too much curvature cannot be localized at a single point.

Here are two examples of semiclassical correlators:\hfill\break
{\bf Example 1}: Consider the three-point function on the sphere,
\eqn\thrpt{
\bigl< \ee{{\theta_1}\phi/\gamma}(z_1,\zb_1)
\,\ee{{\theta_2}\phi/\gamma}(z_2,\zb_2)
\,\ee{{\theta_3}\phi/\gamma}(z_3,\zb_3)\bigr>\ ,
}
where $\theta_i<1$ are considered to be $\CO(1)$ as
$\gamma\to 0$ and $\sum_i \theta_i>2$, so $s<0$.
The classical solution is known in this case
and is M\"obius invariant. It follows immediately from
\regact\ and the transformation properties of circles under
M\"obius transformations that
\eqn\thrptii{
\bigl< \ee{{\theta_1}\phi/\gamma}(z_1,\zb_1)
\,\ee{{\theta_2}\phi/\gamma}(z_2,\zb_2)
\,\ee{{\theta_3}\phi/\gamma}(z_3,\zb_3)\bigr>
\sim {C[\theta_i]\over
\bigl|z_{12}^{\Delta_{123}}z_{13}^{\Delta_{132}}
z_{23}^{\Delta_{231}}\bigr|^2}\ ,
}
where $\Delta_{123}=\Delta_1+\Delta_2-\Delta_3$, etc. The coefficient function
$C$ is generically nonzero. Sadly, this example cannot be extended to higher
point functions because the classical solutions to Liouville theory are not
known in explicit form, except in special cases which have the punctures
symmetrically located \hempel.

\smallskip\noindent
{\bf Example 2}: Consider now the three-point function on
the sphere, but with $s\geq 0$ so we must fix the area.
Using the M\"obius invariance of \avsl\ gives
\eqn\thrptiii{\eqalign{&\bigl< \ee{\alpha_1\phi}(z_1,\zb_1)\,
\ee{\alpha_2\phi}(z_2,\zb_2)\,
\ee{\alpha_3\phi}(z_3,\zb_3)\bigr>_{A=1}^{\rm s.c.} \sim {C[\alpha_i]\over
|z_{12}^{\Delta_{123}}z_{13}^{\Delta_{132}}z_{23}^{\Delta_{231}}|^2}\cr
C[\alpha_i]&=\int_0^\infty {\d\lambda\over\lambda}\,
\lambda^{4(\alpha_1+\alpha_2-\alpha_3)/\gamma}
\int_{\IC} \d^2 w\, {1\over \bigl(|w|^2+1\bigr)^{2 \alpha_1/\gamma}}
{1\over \bigl(|w+\lambda^2|^2+1\bigr)^{2 \alpha_2/\gamma}}\cr
&={\pi\over 4}{\Gamma(j_1-j_2-j_3)\,
\Gamma(j_2-j_3-j_1)\,\Gamma(j_3-j_1-j_2)\over
\Gamma(-2 j_1)\,\Gamma(-2 j_2)\,\Gamma(-2 j_3)}\,\Gamma(-j_1-j_2-j_3-1)\ ,\cr}
}
where $j_k\equiv -\alpha_k/\gamma$.
Strictly speaking, the integrals above converge only for
ranges of $j_k$ for which the arguments of all the $\Gamma$--functions in
\thrptiii\ are positive. As in ordinary string theory,
we define the amplitude at other values of $j$ by analytic continuation.

\noindent {\bf Remarks}:
\item{1)} The formula \avsl\ has an interesting
group-theoretical meaning. Defining $j\equiv -\alpha/\gamma$,
the field $\ee{\alpha \phi}$ transforms under $SL(2,\IC)$ in the
$(j,j)$ representation of $SL(2,\IC)$, hence the suggestive notation.
Group theoretically, the integral
\thrptii\ computes the overlap between a product of vectors in
infinite-dimensional highest weight representations and the
trivial representation.
\item{2)} Analytically continuing $j_3$ to pure imaginary values, and
taking the limit $j_3\to 0$, gives the two-point function
\eqn\twoptfa{
\langle \ee{-j_1 \phi}(z_1,\zb_1)\, \ee{-j_2 \phi}(z_2,\zb_2)
\rangle =-{i \pi^2\over 4}{1\over 2j_1+1}\, \delta(j_1-j_2)
\,{1\over|z_{12}|^{4\Delta_1}}\ ,}
obtained in \nati\ by analytic continuation of
the integration over the dilation subgroup $\IR^*_+\subset SL(2,\IC)$.
For a further discussion of the subtleties of one- and
two-point functions, and their relations to the regularized
volumes of $\IR^*_+$ and $SL(2,\IC)$, see \nati.\foot{That Liouville
two-point functions are
{\it diagonal\/} in Liouville charges might have important
implications for the implementation of string field theory identities \bz.}
\item{3)} Note that fields with $j\in\half \IZ_+$ generically
decouple. If all three fields satisfy $j\in\half \IZ_+$,
then we recover the $SU(2)$ fusion rules.
\item{4)} The two above examples make it clear that the Liouville
correlators are entirely different from the correlators of
a Coulomb gas with a charge at infinity. In particular, there
is no Liouville charge conservation.
\item{5)} It is very unusual to have short distance
singularities in the correlators of the {\it classical\/}
theory as we do in \thrptiii\ and \twoptfa. This is
due to the average over the noncompact group
$SL(2,\IC)$, so we see again that the geometrical origin of the
Liouville field distinguishes it from an ordinary scalar field.
Note that for the $n\geq 4$ -point functions, the operator
products are typically smooth in the semiclassical correlator.

\subsec{Operator Products in Liouville Theory}
\subseclab\ssoplt

The existence of the Seiberg bound \seibound\ dictates that the operator
product expansion in Liouville theory will be rather different from that in
free field theory. Indeed it is not clear that the notion of the operator
product expansion is the correct language to use when discussing Liouville
correlators. Already in the semiclassical theory, we shall see that the
putative short distance behavior of correlators depends on global properties of
the surface and the ``operator product expansion'' appears to be nonlocal
\refs{\nati,\joetalk}.

\ifig\ftwoltp{Left: Insertion of operators $\ee{\alpha \phi}$ and
$\ee{\beta \phi}$ in a surface with boundary. Right: the same operation on a
higher genus surface.}
{\epsfxsize3.5in\epsfbox{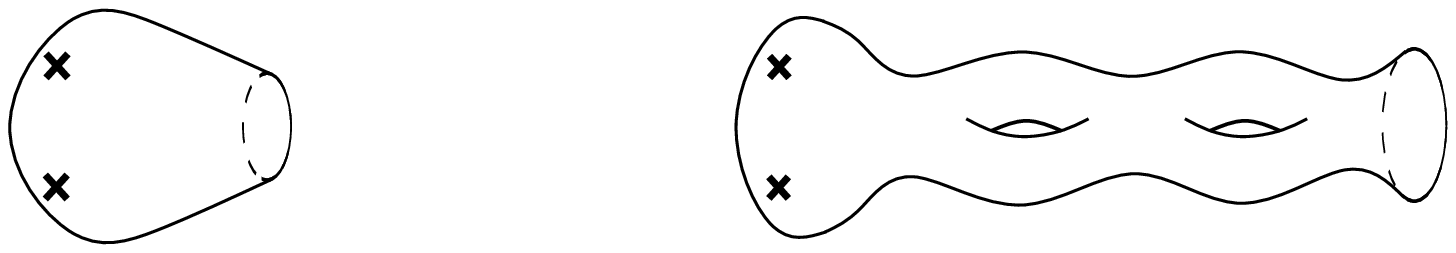}}

Consider the state arising from the insertion of two vertex operators
$\ee{\alpha \phi}$ and $\ee{\beta \phi}$ on a surface with
boundary $\CC$ as in the l.h.s.\ diagram of \ftwoltp, or, more generally, on a
higher genus surface as in the r.h.s.\ diagram of \ftwoltp.
In the semiclassical approximation, the state created by the surface
on the boundary $\CC$ has a wavefunction that depends
on the zero-mode $\phi\dup_0$ of $\phi$ as
\eqn\smiwvfn{\psi(\phi\dup_0)\sim \ee{\alpha \phi\dup_0
+ \beta \phi\dup_0 -\half Q \chi \phi\dup_0}\ ,}
where $\chi=1-2h$ is the Euler character for the surface
(with a single boundary).
If $\alpha+\beta<\half Q \chi$, then the wavefunction
\smiwvfn\ is a real exponential diverging at $\phi\dup_0\to -\infty$
(short distance).  Then we may expect to replace the holes in
the diagrams in \ftwoltp\
each by a sum of local (``microscopic'') operators, as in ordinary
conformal field theory. Note that since the three-point
functions are generically nonzero, we would naively
expect a disastrous sum over operators with conformal
dimensions unbounded from below.

If $\alpha+\beta>\half Q \chi$, on the other hand, then the state is
normalizable and we certainly cannot expand it in an operator product
expansion of local operators. Instead the state must be expanded
in the normalizable macroscopic state operators. Thus, by
sewing, the surface amplitude must have the form
\eqn\softope{
\bigl< \ee{\alpha \phi(z,\zb)}\,\ee{\beta \phi(z,\zb)}\cdots\bigr>
\sim
\int_0^\infty \d E\, c\dup_{\alpha \beta E}\,
|z-w|^{2(E^2/2 + {Q^2/8}-\Delta_\alpha - \Delta_\beta)}
\,\langle E|\Sigma\rangle\ ,
}
where $|\Sigma\rangle$ is a state created by
the rest of the surface (as is standard in discussions of
the ``operator formalism'' \refs{\rBPZ\FMS{--}\opform}).
If we interpret the integral in \softope\ as an
OPE over macroscopic vertex operators $V_E$, then since we sum
over operators with weights $\Delta\geq Q^2/8$, we see that
the OPE is much {\it softer\/} than in ordinary CFT.
This discussion can be generalized.

The essential message is that while we may insert microscopic operators
on a surface we should only do so ``externally.'' We must
factorize on
macroscopic states. The factorization on macroscopic states also
ameliorates the disastrous sum noted above for
$\alpha+\beta<\half Q\chi$.

The essentially non-free field nature of the operator product
expansion in Liouville theory accounts for some unusual
properties of the theory. As one example, note that since
the Liouville theory is conformal for all $\mu>0$,
the cosmological constant $\ee{\gamma \phi}$ is
an exactly marginal operator. This appears to conflict with
the fact that its $n$-point correlation functions are nonvanishing,
since the standard obstruction to exact marginality is
the existence of a ``potential'' for such couplings.
However,
the standard discussion of the obstruction to exact marginality
does not apply because of the strange nature of the
operator product expansion. In later sections on
string theory (sec.~{\it\sbsap}\/) we will see that the unusual OPE of
Liouville also has important consequences for the finiteness
of the theory and for the existence of an infinite
dimensional space of background deformations.

\subsec{Liouville Correlators from Analytic Continuation}
\subseclab\slcfac

In the past two years there has been very interesting progress in
understanding Liouville correlation functions via
``analytic continuation in the number of operators.''
The first step in the calculation of continuum correlators was provided in
\rGTW, where the free field formulation by zero mode integration of the
Liouville field was established.
The essential idea is to treat the Liouville path integral
measure as a free field measure and separate out a zero-mode
$\phi\dup_0$ via $\phi=\phi\dup_0+\hat \phi$, so that
$[\d\phi]=[\d\hat\phi]\,\d\phi\dup_0$. The integral over the
zero mode is
\eqn\zminti{\eqalign{
&\int_{-\infty}^\infty \d\phi\dup_0\ \ee{\sum \alpha_i \phi\dup_0}
\, \ee{-Q \chi \phi\dup_0/2 - B \ee{\gamma \phi\dup_0}}
={1\over \gamma}\, \Gamma(-s)\, B^s\cr
&s\equiv {1\over \gamma}\Bigl(\half Q \chi-\sum \alpha_i\Bigr)\qquad
B\equiv {\mu\over 8 \pi \phi^2}\int\sqrt{\hat g}\,\ee{\gamma \hat \phi}\ .\cr}}
In references \refs{\rGLi\rfftech\kdf{--}\polss},
it is proposed that when $s\in
\IZ_+$ (so there is no negative curvature solution) the $\hat \phi$ integral
can be done using free field techniques. One then obtains a class of amplitudes
as ``functions of $s$,'' manipulates the $s$-dependence to reside solely in the
arguments of $\Gamma$-functions (through factorials) and then ``analytically
continues'' to all values of $s$ using $\Gamma(x+1)$ as an analytic
continuation of $x!$ .

This curious procedure has scored many impressive successes.
In particular \rGLi, the incorporation of the Liouville mode was
shown to cancel the ghastly assemblage of $\Gamma$-functions familiar from the
conformal field theory result and reproduce the relatively simple matrix model
result for many continuum correlation functions.
Additional genus zero correlation functions for $D\le1$ were computed in \kdf.
The genus one partition function for the AD series was
calculated via KdV methods in \rDIFK, and was confirmed from the continuum
Liouville approach in \rBerKl.

Attempts to justify the technique on physical grounds are based on arguments
that for $s\in \IZ_+$, the coefficient of $\log \mu$ in the correlation
function is dominated by the regions where $\phi\to-\infty$ (short distance).
In these regions the Liouville interaction is small and the theory can be
treated as free field theory. See \kdf\ for more detailed discussion. Another
justification, using the quantum group approach to Liouville theory, has been
proposed in \gervcorr.

Nevertheless, these results remain to be better understood.
The proper and complete calculation of correlation functions
in Liouville theory remains the most important open problem in the subject.

\subsec{Quantum Uniformization} 
\subseclab\slqu

The most ambitious approach to the evaluation of Liouville
correlators proceeds by attempting to generalize the original
uniformization program of Klein and \poin\ (described in
sec.~{\it\slcu\/}) to quantum field
theory. This program was one of the central motivations that
led Belavin, Polyakov, and Zamolodchikov to the study of the
minimal models of conformal field
theory \rBPZ. This program has also been pursued in a
series of papers by Gervais and collaborators \refs{\gervais,\gervcorr}\
using the operator formalism. In this section we shall try to clarify
the relation of the original uniformization program to the quantum Liouville
theory, and in particular elucidate
the role played by what we now interpret as quantum Liouville correlators.

We begin by recalling the classical theory of \poin\ \fuchsprs.\foot{We do
not adhere here to the historical development.}
We restrict attention
to the $n$-punctured sphere $X=\hat{\IC}-\{z_1,\dots z_n\}$.
Let us try to solve the classical Liouville equation with
sources, \lqsrc. We set $\theta_i\equiv \gamma\alpha_i$ and
consider the case with $s \gamma^2 =2-\sum \theta_i<0$. The
metric $\ee{\gamma \phi}|\d z|^2$ on $X$ will be obtained as a
pullback of the Poincar\'e metric on the unit disk $D$,
\eqn\dskmet{\d s^2={16\over\mu}\,{|\d w|^2\over (1-|w|^2)^2}\ ,}
via a uniformization map $w=f(z)$, where $f:X\to D$.
(The metric \dskmet\ is related to the metric \pomet\ on the
upper half plane via the Cayley map $w=(z-i)/(z+i)$ from the upper half plane
to the disk.)

The main observation is that $f(z)$ can be obtained as a
ratio of solutions of a {\it linear\/} differential
equation of Fuchsian type (i.e.\ having only regular singular points).
 To see this, recall the result \enrgmo,
\eqn\teeschw{T(z)(\d z)^2 ={1\over \gamma^2}S\bigl[f(z);z\bigr]\,(\d z)^2\ ,}
where $S$ is the Schwartzian derivative.
$T(z)$ is analytic and has second order poles at the sources of
curvature so we may write a partial fraction decomposition,
\eqn\quadiff{\omega\dup_X\equiv{\gamma^2\over 2}T(z)
=\sum_i\Bigl( {h_i\over (z-z_i)^2}+ {c_i\over z-z_i}\Bigr)\ ,}
where
\eqn\thwts{h_i={1\over 4}\bigl(1-(1-\theta_i)^2\bigr)\ ,}
and the $c_i$ are constants known as accessory parameters.

In order to find the map $f$, one might first turn
to solve the nonlinear differential equation
\eqn\omschw{S[f;z]=2\, \omega\dup_X(z)\ .}
This problem may be linearized by considering the Fuchsian
differential equation
\eqn\fuchs{{\d^2 y\over \d z^2}+\omega\dup_X\,y=0}
since, if $y\dup_1$, $y\dup_2$ are any two linearly independent solutions
of \fuchs\ then $f(z)=y\dup_1/y\dup_2$ satisfies \omschw.

\exercise{}

Check \omschw\ for $f=y\dup_1/y\dup_2$. Note that the transformation properties
of $S[f;z]$ under the M\"obius group insure that we can take {\it any\/} two
solutions $y\dup_1$, $y\dup_2$.

\endexercise

The differential equation \fuchs\ has regular singular points
at $z=z_i$, so if $z$ is continued around $z_i$ the solutions
$y\dup_1$, $y\dup_2$ will have monodromy
\eqn\monod{\eqalign{
y\dup_1&\to M_{11}\, y\dup_1 + M_{12}\, y\dup_2\cr
y\dup_2&\to M_{21}\, y\dup_1 + M_{22}\, y\dup_2\ ,\cr}
}
inducing a M\"obius transformation on $f$.
Thus, if $X=D/\Gamma$ where $D$ is the \poin\ disk and $\Gamma$
is a discrete subgroup of $SU(1,1)/\IZ_2\cong PSL(2,\IR)$, then
there exist $y\dup_1,y\dup_2$ such that $f:X\to D$ is a multivalued mapping,
inverting locally the projection $D\to X$. The different values of
$f(z)$ are obtained by applying M\"obius transformations in $\Gamma$.
Thus the $n$-punctured sphere $X$ and its accompanying uniformization map $f$
have led to a Fuchsian differential equation \fuchs\ with the property that
the monodromy around regular singular points forms a {\it discrete\/} subgroup
$\Gamma\subset SU(1,1)/\IZ_2$.

Klein and \poin\ tried to show the converse of the above chain of logic.
Namely, {\it if\/} the parameters $c_i$ in \quadiff\
are appropriately chosen, then the monodromy group of the
differential equation \fuchs\ would be a discrete subgroup
$\Gamma$ of $SU(1,1)$, and $f=y\dup_1/y\dup_2$ could be normalized
so that the images of $f(X)$ under $\Gamma$ tesselate $D$.
(In general the $c_i$ have to be highly non-trivial functions of
the $z_i$ and $h_i$ to result in discrete monodromy and hence in
reasonable surfaces $X=D/\Gamma$.) By suitable choice of the $c_i$, in
principle any surface $X$ could be obtained.
This original approach to uniformization foundered on the
inability to calculate, or even show existence of, appropriate
parameters $c_i$. As we shall see, Klein and \poin\ got
stuck on the problem of computing Liouville correlators.

In the cases where $f$ is a uniformizing map, we may obtain a
solution to the Liouville equation simply by pulling
back the \poin\ metric,
\eqn\pullbacki{
\ee{\gamma \phi}|\d z|^2=f^*\Bigl({16\over\mu}{|\d w|^2\over (1-|w|^2)^2}\Bigr)
={16\over\mu}{|C|^2\,|\d z|^2\over \bigl(|y\dup_2|^2-|y\dup_1|^2\bigr)^2}\ ,
}
which yields
\eqn\pullbackii{\ee{-\ha \gamma \phi}={\sqrt\mu\over 4|C|}
\bigl(|y\dup_2|^2-|y\dup_1|^2\bigr)\ ,}
where the constant $C$ is the Wronskian of $y\dup_1$, $y\dup_2$. Note that
although $y\dup_1,y\dup_2$ have monodromy \monod, $\ee{-\ha\gamma\phi}$
is single-valued. Finally, we combine
\eqns{\teeschw{,\ }\quadiff{,\ }\fuchs{,\ }\pullbackii}\
to obtain the important result
\eqn\nulli{\p_z^2 \ee{-\ha\gamma \phi}+{\gamma^2\over 2}
T(z)\,\ee{-\ha\gamma\phi}=0\ .}

\exercise{}

Show that \nulli\ is an identity by working out the
second derivative of the exponential and using the
formula for $T$ in terms of $\phi$.

\endexercise

\noindent{\bf Example}: The triangle functions.\nl
The uniformization of the three-punctured sphere is explicitly known \nehari.
In this case, \fuchs\ has three regular singular points and can
therefore be transformed to the Gauss hypergeometric equation. The mapping is
given by
\eqna\ratofeff
$$f(z)=N\, {\CF_2(x) \over\CF_1(x)}\ ,\eqno\ratofeff a$$
where
$$\eqalign{\CF_2(x)&=x^{1-\t_1} {}_{2}F_{1}\bigl(2-\half(\t_1+\t_2+\t_3),\
    1+\half(-\t_1+\t_2-\t_3);\ 2-\t_1;\ x\bigr)\cr
\CF_1(x)&= {}_{2}F_{1}\bigl(1+\half(\t_1-\t_2-\t_3),\
  \half(\t_1+\t_2-\t_3);\ \t_1;\ x\bigr)\cr
N^2&=(1-\t_1)^2 \bigl(\Delta(\t_1-1)\bigr)^2\,
{\Delta\bigl(2-\half(\t_1+\t_2+\t_3)\bigr)
  \,\Delta\bigl(\half(-\t_1+\t_2+\t_3)\bigr)\over
  \Delta\bigl(\half(\t_1+\t_2-\t_3)\bigr)\,
  \Delta\bigl(\half(\t_1-\t_2+\t_3)\bigr)}\cr
x&={z-z_1\over z - z_2}\,{z_{32}\over z_{31}}\qquad\qquad
\Delta(y)={\Gamma(y)\over \Gamma(1-y)}\ .\cr}\eqno\ratofeff b$$

\ifig\ftessel{A tesselation of the Poincar\'e disk:
A copy of an adjacent white and black triangle maps to the thrice-punctured
sphere. The images of the triangles under the monodromy group of the associated
Fuchsian differential equation tesselate the Poincar\'e disk.}
{\epsfxsize 2.5in\epsfbox{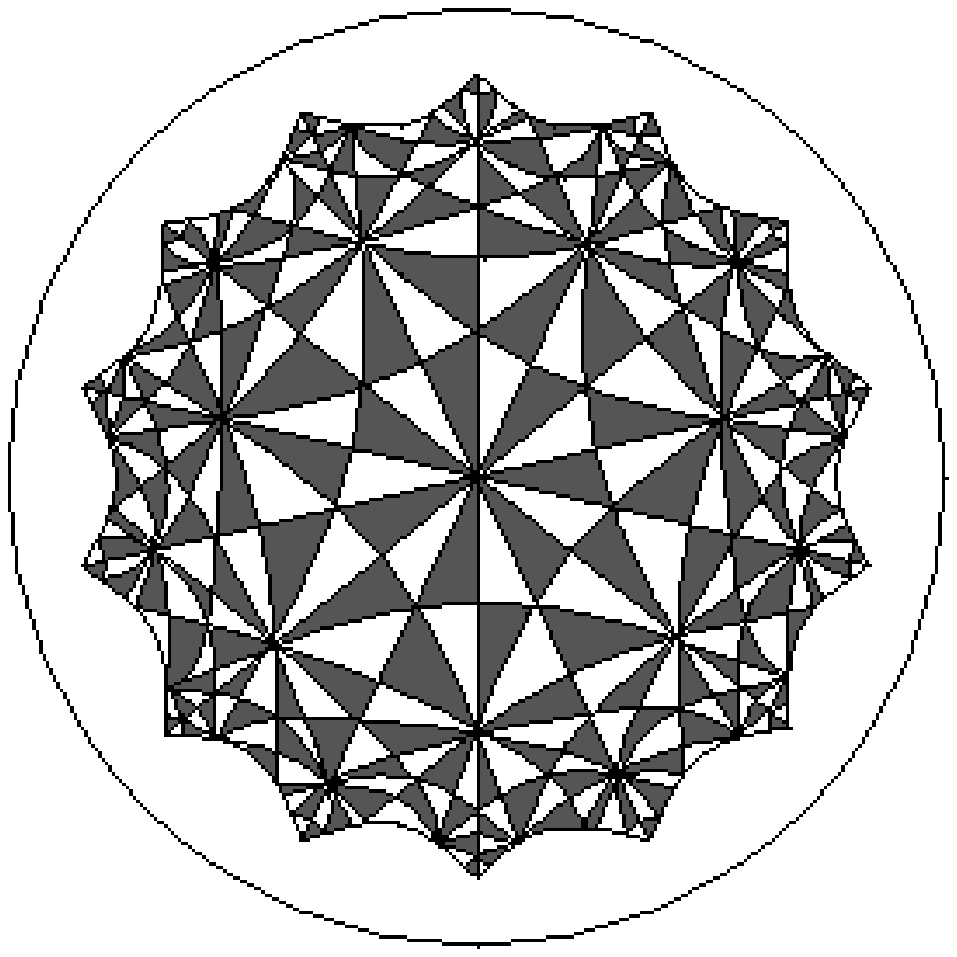}}

The mapping \ratofeff{}\ carries a circle through the points $z_1,z_2,z_3$ to a
curvilinear triangle in $D$ with opening angles $\pi(1-\t_i)$. Note the
geometrical conditions
\eqn\seibd{\t_i\leq 1\ ,
}
since the opening angles are $\geq 0$
(recall eqs.~\eqns{\micmeti{--}\monaa}), and
\eqn\xagain{\t_1+\t_2+\t_3-2\geq 0}
since a hyperbolic triangle must have its sum of interior angles
less than or equal to $\pi$.
If the $\t_i$ are reciprocals of integers, then the triangles
tesselate the disk $D$ as in \ftessel.

Finally we write the classical answer for the
monodromy-invariant solution to \nulli.
Combining \pullbackii\ and \ratofeff{}\ we obtain
\eqn\clsscori{\eqalign{\ee{-\ha\gamma \phi}
&={\sqrt\mu\over2}{1\over (1-\t_1)|N|}
\biggl|{z-z_2\over z - z_3}{z_{31}\over z_{21}}\biggr|^{\t_3}
\,\biggl|{z-z_1\over z - z_2}{z_{32}\over z_{31}}\biggr|^{\t_1}
\,\biggl|{(z-z_3)^2 z_{21}\over z_{31}z_{23}}\biggr|\cr
&\qquad\qquad\cdot\Bigl( |\CF_1|^2 - N^2 | \CF_2|^2\Bigr)\ .\cr}}
Using the transformation properties of hypergeometric
functions and identities on $\Gamma$--functions,
it can be shown that \clsscori\ is fully symmetric in
$(z_1,\theta_1)$, $(z_2,\theta_2)$, and $(z_3,\theta_3)$.

We now interpret the above equations in terms of conformal field theory.
The vertex operator
$\Psi=\ee{-\ha\gamma \phi}$ has conformal weight
$\Delta=-\half-{3\over 8} \gamma^2$. The
central charge is $c=1+3 Q^2$ and therefore we have
\eqn\kacvl{
\Delta={1\over 16}\Bigl(c-5+\sqrt{(c-1)(c-25)}\Bigr)\ .}
It immediately follows, as discussed in \rBPZ\ (see also \rcftrev), that
$\bigl(L_{-1}^2-{2(2\Delta+1)\over 3} L_{-2}\bigr)|\Delta\rangle$
is a singular vector in the Verma module built on $|\Delta\rangle$,
and therefore
\eqn\nullfld{\p_z^2 \ee{-\ha \gamma \phi}+{\gamma^2\over 2}\,
:T(z)\,\ee{-\ha\gamma\phi}:}
is a null field (where we use conformal normal-ordering in the
second term). Now, {\it if\/} the null field decouples in
correlation functions,\foot{The Liouville theory is sufficiently subtle that
this is an open question.} we may put
\eqn\nullii{\Bigl<\bigl(\p_z^2 \ee{-\ha \gamma \phi}+{\gamma^2\over 2}
T(z) \ee{-\ha \gamma \phi}\bigr)
\prod_i \ee{{\t_i}\phi/\gamma}(z_i,\zb_i )\Bigr> =0\ .}

In view of these observations, the classical uniformization
theory takes on new meaning: the classical solution in
the presence of sources $\ee{-\ha\gamma \phi\dup_{\rm cl}}$
corresponds to the semiclassical correlator
$\langle \ee{-\ha \gamma \phi}
\prod \ee{{\t_i\over \gamma}\phi}\rangle_{\rm s.c.}$. The classical
equation \nulli\ is the null-vector decoupling equation,
while \pullbackii\ becomes the decomposition of the correlation
function into ``conformal blocks'' $y\dup_1,y\dup_2$. These blocks
are assembled into monodromy--invariant combinations. The geometrical
conditions \seibd\ and \xagain\ become respectively the Seiberg bound and
the condition for the existence of a classical solution.
Finally, the partial fraction decomposition \quadiff\ is the familiar
Ward identity for the insertion of an energy-momentum tensor
in a correlator of primary fields:
\eqn\accessi{
\omega\dup_X=\half \gamma^2\biggl\langle T_{zz} \prod
\ee{\t_i/\gamma \phi(z_i)}\biggr\rangle_{\rm s.c.}\ .
}
In particular, the accessory
parameters $c_i$ are given by
\eqn\semiclass{
c_i=\half \gamma^2 {\p\over \p z_i} \log \biggl\langle \prod
\ee{\t_i/\gamma \phi(z_i)}\biggr\rangle_{\rm s.c.}
}
When combined with \leadact,
this last formula for the accessory parameters makes sense
independently of the existence of a quantum Liouville
theory and has been rigorously proven recently by Takhtadjan
and Zograf \takht.

\noindent

\danger{Some four-point functions.}
Let us assume that the null-field decouples as in \nullii. Then,
it follows directly from the $SL(2,\IC)$ Ward identities that
\nullii\ reduces to an ODE related to Riemann's differential equation (as in
\rBPZ). A straightforward calculation shows that
\eqn\classcorii{\eqalign{
\Bigl< \ee{-\ha \gamma \phi}(z,\zb) \prod_{i=1}^3
\,\ee{{\t_i\over \gamma}\phi}(z_i,\zb_i)\Bigr> &=
\bigl({\mu\over4}\bigr)^{\ha-\sum \theta_i/\gamma^2}
{1\over (1-\hath_1)|\hat N|}
|z_{12}^{\Delta_{123}}z_{13}^{\Delta_{132}}z_{23}^{\Delta_{231}}|^{-2}\cr
\cdot\Bigl|{z-z_2\over z - z_3}{z_{31}\over z_{21}}\Bigr|^{\hath_3}
\Bigl|{z-z_1\over z - z_2}{z_{32}\over z_{31}}\Bigr|^{\hath_1}
&\Bigl|{(z-z_3)^2 z_{21}\over z_{31}z_{23}}\Bigr|
\Bigl|{(z-z_1)^2 (z-z_2)^2(z-z_3)^2 \over z_{21}
z_{31}z_{23}}\Bigr|^{\gamma^2/4}\cr
&\cdot\Bigl( |\hat\CF_1(x)|^2- \hat{N}^2  |\hat\CF_2(x)|^2\Bigr)\cr
 \hat\CF_1(x)= {}_{2}F_{1}\bigl(1+ & \half(\hath_1-\hath_2-\hath_3),
\half(\hath_1+\hath_2-\hath_3);\hath_1;x)\cr
\hat\CF_2(x)=x^{1-\hath_1}\,
 {}_2 F_1\bigl(2-\half(\hath_1+ & \hath_2+\hath_3),
 1+\half(-\hath_1+\hath_2-\hath_3);2-\hath_1;x)\ ,\cr}
 }
where the quantum and classical expressions are related by the
simple shift $\hath=\t-\half \gamma^2$.
Of course, conformal invariance only determines the
correlator up to an overall function $n(\t_1,\t_2,\t_3)$ which
is totally symmetric in the $\t_i$. The prefactor
$\bigl((1-\hath_1)|\hat N|\bigr)\inv$ in
\classcorii\ is obtained by comparing with
the semiclassical answer \clsscori, where the overall normalization
{\it is\/} determined. Since the rest of the terms in the expression
satisfy the substitution rule $\t\to\hath$ relating classical and
quantum expressions, it is a fair guess that the prefactor
$\bigl((1-\hath_1)|\hat N|\bigr)\inv$ in \classcorii\ is exact.

The fully quantum correlator \classcorii\ is a new result. As opposed to the
matrix model results we will describe in later chapters, \classcorii\
gives the Liouville correlator as a function of the moduli of the
4-punctured sphere --- if properly understood, \classcorii\ could be integrated
over the positions of the punctures to derive the
(already automatically integrated)
matrix model results for pure gravity.
The correlator \classcorii\ has many strange properties possibly
illustrating the strange nature of the OPE in Liouville theory.
Of particular note is the case where some of the operators saturate the Seiberg
bound $\alpha=Q/2$, which, classically, corresponds to sources
producing triangles with corner angle $=0$. For example,
if all three $\t_i/\gamma = Q/2$ then the prefactor in
\classcorii\ develops a pole and the difference
of hypergeometric functions vanishes. A short calculation
shows that the limit $\alpha_i\to Q/2$ is
smooth and
\eqn\spccorr{\eqalign{
\Bigl< \ee{-\ha \gamma \phi} \prod_{i=1}^3\,\ee{\ha Q\phi}(z_i,\zb_i)\Bigr>&=
\bigl({\mu\over4}\bigr)^{\ha-3Q/2\gamma}
|z_{12}^{\Delta_{123}}z_{13}^{\Delta_{132}}z_{23}^{\Delta_{231}}|^{-2}\cr
\cdot\biggl|{(z-z_1)(z-z_3)\over z_1 -
z_3}\biggr|&\biggl|{(z-z_1)(z-z_2)^{1/2}(z_{23})^{1/4} \over
z_{31}^{1/4}z_{21}^{3/4}}\biggr|^{\gamma^2}\cr
&\cdot\pi\Bigl( F(1-x)\bar{F}(x)+ F(x)\bar{F}(1-x)\Bigr)\ ,\cr}}
where $F(x)=F(\ha,\ha;1,x)$ is an elliptic integral of
the first kind. In particular, $F$ has {\it logarithmic}
singularities $F(x)\sim {1\over \pi}\log({1\over 1-x})$
as $x\to 1^-$, resulting in logarithmic
short--distance singularities in the correlator \spccorr.\foot{It is sometimes
suggested in the literature that
the subleading logarithms indicate that the correct vertex operator is
$\phi\, \ee{(Q/2) \phi}={\p\over \p \alpha}\,\ee{\alpha \phi}|_{\alpha=Q/2}$,
with the derivative corresponding to the limiting procedure needed above.}

\noindent{\bf Remarks}:\nl
\item{1)} The formula \classcorii\ probably only applies when
$s\leq 0$, otherwise there are paradoxes.
\item{2)} The operator $\ee{-\ha \gamma \phi}$ is by no means
the only null vector in the Liouville theory. Using the
Kac determinant formula, one may ask for the set of all
operators $\ee{\alpha \phi}=\ee{-j \gamma \phi}$ which
weights $\Delta=\Delta_{p,q}(c)$ where $p,q\geq 1$
are integer. The result is
$$j_{p,q}=\half (p-1) + {1\over \gamma^2}(q-1)\ .$$
In principle this allows one to extend the above
example to an infinite set of correlators.

\subsec{Surfaces with boundaries}
\subseclab\sswb

The final method for extracting Liouville correlators,
and the one which is most closely connected to
matrix model methods, is the computation of macroscopic
loop amplitudes. These are amplitudes in Liouville theory
for manifolds with boundary, for which the Liouville action picks up
the extra boundary contribution
\eqn\liouvbdry{
S\to S_{\rm Bulk}+{Q\over 8\pi}\oint_{\p\Sigma}\d\hat s\,\phi\,\hat k  +
{\rho\over 4\pi\gamma^2}\oint_{\p\Sigma}\d\hat s\, \ee{\ha\gamma\varphi}\ ,}
where $\hat k$ is the extrinsic curvature of the boundary,
$\d\hat s$ is the reference line element, and $\rho$ is the
boundary cosmological constant.

We have a well-defined variational principle
if we choose
Dirichlet boundary conditions ($\delta\varphi|_{\p\Sigma}=0$), or Neumann
boundary conditions:
\eqn\neu{
{\p(\gamma\varphi)\over\p n}+\hat k+{\rho\over 2}
 \ee{\ha\gamma\varphi}=0\ ,
}
where the first term is the normal derivative.

Just as we can introduce amplitudes at fixed area using the
operator \area, when using Neumann boundary conditions
we can introduce amplitudes at fixed length by introducing
the length operator of a boundary loop $\CC$, given by
\eqn\lenth{\ell=\oint_{\CC} \d\hat s\, \ee{\ha \gamma \phi} \ .}
While this is obvious in the classical theory,  surprisingly it
continues to hold exactly in the quantum theory \mms.

\exercise{Boundary operators}

a) Assuming $\phi$ has free field Neumann short distance
singularities near the boundary,
$$\bigl< \phi(z)\,\phi(w)\bigr> \sim -\log|z-w|^2-\log|\bar z-\bar{w}|^2$$
(where we think of the boundary as the $x$--axis for the upper
half plane), show that the vertex operator $\ee{\alpha \phi}$,
when inserted on the boundary has boundary conformal
weight $\Delta_b=-2 \alpha^2+Q \alpha$ and thus \lenth\ is
well-defined. A discussion of boundary operators in conformal
field theory may be found in \cardy.

b) Show that the argument analogous to \ngltrm\ suggests the
bound
\eqn\bdryseib{\alpha \le {Q\over 4} }
for boundary operators.\foot{In the dense phase of the $O(n)$ model
coupled to gravity,
Kostov and Staudacher \kostaudon\ have given examples of loop operator
exponents which appear to give counterexamples to the bound \bdryseib.}

\endexercise

{}From our experience with conformal field theories in
chapt.~\slascft, we may expect that if we insert a ``macroscopic
loop operator''
\eqn\liuloop{W_\CC(\ell)
= \delta\Bigl(\ell-\oint_{\CC}\d\hat s\, \ee{\ha \gamma \phi} \Bigr)}
in the Liouville path integral and then shrink the
circumference to zero, then $W_\CC(\ell)$ may be replaced by an infinite sum
\eqn\liuloco{
W(\ell)\sim\sum_{j\geq 0} \ \ell^{\,x_j}\, \sigma_j\ ,}
where $\sigma_j$ are local operators which can couple to the boundary
state created by \liuloop.

\exercise{Exponents in Loop Expansion}

Suppose that there is an expansion like \liuloco\
in which the operators $\sigma_j$ have Liouville charge
$\alpha_j$, i.e.\
$$ \sigma_j\sim \CP(\p^* \phi,\pb^* \phi)\, \ee{\alpha_j \phi}\ ,$$
where $\CP$ is a polynomial. Show that the
exponents $x_j$ of \liuloco\ can then be found by a variant of the
simple scaling argument we have used in \fixai\ and earlier.
Consider a Liouville path integral with the operator $W(\ell)$ inserted, and
shift $\phi$, remembering to take into
account the change in Euler character from shrinking the hole.
Show that the path integral scales as
$$ \ee{-\ha Q\,\delta \phi
+ \alpha_j\, \delta\phi + \ha \gamma\, x_j\, \delta \phi}\ ,$$
from which follows
\eqn\exfrsj{x_j=Q/\gamma- 2 \alpha_j/\gamma
={2\over \gamma}(\half Q-\alpha_j)\ .}
Note that $x_j\geq 0$.

\endexercise

It turns out that, because of the geometrical nature of
Liouville theory, the expansion \liuloco\ is only
valid under certain circumstances. This may be seen by a semiclassical
study of amplitudes with loops \mss, analogous to the
semiclassical considerations above.
The main results of this study are the following:
\item{1)} Let $-s \gamma=\sum \alpha_i-\half Q \chi$, where $\chi=2-2h-B$ on a
surface with $h$ handles and $B$ boundaries. If $-s \gamma>0$, the $\ell\to 0$
behavior of $W(\ell)$ is equivalent to a sum of local operators. In particular,
this is always the case if there are two or more loops on the surface
(including the one that shrinks).
\item{2)} As noted in the above exercise, $x_j\geq 0$ for local
operators. Coefficients of negative powers of $\ell$ as
$\ell\to 0$ arise from small area divergences and are analytic
in $\mu$ (and other coupling constants, in the context of
2D gravity).
Therefore they are interpreted as arising from infinitesimal size surfaces, and
such terms are classified as non-universal contributions when comparing with
matrix model answers.

\newsec{2D Euclidean Quantum Gravity II: Canonical Approach} 
\seclab\stdeqg

It will be useful to work with the canonical approach to two dimensional
gravity. In this chapter, we are led to introduce in particular some
of the details of the algebraic (BRST) point of view (to be pursued
further in secs.~{\it\ssastdcc, \ssascsc\/}).  We hope that providing
a common language will help bridge the schism between the algebraicists
and the matrix model theorists, who are after all employing
two complementary approaches to study the same subject.

\subsec{Canonical Quantization of Gravitational Theories}

For a review of the canonical approach to Einsteinian general
relativity, see \refs{\MTW,\Wald}. Since gravitational theories are gauge
theories, we are immediately led to study constrained dynamics.

The canonical approach applies to spacetimes $M$ which admit a space-time
foliation: we assume there is a diffeomorphism $\Sigma\times \IR\to M$, where
$\Sigma$ is a $D$-dimensional spacelike manifold. Choosing a unit normal
$n^\mu$ to the surface $\Sigma$, we may project the metric onto components
parallel and perpendicular to the surface. The restriction $^{(D)}g$ of the
metric $g$ to the spatial surface defines the canonical coordinates, while the
time--time and time--space components of the metric are expressed in terms of
Lagrange multipliers for constraints of the theory (the ``lapse'' and
``shift''). In Einsteinian general relativity, the constraints associated with
the lapse and shift are the time--time and time--space components of the
Einstein equations: $G_{00}=0, G_{0i}=0$. In the canonical theory, these are
the generators of time and space diffeomorphisms.

In the canonical quantization of gravity, wavefunctions are functions of the
spatial metric (and other fields in the theory): $\Psi=\Psi[^{(D)}g,{\rm
matter}]$. The requirement of gauge invariance states that wavefunctions are
required to obey operator versions of the space and time diffeomorphism
constraints. The Wheeler--DeWitt equation is the equation expressing the
invariance under the generator of time-diffeomorphisms
\refs{\wheeler\dewitt{--}\laflamme}\ and plays a fundamental role in the
theory.

In non-Einsteinian metric theories of gravity, including for example
gravity in one and two spacetime dimensions,
a completely analogous formulation may still be obtained
by performing constrained quantization of a theory with
diffeomorphism invariance.

\subsec{Canonical Quantization of 2D Euclidean Quantum Gravity} 
\subseclab\sscqotd

Diffeomorphisms are generated by the energy-momentum tensor $T_{\alpha \beta}$.
In the canonical approach the diffeomorphism constraints of quantum gravity
become the statements that the tensor product theory $Liouville\otimes matter$
is a conformal field theory of central charge $c=0$ (including the ghosts) with
a BRST complex, and moreover the states in the theory lie in the BRST
cohomology of the theory.

If massive matter is coupled to gravity, then the realization of the Virasoro
algebra on the full Hilbert space is far from obvious. In the special case
where the Hilbert space is a tensor product $\CL\otimes \CM$ of Liouville and
matter $\CM$ conformal field theories (e.g., $\CM=M(p,q)$ minimal conformal
field theory, is frequently considered), however, the situation simplifies
dramatically. Naively the wavefunctions are now functions of the spatial
metric, parametrized by $\phi(\sigma)$ and the matter degrees of freedom. When
formulating 2D quantum gravity in the context of conformal field theory,
however, the diffeomorphism constraints are properly enforced through the
calculation of BRST cohomology with respect to the Virasoro algebra. The
condition that nontrivial cohomology exists immediately implies that the total
central charge is zero (see comment after \eQ) so that
\eqn\totalc{
c+1+3 Q^2 - 26 =0 \quad\Longrightarrow\quad Q^2 = {25-c\over 3}\ ,
}
where $c$ is the central charge of $\CM$.

\noindent{\bf Remarks}:
\item{1)} There are three kinds of cohomology problems we can study, depending
on how we treat the zero modes of $b(z), \bar{b}(\zb)$. In ``relative
cohomology'' we require $b_0=\bar{b}_0=0$ on states and gauge parameters. In
``semi-relative cohomology'' we impose the condition $b_0^-=b_0-\bar{b}_0=0$ on
states and gauge parameters. In ``absolute cohomology'' we impose no conditions
pertaining to the $b,\bar{b}$ zero modes.
\item{2)} In 2D gravity (and string theory) there is an important duality on
the cohomology spaces. If $\Phi_a$ forms a basis for the semi-relative
cohomology then there will be a dual basis defined such that the BPZ inner
product $\langle\Phi^b,\,\Phi_a\rangle
\equiv\langle0|\Phi^b(\infty)\Phi_a(0)|0\rangle$ is diagonalized.
If $\Phi_a$ is in the semi-relative
cohomology then $\Phi^a$ will not be in the semi-relative cohomology. One can
define $\tilde \Phi^b\equiv b_0^-\Phi^b$ which will be in the semi-relative
cohomology. This conjugation $\Phi_b\to \tilde \Phi^b$ which exchanges states
of ghost number $G$ and $5-G$ plays a crucial role in string field theory, and
will be important in the considerations of chapt.~\svoccm.
\item{3)} In the literature, not much attention is devoted to defining
precisely the boundary conditions on field space (i.e., spacetime) for the
cohomology problem. However, such boundary conditions are very important
physically, as we shall see.

\subsec{KPZ states in 2D Quantum Gravity}
\subseclab\sskpztd

KPZ states refer to a special class of BRST cohomology
classes associated to the primary fields of the $(p,q)$
minimal models which result when these theories are
coupled to gravity.

For states with a trivial ghost structure the Wheeler--DeWitt
constraint, implementing invariance under time diffeomorphisms, becomes
\eqn\wdwcon{
\bigl(L_0 + \overline L_0 -2\bigr)\Psi=0
}
where $L_n$ are the modes of the total stress energy tensor
for $\CL\otimes \CM$, and the $-2$ is the ghost contribution.
For $\CM=M(p,q)$, the KPZ operators
$\CO=\ee{\alpha \phi}\Phi_0$, where $\Phi_0$ is a primary
in $M(p,q)$ (as described in sec.~{\it\spia\/}), the wavefunction
$\Psi_\CO$ further factorizes
\eqn\facwvfn{\Psi_\CO=\Psi_\CO^{\rm matter}\otimes \Psi_\CO^{\rm gravity}\ ,}
and is separately an eigenstate of $(L_0+\overline L_0)_{\rm matter}$.
In this case the WdW equation becomes
\eqn\wdwi{
\Bigl((L_0+\overline L_0)_{\rm Liouv} + \Delta_X + \overline\Delta_X -2\Bigr)
\Psi^{\rm gravity}=0\ .
}

\ifig\fdecbess{Solution to minisuperspace Wheeler--DeWitt equation
decaying at large lengths.}
{\epsfxsize2.75in\epsfbox{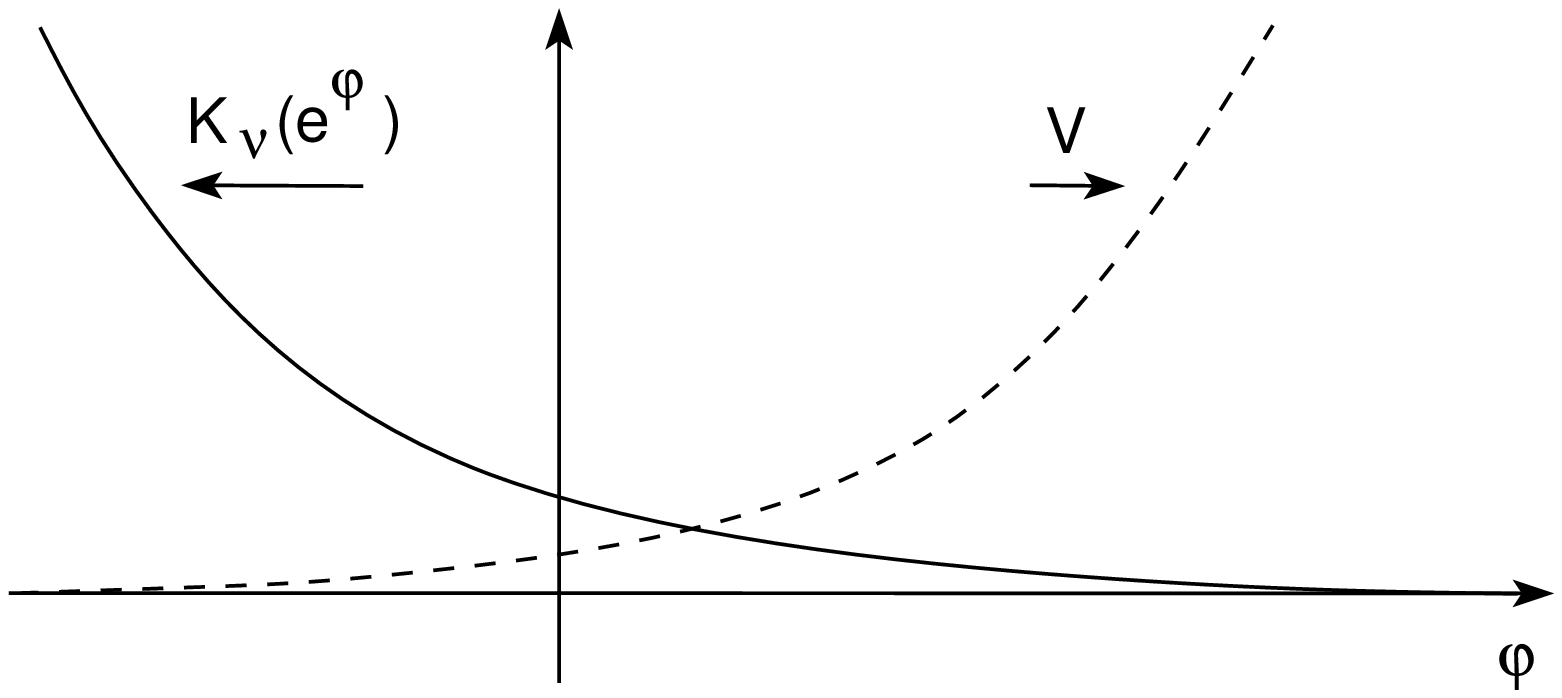}}

If matter boundary conditions are separately diffeomorphism invariant,
we expect $\Psi$ to depend on only the diffeomorphism
invariant information in $\phi(\sigma)$, namely, on the
length $\ell=\oint \ee{\ha \gamma \phi(\sigma)}$. In any case,
in the minisuperspace approximation we replace
\eqn\repalce{
\half (L_0+\overline L_0)_{\rm Liouv}\quad\rightarrow\quad
{\gamma^2\over 4}\Bigl(-(\ell {\p\over \p \ell})^2+4 \mu \ell^2 \Bigr)
+{1\over 8}Q^2\ .}
Using the KPZ formula \embeta\ written as
\eqn\kpzi{\Delta_X-1+{1\over 8}Q^2=\ha\bigl(\alpha-{\half Q}\bigr)^2}
(as suggested after \qandel),
we obtain the minisuperspace Wheeler--DeWitt equation
\eqn\msswdw{\Bigl(-(\ell {\p\over \p \ell})^2
+4 \mu \ell^2 +\nu^2\Bigr)\psi(\ell)=0\ ,\qquad
\nu=\pm {2\over \gamma}\bigl(\alpha-\half Q\bigr)\ .}
The solution decaying at large lengths is
the non-normalizable wavefunction
\eqn\mswvfn{\psi\dup_{\CO}(\ell)\propto  K_{\nu}(2\sqrt{\mu}\ell)\ ,}
illustrated in \fdecbess.

As promised in sec.~{\it\ssclst\/}, the wavefunctions corresponding
to geometries \micmet{}\ appear naturally in the theory. From the geometrical
picture of chapt.~\sbrlt, it is natural to associate this
geometry with the insertion of a local operator at $t=-\infty$.
In \nati, Seiberg has further interpreted the blowup of the wavefunction
at short distances as being physically appropriate. The idea is
that the wavefunctions associated to local operators in quantum gravity
should have support on metrics which are infinitesimally
small in the physical metric $\ee{\gamma \phi}\hat g$ (because
they are {\it local\/}).

{\bf Remark}: In \repalce\ we appear to have made an approximation.
Astonishingly, matrix model calculations (for example eq.~\wvfnht\ below)
confirm that \mswvfn\ is exact.
It is not understood why this should be so.

\subsec{LZ states in 2D Quantum Gravity}
\subseclab\sslzs

So far we have discussed only the KPZ states in which the
ghost modes are not excited. These
form only part of the full spectrum of the theory, as
demonstrated in the continuum formulation in the
work of Lian--Zuckerman \refs{\lzi,\rbong}.
Treating the Liouville field as free,
they calculated the semi-infinite
(BRST) cohomology of $\CL\otimes M(p,q)$, and found that the cohomology is
spanned by operators of the form
\eqn\lzops{\CO_{n}\, \ee{\alpha_{n} \phi}\ ,\qquad
{\alpha_{n}\over\gamma}={p+q-n\over 2 q}\qquad
n\geq 1,\ \not= 0\,{\rm mod}\,p,\ \not=0\,{\rm mod}\,q\ ,}
and $\gamma$ is determined as in \ealph. The operator $\CO_n$ is
made of ghosts, matter, and derivatives of $\phi$.
The ghost number of $\CO_n$ depends linearly on $n$.

In the KP formalism of the matrix model to be described in
sec.~{\it\sscsotmc\/}, on the other hand,
scaling operators formed from fractional
powers of Lax operators (which have known lattice analogs)
will be constructed and {\it scale\/} like
Liouville operators of the form
\eqn\lzopsii{\CO_n\, \ee{\alpha_n \phi}\ ,\qquad
{\alpha_n\over\gamma}={p+q-n\over 2 q}\quad n\geq 1,\ \not= 0\,{\rm mod}\,q\ ,}
where $q<p$ (but the $n\not= 0\,{\rm mod}\,p$ restriction is lifted).
In sec.~{\it\sscsotmc\/}, we will see how these operators arise in
the matrix model formulation.

Let us now consider the discrepancies between the calculations.
First, in the LZ computation there is no reason to restrict
attention to states satisfying $\alpha\le Q/2$. This is quite appropriate,
since the computation applies equally well when $\mu=0$, in which
case there is no wall to induce total reflection of the wavefunctions
and hence identify states with $\pm E$ or $\pm \nu$.
There is a further
discrepancy of operators with $n=0\,{\rm mod}\,p$. 
This has been partially explained with boundary operators \mms.
Apart from this, the two calculations are in remarkable agreement.
Nevertheless it is an important open problem to understand
better the physical meaning of the Lian--Zuckerman states and
their relationship, if any, to the infinite tower of scaling
operators in the matrix model.

In the case of the one-matrix model, the infinite tower of
operators corresponding to $K^{j-\ha}|_+$ (in the notation of
sec.~{\it\sscsotmc\/}) are denoted by $\sigma_j$, and will be studied in
more detail in sec.~{\it\sltlo\/} below.

\subsec{States in 2D Gravity Coupled to a Gaussian Field: more BRST}
\subseclab\ssstdgmb

Consider now the coupling of Euclidean gravity to a massless
Euclidean scalar field in two dimensions:
\eqn\liugausii{
S=\int \d^2 z\,  \sqrt{\hat g}
\biggl({1\over 8 \pi} (\hat\nabla \phi)^2
+{Q\over 8 \pi}\phi R(\hat g)+
{\mu\over 8 \pi \gamma^2} \ee{\gamma \phi}\biggr)+
\int \d^2 z\,  \sqrt{\hat g}
{1\over 8 \pi}(\hat\nabla X)^2\ ,}
where $X$ is the real massless boson.
The KPZ equations \eQ\ and \ealph\ for $D=1$
imply that $Q=\sqrt{8}$ and $\gamma=\sqrt{2}$.

\danger{Cosmological constant operator at $c=1$}
According to some authors, the correct quantum effective action must have a
cosmological constant term given by $\phi\, \ee{\gamma \phi}$. Many
confusing issues related to this point are not well understood
(as of Sep.~92).
The argument in favor of this identification is that the usual relation
between the wavefunction and vertex operator, together with the wavefunction
behavior $K_0(\ell)\sim \log\ell$, suggests an extra factor of $\phi$. A second
argument is based on the $p\to 0$ behavior of amplitudes studied in
sec.~{\it\ssposme\/} below,
and a third is based on the relation between bare and
renormalized cosmological constants at $c=1$ given in sec.~{\it\ssmlft\/}.
We find none of these arguments entirely convincing.

\danger{Spectrum of $\mu=0$ versus $\mu>0$ }
The BRST cohomology of the theory \liugausii\ was calculated in the $\mu=0$
theory by Lian and Zuckerman. Their results were simplified and extended in
\refs{\bouwknegt,\witzwie,\grndrng}.
In this section we describe some of these results.
The following argument, based on the string-theoretic/spacetime interpretation
of these theories described in chapt.~\stdcst, suggests that, except for the
imposition of the Seiberg bound, the physical states should be the same: the
Liouville interaction disappears for $\phi\to -\infty$. Thus, states that have
wavefunctions concentrated in this region must behave like states in the free
theory, in particular, the interaction is arbitrarily weak in this region and
``ought not'' create or destroy extra states. This is not true of states
concentrated at $\phi\to +\infty$, which is why we must impose the Seiberg
bound. This argument is surely correct for the tachyon cohomology classes, but
is not obviously correct for the global modes associated to the discrete
states.

The nature of the cohomology depends strongly on the value of $q$,
the $X$-field momentum as measured by $\sqrt{2}\,\p X$.
For generic $q\notin \IZ$ there are states in the BRST
cohomology of ghost number $G=2$ and dimension zero.\foot{Ghost number
$G$ always
refers to the total left+right moving ghost number in the closed string case.}
These are the  gravitationally dressed
vertex operators
\eqn\eucvert{\eqalign{
V_q&= c \bar{c}\, \ee{i q X/\sqrt{2}}\, \ee{\sqrt{2}(1-\ha|q|)\phi}\cr
\overline V_q&= c \bar{c}\, \ee{i q X/\sqrt{2}}
\, \ee{\sqrt{2}(1+\ha|q|)\phi}\ .\cr}
}
The operators $\overline V_q$ violate the
condition $\alpha\le Q/2$ discussed in sec.~{\it\sssb}.
We will confirm below that they do not
appear in the matrix model computations.
As in sec.~{\it\sskpztd\/}, we expect 2D gravity wavefunctions
associated to the operators $V_q$ to be $\mu^{|q|/2}K_q(2 \sqrt{\mu}\ell)$.
We will confirm this in chapt.~\slascomm. As discussed in sec.~{\it\sscqotd\/}
above we should distinguish between absolute, relative,
and semi-relative cohomology. If we are working with
the absolute cohomology,
we must introduce the operator \witzwie
\eqn\aoper{a=[Q,\phi]=c\, \p \phi + \sqrt{2}\,\p c\ ,}
and its holomorphic conjugate.
Then we have extra states:
$a V_q, \bar{a} V_q, a\bar{a} V_q$. In the semi-relative
cohomology, we must include the extra state $(a+\bar{a})\,V_q$.

In fact, the $c=1$ model has much more cohomology.
First of all, there are many more primary fields in the theory
which may be gravitationally dressed by Liouville
exponentials. This is most elegantly seen by
considering the chiral $SU(2)$ current algebra that
arises when a Gaussian field $X$ is compactified on a self-dual
radius \rcftrev. The currents are given by
\eqn\sutwo{
J^{(\pm)}(z)=
\ee{\pm i\sqrt{2} X} \qquad \qquad J^{(3)}(z)={i\over \sqrt{2}}\,\p X\ .
}
Then, for $s=0,1/2,1,\dots$, we have highest weight fields
$\psi\dup_{s,s}=\ee{i s \sqrt{2} X}$ for the global $SU(2)$.
We can thus make chiral weight $(1,0)$ Virasoro highest weight
fields from
\eqn\virhwt{\psi\dup_{j,m}(z)=\sqrt{(j+m)!\over (j-m)!\,(2j)!}\
\Bigl(\oint {\d z\over 2 \pi i}
\,\ee{-i\sqrt{2} X}\Bigr)^{j-m}\ \psi\dup_{j,j}\ .}
where $m\in\{-j,-j+1,\dots, j-1,j\}$.

\exercise{Characters of Fock modules}

When $q\in \IZ$, the Fock module has a highest weight
vector with Virasoro weight $\Delta=q^2/4$. In this case it
is known from Virasoro representation theory that the
Fock space $\CF_q$ becomes infinitely reducible, i.e., that
$\CF_q$ contains infinitely many Virasoro primaries.

The characters of the irreducible $c=1$ representations of
the Virasoro algebra with weight $\Delta$ are
\refs{\rcftrev,\kacbook}:
\eqn\irredchrs{\eqalign{
\chi\dup_\Delta&= {q^\Delta\over \eta} \qquad \sqrt{4 \Delta}\notin \IZ\cr
\chi\dup_{\Delta_n}&= {q^{n^2/4}-q^{(n+2)^2/4}\over \eta} \qquad \Delta=n^2/4,
n\in  \IZ\ .\cr}}

a) Using these characters and the fact that $\CF_q$ contains no singular
vectors, show that when $q\in \IZ$ the Fock module can be written as
\eqn\decomfock{\CF_{n/\sqrt{2}}=\oplus_{r=0}^\infty
L\bigl(c=1,\Delta={(n+2r)^2\over 4}\bigr)\ ,}
where $L$ is the irreducible representation with highest weight
$\Delta$.

b) Show that the state $\alpha_{-1}\overline{\alpha}_{-1}|0\rangle$
corresponding to $\p X\pb X$ is an
example of a nontrivial Virasoro primary in the Fock module with
$q=0$.

\endexercise

Therefore the chiral cohomology contains the fields
\eqn\openstr{\eqalign{Y_{j,m}^+(z)
&=c\, \psi\dup_{j,m}(z)\, \ee{\sqrt{2}(1-j)\phi}\cr
&= c\, \CP_{n,r}(\partial X)\,\ee{i{1\over \sqrt{2}} n X}
\,\ee{{\sqrt{2}\over2}(2 - (n+2r)) \phi }\cr}}
with ghost number $G=1$ and dimension zero.
In the second line of \openstr, we set $n=2m$ and we have emphasized the
description of the exercise: the highest weight in the $r^{\rm th}$
term of \decomfock\ is generated by the highest weight state
$\CP_{n,r}(\partial X)\,\ee{i{1\over \sqrt{2}} n X}$,
where $ \CP_{n,r}(\partial X)$ is a polynomial in
derivatives of $X$ of dimension $nr+r^2$,
and $s=r+n/2$. This state has Liouville momentum
$-ip_\phi/\sqrt{2}=-\alpha/\gamma+Q/2 \gamma=(n+2r)/2=s$.

Although we have constructed these states by appealing
to the symmetry structure at the self-dual radius, they
will give rise to BRST cohomology classes at other radii
by combining left-- and right--movers. In particular, at
infinite radius we may form the states
\eqn\dimzrgh{\CS_{j,m}= Y^+_{j,m}\,\overline{Y}^+_{j,m}}
with ghost number $G=2$ and dimension zero.
In the absolute cohomology we must include the states $a Y_{j,m}^+$, and so on.

\ifig\fydis{A plot of the quantum numbers of the special states in the
$(p\dup_X,-ip\dup_\phi)/\sqrt{2}$ plane. The special states intersecting the
tachyon dispersion line at $|m|=j$ are called ``special tachyons.'' Note that
if one works at $\mu=0$, the Seiberg bound does not hold and one should include
the other states $Y^-_{j,m}$. These constitute an identical plot obtained by
reflecting $p\dup_\phi\to-p\dup_\phi$.}{\epsfxsize4in\epsfbox{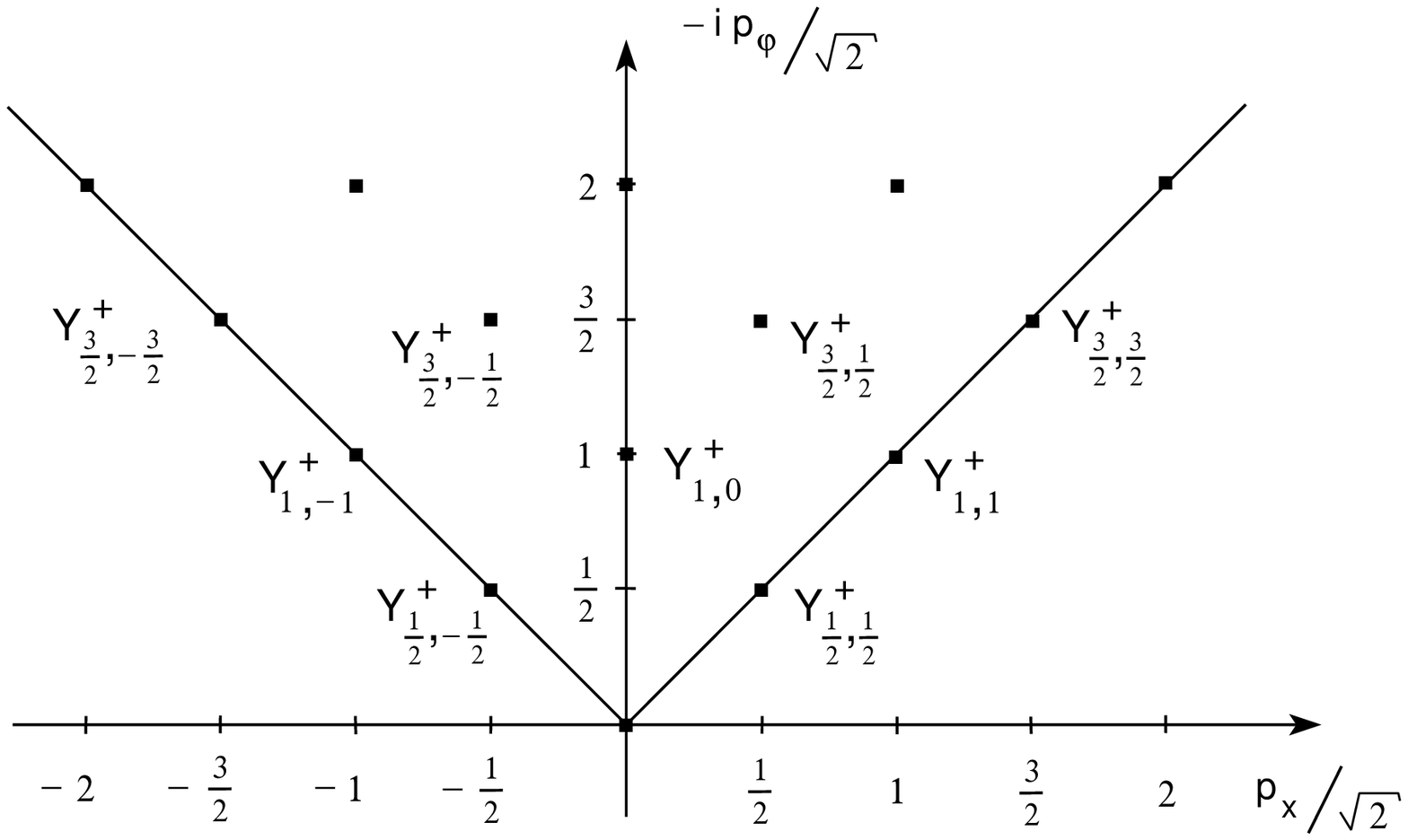}}

We may plot the quantum numbers of these states as in \fydis.
The big surprise, discovered by Lian and Zuckerman, is
that at the points in \fydis\ interior to the wedge
there are extra cohomology classes. The above-mentioned classes only
account for half of the BRST cohomology. For every class
$Y_{j,m}^+$ with $j=1,2,...$, and $|m|<j$ there is a corresponding
class $\CO_{j-1,m}$ with the same $X,\phi$ momenta but with
ghost number zero. The first three examples are
\eqn\grgens{\eqalign{\CO_{0,0}&=1\cr
\CO_{1/2,1/2}&=\Bigl(bc-{1\over \sqrt{2}}
(\p \phi+i\p X)\Bigr)\ee{-(\phi-iX)/\sqrt{2}}\cr
\CO_{1/2,-1/2}&=\Bigl(bc-{1\over \sqrt{2}}
(\p \phi-i\p X)\Bigr)\ee{-(\phi+iX)/\sqrt{2}}\ .\cr}
}

\ifig\fodis{The wedge of \fydis, with the chiral ground ring states
enumerated.}
{\epsfxsize4in\epsfbox{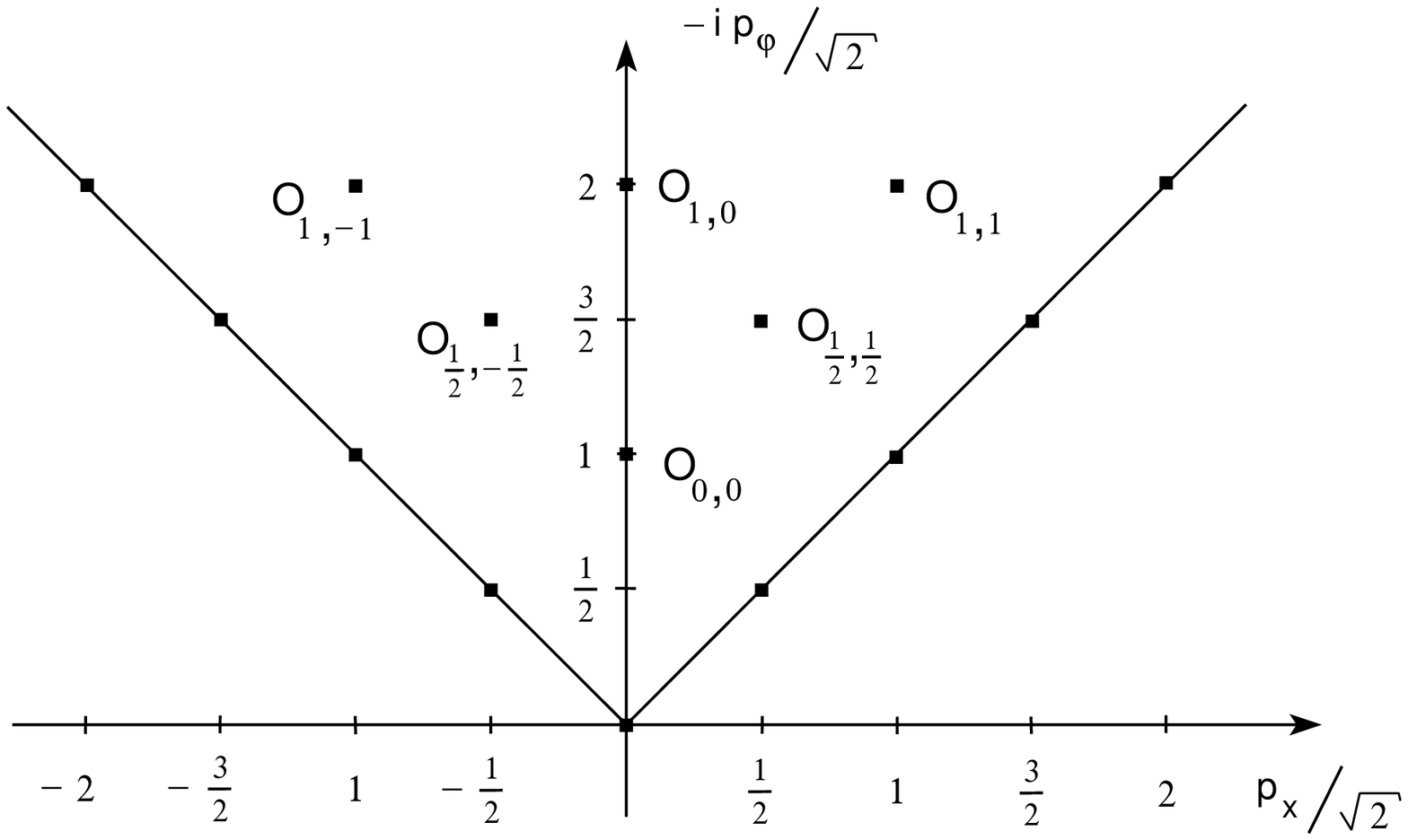}}

A plot of these ``ground ring'' states is shown in \fodis. Lian and Zuckerman
show that there are no other chiral cohomology classes. The full closed string
cohomology is formed by combining the above classes subject to constraints on
left-- and right--moving momenta. Since we do not compactify the Liouville
field, we must impose $p^L_\phi=p^R_\phi$. The conditions on $p\dup_X$ depend
on the radius of compactification \rcftrev. For the $X$-field with infinite
radius $R=\infty$ (our usual case), we have $p_X^R=p_X^L$. When $X$ is
compactified at special radii, e.g.\ the self-dual radius, this condition may
be relaxed and there will be more BRST cohomology classes.

{\bf Remark}: In general there will be ``special states'' when the $X$-field is
compactified on a circle of radius $r={1\over \sqrt{2}} {p\over q}$ where $p,q$
are relatively prime integers. The special states must have the
$(p^L_X,p^R_X)\sqrt{2}=(kq+lp,kq-lp)$ where $k,l$ are arbitrary integers. Note
that the zero momentum special states are present at every radius.

The appearance and disappearance of special states as the radius is varied is a
puzzling phenomenon. It has been discussed in \vermster.

Thus we may finally summarize the relative closed string cohomology at
$R=\infty$: In addition to the tachyon states \eucvert\ we have four states at
ghost numbers $G=0,1,2$ in the relative cohomology:
\eqn\clsdcoho{\eqalign{G=0:\quad & \CR_{j,m}
=\CO_{j,m}\bar{\CO}_{j,m}\qquad\qquad\quad j=0,1/2,\dots;\ |m|\leq j\cr
G=1:\quad & \CJ_{j,m}=Y^+_{j,m}\,\bar{\CO}_{j-1,m}\quad
\bar\CJ_{j,m}=\CO_{j-1,m}\bar{Y}^+_{j,m}\qquad
j=1,3/2,\dots; |m|< j\cr
G=2:\quad &\CS_{j,m}
= Y^+_{j,m}\,\bar{Y}^+_{j,m}\qquad\qquad\qquad j=1,3/2,\dots; |m|< j\ .\cr}}
As pointed out in \witzwie, the semi-relative cohomology is
more appropriate for comparison with closed string field
theory (see \bz). The semi-relative cohomology has 4 more
states at ghost numbers 1,2,3 obtained by multiplying
the above operators by  $a+\bar{a}$.
Explicit formulae for special state representatives, as
well as an alternative proof of the  Lian--Zuckerman theorem
has been given in \bouwknegt.

\noindent{\bf Remark: Conjugate States}\foot{We thank N. Seiberg
for clarifying this point.}\par\nobreak
The tilde conjugation $\Phi_s\to \tilde \Phi^s$ described in
sec.~{\it\sscqotd\/} above is important for understanding the factorization
properties of amplitudes. The behavior of this conjugation is
rather different at $\mu=0$ and $\mu>0$. At $\mu=0$ we have
standard free-field formulae. In particular
$\Phi_s\to \tilde \Phi^s$ exchanges states with ghost number
$G$ and $5-G$. It also exchanges $(+)$--states with $(-)$--states.
At $\mu>0$, there are no $(-)$ states and it might appear that
a fundamental axiom for constructing string field theory
has broken down. This is not the case, since the Liouville
2-point function has a geometrical divergence coming from
the volume of the dilation group $\IR^*_+$ (see sec.~{\it\sssa\/}).
This divergent numerator is precisely what is needed to
cancel the division by the volume of the conformal Killing
group that results if we only insert $4$ out of $6$
$\,c,\bar c\,$--zero modes. Thus we can have a nonzero 2-point
function:
\eqn\tldecji{
\Bigl< c \bar c\, \ee{i p_1 X/\sqrt{2}}\, \ee{\sqrt{2}(1-\half |p_1|)\phi}
\,c \bar c \,\ee{i p_2 X/\sqrt{2}}\, \ee{\sqrt{2}(1-\half |p_2|)\phi} \Bigr>
\sim \delta(p_1+p_2)\ .}
On the RHS we have one, rather than two, $\delta$--functions in
the momenta of the problem. In general, we see that the
conjugation $\Phi_s\to \tilde \Phi^s$ at $\mu>0$ exchanges
ghost numbers $G$ and $4-G$, and preserves the $(+)$--states
satisfying the Seiberg bound.

As emphasized in \grndrng, the existence of
the ghost number one BRST classes
implies the   existence of a large symmetry
algebra. Indeed, quite generally,
given   a dimension zero BRST class $\Omega^{(0)}$
we may associate with it a
descent multiplet $(\Omega^{(0)},\Omega^{(1)},\Omega^{(2)})$
consisting of
$0,1,2$ forms defined by the descent equations:
\eqn\descent{\eqalign{ 0&=\{Q,\Omega^{(0)}\}\cr
\d \Omega^{(0)}&=\{Q,\Omega^{(1)}\}\cr
\d \Omega^{(1)}&=\{Q,\Omega^{(2)}\}\cr}
}

\exercise{Descent Equations}

a) Using $\{Q,b_{-1}\}= L_{-1}$, show that in terms
of states associated to the operators the descent equations
read:
\eqn\descenti{\eqalign{
|\Omega^{(1)}_z \rangle &=b_{-1}|\Omega^{(0)}\rangle \cr
|\Omega^{(1)}_\zb \rangle &=\bar{b}_{-1}|\Omega^{(0)}\rangle \cr
|\Omega^{(2)}_{z\zb} \rangle &=b_{-1}\bar{b}_{-1}|\Omega^{(0)}\rangle\ .\cr}
}

\endexercise

The significance of the descent multiplet is that to any
BRST invariant dimension zero operator $\Omega^{(0)}$, we may
associate 1) a corresponding charge
\eqn\corrchrge{\CA(\Omega^{(0)})\equiv \oint \Omega^{(1)}\ ,}
conserved up to BRST exact operators, and 2)
a corresponding modulus, by which we can deform the
action,
\eqn\corrmdls{\Delta S=\int_\Sigma \Omega^{(2)}\ ,}
while preserving BRST symmetry.

\exercise{Tachyon descent multiplet}

Show that the descent multiplet for the tachyon vertex
operator is
\eqn\tchmult{ \eqalign{
G=2:\quad V_p^{(0)} & = c\overline c\, \ee{ipX/\sqrt{2}}
\,\ee{(\sqrt 2(1-\half|p|)\phi} \cr
G=1:\quad V_p^{(1)} & = \left(\d z \,\overline c-\d\overline z\,c\right)
\,\ee{ipX/\sqrt{2}}\,\ee{(\sqrt 2(1-\half|p|)\phi} \cr
G=0:\quad V_p^{(2)} & = \d z\wedge \d\overline z\,\, \ee{ipX/\sqrt{2}}
\,\ee{(\sqrt 2(1-\half|p|)\phi}\ .\cr}}

\endexercise

\ifig\fadis{Closed string symmetry charges $\CA_{j,m}\equiv \oint \d z\,
\Omega^{(1)}_{j,m}$. Note there are also conjugate charges
$\bar{\CA}_{j,m}$ at the same values of $(p_x,p_\ph)$.}
{\epsfxsize4in\epsfbox{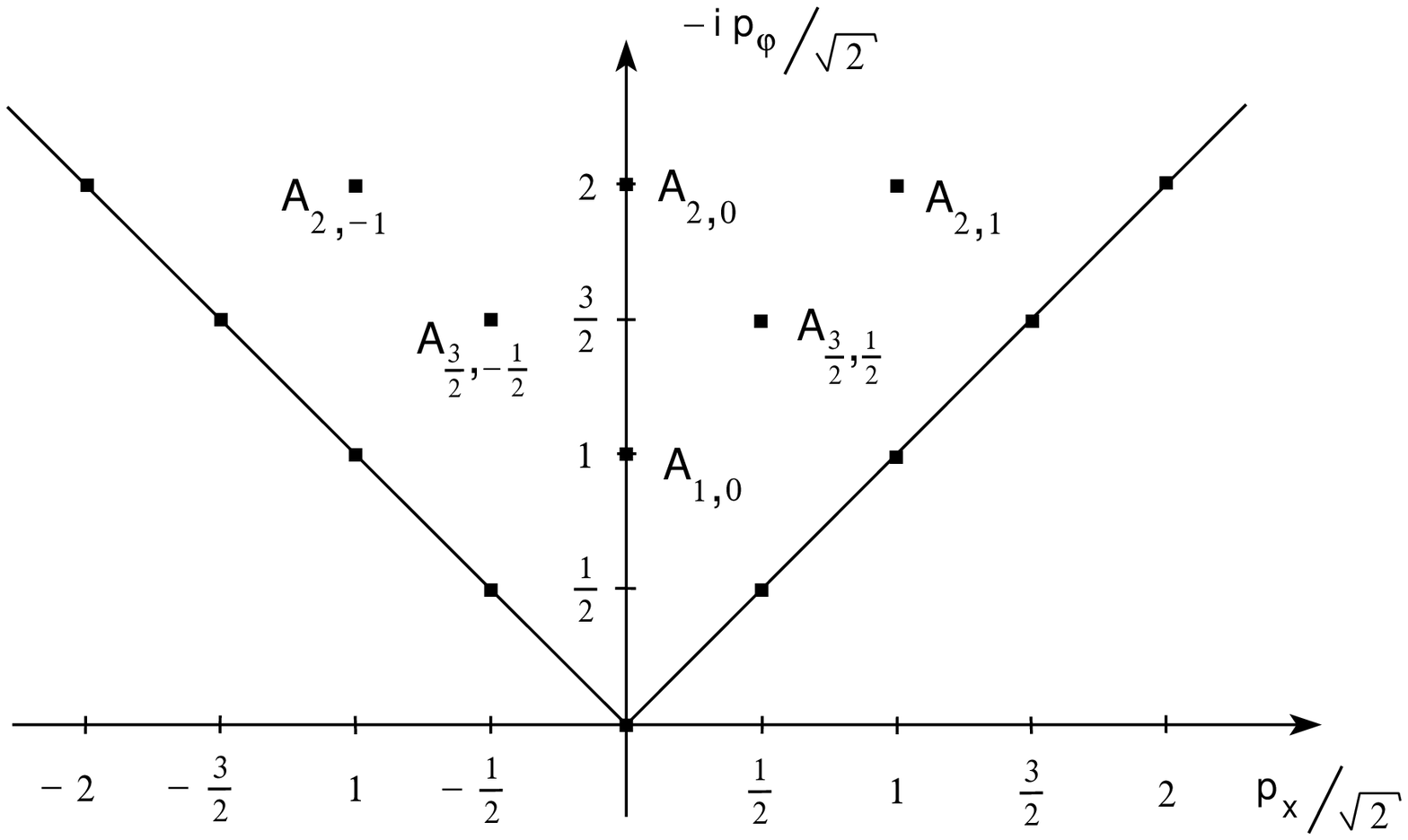}}

The descent multiplet turns out to be nontrivial for the ghost number
$G=1$ states in \clsdcoho:
$\CJ^{(0)}_{j,m}=Y^+_{j,m}\bar{\CO}_{j-1,m}$
and its holomorphic conjugate. Therefore, there are
corresponding currents $\Omega^{(1)}_{j,m}$, conserved up
to BRST exact operators, which produce ``discrete charges''
\eqn\dischrge{\CA_{j,m}\equiv \oint \d z\, \Omega^{(1)}_{j,m}\ ,}
and their holomorphic conjugates $\bar{\CA}_{j,m}$, which are
conserved up to BRST exact operators.
As described in \witzwie\ and in chapt.~\svoccm\ below,
the existence of these
charges have nontrivial consequences for correlation functions
computed in the $\mu=0$ theory. The quantum numbers of the
charges are plotted in \fadis.

As at $c<1$, an important open problem is to understand better
the role of these states in quantum gravity.
Moreover, an important open problem is to find matrix model
techniques for investigating the $\CO_{u,n}$.

\newsec{2D Critical String Theory} 
\seclab\stdcst

Further insight into the spectrum of 2D gravity is obtained when we consider
the string--theory/target space point of view, in which we regard $\phi$ as
a spacetime coordinate. The KPZ formula is now interpreted as the on-shell
condition for Euclidean target space.

\subsec{Particles in $D$ Dimensions: QFT as 1D Euclidean Quantum Gravity.}

In chapters \sqliouv\ and \stdeqg, we have discussed 2D Euclidean
quantum gravity. In this section, we apply the same techniques
to 1D Euclidean Quantum Gravity.
While the theory is trivial as a theory of quantum gravity,
it has an important and obvious reinterpretation in terms of target space
Euclidean quantum field theory.

\noindent{\bf Path Integral Approach}\par\nobreak
An example which will illuminate our later considerations is that of a particle
moving through {\it Euclidean\/} spacetime. This may be thought of as $1D$
quantum gravity since the system is described by the action
\eqn\prta{S=\ha \int \d \tau\, \sqrt{g(\tau)}\Bigl(
g^{\tau \tau}\bigl({\d X^\mu\over \d\tau}\bigr)^2 - m^2\Bigr)\ .}
We consider the path integral
\eqn\ptpth{\CA(X_I,X_f)=\int {\d g\, \d X\over {\it Diff}}\, \ee{S}\ ,}
with boundary conditions $X_i^\mu,X_f^\mu$ on $X^\mu$. We can fix the
gauge by transforming the einbein to a constant,
$f^* e = s$, where $s$ is the single coordinate invariant
quantity (i.e.\ modulus), namely the length. The path integral becomes
\eqn\ptptpth{\eqalign{\CA(X_i,X_f)&=\int_0^\infty {\d s\over s^{1/2}}\,
\bigl(\det{}'(-s^{-2}\p_t^2)\bigr)^{(1-D)/2}\,\ee{-(\Delta X)^2/2s-m^2 s/2}\cr
&\propto\int_0^\infty {\d s\over s^{D/2}}\, \ee{-(\Delta X)^2/2s - m^2 s/2}
\propto\int {\d^D p\over(2 \pi)^D}
\,{\ee{i p \Delta X}\over p^2+m^2}\ ,\cr}}
since the determinant is proportional to $s$.

\noindent{\bf Canonical Approach}\par\nobreak

Turning to the canonical approach, the action \prta\
has a gauge invariance:
\eqn\diffact{
\delta X=\epsilon(\tau) X' \qquad\qquad \delta e(\tau)=\epsilon'(\tau)
e(\tau)+\epsilon(\tau) e'(\tau)\ .
}
We can fix the gauge by putting $e=1$ at the price of
imposing a constraint.
The Wheeler--DeWitt operator, which generates $\tau$ diffeomorphisms,
is simply  $H=p^2 + m^2$, where $p^\mu(\tau)$ is the
field canonically conjugate to $x^\mu(\tau)$.
The Wheeler--DeWitt equation is
the Euclidean Klein-Gordon equation:
\eqn\euckg{H \psi(x)=\Bigl(-{\p^2\over \p x^2}+m^2\Bigr)\psi=0 \ .}

If we isolate one Euclidean coordinate,
call it $\phi$, as a special coordinate, then we can
write the Euclidean on-shell wavefunctions
as $\ee{i px }\, \ee{\pm \sqrt{p^2+m^2}\phi}$. As long as
there are no tachyons in the theory, these wavefunctions
have exponential growth and are not normalizable.
Conversely, the existence of Euclidean on-shell normalizable
wavefunctions is a signal of tachyons in the theory.

In order to describe off-shell physics, we introduce the
normalizable states which diagonalize the Wheeler--DeWitt
operator: $\ee{ipx + i E \phi}$, with eigenvalue
$E^2+p^2 + m^2$.  For example, in a mixed
position-space/momentum-space representation
where we Fourier transform  with respect to all other
coordinates, we may describe the propagator as
\eqn\fftpr{\eqalign{G(\varphi_1,p;\varphi_2,-p)&=
\int_{-\infty}^{\infty} \d E\,
{\ee{-iE\varphi_1}\,\ee{iE\varphi_2}\over E^2+\vec p^{\,2}+m^2}\cr
&=\theta(\phi_1-\phi_2) {1\over \sqrt{\vec p^{\,2}+m^2}}
\,\ee{-\sqrt{\vec p^{\,2}+m^2}| \varphi_1-\varphi_2| }
+[1\leftrightarrow 2]\ .\cr}}

\exercise{Back to the wall}

What happens if $\phi$ is restricted to be semi-infinite?
Put a boundary condition that the wavefunctions vanish
at $\phi=\log \mu$ and calculate the analog of \fftpr.

\endexercise

\noindent{\bf Interactions and Topology-Change}\par\nobreak
One-dimensional quantum gravity from the target space viewpoint
provides a useful insight into the origin of
the violation of the Wheeler--DeWitt constraint in
topology-changing processes. In this case, a topology-changing
process corresponds to one 0-dimensional space splitting
into two as in
\eqn\etcpr{\vcenter{\figins{\epsfxsize1.75in\epsfbox{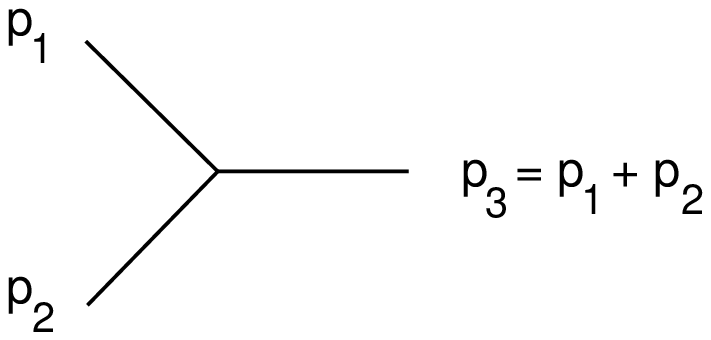}}}\ \ .}
The ``violation of the Wheeler--DeWitt constraint'' is simply the
familiar fact that if $p_1^2=p_2^2=-m^2$ are on-shell momenta then
in general $p_3^2 =(p_1+p_2)^2\not=-m^2$ will not be on-shell.

This above basic
phenomenon can also be realized as the result of a contact
term arising from a singularity at the boundary of ``moduli space.''
Consider the wavefunction of a particle that interacts
with an external potential $V$ so that the wavefunction becomes
\eqn\wvfni{\tilde{\psi}(\tau)
=\int^\tau_{-\infty} \d \tau'\, \ee{-H(\tau-\tau')}V \psi\ ,}
where $H$ is the Wheeler--DeWitt operator. Note that
\eqn\violi{H \tilde{\psi}=
\int^\tau_{-\infty} \d \tau' {\p\over \p \tau'}
\Bigl(\ee{-H(\tau-\tau')}\,V \psi\Bigr)=V \psi \ne 0\ ,}
so the condition $H\psi=0$ is not preserved under time evolution.

\subsec{Strings in $D$ Dimensions: String Theory
as 2D Euclidean Quantum Gravity}
\subseclab\sbsdd

{\bf Nonlinear $\sigma$-Model Approach}.
The particle Lagrangian \prta\ can be generalized to a
string Lagrangian, which we recognize as a 2D nonlinear $\sigma$-model,
and the quantum theory involves a path integral over surfaces.
To describe
strings propagating in general manifolds we should in principle
consider {\it arbitrary\/} 2d quantum field theories:
\eqn\pertaction{ S_\sigma={1\over 4\pi}\int \d^2 z\,\sqrt g\,
\Bigl(T(X)+R^{(2)}D(X)+g^{ab}\,\p_aX^\mu\,\p_b X^\nu \,G_{\mu\nu}(X)
+\cdots \Bigr)\ ,}
where $X^{\mu=1,...,D}$ parametrize a $D$-dimensional spacetime target space
and the ellipsis indicates a sum over a possibly infinite
set of irrelevant operators.

\danger{Pertinent operators?}
We are expanding here around the Gaussian fixed point,
since we think of each coordinate $X^\mu$ as a Gaussian
field. Including arbitrary interactions is a very formal
procedure which must be made well-defined. An infinite
sum of irrelevant operators might not be irrelevant at all,
but might be the effect of expanding
around the wrong fixed point.

In standard treatments of string theory \grschwtt, it is shown that
a consistent string theory can be formulated from
models of the above type when
they are conformally invariant (more precisely, BRST invariant).
The model is conformally invariant when the $\beta$--functions
vanish, that is, when the {\it spacetime\/} equations of motion,
\eqn\betafn{\eqalign{\beta^G_{\mu\nu}=&R_{\mu\nu}+2\nabla_\mu\nabla_\nu D
	-\nabla_\mu T\,\nabla_\nu T+\cdots =0\cr
	\beta^D=&{26-d\over 3}-R+4(\nabla D)^2
	-4\nabla^2 D+(\nabla T)^2-2 T^2 +\cdots =0\cr
	\beta^T=&-2\nabla^2T+4\nabla D\,\nabla T -4 T +\cdots =0\ ,\cr}}
are satisfied. The dots indicate higher order (in the string tension $\alpha'$)
corrections, including tachyon interactions. These $\beta$--function equations
themselves follow from an action\foot{The nonderivative dependence on $T$
follows from very general considerations \bkstach.} \rcmpf
\eqn\stact{S={1\over 2 \pi \kappa^2}\int \d^dx\, \sqrt{G}\, \ee{-2 D}
\Bigl(R+4(\nabla D)^2 + {26-d\over 3} - (\nabla T)^2 + 2 T^2\Bigr) + \cdots
\ ,}
where $\kappa$ is the string coupling.

Consider the case when the matter conformal field theory
$S_{\rm CFT}$ is a product of Gaussian models,
\eqn\prodgss{S_{\rm CFT}=\int \d^2 z\,  \sqrt{\hat g}
{1\over 8 \pi}(\hat\nabla X^\mu)^2\ ,}
together with one CTFF field $\phi$
(the Chodos--Thorn/Feigen--Fuks field described in sec.~{\it\ssctff\/}).

Identifying $\phi$ with a spacetime coordinate in \pertaction,
we read off from comparison of \liouvs\ with \pertaction:
\eqn\soln{\langle T\rangle =0
\,,\qquad\qquad\langle D\rangle ={ Q\over 2}\phi
\,,\qquad\qquad\langle G_{\mu\nu}\rangle =\delta_{\mu\nu}\ .}
Substituting \soln\ into \betafn\ and working to lowest order
in $\langle T \rangle$ shows that $\beta=0$ is satisfied
provided the KPZ formulae described in chapt.~\sqliouv\ are satisfied,
so in particular $Q={ 2\over \gamma}+\gamma=\sqrt{(26-d)/3}$ (where
$d=c+1$ in the critical string interpretation).

Now let us replace the CTFF field by a Liouville field, i.e.\ instead of a free
field we now have the Liouville interaction term.
Comparing actions \liouvs\ with \pertaction, we find the same dilaton and
metric expectation values as in \soln, but a new tachyon expectation value:
\eqn\solnii{\langle T\rangle ={\mu\over 2 \gamma^2}\, \ee{\gamma \phi}
\,,\qquad\qquad\langle D\rangle ={ Q\over 2}\phi
\,,\qquad\qquad\langle G_{\mu\nu}\rangle =\delta_{\mu\nu}\ .}

\danger{Conformal background?}
The background \solnii\ no longer solves the lowest order $\beta$-function
equations \betafn. This has been blamed either on the possibility of field
redefinitions \mart, or on the fact that the above equations are only the
lowest order terms in the $\beta$-function. We nevertheless continue with
this review, since the Liouville theory {\it is\/} conformal.

\danger{More subtleties}
There are many other subtleties and caveats associated with these assertions.
For example, due to the difficulties of treating theories with matter
central charge $c>1$ for $\mu>0$, we can really understand only the case
of a single gaussian model in \prodgss.

The construction of a consistent string theory can be carried out for any
conformal field theory with total central charge $c=26$. In the case of a
tensor product of Gaussian models, we identify each Gaussian model field with a
macroscopic spacetime dimension. An arbitrary CFT is an abstract version of
target spaces made from products of Gaussian models. The minimal models with
$c<1$, for example, can be thought of as generalized Euclidean signature
spacetimes. They can be augmented to $c=26$ and converted to consistent target
spaces for string propagation by coupling to a Liouville theory since the
Liouville mode has a tunable central charge.\foot{But it always has the same
number of field theoretic degrees of freedom. This remarkable aspect of
Liouville theory has been explored in detail in \refs{\nati,\kutseib}.}
For example, by introducing a free CTFF field we can tune to lower
dimensional critical string theories \rmabetal.
We have already discussed some aspects of tensor products of Liouville and
matter sectors in sec.~{\it\sstdglt\/}, and pointed out the relation between
critical strings in $d=D+1$ dimensions and
``non-critical strings'' in $D$ dimensions (when the latter interpretation
exists, see footnote after \egam).

\subsec{2D String Theory: Euclidean Signature}
\subseclab\stdstes

It is useful to recall at this point the dual interpretations of the
theories we consider:
\item{i)} matter coupled to 2D quantum gravity.
\item{ii)} critical strings moving through specific background geometries.

\noindent In particular, as described in the previous section,
gravity coupled to a $c=1$ Gaussian model can be interpreted as
a $d=2$ critical string theory.
The critical string interpretation of the $c=1$ matrix model is subtle and
still changing\foot{June 1992}.
Our specific action \liugausii\
describes strings moving in two Euclidean spacetime dimensions $(X,\phi)$,
and in the next section we shall consider its Minkowskian continuations.

In general, the KPZ formula \eqns{\embeta{,\ }\kpzi}\ that determines the
gravitational dressing for an operator coupled to 2d gravity has a dual
interpretation as the Euclidean
on-shell condition for string propagation in the critical string target space
picture. Recall that for $c=1$, we have $Q=2\sqrt2$.
Thus the operators in \eucvert,
$$\ee{i q X/\sqrt{2}}\, \ee{\sqrt{2}(1\mp\ha|q|)\phi}
=\ee{i p\dup_X X}\, \ee{\ha Q\phi+iE\phi}\ ,$$
create states that satisfy the Euclidean on-shell condition
$$E^2 + p_X^2 =0$$
for a massless particle (where $p\dup_\phi=E$ and $p\dup_X=q/\sqrt2$ are the
$\phi$ and $X$ momenta). We recognize that the KPZ formula (written in the form
\kpzi) is the dispersion relation for massless propagation.

It should come as no surprise to find massless propagation in $d=2$ critical
string theory. In the light cone gauge approach to string theory, there are
physical excitations associated with the motion of the string center of mass
and with the transverse oscillations of the string. In two spacetime
dimensions, there are no transverse oscillations so we expect to find a single
field theoretic degree of freedom. The center of mass degree of freedom, which
is identified with the tachyon field $T(X^\mu)$ of the $26$-dimensional
critical string, has mass-squared $m^2=(2-d)/12$ in $d$ dimensions. Thus, in
two dimensional string theory we expect to find one massless field theoretic
excitation.

One way to confirm the lightcone statements is to
consider the $\beta$-function derived spacetime action \stact\ in a general
linear dilaton background in $d$ dimensions, i.e.\ \soln\ or \solnii\
with $Q^2=(26-d)/3$ (again $d=c+1$ in the critical string interpretation).
Changing variables in \stact\
to $T=\ee{D}\,b(x,\phi)$, we find that the tachyon field has action
\eqn\tachact{\eqalign{S_T&={1\over 2 \pi \kappa^2}\int \d x\, \d\phi\,
\ee{-2 D}\Bigl((\nabla T)^2-2 T^2+\cdots \Bigr) \cr
&={1\over 2 \pi \kappa^2}\int \d x\,\d\phi\,\Bigl((\nabla b)^2
+\bigl((\nabla D)^2-\nabla^2 D-2\bigr)\, b^2+\cdots \Bigr)\cr
&={1\over 2 \pi \kappa^2}
\int \d x\,\d\phi\Bigl((\nabla b)^2 - {2-d\over 12}\, b^2 + {\rm
interactions}\Bigr)\ .\cr}}
In particular for $d=2$, the field $b$ is massless.

{\bf Remark:} We can view the KPZ formula as the
on-shell condition for the Euclidean target space propagator
as well for $c\ne1$. Indeed from \kpzi\ we have
\eqn\kpzfor{
-\ha\Bigl(\alpha-{Q\over 2}\Bigr)^2 + \Delta_X + {1-c\over 24} =0\ ,}
which we read as the Euclidean on-shell condition:
\eqn\onshell{\half E^2 + \half p^2 + \half m^2 =0\ .}
In the $c=1$ model, we have seen just above that the analogy
\eqn\eanal{\Delta_X\ \longleftrightarrow\ \half p^2 \qquad
\qquad {1-c\over 24}\ \longleftrightarrow\ \half m^2}
is exact, with $m^2=0$.

Following the particle example we can --- in the minisuperspace
approximation --- immediately discuss the propagator
\eqn\stanprop{\eqalign{G(\ell_1,p;\ell_2,-p)
=&\int_{0}^\infty \d E\, {1\over E^2+p^2+m^2}
\,\psi\dup_E(\ell_1)\,\psi\dup_E(\ell_2)\cr
&=\theta(\ell_2-\ell_1)\,I_{\omega_p}(2\sqrt{\mu}\,\ell_1)
\,K_{\omega_p}(2\sqrt{\mu}\,\ell_2) + [1\leftrightarrow 2]\ ,\cr}}
where $\omega_p\equiv + \sqrt{p^2+m^2}$. This is the 2D gravity analog of
\fftpr\ in the minisuperspace approximation. From the point of view of 2D
quantum gravity, this is the universe--universe propagator of third
quantization \refs{\joei\dj{--}\wadia}.

\subsec{2D String Theory: Minkowskian Signature}
\subseclab\stdstms

Now we consider the possible Minkowskian continuations of our
Euclidean action \liugausii.

\noindent {\bf A)} $X$ is Euclidean time.\par\nobreak
The $c=1$ model has the clearest target space interpretation
of the models we have studied. In particular if we rotate
$X\to i t$, we can consider $t$ as a Minkowskian time coordinate.
Taking account of the tachyon condensate, we have seen how to get the target
space wavefunctions (e.g.\ \mswvfn).
Then the on-shell wavefunctions are, at tree level,
\eqn\onshmn{\ee{i E t}\, K_{i E}(\ell)
=\ee{i E t}\Bigl( {\ell^{i E}\over \Gamma(1+iE)}
-{\ell^{-i E}\over \Gamma(1-iE)}\Bigr) +\CO(\ell^2)\ .}
Physically these wavefunctions describe the reflection
scattering of an incoming tachyon by the Liouville wall.
Since the Bessel function is a sum of incoming and outgoing
waves we may, without further ado,
read off the genus zero $1\to 1$ scattering amplitude
in the theory:
$$S(E)=- {\Gamma(i E)\over \Gamma(-i E)}\ .$$
In sec.~{\it\ssnpsm\/} below, we will calculate the full nonperturbative
$S$-matrix for this theory.

The scattering cohomology classes are
\eqn\minkvrtx{
V_\omega^\pm = c \bar{c}\, \underbrace{\ee{(\pm i\omega t+i
\omega\phi)/\sqrt{2}}}_{\rm
wavefunction}\overbrace{\ee{\sqrt{2} \phi}}^{\rm coupling\; constant}
\ ,}
where $\omega>0$. Note that

\item{i)} In quantum mechanics wavefunctions depend on time as
$\psi\sim \ee{-i E t}$ where $E\geq 0$ is a positive
energy. When calculating scattering matrices in a
path integral formalism \faddeev\ we insert
$\psi_{\rm out}^*$ and $\psi\dup_{\rm in}$ respectively for outgoers and
incomers. Therefore the vertex operators create scattering
states according to:
\eqn\scatinti{\eqalign{&V_\omega^-: {\rm\ incoming\ rightmover}\cr
&V_\omega^+: {\rm\ outgoing\ leftmover}\ .\cr}}
Since we are effectively discussing scattering theory in a half-space,
incomers are rightmovers and outgoers are leftmovers.
This is the spacetime version of the Seiberg bound \seibound.

\item{ii)} We must work with macroscopic states to have
(plane-wave) normalizable wavefunctions in
Minkowski space, required to set up a sensible scattering theory.

\noindent{\bf B)} $\phi$ is Euclidean time.\par\nobreak
In this case we must rotate $\phi\to i t$ to
obtain a Minkowskian interpretation. Unfortunately,
the rotation is problematic for $\mu>0$
\joetalk. The reason is evident from the zero-mode
part of the Liouville path integral \zminti. If $\mu>0$, then
in the complex $\phi_0$ plane (i.e.\ zero modes of $\phi$)
there is a series of
``ridges'' along the lines ${\rm Im}(\phi_0)=(2n+1) \pi/\gamma$, $n\in \IZ$,
which invalidate any contour rotation:
the right answer cannot be obtained by rotating $\phi\to i t$ and
expanding in a series of $\delta$-functions (except, perhaps, by dumb luck).

These objections disappear if we consider the ``free Liouville
theory'' with $\mu=0$. There is no obstruction to
rotating $\phi\to i t$, where $t$ is a timelike coordinate.
The natural BRST classes are
\eqn\bsvrtx{
T^\pm_k=c \bar{c}\, \ee{ik(X\pm t)/\sqrt{2}}\, \ee{i \sqrt{2} t}\ ,
}
which now have the interpretation
\eqn\scatintii{\eqalign{
&T_k^+ : {\rm\ incoming\ leftmover}\quad k<0 \cr
&T_k^+ : {\rm\ outgoing\ leftmover}\quad k>0 \cr
&T_k^- : {\rm\ outgoing\ rightmover}\quad k<0 \cr
&T_k^- : {\rm\ incoming\ rightmover}\quad k>0\ . \cr}}
Since there is no wall at $\mu=0$, we can have both
leftmovers and rightmovers. Moreover, the string
coupling becomes time-dependent, $\kappa(t)=\kappa_0\, \ee{i \sqrt{2} t}$,
and the dilaton field is purely imaginary.\foot{In conventional closed string
field theory \bz, one imposes reality conditions on the string
field forcing the dilaton to be real.}
Clearly the physics of this model is rather different
from case {\bf A)} and any relation between the models is
only mathematical. We will return to this world
briefly in sec.~{\it\srabsm\/}.

\subsec{Heterodox remarks regarding the ``special states''}
\subseclab\sshrrss

There are three reasons why the infinite class of special states is
exciting and interesting:
\item{1)} They correspond to a large unbroken symmetry group of
the string gauge group.
\item{2)} The only difference in degrees of freedom between
strings and fields in 2D is in the special states.
The spacetime meaning of the special states
is not understood and should be stringy and interesting.
\item{3)} They enter non-trivially into the 2d black hole metric.

\noindent Let us elaborate on these three points:
\item{1)} In the 26-dimensional bosonic string with Minkowski
space background there is an analog of the special states.
They are all at zero momentum and their physical interpretation
is clear. The linearized gauge symmetry of
string field theory is
\eqn\symmact{\Psi\to \Psi+Q\Lambda+\kappa [\Psi,\Lambda]+\cdots\ ,}
where the last term is the string product described in \bz, and
the infinitesimal symmetry generator $\Lambda$ has
ghost number $G=1$. $\Psi$ represents deviations of the fields from background
values, so a symmetry of the background should take $\Psi=0\to\Psi=0$ and
therefore satisfy
$Q \Lambda=0$, i.e., the symmetry should act linearly on small deviations from
the
background, as follows from \symmact. Moreover, modifying $\Lambda\to \Lambda +
Q \epsilon$ doesn't change the linearized action on the on-shell fields.
Therefore the nontrivial BRST classes of ghost number $G=1$ correspond to
on-shell symmetries of the string background \witzwie. In the case of the
``special states'' of Minkowski space, they correspond to the unbroken
translation symmetries of the vacuum defined by Minkowski space.\foot{Together
with dual symmetries for the $B$-field.} Reasoning by analogy, it would seem
that the infinite number of special states in the 2D string correspond to a
much larger symmetry group. It has been suggested in \vermster\ that this is
also related to the fact that in the 2d string there are far fewer states in
the theory.
\item{2)} The vertex operators representing small changes in the tachyon
background are just those given in \eucvert. The question thus arises as to the
spacetime meaning of the special state operators. It has been suggested in
\polss\ that these represent global modes of spacetime fields which have no
propagating degrees of freedom. The basic idea can be seen by considering $1+1$
dimensional gauge theories of electromagnetism and gravity. In $1+1$
dimensional (classical) electromagnetism and gravitation, for example, the
fields $A_\mu(x)$ and $G_{\mu \nu}(x)$ have no propagating modes, yet the
background electric holonomy $\oint \d\sigma A_{\sigma}$ and the circumference
of the world $\oint \d X\, \sqrt{G_{XX}}$ are gauge-invariant observables
when $X$ is compactified.
%
\item{3)} There are indications that understanding
special state correlators would aid in the search for a model with both the
black hole mass and the cosmological constant turned on.

\noindent For these reasons the ``special states'' have been the subject
of intense investigation for the past year and a half.
Sadly, some of these investigations have been rather misguided.

When we compute BRST cohomology, we must pay proper attention to the boundary
conditions of the fields representing BRST cohomology. In
electromagnetism in four dimensions, for example,
BRST cohomology will be represented by
plane-wave states of the gauge field $A^\mu\sim \xi^\mu\, \ee{i k\cdot x}$,
$k^2=k\cdot \xi=0$, representing transverse photons. Of course, $k$ is real
because we want only to consider plane-wave normalizable states. In addition
there are other BRST invariant field configurations which are not plane-wave
normalizable. For example, in $1+1$ electrodynamics on $\IR^2$ we can work in
$A_1=0$ gauge, but then $A_0=E x$ for $E$ constant is not normalizable. This
corresponds to the Coulomb force. We should therefore distinguish the {\it
scattering cohomology} representing states for which one can scatter and
compute an $S$-matrix, from the {\it background cohomology} which represents
gauge-invariant global information which cannot be changed by small wavelike
field perturbations.

This discussion applies to the 2D string.
As we have seen, when rotating the coordinate $X$ to Minkowskian
time, the primary matter fields have negative conformal weight.
Thus, since $\omega$ must be real to provide plane wave normalizable
incoming and outgoing wavefunctions, the {\it only\/} BRST
cohomology classes in the Minkowskian theory with $\omega > 0$
are those in \minkvrtx.
This reasoning breaks down for the case of zero $t$-momentum.
On the other hand before looking for the effects of special
state operators like
\eqn\othsst{c \bar{c}\,\CP_{0,r}(\p^*t)\,\overline{\CP}_{0,r}(\pb^*t)
\,\ee{\sqrt{2}(1\pm r) \phi}\ ,}
(where the Seiberg bound implies we must take $1-r$), we must require that the
wavefunctions in question do not change the asymptotic behavior of the
lagrangian of the theory. In fact this is only the case for the operator
$\p t\,\pb t$. The
other states have non-normalizable wavefunctions and thus belong to the
background cohomology groups. We cannot form well-defined wavepackets for them
and they will not be changed by scattering processes since such processes
involve wavepacket normalizable quanta from the scattering cohomology.

The special states are very interesting for string theory, but they have no
place in the wall $S$-matrix of the 2D string. To paraphrase a warning to
previous generations \sidney : Those who look for special states in the
singularities of the $c=1$ $S$-matrix are like the man who settled in
Casablanca for the waters. They were misinformed.

The situation is rather more confused for the bulk--scattering
matrix described in chapt.~\svoccm.

\subsec{Bosonic String Amplitudes and the ``$c>1$ problem''}
\subseclab\sbsap

In this section we consider some of the ``tachyonic'' divergences that occur
in bosonic string theories.
\smallskip

\noindent
{\it First Description\/}
\par\nobreak
Let us return to the operator formalism description of string
amplitudes.
In general, the amplitudes $\CA_{h,n}$ are meaningless because of the
singularities of the string density $\Omega$ on the boundaries of moduli space.
A traditional way of avoiding this problem has been the introduction
of supersymmetry. An alternative way around the problem is provided
by low dimensional string theory\rDS, since in low dimensions the
tachyon (which causes the divergences) becomes massless or massive
as we have seen in \tachact.
We can see how this comes about by considering the
one-loop partition function in the example of
a general linear dilaton background (i.e.\ non-zero $Q$ in \solnii)
coupled to some matter conformal field theory $\CC$,
\eqn\oneloopi{
\Omega_{1,0}\sim \d \tau\, \eta^2 \wedge \d \bar{\tau}\, \bar{\eta}^2\
Z_{{\rm Liouville}\otimes \CC}(q,\bar q)\ .}
(Note the leading $\eta^2\bar{\eta}^2$ is from the ghosts.)
The behavior of the partition function as $q\to 0$, which accounts
for the tachyon divergences of the theory, is obtained by writing
the partition function
as a sum over eigenstates of $L_0,\overline L_0$:
\eqn\eigns{
Z_{{\rm Liouville}\otimes \CC}
=\sum_i\int_0^\infty \d E\, f_i(E)\, (q\bar{q})^{\ha E^2
+{1\over8}Q^2+ \Delta_i -26/24}\ ,}
where $f_i(E)$ represents the density of Liouville states in \curtthr.
Including also the leading $(q\bar q)^{2/24}$ from the ghosts in \oneloopi,
we arrive at the condition \refs{\nati,\kutseib}\ for
{\it no\/} tachyonic divergences:
\eqn\notachs{
{\rm min}_{\Delta_i\in\CC}
\Bigl\{\half E^2 +{\textstyle{1\over 8}}Q^2 + \Delta_i -1\Bigr\}\geq 0
\quad\Longrightarrow\quad
c_{\rm eff}(\CC)\equiv  c-24\,{\rm min}\Delta_i \leq 1\ .}
{}From this point of view, we see that the problem is not necessarily that
$c>1$ {\it per se\/}, but is rather an issue involving the value of $c$
together with the spectrum of the theory.

The condition \notachs\
is of course only a necessary condition. We should also worry
about the existence of divergences when operators approach each other.
In this case the softening of the Liouville operator product
expansion discussed above explains the lack of divergences
on the boundaries of punctured moduli space. In particular,
if we look at the operator product of two dressed matter
primaries $\Phi_1\, \ee{\alpha \phi}$ and $\Phi_2\, \ee{\beta\phi}$,
then from \softope\ we have
\eqn\sftopeii{\eqalign{
&\Phi_1\ee{\alpha \phi}(z,\zb)\, \Phi_2\ee{\beta \phi}(w,{\bar{w}})\cr
&\qquad\sim\sum_{\Delta_X} \int_0^\infty
\d E\ c_{1,2,(X,E)}\, |z-w|^{2(\ha E^2+{1\over 8}Q^2 + \Delta_X -2)}
\ \Phi_{\Delta_X}\, V_E(w,\bar{w})\cr}}
(where $c_{1,2,(X,E)}$ is the coefficient of the field $\Phi_{\Delta_X}$
and its
gravitational dressing $V_E(w,\bar{w})$ in the operator product expansion of
the two above operators).
The worst singularity at $z=w$ comes from the contribution
near $E=0$,
$$ {1\over |z-w|^2} |z-w|^{{1\over12}(1-c_{\rm eff}(\CC))}\ ,$$
and is integrable when the condition \notachs\ is satisfied.
(The case $c_{\rm eff}(\CC)=1$ is a borderline case. In the $c=1$
model, it turns out that $c_{1,2,E}\to 0$ as $E\to 0$.)

Based on these two examples, we may guess that all bosonic
string amplitudes in fact do exist when \notachs\ is satisfied.
The matrix model approach to 2D string theory has the great virtue of
confirming this,
and moreover gives
an infinite dimensional space of background perturbations.

\smallskip\noindent{\it Second Description\/}\par\nobreak
We can also describe these divergences from
the point of view of the spacetime theory by interpreting
the norm of the plumbing fixture coordinate $q$ as
$|q|=\ee{-s}$, where $s$ is a proper time coordinate such
as introduced following \ptptpth\ for the field theory propagator.
{}From this point of view, we see that the divergences are due to
on-shell tachyons and massless particles. When \notachs\
is satisfied as a strict inequality, we see that the amplitudes
are finite because only zero-momentum massive particles
flow. As usual, the massless particles present a special
case at $c=1$, but they are derivatively coupled.

\ifig\fexpws{The case of the exploding worldsheet. Since every order in
perturbation theory adds a hole to the surface, this is
an overly optimistic rendering. Summing up such a perturbation
expansion, the worldsheet on scales larger than the cutoff is
all holes \nati.}{\epsfbox{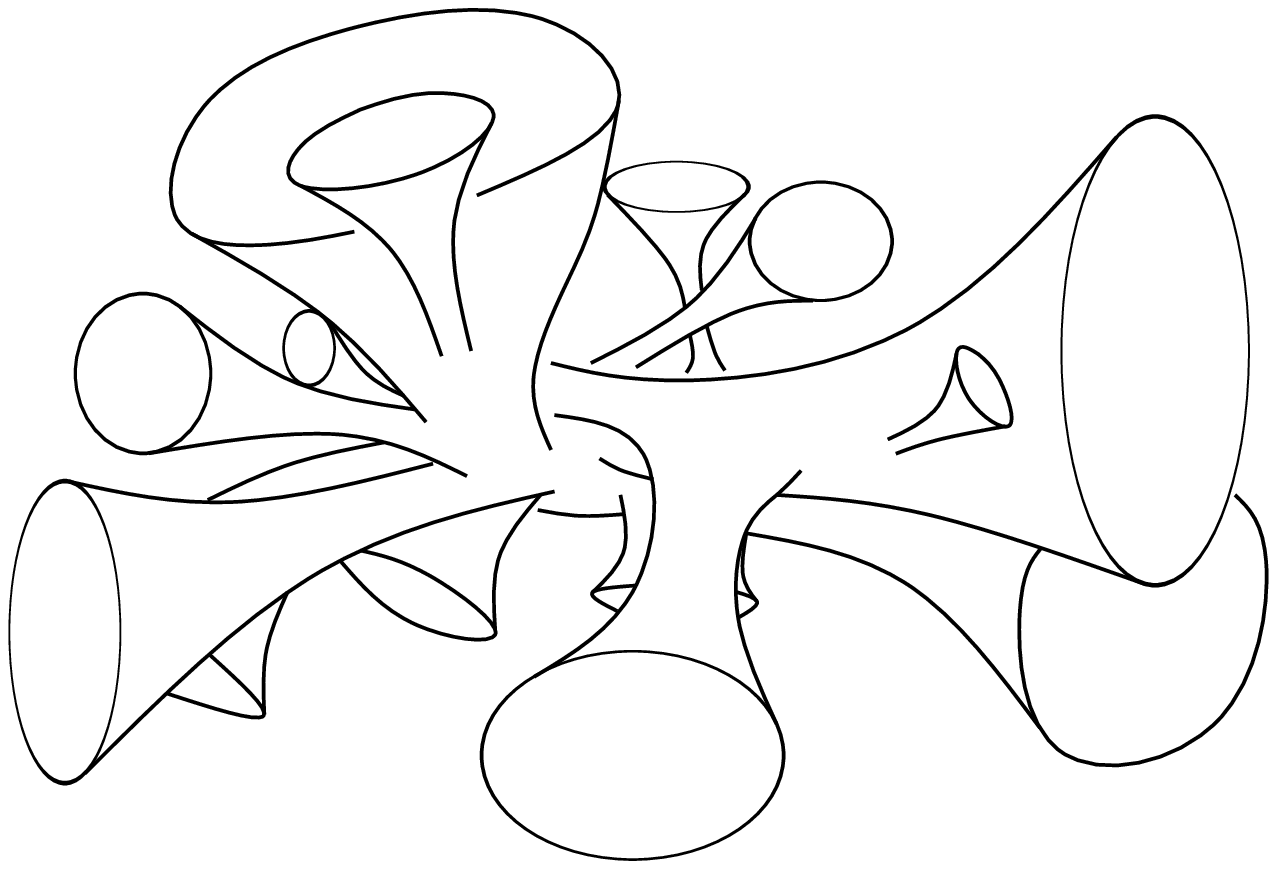}}

\smallskip\noindent{\it Third Description\/} \nati\par\nobreak
We may also consider the above phenomenon from the worldsheet
point of view.
We consider the Liouville theory coupled to some conformal
field theory $\CC$ such that the total central charge is 26.
The conformal field theory $\CC$ is assumed to have a spectrum
bounded from below: that is, we are considering strings
in Euclidean space.
In general we expect Euclidean propagators in Liouville theory to
have the form
\eqn\genprofrm{\int \d E\, {f(E)\over E^2+p^2+m^2}
\, \psi\dup_E(\ell_1)\,\psi\dup_E(\ell_2)\ ,}
where as explained in sec.~{\it\stdstes\/} we identify
\eqn\keyana{p^2+m^2=\Delta_X + {1-c\over 24}\ .}
Suppose the unit operator flows through the loop and $c>1$. Then
there is a zero in the propagator for $E$ real. That is, there
exists an on-shell, normalizable (macroscopic) state in Euclidean space.
As in 1D, we should suspect that there are tachyons in the theory.
Recalling the semiclassical Liouville pictures discussed in chapt.~\sbrlt,
the troubles caused by these states have a graphical worldsheet illustration.
Insertion of an operator dressed by a macroscopic Liouville state
is not a local disturbance to the surface: it creates a macroscopic hole and
tears the surface apart.
In any lattice description of a $c>1$ model, unless we fine-tune
there will be nonzero couplings to the operator that creates
the on-shell macroscopic state whose existence we have
established. In particular, using the KPZ dressing formulae of
sec.~{\it\spia\/}, we see that the cosmological constant operator
itself becomes a macroscopic state.
Bringing down any such operators from the exponential in a perturbative
expansion of the path integral,
we see that the typical resulting ``worldsheet'' would look as depicted in
\fexpws. Evidently a worldsheet description of the physics is no longer
most appropriate.  Once more, the condition that
would prevent this explosion is \notachs.

\newsec{Discretized surfaces, matrix models, and the continuum limit}
\seclab\sDsMmCl

Now that we have some idea of the physics we are looking for, we will
study the  ``experimental'' results of the matrix model.
The next four chapters are devoted to defining the
continuum limit for the models of $c<1$ matter coupled to
gravity associated with the one matrix model. We mention
matrix chains briefly.
We will emphasize both the role of macroscopic loops and also
the fermionic formulation of the matrix model, which
lies at the heart of the exactly solvable nature of these models.

\subsec{Discretized surfaces}
\subseclab\sdsmmcl

We begin by considering a ``$D=0$ dimensional string theory'', i.e.\ a pure
theory of surfaces with no coupling to additional ``matter'' degrees of freedom
on the string worldsheet. This is equivalent to the propagation of strings in a
non-existent embedding space. For partition function we take
\eqn\eZdo{Z=\sum_h\int\CD g\ \e{-\beta A + \gamma \chi}\ ,}
where the sum over topologies is represented by the summation over $h$, the
number of handles of the surface, and the action consists of couplings to the
area $A=\int\sqrt g$, and to the Euler character
$\chi={1\over4\pi}\int\sqrt g\,R=2-2h$.

\ifig\frandtri{A piece of a random triangulation of a surface.
Each of the triangular faces is dual to a three point vertex of a quantum
mechanical matrix model.}{\epsfxsize3.25in\epsfbox{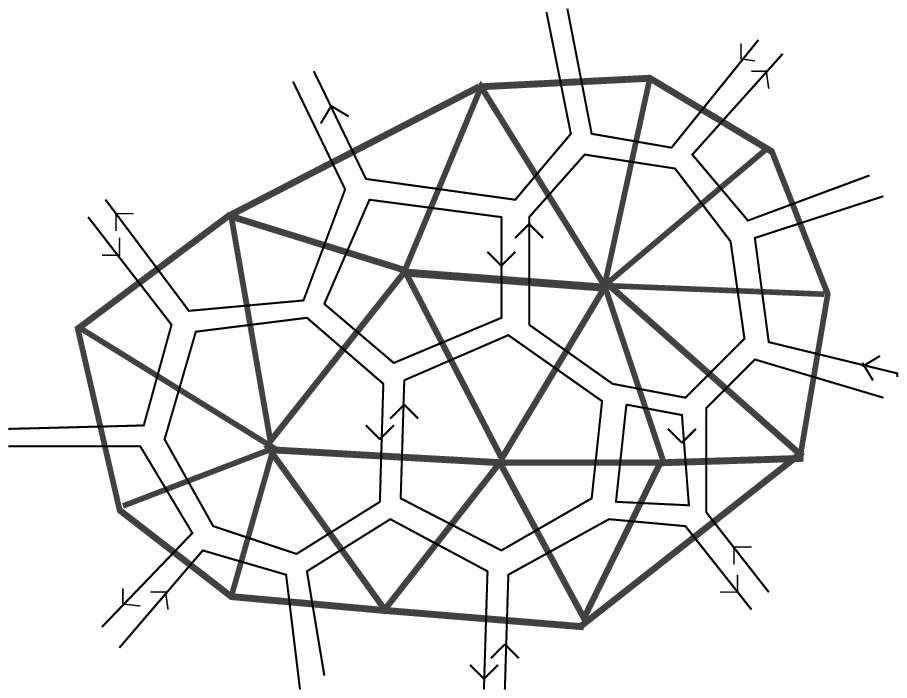}}

The integral $\int\CD g$ over the metric on the surface in \eZdo\ is
difficult to calculate in general. The most progress in the continuum has
been made via the Liouville approach which we briefly reviewed in
chapt.~\sqliouv.
If we discretize the surface, on the other hand, it turns out that \eZdo\
is much easier to calculate, even before removing the finite cutoff. We
consider in particular a ``random triangulation'' of the surface
\rDavidetal, in which the surface is constructed from triangles, as in
\frandtri. The triangles are designated to be equilateral,\foot{We point out
that this constitutes a basic difference from the Regge calculus, in which
the link lengths are geometric degrees of freedom. Here the geometry is
encoded entirely into the coordination numbers of the vertices.
This restriction of degrees of freedom roughly corresponds to fixing a
coordinate gauge, hence we integrate only over the gauge-invariant moduli
of the surfaces.} so that
there is negative (positive) curvature at vertices $i$ where the number
$N_i$ of incident triangles is more (less) than six, and zero curvature
when $N_i=6$. The summation over all such random triangulations is thus
the discrete analog to the integral $\int \CD g$ over all possible geometries,
\eqn\ediscr{\sum_{{\rm genus}\ h}\ \int \CD g \quad
\to \ \sum_{\scriptstyle\rm random \atop \scriptstyle\rm triangulations}\ .}

The discrete counterpart to the
infinitesimal volume element $\sqrt g$ is $\sigma_i=N_i/3$,
so that the total area $|S|=\sum_i \sigma_i$ just counts the total number of
triangles, each designated to have unit area. (The factor
of $1/3$ in the definition of $\sigma_i$ is because
each triangle has three vertices and is counted three times.)
The discrete counterpart to the
Ricci scalar $R$ at vertex $i$ is $R_i=2\pi(6-N_i)/ N_i$, so that
$$\int\sqrt g\,R\to \sum_i 4 \pi(1-N_i/6)
=4 \pi(V-\half F)=4 \pi(V-E+F)=4 \pi \chi\ .$$
Here we have used the simplicial definition which gives the Euler
character $\chi$ in terms of the total number of vertices, edges, and
faces $V$, $E$, and $F$ of the triangulation (and we have used the
relation $3F=2E$ obeyed by triangulations of surfaces, since each
face has three edges each of which is shared by two faces).

In the above, triangles do not play an essential role and may be replaced
by any set of polygons. General random polygonulations of surfaces
with appropriate fine tuning of couplings may, as we shall see, have more
general critical behavior, but can in particular always reproduce the
pure gravity behavior of triangulations in the continuum limit.

\subsec{Matrix models}
\subseclab\ssMmdls

We now demonstrate how the integral over geometry in \eZdo\ may be
performed in its discretized form as a sum over random triangulations. The
trick is to use a certain matrix integral as a generating functional for
random triangulations. The essential idea goes back to work \rthooft\ on
the large $N$ limit of QCD, followed by work on the saddle point approximation
\rBIPZ.

We first recall the (Feynman) diagrammatic expansion of the (0-dimensional)
field theory integral.
%
\eqn\esft{\int_{-\infty}^\infty
{\d\ph\over\sqrt{2\pi}}\, \e{-\ph^2/2+ \lambda\ph^4/4!}\ ,}
where $\ph$ is an ordinary real number.\foot{The integral is understood to
be defined by analytic continuation to negative $\lambda$.} In a formal
perturbation series in $\lambda$, we would need to evaluate integrals such as
\eqn\epel{{\lambda^n\over n!}
\int_\ph\e{-\ph^2/2}\left({\ph^4\over4!}\right)^n\ .}
Up to overall normalization we can write
\eqn\essft{\int_\ph \e{-\ph^2/2}\ph^{2k}=
\left.{\del^{2k}\over\del J^{2k}}\int_\ph\e{-\ph^2/2+J\ph}\right|_{J=0}
=\left.{\del^{2k}\over\del J^{2k}}\,\e{J^2/2}\right|_{J=0}\ .}
Since ${\del\over\del J}\ee{J^2/2}=J \ee{J^2/2}$, applications of
$\del/\del J$ in the above need to be paired so that any factors of $J$
are removed before finally setting $J=0$. Therefore if we represent each
``vertex'' $\lambda\ph^4$ diagrammatically as a point with four emerging
lines (see fig.~\the\figno b), then \epel\ simply counts the number of ways
to group
such objects in pairs. Diagrammatically we represent the possible pairings
by connecting lines between paired vertices. The connecting line is known
as the propagator $\langle\ph\,\ph\rangle$ (see fig.~\the\figno a) and the
diagrammatic rule we have described for connecting vertices in pairs is
known in field theory as the Wick expansion.

$$\vbox{\hrule height .7pt width 30pt\vskip30pt
\hbox{\quad(a)}}\qquad\qquad\qquad
\vbox{\hbox{\hskip20pt\vrule width .7pt height 40pt}\vskip-20pt\hrule width
40pt height .7pt\vskip30pt\hbox{\quad\ (b)}}$$
\vglue5pt\nobreak
\centerline{\footnotefont{\bf Fig.~\the\figno:}
(a) the scalar propagator.   (b) the scalar four-point vertex.}
\bigbreak\xdef\fscpv{fig.~\the\figno}\global\advance\figno by1

When the number of vertices $n$ becomes large, the allowed diagrams begin
to form a mesh reminiscent of a 2-dimensional surface. Such diagrams
do not yet have enough structure to specify a Riemann surface. The
additional structure is given by widening the propagators to ribbons (to
give so-called ``fat'' graphs). From the standpoint of \esft,
the required extra structure is given by replacing the scalar
$\ph$ by an $N\times N$ hermitian matrix $M^i{}_j$. The analog of \essft\
is given by adding indices and traces:
\eqn\esmft{\eqalign{\int_M\e{-\tr M^2/2} M^{i_1}{}_{j_1}\cdots M^{i_n}{}_{j_n}
&=\left.{\del\over\del J^{j_1}{}_{i_1}}
\cdots{\del\over\del J^{j_n}{}_{i_n}} \e{-\tr M^2/2+\tr J M}\right|_{J=0}\cr
&=\left.{\del\over\del J^{j_1}{}_{i_1}}
\cdots{\del\over\del J^{j_n}{}_{i_n}}\, \e{\tr J^2/2}\right|_{J=0}\ ,\cr}}
where the source $J^i{}_j$ is as well now a matrix. The measure in \esmft\ is
the invariant
$\d M=\prod_i\d M^i{}_i\,\prod_{i<j}\d{\rm Re} M^i{}_j\,\d{\rm Im} M^i{}_j$,
and the normalization is such that $\int_M \ee{-\tr M^2/2}=1$.
To calculate a quantity such as
\eqn\empel{{\lambda^n\over n!}\int_M \e{-\tr M^2/2}(\tr M^4)^n\ ,}
we again lay down $n$ vertices (now of the type depicted in fig.~\the\figno b),
and connect the legs with propagators $\langle
M^i{}_j\,M^k{}_l\rangle=\delta^i_l\,\delta^k_j$ (fig.~\the\figno a).
The presence of upper and lower matrix indices is represented in
fig.~\the\figno\ by the
double lines\foot{This is the same notation employed in the large $N$ expansion
of QCD \rthooft.} and it is understood that the sense of the arrows is to be
preserved when linking together vertices. The resulting diagrams are similar to
those of the scalar theory, except that each external line has an associated
index $i$, and each internal closed line corresponds to a summation over an
index $j=1,\ldots,N$. The ``thickened'' structure is now sufficient to
associate a Riemann surface to each diagram, because the closed internal loops
uniquely specify locations and orientations of faces.

\font\bigarrfont=cmsy10 scaled\magstep 3
\def\extarr{\mathord-\mkern-6mu}
\def\vuline{\raise7.5pt\hbox{\textfont2\bigarrfont$\uparrow$}\hskip-4.75pt
\hbox{\vrule width.7pt depth -18 pt height 27pt}}
\def\vdline{\raise15pt\hbox{\textfont2\bigarrfont$\downarrow$}\hskip-4.75pt
\vrule width.7pt depth -4 pt height 27pt}
$$\hbox{\vbox{\hbox{\textfont2\bigarrfont$\extarr
\mathord\rightarrow\mkern-6mu\extarr\extarr$}
\vskip-10pt\hbox{$\textfont2\bigarrfont\extarr\extarr
\mathord\leftarrow\mkern-6mu\extarr$}\vskip21pt\hbox{\quad\ (a)}}
\qquad\qquad\qquad
\vbox{\hbox{$\textfont2\bigarrfont\mathord\rightarrow\mkern-6mu\extarr
\vuline\hskip2.5pt\vdline\mkern-1mu\mathord\rightarrow\mkern-6mu\extarr$}
\vskip-10pt
\hbox{$\textfont2\bigarrfont\extarr\mathord\leftarrow\mkern-6mu
\lower23pt\hbox{\vuline}\hskip2.5pt
\lower23pt\hbox{\vdline}\mkern-2mu\extarr\mathord\leftarrow$}
\vskip7pt\hbox{$\,$\qquad(b)}}}$$
\nobreak
\centerline{\footnotefont{\bf Fig.~\the\figno:}
(a) the hermitian matrix propagator.
(b) the hermitian matrix four-point vertex.}
\bigbreak\xdef\fmapv{fig.~\the\figno}\global\advance\figno by1

To make contact with the random triangulations discussed earlier,
we consider the diagrammatic expansion of the matrix integral
\eqn\ecmm{\e{Z}=\int\d M\ \e{-\ha\tr M^2+{g\over\sqrt N}\tr M^3}}
(with $M$ an $N\times N$ hermitian matrix, and the integral again
understood to be defined by analytic continuation in the coupling $g$.)
The term of order $g^n$ in a power series expansion counts the number of
diagrams constructed with $n$ 3-point vertices. The dual to such a diagram (in
which each face, edge, and vertex is associated respectively to a dual
vertex, edge, and face) is identically a random triangulation inscribed on
some orientable Riemann surface (\frandtri). We see that the matrix integral
\ecmm\ automatically generates all such random triangulations.\foot{Had we used
real symmetric matrices rather than the hermitian matrices $M$, the two indices
would be indistinguishable and there would be no arrows in the propagators and
vertices of \fmapv. Such orientationless vertices and propagators generate an
ensemble of both orientable and non-orientable surfaces, and have
been studied, e.g., in \rnonor.}

Since each triangle has unit area, the area of the
surface is just $n$. We can thus make formal identification with \eZdo\
by setting $g=\ee{-\beta}$. Actually the matrix integral generates both
connected and disconnected surfaces, so we have written $\ee Z$ on the
left hand side of \ecmm. As familiar from field theory, the exponential of
the connected diagrams generates all diagrams, so $Z$ as defined above
represents contributions only from connected surfaces. We see that the
{\it free energy\/} from the matrix model point of view is actually the
{\it partition function\/} $Z$ from the 2d gravity point of view.

There is additional information contained in $N$, the size of the matrix.
If we change variables $M\to M\sqrt N$ in \ecmm, the matrix action becomes
$N\,\tr(-\half\tr M^2+g\tr M^3)$, with an overall factor of
$N$.\foot{Although we could as well rescale $M\to M/g$ to pull out an overall
factor of $N/g^2$, note that
$N$ remains distinguished from the coupling $g$ in the model since it enters
as well into the traces via the $N\times N$ size of the matrix.} This
normalization makes it easy to count the power of $N$ associated to any
diagram. Each vertex contributes a factor of $N$, each propagator (edge)
contributes a factor of $N\inv$ (because the propagator is the inverse of
the quadratic term), and each closed loop (face) contributes a factor of
$N$ due to the associated index summation.
Thus each diagram has an overall factor
\eqn\efN{N^{V-E+F}=N^\chi=N^{2-2h}\ ,}
where $\chi$ is the Euler character of the surface associated to the diagram.
We observe that the value $N=\ee{\gamma}$ makes contact with the coupling
$\gamma$ in \eZdo.  In conclusion, if we take
$g=\ee{-\beta}$ and $N=\ee{\gamma}$, we can formally identify the continuum
limit of the partition function $Z$ in \ecmm\ with the $Z$ defined in \eZdo.
The metric for the discretized formulation is not smooth, but one can imagine
how an effective metric on larger scales could arise after averaging over
local irregularities. In the next section, we shall see explicitly how this
works.

(Actually \ecmm\ automatically calculates \eZdo\ with the measure factor
in \ediscr\ corrected to $\sum_S {1\over|G(S)|}$, where $|G(S)|$ is the
order of the (discrete) group of symmetries of the triangulation $S$. This
is familiar from field theory where diagrams with symmetry result in an
incomplete cancellation of $1/n!$'s such as in \epel\ and \empel. The
symmetry group $G(S)$ is the discrete analog of the isometry group of a
continuum manifold.)

The graphical expansion of \ecmm\ enumerates graphs as shown in \frandtri,
where the triangular faces that constitute the random triangulation are dual to
the 3-point vertices. Had we instead used 4-point vertices as in \fmapv b, then
the dual surface would have square faces (a ``random squarulation'' of the
surface), and higher point vertices $(g_k/N^{k/2-1})\tr M^k$ in the matrix
model would result in more general ``random polygonulations'' of surfaces.
(The powers of $N$ associated with the couplings are chosen so that the
rescaling $M\to M\sqrt N$ results in an overall factor of $N$ multiplying the
action. The argument leading to \efN\ thus remains valid, and the power of
$N$ continues to measure the Euler character of a surface constructed from
arbitrary polygons.) The different possibilities for generating vertices
constitute additional degrees of freedom that can be realized as the coupling
of 2d gravity to different varieties of matter in the continuum limit.

\subsec{The continuum limit}

{}From \efN, it follows that we may expand $Z$ in powers of $N$,
\eqn\elne{Z(g)
=N^2 Z_0(g)+Z_1(g)+N^{-2}Z_2(g) + \ldots=\sum N^{2-2h} Z_h(g)\ ,}
where $Z_h$ gives the contribution from surfaces of genus $h$.
In the conventional large $N$ limit, we take $N\to\infty$ and only $Z_0$,
the planar surface (genus zero) contribution, survives. $Z_0$ itself may be
expanded in a perturbation series in the coupling $g$, and for large order $n$
behaves
as (see \rBIZ\ for a review)
\eqn\eloln{Z_0(g)\sim \sum_n n^{\Gamma_{\rm str}-3} (g/\gc)^n\sim
(\gc-g)^{2-\Gamma_{\rm str}}\ .}
These series thus have the property that they diverge as $g$ approaches
some critical coupling $\gc$. We can extract the continuum limit of these
surfaces by tuning $g\to\gc$. This is because the expectation value of the area
of a surface is given by
$$\langle A\rangle=\langle n\rangle
={\del\over\del g}\ln Z_0(g)\sim {1\over g-\gc}$$
(recall that the area is proportional to the number of vertices $n$, which
appears as the power of the coupling in the factor $g^n$ associated to each
graph).
As $g\to\gc$, we see that $A\to\infty$ so that we may rescale the area of
the individual triangles to zero, thus giving a continuum surface with finite
area. Intuitively, by tuning the coupling to the point where the
perturbation series diverges, the integral becomes dominated by diagrams
with infinite numbers of vertices, and this is precisely what we need to
define continuum surfaces.

There is no direct proof as yet that this procedure for defining continuum
surfaces is ``correct,'' i.e.\ that it coincides with the continuum
definition \eZdo. We are able, however, to compare properties of the
partition function and correlation functions calculated by matrix model
methods with those (few) properties that can be calculated directly in the
continuum, as reviewed in preceding chapters.
This gives implicit confirmation that
the matrix model approach is sensible and gives reason to believe other
results derivable by matrix model techniques (e.g.\ for higher genus) that
are not obtainable at all by continuum methods.
In sec.~{\it\sdotcl\/}, we shall give a more precise formulation of
what we mean by the continuum limit.

One of the properties of these models derivable via the continuum Liouville
approach is a ``critical exponent'' $\Gamma_{\rm str}$,
defined in terms of the area dependence of the partition function for surfaces
of fixed large area $A$ as
\eqn\elpoa{Z(A)\sim A^{(\Gamma_{\rm str}-2)\chi/2-1}\ .}
Recall that the
unitary discrete series of conformal field theories is labelled by an integer
$m\ge2$ and has central charge $D=1-6/m(m+1)$
(for a review, see e.g.\ \rcftrev), where the central charge is normalized such
that $D=1$ corresponds to a single free boson. If we couple conformal field
theories with these fractional values of $D$ to 2d gravity, we see
from \egam\ the continuum Liouville theory prediction for the exponent
$\Gamma_{\rm str}$
\eqn\epbl{\Gamma_{\rm str}
={1\over12}\bigl(D-1-\sqrt{(D-1)(D-25)}\,\bigr)=-{1\over m}\ .}
The case $m=2$, for example, corresponds to $D=0$
and hence $\Gamma_{\rm str}=-\ha$ for
pure gravity. The next case $m=3$ corresponds to $D=1/2$, i.e.\ to a
1/2--boson or fermion. This is the conformal field theory of the critical Ising
model, and we learn from \epbl\ that the Ising model coupled to 2d gravity has
$\Gamma_{\rm str}=-{1\over3}$.

In chapt.~\sMMTI\ we shall present the solution to the matrix model formulation
of the problem, and the value
of the exponent $\Gamma_{\rm str}$ provides a coarse means
of determining which specific continuum model results from taking the continuum
limit of a particular matrix model.
Indeed the coincidence of $\Gamma_{\rm str}$ and
other scaling exponents (defined in chapt.~\sqliouv) calculated from the
two points of view were originally the only evidence that the continuum limit
of matrix models was a suitable definition for the continuum problem of
interest. Subsequently, the simplicity of matrix model results for
correlation functions has spurred a rapid evolution of continuum Liouville
technology so that as well many correlation functions can be computed in both
approaches and are found to coincide.\foot{In particular,
following the confirmation that the matrix model approach reproduced the
scaling results of \rKPZ, some 3-point couplings for order parameters at genus
zero were calculated in \rkade\ from the standpoint of ADE face models on
fluctuating lattices. The connection to KdV (to be
reviewed in sec.~{\it\sKdV\/} here) was made in \rD,
and then general correlations of order parameters (not yet known
in the continuum) were calculated in \rDIFK.
Using techniques described in sec.~{\it\slcfac\/},
continuum calculations of the correlation
functions (when they can be done) have been found to
agree with the matrix model (for a review, see \rDGZ).
For $D=1$, the matrix model
approach of \refs{\bkz\gzj\parisi{--}\GMil}\ was used in \refs{\moore,\msi}\
(also \refs{\rGKn,\gki\rkco}) to calculate a variety of correlation functions.
These were also calculated in the collective field approach
\refs{\dj\wadia\joesea\rColl{--}\jevscatii}\ where up to 6-point amplitudes
were derived, and found to be in agreement with
the Liouville results of \kdf.}

\subsec{A first look at the double scaling limit}
\subseclab\ssfldsl

Thus far we have discussed the naive $N\to\infty$ limit which retains only
planar surfaces. It turns out that the successive coefficient functions
$Z_h(g)$ in \elne\ as well diverge at the same critical value of the
coupling $g=\gc$ (this should not be surprising since the divergence of the
perturbation series is a local phenomenon and should not depend on global
properties such as the effective genus of a diagram).
As we saw in \eZainf, for the higher genus contributions
\eloln\ is generalized to
\eqn\elolnhg{Z_h(g)\sim \sum_n n^{(\Gamma_{\rm str}-2)\chi/2-1} (g/\gc)^n\sim
(\gc-g)^{(2-\Gamma_{\rm str})\chi/2}\ .}
We see that the contributions from higher genus ($\chi<0$)
are enhanced as $g\to\gc$.
This suggests that if we take the limits $N\to\infty$ and $g\to\gc$ not
independently, but together in a correlated manner, we may compensate the
large $N$ high genus suppression with a $g\to\gc$ enhancement. This would
result in a coherent contribution from all genus surfaces \refs{\rDS{--}\rGM}.

To see how this works explicitly,
we write the leading singular piece of the $Z_h(g)$ as
$$Z_h(g)\sim f_h(g-\gc)^{(2-\Gamma_{\rm str})\chi/2}\ .$$
Then in terms of
\eqn\ekappa{\kappa\inv\equiv N(g-\gc)^{(2-\Gamma_{\rm str})/2}\ ,}
the expansion \elne\ can be rewritten\foot{Strictly speaking the first two
terms here have additional non-universal pieces that need to be subtracted
off.}
\eqn\elner{Z=\kappa^{-2}f_0+f_1+\kappa^2 f_2+\ldots
=\sum_h \kappa^{2h-2}\,f_h\ .}
The desired result is thus obtained by taking the limits $N\to\infty$,
$g\to\gc$ while holding fixed the ``renormalized'' string coupling $\kappa$
of \ekappa. This is known as the ``double scaling limit.''
\elner\ is an asymptotic expansion for $\kappa\to0$. In
secs.~{\it\sstagpf, \sKdV\/} below, we show how to find a function
$Z(\kappa)$ with identically that asymptotic expansion.

\newsec{Matrix Model Technology I: Method of Orthogonal Polynomials} 
\seclab\sMMTI

The large $N$ limit of the matrix models considered here was originally solved
by saddle point methods in \rBIPZ. In this chapter we shall instead present the
orthogonal polynomial solution to the problem (\rBIZ\ and references therein)
since it extends readily to
subleading order in $N$ (higher genus corrections).

\subsec{Orthogonal polynomials}
\subseclab\ssOP

In order to justify the claims made at the end of sec.~{\it\ssfldsl\/},
we introduce some formalism to solve the matrix models.
We begin by rewriting the partition function \ecmm\ in the form
\eqn\egpf{\e{Z}=\int \d M\ \e{-\tr V(M)}=\int\prod_{i=1}^N
\d \lambda_i\, \Delta^2(\lambda)\ \e{-\sum_i V(\lambda_i)}\ ,}
where we now allow a general polynomial potential $V(M)$. In \egpf, the
$\lambda_i$'s are the $N$ eigenvalues of the hermitian matrix $M$, and
\eqn\eVand{\Delta(\lambda)=\prod_{i<j}(\lambda_j-\lambda_i)}
is the Vandermonde determinant.\foot{\egpf\ may be derived via the usual
Fadeev-Popov method: Let $U_0$ be the unitary matrix such that
$M=U_0^\dagger \Lambda' U_0$, where $\Lambda'$ is a diagonal matrix with
eigenvalues $\lambda'_i$. The right hand side of \egpf\ follows by
substituting the definition
$1=\int \prod_i\d \lambda_i\,\d U\,\delta(U M U^\dagger-\Lambda)
\,\Delta^2(\lambda)$ (where $\int\d U\equiv1$).
We first perform the integration over $M$, and then $U$ decouples due to
the cyclic invariance of the trace so the integration over $U$ is trivial,
leaving only the integral over the eigenvalues $\lambda_i$ of $\Lambda$.
To determine $\Delta(\lambda)$, we note that only the infinitesimal
neighborhood $U=(1+T)U_0$ contributes to the $U$ integration, so that
$$1=\int \prod_{i=1}^N\d \lambda_i\,\d U\,
\delta^{N^2}\!\bigl(U M U^\dagger-\Lambda\bigr)\,\Delta^2(\lambda)
=\int\d T\ \delta^{N(N-1)}\bigl([T,\Lambda']\bigr)\Delta^2(\lambda')\ .$$
Now $[T,\Lambda']_{ij}=T_{ij}(\lambda'_j-\lambda'_i)$, so \eVand\ follows
(up to a sign) since the integration $\d T$ above is over
real and imaginary parts of the off-diagonal $T_{ij}$'s.}
Due to antisymmetry in interchange of any two eigenvalues, \eVand\ can be
written $\Delta(\lambda)=\det\,\lambda^{j-1}_i$
(where the normalization is determined by comparing leading terms).
In the case $N=3$ for example we have
$$(\lambda_3-\lambda_2)(\lambda_2-\lambda_1)(\lambda_3-\lambda_1)=
\det\pmatrix{1&\lambda_1&\lambda_1^2\cr
1&\lambda_2&\lambda_2^2\cr
1&\lambda_3&\lambda_3^2\cr}\ .$$

The now-standard method for solving \egpf\ makes use of an infinite set of
polynomials $P_n(\lambda)$, orthogonal with respect to the measure
\eqn\eopoly{\int_{-\infty}^\infty
\d \lambda\ \ee{-V(\lambda)}\,P_n(\lambda)\,P_m(\lambda)
=h_n\,\delta_{nm}\ .}
The $P_n$'s are known as orthogonal polynomials and are functions of a single
real variable $\lambda$. Their normalization is given by having leading term
$P_n(\lambda)=\lambda^n+\ldots$, hence the constant $h_n$ on the r.h.s.\ of
\eopoly. Due to the relation
\eqn\evand{\Delta(\lambda)=\det\,\lambda^{j-1}_i=
\det\,P_{j-1}(\lambda_i)}
(recall that arbitrary polynomials may be built up by adding
linear combinations of preceding columns, a procedure that leaves the
determinant unchanged),
the polynomials $P_n$ can be employed to solve \egpf. We substitute the
determinant $\det\,P_{j-1}(\lambda_i)=\sum(-1)^{\pi}
\prod_k P_{i_k-1}(\lambda_k)$
for each of the $\Delta(\lambda)$'s in \egpf\
(where the sum is over permutations $i_k$ and $(-1)^{\pi}$ is the parity of
the permutation).  The integrals over individual
$\lambda_i$'s factorize, and due to orthogonality the only contributions are
from terms with all $P_i(\lambda_j)$'s paired. There are $N!$ such terms so
\egpf\ reduces to
\eqn\egrt{\e{Z}
=N!\prod_{i=0}^{N-1}h_i=N!\, h_0^N\,\prod_{k=1}^{N-1}f_k^{N-k}\ ,}
where we have defined $f_k\equiv h_k/h_{k-1}$.

In the naive large $N$ limit (the planar limit), the rescaled index $k/N$
becomes a continuous variable $\xi$ that runs from 0 to 1, and $f_k/N$ becomes
a continuous function $f(\xi)$. In this limit, the partition function
(up to an irrelevant additive constant) reduces to a simple one-dimensional
integral:
\eqn\enlnl{{1\over N^2} Z={1\over N}\sum_k(1-k/N)\ln f_k
\sim\int_0^1\d \xi(1-\xi)\ln f(\xi)\ .}

To derive the functional form for $f(\xi)$, we assume for simplicity that the
potential $V(\lambda)$ in \eopoly\ is even.
Since the $P_i$'s form a complete set of basis vectors in the space of
polynomials, it is clear that $\lambda P_n(\lambda)$ must be expressible as a
linear combination of lower $P_i$'s, $\lambda P_n(\lambda)=\sum_{i=0}^{n+1}
a_i\,P_i(\lambda)$ (with $a_i=h_i\inv\int\ee{-V}\lambda P_n\,P_i$). In fact,
the orthogonal polynomials satisfy the simple recursion relation,
\eqn\elpn{\lambda P_n=P_{n+1}+r_n\,P_{n-1}\ ,}
with $r_n$ a scalar coefficient independent of $\lambda$. This is because any
term proportional to $P_n$ in the above vanishes due to the assumption that the
potential is even, $\int\ee{-V}\lambda\,P_n\,P_n=0$. Terms proportional to
$P_i$ for $i<n-1$ also vanish since $\int \ee{-V}P_n\,\lambda\,P_i=0$ (recall
$\lambda P_i$ is a polynomial of order at most ${i+1}$ so is orthogonal to
$P_n$ for $i+1<n$).

By considering the quantity $P_n\lambda P_{n-1}$ with $\lambda$
paired alternately with the preceding or succeeding polynomial, we derive
$$\int \ee{- V} \,P_n\, \lambda\, P_{n-1}=r_n\,h_{n-1}=h_n\ .$$
This shows that the ratio $f_n=h_n/h_{n-1}$ for this simple case\foot{In
other models, e.g.\ multimatrix models, $f_n=h_n/h_{n-1}$ has a more
complicated dependence on recursion coefficients.} is identically the
coefficient defined by \elpn, $f_n=r_n$.
Similarly if we pair the $\lambda$ in $P_n'\,\lambda\,P_n$ before and
afterwards, integration by parts gives
\eqn\enhn{n h_n=\int \ee{-V}\,P_n'\,\lambda\,  P_n
= \int \ee{-V}\,P_n'\, r_n\, P_{n-1}=r_n\int\ee{-V}\, V'\,P_n\,P_{n-1}\ .}
This is the key relation that will allow us to determine $r_n$.

\subsec{The genus zero partition function}

Our intent now is to find an expression for $f_n=r_n$ and substitute into
\enlnl\ to calculate a partition function.
For definiteness, we take as example the potential
\eqn\epex{\eqalign{&V(\lambda)={1\over 2g}\Bigl(\lambda^2+{\lambda^4\over N}
+b{\lambda^6\over N^2}\Bigr)\ ,\cr
\llap{\rm with\ derivative\quad\qquad}
g &V'(\lambda)
=\lambda+2{\lambda^3\over N}+3b{\lambda^5\over N^2}\ .\cr}}
The right hand side of \enhn\ involves terms of the form
$\int\ee{-V}\, \lambda^{2p-1}\,P_n\,P_{n-1}$. According to \elpn, these may be
visualized as ``walks'' of $2p-1$ steps ($p-1$ steps up and $p$ steps down)
starting at $n$ and ending at $n-1$, where each step down from $m$ to $m-1$
receives a factor of $r_m$ and each step up receives a factor of unity. The
total number of such walks is given by ${2p-1\choose p}$, and each results in a
 final factor of $h_{n-1}$ (from the integral
$\int\ee{-V}\,\,P_{n-1}\,P_{n-1}$) which combines with the $r_n$ to cancel the
$h_n$ on the left hand side of \enhn.
For the potential \epex, \enhn\ thus gives
\eqn\egnpx{g n=r_n+{2\over N}r_n(r_{n+1}+r_n+r_{n-1})+
{3 b\over N^2}(10 \ rrr \ {\rm terms})\ .}
(The 10 $rrr$ terms start with $r_n(r_n^2 +r_{n+1}^2+r_{n-1}^2+\ldots)$
and may be found e.g.\ in \rIYL.)

As mentioned before \enlnl, in the large $N$ limit the index $n$ becomes a
continuous variable $\xi$, and we have
$r_n/N\to r(\xi)$ and $r_{n\pm1}/N\to r(\xi\pm\varepsilon)$,
where $\varepsilon\equiv 1/N$. To leading order in $1/N$, \egnpx\ reduces to
\eqn\egxw{\eqalign{g \xi=r + 6 r^2+30 b r^3&=W(r)\cr
&=\gc+\half W''|_{r=r_c}\bigl(r(\xi)-r_c\bigr)^2+\ldots\ .\cr}}
In the second line, we have expanded $W(r)$ for $r$ near a critical point
$r_c$ at which $W'|_{r=r_c}=0$
(which always exists without any fine tuning of the parameter $b$),
and $\gc\equiv W(r_c)$. We see from \egxw\ that
$$r-r_c\sim(\gc-g \xi)^{1/2}\ .$$
For a general potential $V(\lambda)={1\over2g}\sum_p a_p\,\lambda^{2p}$ in
\epex, we would have
\eqn\eVtoW{ W(r)=\sum_p a_p{(2p-1)!\over (p-1)!^2}\,r^p\ .}

To make contact with the 2d gravity ideas of chapt.~\sDsMmCl, let us
suppose more generally that the leading singular behavior of $f(\xi)$
$\bigl(=r(\xi)\bigr)$ for large $N$ is
\eqn\efx{f(\xi)-f_c \sim (\gc - g \xi)^{-\Gamma_{\rm str}}}
for $g$ near some $\gc$ (and $\xi$ near 1).
(We shall see that $\Gamma_{\rm str}$ in the
above coincides with the critical exponent $\Gamma_{\rm str}$ defined in
\elpoa.)
The behavior of \enlnl\ for $g$ near $\gc$ is then
\eqn\emc{\lbspace\eqalign{{1\over N^2}Z
\sim\int_0^1\d \xi\,(1-\xi)(\gc- g \xi)^{-\Gamma_{\rm str}}
&\sim(1-\xi)(\gc-g \xi)^{-\Gamma_{\rm str}+1}\Big|_0^1
+\int_0^1\d \xi\,(\gc- g \xi)^{-\Gamma_{\rm str}+1}\cr
&\sim(\gc-g)^{-\Gamma_{\rm str}+2}
\sim \sum_n n^{\Gamma_{\rm str}-3}(g/\gc)^n\ .}}
Comparison with \elpoa\ shows that the large area (large $n$) behavior
identifies
the exponent $\Gamma_{\rm str}$ in \efx\ with the critical exponent defined
earlier. We also note that the second derivative of $Z$ with respect to
$x=\gc-g$ has leading singular behavior
\eqn\elsbz{Z''\sim(\gc-g)^{-\Gamma_{\rm str}}\sim f(1)\ .}

{}From \efx\ and \emc\ we see that the behavior in \egxw\ implies a critical
exponent $\Gamma_{\rm str}=-1/2$. From \epbl, we see that this corresponds to
the case $D=0$, i.e.\ to pure gravity. It is natural that pure gravity should
be present for a generic potential. With fine tuning of the parameter $b$ in
\epex, we can achieve a higher order critical point, with
$W'|_{r=r_c}=W''|_{r=r_c}=0$, and hence the r.h.s.\ of \egxw\ would instead
begin with an $(r-r_c)^3$ term. By the same argument starting from \efx, this
would result in a critical exponent $\Gamma_{\rm str}=-1/3$. With a general
potential $V(M)$ in \egpf, we have enough parameters to achieve an $m^{\rm th}$
order critical point \volodya\ at which the first $m-1$ derivatives of $W(r)$
vanish at $r=r_c$. The behavior is then $r-r_c\sim (\gc-g \xi)^{1/m}$ with
associated critical exponent $\Gamma_{\rm str}=-1/m$. As anticipated at the end
of sec.~{\it\ssMmdls\/}, we see that more general polynomial matrix
interactions provide the necessary degrees of freedom to result in matter
coupled to 2d gravity in the continuum limit.

\subsec{The all genus partition function}
\subseclab\sstagpf

We now search for another solution to \egnpx\ and its generalizations that
describes the contribution of all genus surfaces to the partition function
\enlnl. We shall retain higher order terms in $1/N$ in \egnpx\ so that e.g.\
\egxw\ instead reads
\eqn\egxwp{\eqalign{g \xi
&=W(r)+ 2 r(\xi)\bigl(r(\xi+\varepsilon)+r(\xi-\varepsilon)-2r(\xi)\bigr)\cr
&=\gc+\half W''|_{r=r_c}\bigl(r(\xi)-r_c\bigr)^2
+ 2 r(\xi)\bigl(r(\xi+\varepsilon)+r(\xi-\varepsilon)-2r(\xi)\bigr)+\ldots\
.\cr}}
As suggested at the end of sec.~{\it\ssfldsl\/},
we shall simultaneously let $N\to\infty$
and $g\to\gc$ in a particular way. Since $g-\gc$ has dimension [length]$^2$,
it is convenient to introduce a parameter $a$ with dimension length and
let $g-\gc=\kappa^{-4/5}a^2$, with $a\to0$. Our ansatz for a coherent
large $N$ limit will be to take $\varepsilon\equiv 1/N=a^{5/2}$
so that the quantity $\kappa\inv=(g-\gc)^{5/4}N$ remains finite as $g\to\gc$
and $N\to\infty$.

Moreover since the integral \enlnl\ is dominated by
$\xi$ near 1 in this limit, it is convenient to change variables from $\xi$ to
$z$, defined by $\gc-g\xi=a^2 z$.
Our scaling ansatz in this region is $r(\xi)=r_c+a u(z)$.
If we substitute these definitions into \egxw, the leading terms are of order
$a^2$ and result in the relation $u^2\sim z$. To include the higher derivative
terms, we note that
$$r(\xi+\varepsilon)+r(\xi-\varepsilon)-2r(\xi)
\sim \varepsilon^2{\del^2 r\over \del \xi^2}
=a {\del^2\over \del z^2} a u(z)\sim a^2 u''\ ,$$
where we have used
$\varepsilon(\del/\del \xi)=-g a^{1/2}(\del/\del z)$
(which follows from the above change of variables from $\xi$ to $z$).
Substituting into \egxwp, the vanishing of the coefficient of $a^2$ implies
the differential equation
\eqn\eplv{z=u^2-{\textstyle 1\over3}u''}
(after a suitable rescaling of $u$ and $z$).
In \elsbz, we saw that the
second derivative of the partition function
(the ``specific heat'') has leading singular behavior given by
$f(\xi)$ with $\xi=1$, and thus by $u(z)$ for $z=(g-\gc)/a^2=\kappa^{-4/5}$.
The solution to \eplv\ characterizes the behavior of the partition function of
pure gravity to all orders in the genus expansion. (Notice that the leading
term is $u\sim z^{1/2}$ so after two integrations the leading term in $Z$ is
$z^{5/2}=\kappa^{-2}$, consistent with \elner.)

Eq.~\eplv\ is known in the mathematical literature as the
Painlev\'e I equation.
The perturbative solution in powers of $z^{-5/2}=\kappa^2$ takes the form
$u=z^{1/2}(1-\sum_{k=1}u_k z^{-5k/2})$,
where the $u_k$ are all positive.\foot{The first term, i.e.\ the
contribution from the sphere, is dominated by a regular part which has
opposite sign.  This is removed by taking an additional derivative of $u$,
giving a series all of whose terms have the same sign ---
negative in the conventions of \eplv. The other solution, with leading term
$-z^{1/2}$, has an expansion with alternating sign
which is presumably Borel summable, but not physically relevant.}
This verifies for this model the claims made in
eqs.~\eqns{\elolnhg{--}\elner}.
For large $k$, the $u_k$ go asymptotically as $(2k)!$, so the solution for
$u(z)$ is not Borel summable (for a review of these issues
in the context of 2d gravity, see e.g.\ \rGZaplob).
Our arguments in chapt.~\sDsMmCl\ show only that the
matrix model results should agree with 2d gravity order by order in
perturbation theory. How to insure that we are studying nonperturbative gravity
as opposed to nonperturbative matrix models is still an open question.
Some of the constraints that the solution to \eplv\ should satisfy are
reviewed in \rfrdc. In particular it is known that real solutions to \eplv\
cannot satisfy the Schwinger--Dyson (loop) equations for the theory.

In the case of the next higher multicritical point, with $b$ in \egxw\ adjusted
so that $W'=W''=0$ at $r=r_c$, we have
$W(r)\sim \gc+{1\over6}W'''|_{r=r_c}(r-r_c)^3+\ldots$
and critical exponent $\Gamma_{\rm str}=-1/3$.
In general, we take $g-\gc=\kappa^{2/(\Gamma_{\rm str}-2)}a^2$, and
$\varepsilon=1/N=a^{2-\Gamma_{\rm str}}$ so that the combination
\eqn\ecltsc{(g-\gc)^{1-\Gamma_{\rm str}/2}N=\kappa\inv}
is fixed in the limit $a\to0$.
The value  $\xi=1$ now corresponds to $z=\kappa^{2/(\Gamma_{\rm str}-2)}$, so
the string coupling $\kappa^2=z^{\Gamma_{\rm str}-2}$.
The general scaling scaling ansatz is
$r(\xi)=r_c+a^{-2 \Gamma_{\rm str}}u(z)$,
and the change of variables from $\xi$ to $z$
gives $\varepsilon(\del/\del \xi)=-g a^{-\Gamma_{\rm str}}(\del/\del z)$.

For the case $\Gamma_{\rm str}=-1/3$, this means in particular that
$r(\xi)=r_c+a^{2/3}u(z)$, $\kappa^2=z^{-7/3}$, and
$\varepsilon(\del/\del \xi)=-ga^{1/3}{\del\over \del z}$.
Substituting into the large $N$ limit of \egnpx\ gives
(again after suitable rescaling of $u$ and $z$)
\eqn\etmcp{z=u^3-u u'' -\half(u')^2+\alpha\, u''''\ ,}
with $\alpha={1\over10}$. The solution to \etmcp\ takes the form
$u=z^{1/3}(1+\sum_k u_k\,z^{-7k/3})$. It turns out that the coefficients $u_k$
in the perturbative expansion of the solution to \etmcp\ are positive definite
only for $\alpha<{1\over12}$, so the $3^{\rm th}$
order multicritical point does not
describe a unitary theory of matter coupled to gravity. Although
from \epbl\ we see that
the critical exponent $\Gamma_{\rm str}=-1/3$
coincides with that predicted for the
(unitary) Ising model coupled to gravity,
it turns out \refs{\rstaudyl,\rIYL,\rising}\
that \etmcp\ with $\alpha={1\over10}$
instead describes the conformal field theory of the Yang--Lee edge singularity
(a critical point obtained by coupling the Ising model to a particular value of
imaginary magnetic field) coupled to gravity. The specific heat of the
conventional critical Ising model coupled to gravity turns out (see
sec.~{\it\ssIM\/})
to be as well determined by the differential equation \etmcp, but
instead with $\alpha={2\over27}$.

For the general $m^{\rm th}$ order critical point of the potential $W(r)$,
\eqn\egenW{W(r)=g\dup_c-\alpha(r_c-r)^m\ ,}
we have seen that the associated model of matter coupled to gravity has
critical exponent $\Gamma_{\rm str}=-1/m$.
With scaling ansatz $r(\xi)=r_c+a^{2/m}u(z)$, we find leading behavior
$u\sim z^{1/m}$ (and $Z\sim z^{2+1/m}=\kappa^{-2}$ as expected).
The differential equation that results from substituting the double scaling
behaviors given before \etmcp\ into the generalized version of \egnpx\ turns
out to be the $m^{\rm th}$ member of the KdV hierarchy of differential
equations (of which Painlev\'e I results for $m=2$). In the next section, we
shall provide some marginal insight into why this structure emerges.

The one-matrix models reproduce the $(2,2m-1)$ minimal models (in the
nomenclature mentioned after \egam) coupled to quantum gravity.
The remaining $(p,q)$ models coupled to gravity can be realized in terms of
multi-matrix models (to be defined in sec.~{\it\ssmmm}).

\subsec{The Douglas Equations and the KdV hierarchy}
\subseclab\sKdV

We now wish to describe superficially why the KdV hierarchy of differential
equations plays a role in 2d gravity. To this end it is convenient to switch
from the basis of orthogonal polynomials $P_n$ employed in
sec.~{\it\ssOP\/} to a basis of orthonormal polynomials
$\Pi_n(\lambda) = P_n(\lambda)/\sqrt{h_n}$ that satisfy
\eqn\eonP{\int_{-\infty}^\infty\d \lambda\ \ee{-V}\,\Pi_n\,\Pi_m
=\delta_{nm}\ .}
In terms of the $\Pi_n$, eq.~\elpn\ becomes
$$\eqalign{\lambda\Pi_n&=\sqrt{h_{n+1}\over h_n}\,\Pi_n
+r_n\sqrt{h_{n-1}\over h_n}\,\Pi_{n-1}
=\sqrt{r_{n+1}}\,\Pi_{n+1}+\sqrt{r_n}\,\Pi_{n-1}\cr
&=Q_{nm}\,\Pi_m\ .\cr}$$
In matrix notation, we write this as $\lambda\Pi= Q\Pi$,
where the matrix $Q$ has components
\eqn\eQmn{Q_{nm}=\sqrt{r_m}\delta_{m,n+1}+\sqrt{r_n}\delta_{m+1,n}\ .}
Due to the orthonormality property \eonP, we see that $\int\ee{-V}\lambda
\Pi_n\,\Pi_m=Q_{nm}=Q_{mn}$, and
$Q$ is a symmetric matrix. In the continuum limit, $Q$ will therefore become
a hermitian operator.

To see how this works explicitly \refs{\rD,\bdss},
we substitute the scaling ansatz
$r(\xi)=r_c+a^{2/m}u(z)$ for the $m^{\rm th}$ multicritical model into \eQmn,
$$Q\to(r_c+a^{2/m}u(z))^{1/2}\,\e{\varepsilon{\del\over\del \xi}}
+\e{-\varepsilon{\del\over\del \xi}}(r_c+a^{2/m}u(z))^{1/2}\ .$$
With the substitution
$\varepsilon{\del\over\del \xi}\to -g a^{1/m}{\del\over\del z}$, we find the
leading terms
\eqn\eQmnc{Q=2 r_c^{1/2}+{a^{2/m}\over\sqrt{r_c}}(u+r_c \kappa^2\del_z^2)\ ,}
of which the first is a non-universal constant and the second is a hermitian
$2^{\rm nd}$ order differential operator.

The other matrix that naturally arises is defined by differentiation,
\eqn\edA{{\del\over\del \lambda}\Pi_n=A_{nm}\Pi_m \ ,}
and automatically satisfies $[A,Q]=1$. The matrix $A$ does not have any
particular symmetry or antisymmetry properties so it is convenient to correct
it to a matrix $P$ that satisfies the same commutator as $A$.
{}From our definitions, it follows that
$$0=\int{\del\over\del \lambda}\bigl(\Pi_n\,\Pi_m\,\ee{-V}\bigr)\quad
\quad\Rightarrow\quad A+A^T=V'(Q)\ ,$$
where we have differentiated term by term and used
$\int\ee{-V}\lambda^\ell\, \Pi_n\,\Pi_m=(Q^\ell)_{nm}$.
The matrix $P\equiv A-\ha V'(Q)=\ha(A-A^T)$ is therefore anti-symmetric and
satisfies
\eqn\epqd{\bigl[P,Q\bigr]=1\ .}

To determine the order of the differential operator $Q$ in the continuum
limit, let us assume for example that the potential $V$ is of order
$2\ell$, i.e.\ $V=\sum_{k=0}^{\ell}a_k\,\lambda^{2k}$. For $m>n$, the
integral $A_{mn}=\int \ee{-V}\Pi_n{\del\over\del \lambda}\Pi_m=\int
\ee{-V} V'\,\Pi_n\,\Pi_m$ may be nonvanishing for $m-n\le 2\ell-1$. That
means that $P_{mn}\ne0$ for $|m-n|\le 2\ell-1$, and thus has enough
parameters to result in a $(2\ell-1)^{\rm st}$ order differential operator
in the continuum. The single condition $W'=0$ results in $P$ tuned to a
$3^{\rm rd}$ order operator, and the $\ell-1$ conditions
$W'=\ldots=W^{(\ell-1)}=0$ allow $P$ to be realized as a $(2\ell-1)^{\rm st}$
order differential operator. In \eQmnc, we see that the universal
part of $Q$ after suitable rescaling takes the form $Q=\d^2-u$. For the
simple critical point $W'=0$, the continuum limit of $P$ is the
antihermitian operator $P=\d^3-{3\over4}\{u,\d\}$, and the commutator
\eqn\ert{1=[P,Q]=4R_2'=
\Bigl({\textstyle 3\over4}u^2-{\textstyle 1\over4}u''\Bigr)'}
is easily integrated with respect to $z$ to give an equation equivalent to
\eplv, the string equation for pure gravity. In \ert, the notation $R_2$ is
conventional for the first member of the ordinary KdV hierarchy. The emergence
of the KdV hierarchy in this context is due to the natural occurrence of the
fundamental commutator relation \epqd, which also occurs in the Lax
representation of the KdV equations. (The topological gravity approach has as
well been shown at length to be equivalent to KdV, for a review see
\refs{\rdvv,\rdijklect}.)

In general the differential equations
\eqn\epqcc{[P,Q]=1}
that follow from \epqd\
may be determined directly in the continuum. Given an operator $Q$,
the differential operator $P$ that can satisfy this commutator is constructed
as a ``fractional power'' of the operator $Q$.  This method of
formulating the continuum theory has a beautiful generalization to
a larger class of theories, which is defined in the following two
sections.

\subsec{Ising Model}
\subseclab\ssIM

The first extension of the method of orthogonal polynomials
occurs in the solution of the Ising model. The partition
function of the Ising model on a random surface can be
formulated using the two-matrix model:
\eqn\eIs{\e{Z}=\int \d U\,\d V\
\e{-\tr\bigl(U^2+V^2-2c\, UV+{g\over N}(\ee{H}\,U^4+\ee{-H}V^4)\bigr)}\ ,}
where $U$ and $V$ are hermitian $N\times N$ matrices and $H$ is a constant.
In the diagrammatic expansion of the right hand side, we now have two different
quartic vertices of the type depicted in \fmapv b, corresponding to insertions
of $U^4$ and $V^4$. The propagator is determined by the inverse of the
quadratic term,
$$\pmatrix{1&-c\cr -c&1\cr}\inv={1\over 1-c^2}\pmatrix{1&c\cr c&1\cr}\ .$$
We see that double lines connecting vertices of the same type (either generated
by $U^4$ or $V^4$) receive a factor of $1/(1-c^2)$, while those connecting
$U^4$ vertices to $V^4$ vertices receive a factor of $c/(1-c^2)$.

This is identically the structure necessary to realize the Ising model on a
random lattice. Recall that the Ising model is defined to have a spin
$\sigma=\pm1$ at each site of a lattice, with an interaction $\sigma_i
\sigma_j$ between nearest neighbor sites $\langle ij\rangle$. This interaction
takes one value for equal spins and another value for unequal spins. Up to an
overall additive constant to the free energy, the diagrammatic expansion of
\eIs\ results in the 2d partition function
$$Z=\sum_{\rm lattices}
\sum_{{\scriptstyle\rm spin}\atop{\scriptstyle\rm configurations}}
\e{\beta\sum_{\langle ij\rangle}\sigma_i\,\sigma_j+H\sum_i \sigma_i}$$
where $H$ is the magnetic field. The weights for equal and unequal neighboring
spins are $\ee{\pm \beta}$, so fixing the ratio $\ee{2 \beta}=1/c$ relates the
parameter $c$ in \eIs\ to the temperature $\beta$. It turns out that the Ising
model is much easier to solve summed over random lattices than on a regular
lattice, and in particular is solvable even in the presence of a magnetic
field. This is because there is much more symmetry after coupling to gravity,
since the complicating details of any particular lattice (e.g.\ square) are
effectively integrated out.

We briefly outline the method for solving \eIs\ (see \refs{\rKBK,\rIYL,\rising}
for more details). By methods similar to those used to derive \egpf, we can
write \eIs\ in terms of the eigenvalues $x_i$ and $y_i$ of $U$ and $V$,
$$\e{Z}=\int \Delta(x)\,\Delta(y)\ \prod_i\d x_i\,\d y_i\,\e{-W(x_i, y_i)}
\ .$$
where $W(x_i, y_i)\equiv x_i^2+y_i^2-2c\,x_iy_i
+{g\over N}(\ee{H}x_i^4+\ee{-H}y_i^4)$.
The polynomials we define for this problem are orthogonal with respect to the
bilocal measure
$$\int \d x\,\d y\ \ee{-W(x,y)}\,P_n(x)\,Q_m(y)=h_n\,\delta_{nm}$$
(where $P_n\ne Q_n$ for $H\ne0$). The result for the partition function is
identical to \egrt,
$$\e{Z}\propto\prod_i h_i\propto\prod_i f_i^{N-i}\ ,$$
and the recursion relations for this case generalize \elpn,
$$\eqalign{x\,P_n(x)&=P_{n+1}+r_n\,P_{n-1}+s_n\,P_{n-3}\ ,\cr
y\,Q_m(y)&=Q_{m+1}+q_m\,Q_{m-1}+t_m\,Q_{m-3}\ .\cr}$$
We still have $f_n\equiv h_n/h_{n-1}$, and $f_n$ can be determined in terms of
the above recursion coefficients (although the formulae are more complicated
than in the one-matrix case). After we substitute
the scaling ans\"atze described in sec.~{\it \sstagpf\/}, the formula for
the scaling part of $f$ is derived via straightforward algebra. The result
is that the specific heat $u\propto Z''$ is given by \etmcp\ with
$\alpha={2\over27}$.

\subsec{Multi-Matrix Models}
\subseclab\ssmmm

We now expand slightly the class
of models from single matrix to multi-matrix models.
The free energy of a particular $(q-1)$-matrix model, generalizing \egpf, may
be written \rCMM\
\eqn\eqmm{\eqalign{Z&=\ln\int\prod_{i=1}^{q-1} \d M_i\
\e{-\tr\bigl(\sum_{i=1}^{q-1}V_i(M_i)
-\sum_{i=1}^{q-2} c_i\,M_i M_{i+1}\bigr)}\cr
&=\ln\int \prod_{{\scriptstyle i=1,q-1}\atop {\scriptstyle\alpha=1,N}}
\!\!\d\lambda_i^{(\alpha)}\ \Delta(\lambda_1)\,
\e{-\sum_{i,\alpha}V_i\bigl(\lambda_i^{(\alpha)}\bigr)+
\sum_{i,\alpha} c_i\,\lambda_i^{(\alpha)}\lambda_{i+1}^{(\alpha)}}
\Delta(\lambda_{q-1})\ ,\cr}}
where the $M_i$ (for $i=1,\ldots,q-1$) are $N\times N$ hermitian matrices,
the $\lambda_i^{(\alpha)}$ ($\alpha=1,\ldots,N$) are their eigenvalues, and
$\Delta(\lambda_i)=\prod_{\alpha<\beta}(\lambda_i^{(\alpha)}-
\lambda_i^{(\beta)})$ is the Vandermonde determinant.
The result
in the second line of \eqmm\ depends on having $c_i$'s that couple matrices
along a line (with no closed loops so that the integrations over the relative
angular variables in the $M_i$'s can be performed.)
Via a diagrammatic expansion, the matrix integrals in \eqmm\
can be interpreted to generate a sum over discretized surfaces, where
the different matrices $M_i$ represent $q-1$ different matter states that
can exist at the vertices.
The quantity $Z$ in \eqmm\ thereby admits an interpretation as the partition
function of 2d gravity coupled to matter.

The methods of the previous section generalize to enable
the evaluation of \eqmm\ \rCMM.

\subsec{Continuum Solution of the Matrix Chains}
\subseclab\sscsotmc

Following \rCMM, we can
introduce operators $Q_i$ and $P_i$ that represent the insertions of
$\lambda_i$ and $\d/\d\lambda_i$ respectively in the integral \eqmm.
These operators necessarily satisfy $\bigl[P_i,Q_i\bigr]=1$.
In the $N\to\infty$ limit, we have seen (following \rD) that $P$ and $Q$
become differential operators of finite order, say $p,q$ respectively
(where we assume $p>q$), and these continue to satisfy \epqcc.
\def\QT{K}
In the continuum limit of the matrix problem (i.e.\ the ``double'' scaling
limit, which here means couplings in \eqmm\ tuned to
critical values), $Q$ becomes a differential operator of the form
\eqn\eQx{Q=\d^{q}+\left\{v_{q-2}(z),\d^{q-2} \right\}+\ \cdots\ + 2v_{0}(z)\ ,}
where $\d=\d/\d z$. (By a change of basis of the form $Q\to f\inv(z)Qf(z)$,
the coefficient of $\d^{q-1}$ may always be set to zero.)
The continuum scaling limit of the multi-matrix models
is thus abstracted to the mathematical problem of finding solutions to \epqcc.

The differential equations \epqcc\ may be constructed as follows.
For $p,q$ relatively prime, a $p^{\rm th}$ order
differential operator that can satisfy \epqcc\ is
constructed as a fractional power of the operator $Q$ of \eQx.
Formally, a $q^{\rm th}$ root may be represented within an algebra of
formal pseudo-differential operators (see, e.g.\ \rDrSo) as
\eqn\eQr{Q^{1/q}=\d + \sum_{i=1}^{\infty} \left\{e_i,\d^{-i}\right\}\ ,}
where $\d\inv$ is defined to satisfy
$\d\inv f=\sum_{j=0}^\infty (-1)^j f^{(j)}\,\d^{-j-1}$.
The differential equations describing the $(p,q)$ minimal model coupled to 2d
gravity are given by
\eqn\ecom{\bigl[Q^{p/q}_+,\ Q\bigr]=1\ ,}
where $P=Q^{p/q}_+$ indicates the part of $Q^{p/q}$ with only
non-negative powers of d, and is a $p^{\rm th}$ order differential operator.

To illustrate the procedure we reproduce now the results for the one-matrix
models, which can be used to generate $(p,q)$ of the form $(2l-1,2)$.
{}From \eQmnc, these models are obtained by taking $Q$ to be the hermitian
operator
\eqn\eQom{Q=\QT\equiv\d^2 - u(z)\ .}
The formal expansion of $Q^{l-1/2}=\QT^{l-1/2}$
(an anti-hermitian operator) in powers of $\d$ is given by
\eqn\eDl{\QT^{l-1/2}=\d^{2l-1}- {2l-1\over4}\left\{u,\d^{2l-3}\right\}
+ \ldots\ }
(where only symmetrized odd powers of $\d$ appear in this case).
We now decompose $\QT^{l-1/2}= \QT^{l-1/2}_+ + \QT^{l-1/2}_-$, where
$\QT^{l-1/2}_+=\d^{2l-1}+\ldots$ contains only non-negative powers of $\d$,
and the remainder $\QT^{l-1/2}_-$ has the expansion
\eqn\eQR{\QT^{l-1/2}_-=\sum_{i=1}^\infty\
\bigl\{e_{2i-1},\d^{-(2i-1)}\bigr\}
=\left\{R_{l},\d^{-1}\right\}+O(\d^{-3})+\ldots\ .}
Here we have identified $R_l\equiv e_1$ as the first term in the expansion
of $\QT^{l-1/2}_-$.
For $\QT^{1/2}$, for example, we find $\QT^{1/2}_+=\d$ and $R_1=-u/4$.

The prescription \ecom\ with $p=2l-1$ corresponds here to
calculating the commutator $\bigl[\QT^{l-1/2}_+,\QT\bigr]$.
Since $\QT$ commutes with $\QT^{l-1/2}$, we have
\eqn\eKKc{\bigl[\QT^{l-1/2}_+,\QT\bigr]=\bigl[\QT,\QT^{l-1/2}_-\bigr]\ .}
But since $\QT$ begins at $\d^2$, and since from the l.h.s.\ above
the commutator can have only positive powers of $\d$,
only the leading ($\d\inv$) term from the r.h.s.\ can contribute, which
results in
\eqn\elc{\bigl[\QT^{l-1/2}_+,\QT\bigr]
=\,{\rm leading\ piece\ of}\ \bigl[\QT,2 R_{l}\,\d^{-1}\bigr]
=4R'_{l}\ .}
After integration, the equation $\bigl[\QT^{l-1/2}_+,\QT\bigr]=1$
thus takes the simple form
\eqn\ede{c\,R_{l}[u]=z\ ,}
where the constant $c$ may be fixed by suitable rescaling of $z$ and
$u$.  Such a scaling is enabled by the property that all terms in $R_l$ have
fixed grade, namely $2l$, where the grade of d is defined to be 1 (and $u$
therefore has grade 2).
The grade of $v_{q-\alpha}$ in \eQx\
is $\alpha$ for an operator $Q$ of overall grade $q$.
As we shall see shortly, this
notion of grade is related to the conventional scaling weights of operators.
It can also be used to determine the terms that may appear in many equations,
since only terms of overall equal grade may be related.

The quantities $R_l$ in \eQR\ are easily seen to satisfy a simple recursion
relation. From $\QT^{l+1/2}=\QT\QT^{l-1/2}=\QT^{l-1/2}\QT$, we find
%
$$\QT^{l+1/2}_+=\ha\left(\QT^{l-1/2}_+ \QT
+ \QT\QT^{l-1/2}_+\right)+\bigl\{R_l,\d\bigr\} \ .$$
%
Commuting both sides with $\QT$ and using \elc, simple algebra gives \rGD
\eqn\erec{R'_{l+1}={1\over4}R'''_{l}-uR'_{l}-{1 \over 2}u'R_l\ .}

While this recursion formula only determines $R'_{l}$,
by demanding that the $R_{l}$ ($l\ne0$) vanish at $u=0$, we obtain
\eqn\erex{\eqalign{R_0&=\ha\,,\qquad\qquad
R_1=-{1\over4}u\,,\qquad\qquad
R_2={3\over16}u^2-{1\over16}u''\,,\cr
R_3&=-{5\over32}u^3+{5\over32}\bigl(uu''+\half u'{}^2\bigr)
-{1\over64}u^{(4)}\ .\cr}}
We summarize as well the first few $\QT^{l-1/2}_+$,
\eqn\ekex{\eqalign{\QT^{1/2}_+&=\d\,,\qquad\qquad
\QT^{3/2}_+=\d^3-{3\over4}\{u,\d\}\,,\cr
\QT^{5/2}_+&=\d^5-{5\over4}\{u,\d^3\}+
{5\over16}\left\{(3u^2+u''),\d \right\}\ .\cr}}

After rescaling, we recognize $R_3$ in \erex\ as eq.~\etmcp\
with $\alpha={1\over10}$, i.e. the equation for the (2,5) minimal model
coupled to gravity. In general,
the equations determined by \epqcc\ for general $p,q$ characterize the
partition function of the $(p,q)$ minimal model (mentioned after \egam)
coupled to gravity. To realize these equations in the continuum
limit turns out \refs{\rmrdt,\rtadt} to require only a two-matrix model of the
type \eqmm. The argument given after \epqd\ for the one-matrix case is easily
generalized to the recursion relations for the two-matrix case and shows that
for high enough order potentials, there are enough couplings to tune
the matrices $P$ and $Q$ to become $p^{\rm th}$ and $q^{\rm th}$ order
differential operators.
It is also possible to realize
a $c=1$ theory coupled to gravity in terms of a two-matrix model formulation
of the 6-vertex model on a random lattice
(see e.g.\ \pgrev). 
In \rtmr, it is argued that one can as well realize a wide variety of
$D<1$ theories by means of a one-matrix model coupled
to an external potential.

It is possible to define a larger space of models defined by taking linear
combinations of the above models. In the case where $Q=K=\d^2-u$, for example,
we can consider
\eqn\elcoks{\sum_k t_{(k)}\bigl[\QT^{k-1/2}_+,\QT\bigr]=1\ ,}
which results after suitable rescaling in the ``string equation''
\refs{\rDS,\rGM,\bdss}\
describing a general massive model interpolating between multicritical points,
\eqn\eStrEqn{z=\sum_{k=1}\ \bigl(k+\half\bigr) t_{(k)}\,R_k[u]\ .}

If we consider the higher operators $\QT^{k-1/2}_+$ as perturbations on pure
gravity, $P=K^{3/2}_+ + \sum_j t_{(j)}\QT^{j-1/2}_+$,
then their scaling weights follow from a simple
argument. Since $u\sim z^{1/2}$ for pure gravity and $u$ has grade 2, we see
that a coupling of grade $\alpha$ scales as $[z]^{\alpha/4}$, giving
$\Delta=\alpha/4$ as the gravitationally dressed scaling weight of its
conjugate operator. Now the grade of $t_{(j)}$ is $3-(2j-1)=4-2j$, so
it couples to an operator with weight $1-j/2$.

In the case of unitary minimal models, $z$ couples to the area, so is
proportional to the cosmological constant.
In general, however, $z$ couples to the lowest dimensional operator in the
theory. From the point of view
of a perturbed $(2,2m-1)$ model, we have $u\sim z^{1/m}$, the grade of
$t_{(j)}$ is $(2m-1)-(2j-1)=2m-2j$,
and $\QT^{j-1/2}_+$ scales as $(m-j)/m$ with respect to the lowest dimensional
operator, i.e.\ corresponding to $\alpha/\alpha\dup_0$ rather
than $\alpha/\gamma$ in \ehba.
If we wish to compare to Liouville scaling with respect to the
area, on the other hand, we must multiply by a factor of $\alpha\dup_0/\gamma=
m/2$, which results in
\eqn\ehammj{\half(m-j)}
for the scaling of $\QT^{j-1/2}_+$ viewed
as a perturbation of a $(2,2m-1)$ model (a result we shall use when we
expand macroscopic loops in terms of local operators in these theories).

We can also consider the operators that correspond to these perturbations
from the standpoint of the underlying one-matrix model.
The parameters $t_{(j)}$ in \eStrEqn\ correspond to perturbations
to $j^{\rm th}$ order multicritical potentials of the form \egenW, and are
given in turn by matrix operator perturbations of the form
\eqn\ecritops{\eqalign{&{\rm tr}\,V_{(j)}(M)
={\rm tr}\int_0^1 {\d t\over t}\,W_{(j)}\bigl(t(1-t)M^2\bigr)\ ,\cr
&\hbox{ with}\quad W_{(j)}(r)=g\dup_c-\alpha(r_c-r)^j\ .\cr}}
(Note that the integral in the first line just inverts the expression leading
from $V$ to $W$ in \eVtoW.)

Finally, we note that  for a general $(p,q)$ model,
the grade of the l.h.s.\ of \epqcc\ is $p+q$,
so $z$ will be set equal
(i.e.\ following one integration) to a quantity with grade $p+q-1$.
A coupling with grade $\alpha$ therefore scales as
$[z]^{\alpha/(p+q-1)}$, giving $d=\alpha/(p+q-1)$ as the
gravitationally dressed scaling weight of its conjugate operator.
(The grade of $v_{q-2}=\langle\CP\CP\rangle$ is always 2, where
$\CP$ is the puncture operator whose two-point function calculates the
$2^{\rm nd}$ derivative of the partition function, hence
giving the string susceptibility $\Gamma_{\rm str}=-2/(p+q-1)$.)
If we perturb $P\to P+t\,Q^{|pr-qs|/q-1}_+$, then
$t$ has grade $p+q-|pr-qs|$, and hence couples to an operator
of scaling weight coincident with \edrs\ (after multiplying the latter
by $\gamma/\alpha_{q-1,p-1}=2q/(p+q-1)$ to take into account the coupling
of $z$ to the lowest dimensional operator rather than to area).

\exercise{Scaling of Lax operators}

a) Show that if $Q$ is the Lax operator of order $q$, defining
the $(p,q)$ series, then the operators in \lzopsii\ are
\eqn\frcpwr{Q^{n/q}\big|_+\ .}
Parametrizing $n=k q + \alpha$, $1\leq \alpha\leq q-1$, we
identify these operators with the topological
field theory operators $\sigma_k(\CO_\alpha)$ which
appear in \rdijklect.

b) Calculate the spectrum of indices $\nu$ of \msswdw\ we expect to
find for the Wheeler--DeWitt wavefunctions of the Lian--Zuckerman
states.

\endexercise


\newsec{Matrix Model Technology II: Loops on the Lattice}
\seclab\smmttll

\subsec{Lattice Loop Operators}

\ifig\fphin{Insertion of  $\tr\, \Phi^M$ into the matrix generating functional
results in a vertex emanating $M$ ``spokes''.}
{\epsfxsize2.5in\epsfbox{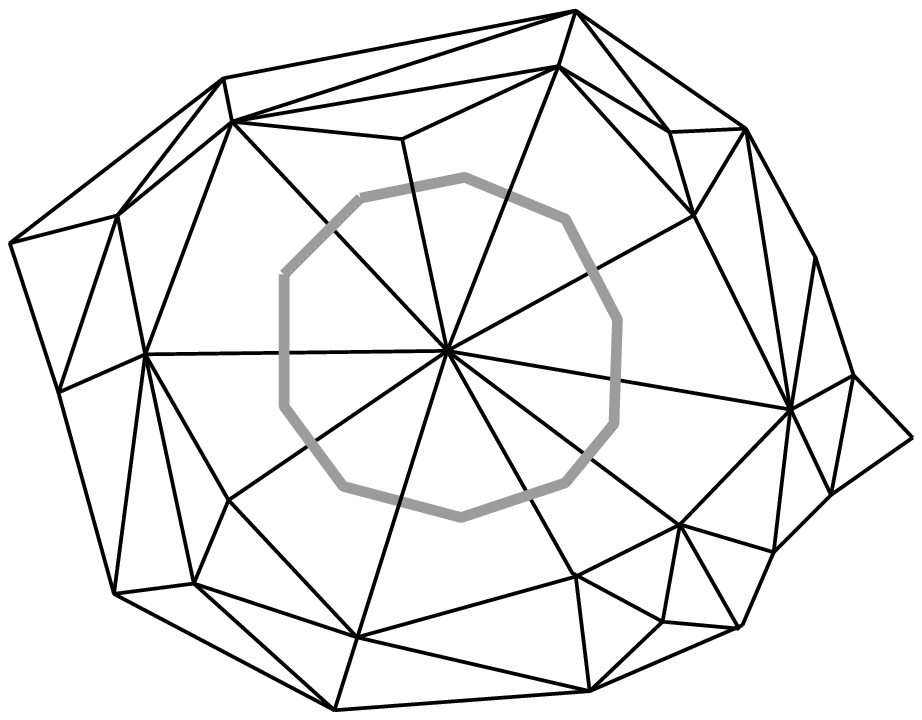}}

Consider the one hermitian matrix model, as discussed in
chapt.~\sMMTI\ (where we now use $\Phi$ rather than $M$ to denote
the $N\times N$ hermitian matrix).
In the Feynman diagram expansion, the
insertion of the operator
\eqn\latloop{{1\over M} \tr\, \Phi^M }
creates a vertex emanating $M$ ``spokes,'' as shown in \fphin.
On the dual triangulated surface, this operator has inserted
a hole with $M$ boundary lengths, and thus has
length $aM$, where $a$ is the lattice spacing.
The factor of $1/M$ in \latloop\ is
needed to take account of the symmetry factor for the Feynman
diagram. In the following we will generally work with
marked loops and discard this factor.

To obtain macroscopic loop amplitudes we must take the
continuum limit $a\to 0$ and to maintain a finite
physical length, we must simultaneously take $M\to \infty$, holding some
combination of $a$ and $M$ fixed. From this point of view,
the local operators of the theory discussed in sec.~{\it\sscsotmc\/},
namely linear combinations of operators of the form \latloop\
(e.g. \ecritops), correspond instead to
``microscopic loops'',  i.e.\ loops only a few lattice spacings in length.

Before proceeding to make precise sense of the continuum
limit, we begin with some heuristic remarks.
Recall that in the orthogonal polynomial formalism
discussed in sec.~{\it\sKdV\/}, the action of $\Phi^M$
was equivalent to $(\Phi_{\rm cr}+Q)^M$, where
\eqn\latceq{Q\to 2 r_c^{1/2}
+a^{2/m}\,r_c^{-1/2}(u+r_c \kappa^2 \p_z^2 )\ .}
Hence if we hold $Ma^{2/m}=2 r_c\,\ell$ fixed, we expect
the loop operator to become the heat kernel operator,
\eqn\limloop{{1\over M} \tr (\Phi^M ) \to \ee{\ell Q}\ .}
After rescaling, we can write $Q=\kappa^2 \d^2/\d z^2 - u(z,\kappa)$,
where $Q$ is the Schr\"odinger operator associated to the model, and $\kappa$
is the topological coupling.
A rigorous discussion of the above limiting procedure
makes use of the free fermion formalism,
implicit in the orthogonal polynomial technique
(and described in chapt.~\smmttff).

An alternative formulation of the lattice loop operator is
\eqn\secloop{W(L)={1\over N} {\rm Tr}(\ee{L \Phi})\ ,}
where $L$ is a ``chemical potential'' for the length. The limiting
form \latceq\ shows that these loop operators have the
same continuum limit, up to a non-universal multiplicative
renormalization. This is a useful observation for making
sense of examples where $r_c=0$.

To analyze macroscopic
loop amplitudes on the lattice, we introduce the resolvent operator
\eqn\lplc{\hat W(\zeta)=\int_0^\infty \d L\, \ee{-\zeta L}\, W(L)\ .}
Defining $\zeta=\ee\rho$, we may interpret $\rho$ as a  bare
boundary cosmological constant, and
\lplc\ should be regarded as the lattice analog of the relation
\eqn\blkcse{Z(\mu_B)=\sum_{A=1}^\infty \ee{-\mu_B A} \,Z(A)\ .}
between fixed area and fixed cosmological constant partition functions.

\subsec{Precise definition of the continuum limit}
\subseclab\sdotcl

Making use of the resolvent operator \lplc,
we can now  present a more technical description of the continuum limit
discussed in chapts.~\sDsMmCl,\sMMTI.
To take the continuum limit of the
lattice expressions, we first study the
$N\to \infty$ asymptotics of the correlation functions
\eqn\latmacam{\eqalign{
\bigl< N|\prod_{i=1}^B \hat W(\zeta_i)|N\bigr>
&\equiv Z^{-1} \int \d \Phi\, \ee{-N\tr\, V(\Phi)}\prod_{i=1}^B
\hat W(\zeta_i)\cr
&\sim \sum_{\chi=2-2h-B} N^\chi\, \CF_\chi[V;\zeta_i]\ ,\cr}}
where $V(\Phi)=\sum_{j\geq 0} T_j\, \Phi^j$ is a polynomial
interaction for $\Phi$.
This is a generating functional for correlation functions
of $B$ operators.
As explained in sec.~{\it\sdsmmcl\/}, at fixed topology the functionals
$\CF_\chi[V;\zeta_i]$ have a lattice expansion
\eqn\lattopexp{\CF_\chi[V;\zeta_i]=\sum_{F_i,L_i\geq 0}
\ \sum_{\CD_\chi[F_i,L_i]}
\prod (T_i/\sqrt{T_2}\,)^{F_i} \prod \zeta_i^{-L_i}\ ,}
%
where $\CD_\chi[F_i,L_i]$ is the set of distinct
``triangulations'' of
a surface into $F_i$ $i$-sided polygons with boundaries of
lattice length $L_i$. Using methods described below (in particular
the loop equations \splsdi), one can show that the functions $\CF_\chi$
have the following mathematical properties, familiar from
the study of phase transitions in statistical mechanical models.

\ifig\fmulticrit{Subspaces of successively higher codimension
in coupling constant space corresponding to multicritical domains.}
{\epsfxsize3.75in\epsfbox{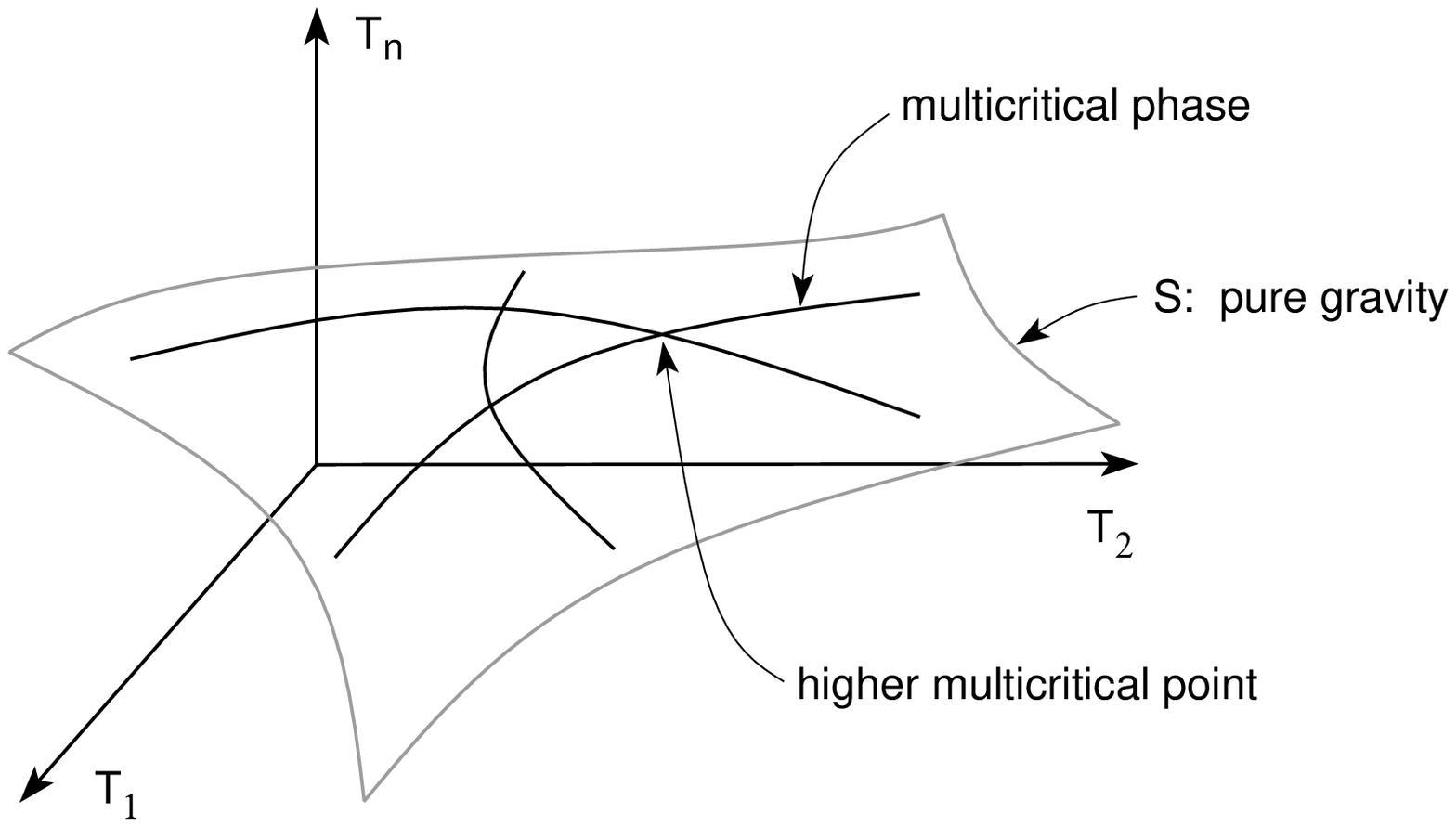}}
\item{1)} The expansions in $V$ and $\zeta\inv_i$ are convergent in a
sufficiently small neighborhood of the origin.
To specify a neighborhood of the origin for the potential $V$,
we first define a filtration on the space of
potentials by requiring that $V$ be in $\CV^{(n)}$, the
space of polynomials of degree $\leq n$ with no constant or linear term.
Considered as a function on the space of polynomials of degree $\leq n$,
we have convergence in a neighborhood of the origin.
\item{2)} The expansions have a finite radius of convergence.
Defining the lattice area $A=\sum F_i$, it can be established that
the asymptotic behavior of the number of distinct triangulations
goes as
$|\CD_\chi|\sim \ee{\mu_* A+ \sum_i \rho_* L_i}\, A^{\theta}\, L_i^{\theta'}$
for $A\to \infty$, $L_i\to \infty$. Thus, the series will
diverge as $V$ approaches a real codimension one subvariety, the
{\it singular subvariety\/} $\CS$ of
$\CV^{(n)}$, and as $\zeta\to \zeta_c$ from above.
\item{3)} A priori $\CS$ and $\zeta_c$ could depend on
the Euler character $\chi$ and the choice
of boundary component, but turn out independent of them.
\item{4)} The variety $\CS$ has subspaces of successively higher
codimension corresponding to multicritical domains, as depicted in
\fmulticrit. For the
space $\CV^{2n}$, the highest multicritical behavior
corresponds to a point, given in \rGM
\eqn\grssmgdl{V_*^{(n)}(\Phi)
=\int_0^1 {\d t\over t}\Bigr(1-\bigl(1-t(1-t)\,\Phi^2\bigr)^n\Bigr)\ .}
(This is just \ecritops\ with normalization $g\dup_c=\alpha=r_c=1$.)
\item{5)} The critical exponents $\theta$, $\theta'$ in (2) are ``universal,''
which means that they only depend on the multicritical domain in
$\CS$. On the other hand, $\mu_*,\rho_*$ are ``non-universal''
which means they can vary from point to point in $\CS$.

\noindent We can use these mathematical properties of the
functions $\CF_\chi$ to learn about the physics of smooth
surfaces as follows.
We would like to distinguish
``universal'' phenomena --- associated with smooth continuum
surfaces and not with the details of lattice decompositions ---
by using the nonanalytic behavior of $\CF_\chi$ as we approach
singular values of $V,\zeta$. The idea is based on the remark
that the contribution of a hole with a finite lattice size, or a surface with a
finite number of polygons, to \latmacam\ will always be analytic in $\zeta,V$.
Thus, the nonanalyticity in $\zeta,V$ must ``arise from holes and surfaces
whose perimeter and area is infinity in lattice units,'' that is, from smooth
continuum surfaces.\foot{The extent to which these surfaces really are smooth
is an interesting question. See \rsmth.}

By turning the above reasoning
on its head, we {\it define\/} the continuum limit by isolating
the nonanalytic dependence on $\zeta_i,V$. More precisely,
we must define scaling functions as we approach singular values
and define the continuum quantities in terms of these scaling
functions. Note that if we use such a definition, continuum
quantities are ambiguous by terms which are
purely analytic in the coupling constants.

\subsec{The Loop Equations}

The correlation functions of $\hat W(\zeta)$ may be
determined by the  ``Schwinger--Dyson'' or
loop equations \refs{\volodya,\fdavid,\fdavidi}\
for the matrix model.

The loop equations are derived by requiring that the matrix
model path integral be independent of a change of variables.
A convenient way to organize arbitrary analytic changes of
variables is to consider the simple transformation
\eqn\chgvrbls{\phi\to \phi + \epsilon {1\over \zeta-\phi}\ .}
Under \chgvrbls, we have
\eqn\sothat{\eqalign{
\tr\, V(\phi)&\to \tr\, V(\phi) + \epsilon\,\tr\, V'(\phi)(\zeta-\phi)^{-1}\cr
\d\phi
&\to \d\phi\Bigl(1+ \epsilon N^2 \bigl(\hat W(\zeta)\bigr)^2\Bigr)\ .\cr}}
Inserting \sothat\ into $\int\d\phi\,\ee{-N\tr V(\Phi)}$ and
equating first order terms in $\epsilon$ gives
\eqn\splsd{\bigl< \bigl(\hat W(\zeta)\bigr)^2\bigr\rangle
= {1\over N}\bigl< \tr\, V'(\phi)(\zeta-\phi)^{-1}\bigr>\ .}

We wish to expand the above in $1/N$.
$\hat W$ is normalized so that the expansion in $1/N$
of $\langle \hat W\rangle$ begins at $\CO(1)$, corresponding
to the disk geometry,
\eqn\expfw{
\bigl< \hat W(\zeta)\bigr>\sim \bigl< \hat W(\zeta)\bigr>_{h=0}+
{1\over N^2} \bigl< \hat W(\zeta)\bigr>_{h=1}+\cdots\ .}
By considering the relevant topologies contributing to
$\bigl< (\hat W(\zeta))^2\bigr>$, we see that the
leading term on the left hand side of \splsd\ has the topology of
two disks. In general, we may separate the contribution of
connected and disconnected geometries:
\eqn\leadenn{
\bigl< \bigl(\hat W(\zeta)\bigr)^2\bigr> =(\bigl< \hat W(\zeta)\bigr> )^2
+{1\over N^2}\bigl< \bigl(\hat W(\zeta)\bigr)^2\bigr>_c\ ,}
where the second term corresponds to connected geometries.
Expanding $V'(\phi)$
as a polynomial in $\zeta-\phi$ with coefficients which are polynomials
in $\zeta$, we see that $\bigl< \hat W(\zeta)\bigr>_{h=0}$ satisfies
a quadratic equation,
\eqn\qdreq{\bigl< \hat W(\zeta)\bigr>_{h=0}^2-V'(\zeta) \bigl<
\hat W(\zeta)\bigr>_{h=0} + Q(\zeta,V)=0\ ,}
where
\eqn\eqfrqz{\eqalign{
Q(\zeta,V)&=\bigl< Q(\zeta,\phi)\bigr>\cr
Q(\zeta,\phi)&=\sum_{k\geq 1} {1\over k!} V^{(k+1)}(\zeta)
{1\over N} \tr(\phi-\zeta)^{k-1}\ .\cr}}
$Q$ is a polynomial in $\zeta$ of degree ${\rm deg}\,(V)-2$
whose coefficients are linear combinations of
$c_j(V)\equiv \bigl<\tr\, \Phi^j\bigr>$, $j\leq {\rm deg}\,(V)-2$.
The disk amplitude is obtained by solving \qdreq.

\ifig\floopd{Pictorial representation of
$V'({\p/\p L})\,\bigl< W(L)\bigr>_c$ .}
{\epsfxsize 4.5in\epsfbox{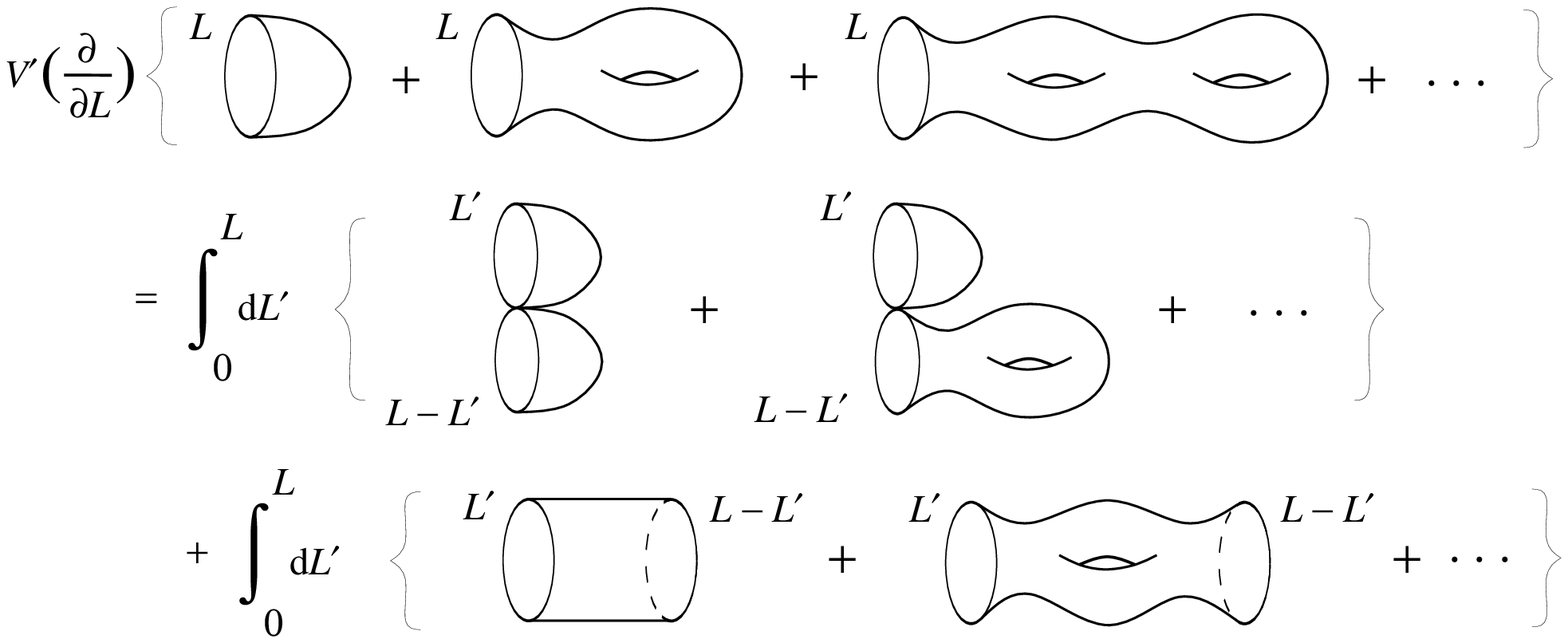}}

\noindent
{\it Geometrical Interpretation: SD Equations as Loop Equations\/}:
\par\nobreak
The loop equations have a beautiful geometrical interpretation
further justifying the identification of $W(L)$ with a loop
operator. Write \splsd\ in the form
\eqn\splsdi{
V'(\zeta) \bigl< \hat W(\zeta)\bigr> + Q(\zeta,V)=
(\bigl< \hat W(\zeta)\bigr>_c)^2+{1\over N^2}
\bigl<(\hat W(\zeta))^2\bigr>_c^2\ .}
Taking an inverse Laplace transform, we obtain
\eqn\loopeq{V'\Bigl({\p\over \p L}\Bigr)\,\bigl< W(L)\bigr>_c
=\int_0^L \d L'\, \Bigl(\bigl< W(L')\bigr>_c \bigl< W(L-L')\bigr>_c
+{1\over N^2}\bigl< W(L')\, W(L-L')\bigr>_c\Bigr)\ ,}
which has the pictorial representation shown in \floopd.

\exercise{}

Derive \loopeq. Note that a polynomial in $\zeta$ does not
have an inverse Laplace transform.

\endexercise

The full set of loop equations may be elegantly summarized
by introducing a ``source'' coupling to the loop operator:
\eqn\zwthsrce{Z[J]\equiv
\int \d\Phi\, \ee{-N\tr\, V(\Phi)+\int_0^\infty \d L'\, J(L')\, W(L')}\ .}
Making the change of variables $\Phi\to \Phi + \epsilon\,\ee{L \Phi}$
and using the above procedures, we obtain
\eqn\fulllpeqs{\eqalign{&V'\Bigl({\p\over \p L}\Bigr)\bigl< W(L)\bigr>_c[J]=
\int_0^L \d L'\, \Bigl(\bigl< W(L')\bigr>_c [J]\ \bigl< W(L-L')\bigr>_c[J]\cr
&\qquad+{1\over N^2}\bigl< W(L')\, W(L-L')\bigr>_c\Bigr)
+{1\over N^2}\int_0^\infty \d L'\, L'\, \bigl< W(L+L')\bigr>_c [J]\ ,\cr}}
for which one may draw a similar pictorial representation.

{\bf Remark:} It is possible, although slightly subtle \refs{\fdavid,\rfknm},
to take the continuum limit of the loop equations we have derived here to write
analogous equations for the continuum amplitudes.
These continuum loop equations have many important applications,
including for example the elimination \fdavid\ of unphysical solutions to the
string equations \eStrEqn.

\newsec{Matrix Model Technology III: Free Fermions from the Lattice} 
\seclab\smmttff

The equivalence of matrix models to theories of free fermions
is the underlying reason for the solvability of matrix models.
In this chapter we describe the free fermion formalism.

\subsec{Lattice Fermi Field Theory}

The free-fermion formalism provides the basis for a rigorous
description of the double-scaling limit of
macroscopic loop operators. The formalism is also a
very efficient way for calculating loops, both on
the lattice and in the continuum. The formalism
was first applied to macroscopic loop amplitudes in \bdss.

In sec.~{\it\ssOP\/}, orthogonal polynomials were introduced and
it was shown that correlation functions are integrals of
powers of $\lambda$ multiplying a Vandermonde determinant.
Interpreting this determinant as a Slater determinant
for a theory of free fermions, we introduce the second-quantized
Fermi field
\eqn\latfermi{\Psi(\lambda)=\sum_{n=0}^\infty a_n \psi\dup_n(\lambda)\ ,}
where $\psi\dup_n$ are the orthonormal wavefunctions built
from the orthogonal polynomials:
\eqn\wvfnii{\psi\dup_n(\lambda)
\equiv {1\over \sqrt{h_n}}P_n(\lambda)\, \ee{-\half N V(\lambda)}\ ,}
and $\{a_n,a^\dagger_m\}=\delta_{n,m} $.

Correlation functions in the matrix model
with $N\times N$ matrices are obtained by calculating
correlation functions in the Fermi sea defined by
\eqn\fermsea{\eqalign{
a_n |N\rangle =0 & \qquad n\geq N\cr
a_n^\dagger |N\rangle =0 & \qquad n<N\ .\cr}}
To see this, introduce the second-quantized operator for multiplication
by $\lambda^n$,
\eqn\oper{\Psi^\dagger \hat \lambda^n\Psi
=\int \d \lambda\,\Psi^\dagger(\lambda)\,\lambda^n\,\Psi(\lambda)\ ,}
and the main observation is
\eqn\mfreq{\eqalign{
\Bigl< \prod_i \tr\, \Phi^{n_i}\Bigr>_{\rm matrix\ model} &\equiv
Z^{-1}\int \d\Phi \prod_i \tr\, \Phi^{n_i}\,\ee{ -N\tr\, V(\Phi)}\cr
&=\bigl< N|\prod_i\bigl(\Psi^\dagger
 \hat \lambda^{n_i}\Psi\bigr)|N\bigr>\ .\cr}}
where $V=\sum_{i\geq 2} g_i\, \Phi^i$ and all but finitely
many $g_i=0$.
The proof of this identity uses the orthogonal polynomial
techniques, e.g.\ for the one-point function:

\eqn\expl{\eqalign{\langle \tr\, \Phi^{n}\rangle
&={\int \prod \d \lambda_i\,
\Delta^2(\lambda_i)(\sum \lambda_i^{n})\prod_i
\ee{-N V(\lambda_i)}\over N! \prod h_i}\cr
&={N\over  N! \prod h_i} \int \prod \d \lambda_i\,
\bigl(\det P_{j-1}(\lambda_i)\bigr)^2 \lambda_1^{n}
\prod_i \ee{-N V(\lambda_i)}\cr
&={N\over  N! \prod h_i}(N-1)!\sum_{j=0}^{N-1}{\prod h_i\over h_j}
\int \d \lambda\, \bigl(P_j(\lambda)\bigr)^2 \,\lambda^{n}\,
\ee{-N V(\lambda)}\cr
&=\sum_{j=0}^{N-1} \langle \psi\dup_j|\lambda^{n}|\psi\dup_j\rangle
=\langle N|\,\Psi^\dagger \hat \lambda^{n}\Psi\,|N\rangle\ .\cr}}

\exercise{}

a) Prove \mfreq\ for two point functions using
the same steps as in \expl.

b) Find a general proof of \mfreq.

c) Show that the lattice loop operator \secloop\ and resolvent
may be realized
in the fermion formalism as
\eqn\frmloop{\eqalign{
W(L)&={1\over N} \Psi^\dagger\,\ee{L \hat \lambda}\Psi\cr
\hat W(\zeta)&=\sum_{n=0}^\infty \zeta^{-n-1}\,
\Psi^\dagger \hat \lambda^n\Psi
=\Psi^\dagger {1\over \zeta-\hat \lambda} \Psi\cr}}

\endexercise

\subsec{Eigenvalue distributions}

We will now justify some of the statements made in
sec.~{\it\sdotcl\/} and indicate why the lattice (and hence continuum)
correlation functions are computable.

By Wick's theorem, we can express all amplitudes in terms of
the fermion two-point function
\eqn\twpt{
K_N(\lambda_1,\lambda_2)\equiv \langle N|\Psi^\dagger(\lambda_1)
\Psi(\lambda_2)|N\rangle\ .}
Therefore, in order to define the double scaling limit we must study the
$N\to \infty$ asymptotic behavior of the kernel $K_N$.
As explained in sec.~{\it\ssMmdls\/}, matrix model correlation
functions have an asymptotic expansion in $1/N$, and
are obtained from the asymptotic expansion:
\eqn\asymptots{K_N\sim
\sum_{j\geq 0} N^{1-j} K_j(\lambda_1,\lambda_2)}
The functions $K_j$ have support on an interval\foot{In more complicated cases
the support can be on unions of intervals.} $I$ which is independent of $j$.
As a special case, note in particular that the diagonal of this kernel is the
eigenvalue density:
\eqn\evdns{
\rho(\lambda)= K_N(\lambda,\lambda)\ .}
By \evdns\ we may identify the interval $I$ with support of
the eigenvalue density in perturbation theory.

\exercise{Eigenvalue Density}

Show that $\rho(\lambda)$ is the probability for finding
an eigenvalue with value $\lambda$ in a random matrix
ensemble described by $V(\lambda)$. That is, show that
it is the matrix expectation value of
\eqn\matexpvl{
{1\over N} \sum_{i=1}^N \delta(\lambda-\lambda_i)\ .}

\endexercise

The easiest way to prove our assertions about the nature of the eigenvalue
densities proceeds by studying the correlation functions of the resolvent
operators $\hat W(\zeta)$. Note that $\hat W(\zeta)$ is only defined for
$\zeta$ off the real axis since $\phi$ has real eigenvalues. Moreover, the
discontinuity of $\hat W(\zeta)$ across the real axis is equal to the
eigenvalue density.

Solving the quadratic equation we see that the roots of the polynomials
define several branch points for $\bigl< \hat W(\zeta)\bigr>_{h=0}\,$,
and since $\rho(\lambda)$ is the discontinuity of
$\bigl<\hat W(\zeta)\bigr>_{h=0}\,$, the support of the genus zero
eigenvalue density must lie on an interval or finite union of
intervals.

\ifig\fsemi{The Wigner semicircle distribution.}
{\epsfxsize2.5in\epsfbox{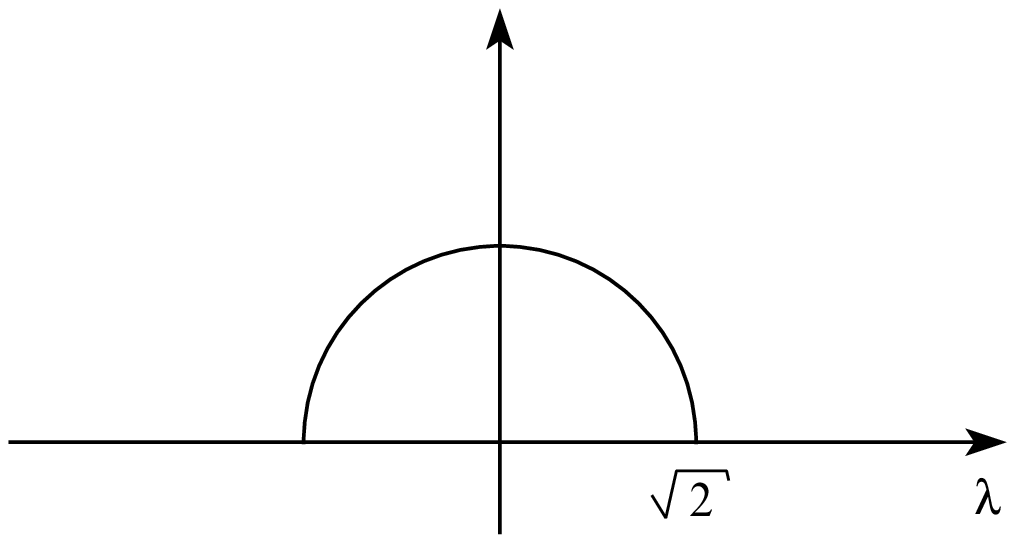}}

\exercise{Derivation of the Wigner Distribution}

As an example of an eigenvalue distribution, we consider the
Gaussian matrix model.
The leading term in the large $N$ asymptotics of the eigenvalue
distribution is the famous Wigner distribution
\eqn\gmev{
K_1(\lambda,\lambda)={1\over 4 \pi}\sqrt{2-\lambda^2}
\ \theta(2-\lambda^2)\ ,
}
shown in \fsemi.

Derive the Wigner distribution from the ``Schwinger--Dyson''
equations of the matrix model using the above procedure.
First show that for the Gaussian potential we have
\eqn\gsspot{\bigl< \hat W(\zeta)\bigr>_{h=0}
=\ha\bigl(\zeta-\sqrt{\zeta^2-2}\bigr)\ ,}
and from this obtain the Wigner distribution \gmev.

\endexercise

The finiteness of the support of the kernels has important implications
for the nonanalyticity in $\zeta$. Consider for example the one-point
function
\eqn\rslvnt{\eqalign{
\bigl< \hat W(\zeta)\bigr> \equiv &
{1\over N}\Bigl< \tr\, {1\over \zeta-\phi}\Bigr>=
{1\over N}\int \d \lambda\,{K(\lambda,\lambda)\over \zeta-\lambda}\cr
&\sim \sum_\chi N^\chi \int_I \d \lambda\, {K_\chi(\lambda,\lambda)\over
\zeta-\lambda}\ .\cr}
}

The nonanalytic dependence on $\zeta$ we are looking for
comes from the contributions in the $\lambda$ integrals
from the integrals near the {\it edge\/} of the support
$I$ of the eigenvalue distribution. In the last expression
we may take $\zeta$ real and $\zeta>\zeta_c$. We encounter
nonanalytic behavior as $\zeta$ hits the edge of the
eigenvalue distribution.

{\bf Example}: Let us verify the statements about analytic
dependence on $\zeta$ in the example of a Gaussian
potential. Expanding $\zeta=\zeta_c+\delta \zeta
=\sqrt{2} + \delta \zeta$, or equivalently, expanding $\lambda$
around the edge of the eigenvalue distribution,
we obtain a nonanalytic function of $(\delta \zeta)^{1/2}$
corresponding (formally) to the one-loop amplitude
$\bigl< W(\ell)\bigr> =\ell^{-3/2}$.

\subsec{Double--Scaled Fermi Theory}
\subseclab\ssDSFT

More generally, to prove \asymptots\ and to investigate the
scaling limit of $K$ near the edge of $I$ more thoroughly, note that using the
recursion relation
\eqn\recrl{\lambda \psi\dup_n
=\sqrt{r_{n+1}}\,\psi\dup_{n+1}+\sqrt{r_n}\, \psi\dup_{n-1}\ ,}
we may write
\eqn\extpt{\eqalign{
\langle N|\Psi^\dagger(\lambda_1)\Psi(\lambda_2)| N\rangle
&=\sum_{n=0}^{N-1} \psi\dup_n(\lambda_1)\psi\dup_n(\lambda_2)\cr
&=\sqrt{r_{N+1}}{\psi\dup_{N+1}(\lambda_1)\psi\dup_N(\lambda_2)-
\psi\dup_{N+1}(\lambda_2)\psi\dup_N(\lambda_1)\over
\lambda_1-\lambda_2}\ ,\cr}
}
and therefore we should study the scaling limit of the
orthonormal wavefunctions themselves.

As discussed in sec.~{\it\sdotcl\/},
the recursion functions $r_n[V]$ have singular
behavior as $V\to \CS$. Moreover, using the recursion relations
for orthogonal polynomials, as elegantly summarized in the
statement $[P,Q]=1$, the large $n$ asymptotics determines
a consistent ansatz for the following behavior.
If $V_*^{(m)}(\lambda)$ is the $m^{th}$ multicritical potential,
then we approach criticality by taking the limit:
\eqn\limtqs{V= \ee{a^2 \mu}\, V_*^{(m)}\quad a\to 0\qquad
\eqalign{&n/N=1-a^2 (z-\mu)\cr
&N a^{2+1/m}=\kappa^{-1}\cr
&r_n[V]\rightarrow r_c + a^{2/m} u(z)\ .\cr}}
The recursion relation \recrl\ implies that if $\psi$ has well-behaved
limiting behavior near the edge of the eigenvalue distribution $\lambda_c$,
\eqn\limpsi{\psi\dup_n(\lambda_c+a^{2/m} \tilde \lambda)
\rightarrow a^\theta \psi(z,\tilde \lambda)
}
(here $a^\theta$ is a normalization factor),
then the $\psi$'s are eigenfunctions of the Lax operator:\foot{In the theory of
the KdV hierarchy, such functions are known as Baker-Akhiezer functions.}
\eqn\lmspii{Q \psi =
\bigl(\kappa^2\,\d^2/\d z^2 - u(z,\kappa)\bigr)\psi= \tilde \lambda \psi\ .}
The limiting form of $\psi$ will be an eigenfunction of $Q$.\foot{The
eigenfunctions of $Q$ are obtained by demanding
appropriate asymptotic behavior in $z$ and $\lambda$,  insuring
convergence of integrals as $z\to \infty$ and exponential
decay of eigenvalue density off the perturbative cut as $\lambda\to \infty$.}


\noindent{\bf Example}. {\it The Gaussian potential\/}
\par\nobreak
We will work through in detail the fermionic formulation
of the double scaling limit for the simplest matrix
potential of all, the Gaussian potential
$V(\Phi)=\Phi^2$. The Gaussian potential
corresponds to the so-called
topological point or the $(1,2)$ point in the Lax
operator classification described in sec.~{\it\sscsotmc\/}.
Perturbations about this point define the correlation functions of
topological gravity, described from the point of view of
topological field theory in \rdijklect.

\ifig\fcoal{As $\lambda\to \lambda_c$,
two stationary phase points coalesce at $i/\sqrt2$.}
{\epsfxsize2.5in\epsfbox{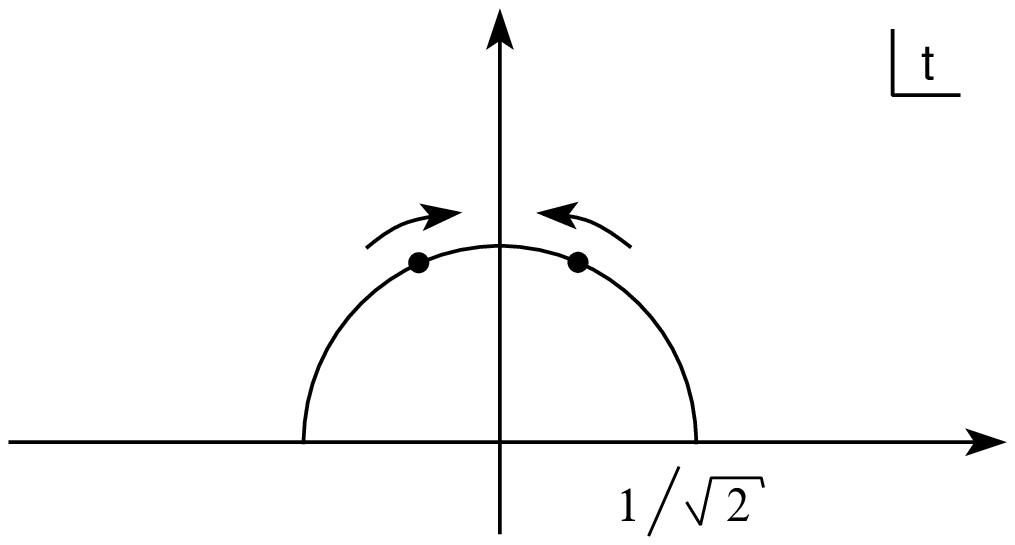}}

We now obtain the full asymptotics in $1/N$ of the contributions
to \mfreq\ of the integrals from the edge of the eigenvalue distribution.
We do this first explicitly for the case of the Gaussian potential.
In the case of a gaussian matrix potential $\ee{-N\tr\phi^2}$,
the orthonormal wavefunctions are simply
\eqn\herfun{\psi\dup_n(\lambda)
={N^{1/4}\over 2^{n/2}\pi^{1/4}\sqrt{n!}}\, H_n(\sqrt{N}\lambda)\,
\ee{-N\lambda^2/2}\ , }
where $H_n$ is a Hermite polynomial, and has the integral representation
\eqn\intrep{
H_n(x)= {2^n\over \sqrt{\pi}}\int_{-\infty}^{\infty}\d t\,(x+it)^n
\,\ee{-t^2}\ .}
Using the stationary phase approximation, one finds two
stationary points for $\lambda^2\not=2$. For $\lambda^2<2$
we find an oscillatory function while for $\lambda^2>2$, the wavefunction
is zero to all orders of the $1/N$ expansion.
We are most interested in the behavior of the wavefunctions
for $\lambda$ infinitesimally close to $\pm \sqrt{2}$, the
edge of the eigenvalue distribution. At this point, the
two stationary phase points coalesce as in \fcoal, and by simultaneously
scaling $N\to \infty$ and $\lambda\to \lambda_c=\sqrt{2}$ one
can obtain a well-defined limit whose asymptotics captures the
contribution of the edge of the eigenvalue distribution to the
entire perturbation series in $1/N$. This simultaneous scaling
is the fermionic version of the double scaling limit.

In detail, let
\eqn\psstolim{Na^3=\kappa^{-1}\qquad
n/N=1-a^2 z\qquad
\lambda=\sqrt{2}(1+ a^2 \tilde \lambda)\ ,}
and let $a\to 0$ holding $z,\kappa,\tilde \lambda$ fixed.
Then we have
\eqn\getairy{\eqalign{\lim_{a\to 0} a^{1/2} \psi\dup_n(\lambda)
&={\kappa^{-1/6}\over \pi 2^{3/4}}\int_{-\infty}^{\infty}
\d t\, \ee{it \kappa^{-2/3}(\tilde \lambda +z)+it^3/3}\cr
&={1\over 2^{1/4} \kappa^{1/6}}\,
{\rm Ai}\Bigl({z+ \tilde \lambda \over \kappa^{2/3}}\Bigr)\ .\cr}}
That is, the double scaling limit of the Hermite functions of the Gaussian
model are Airy functions.\foot{The appearance of these functions is directly
related to the Airy functions which play a key role in the Kontsevich matrix
model.
}

\ifig\fmag{A magnified view of the eigenvalue distribution near the endpoint.
Note the nearby exponential falloff and squareroot growth far from the
endpoint.}{\epsfxsize4in\epsfbox{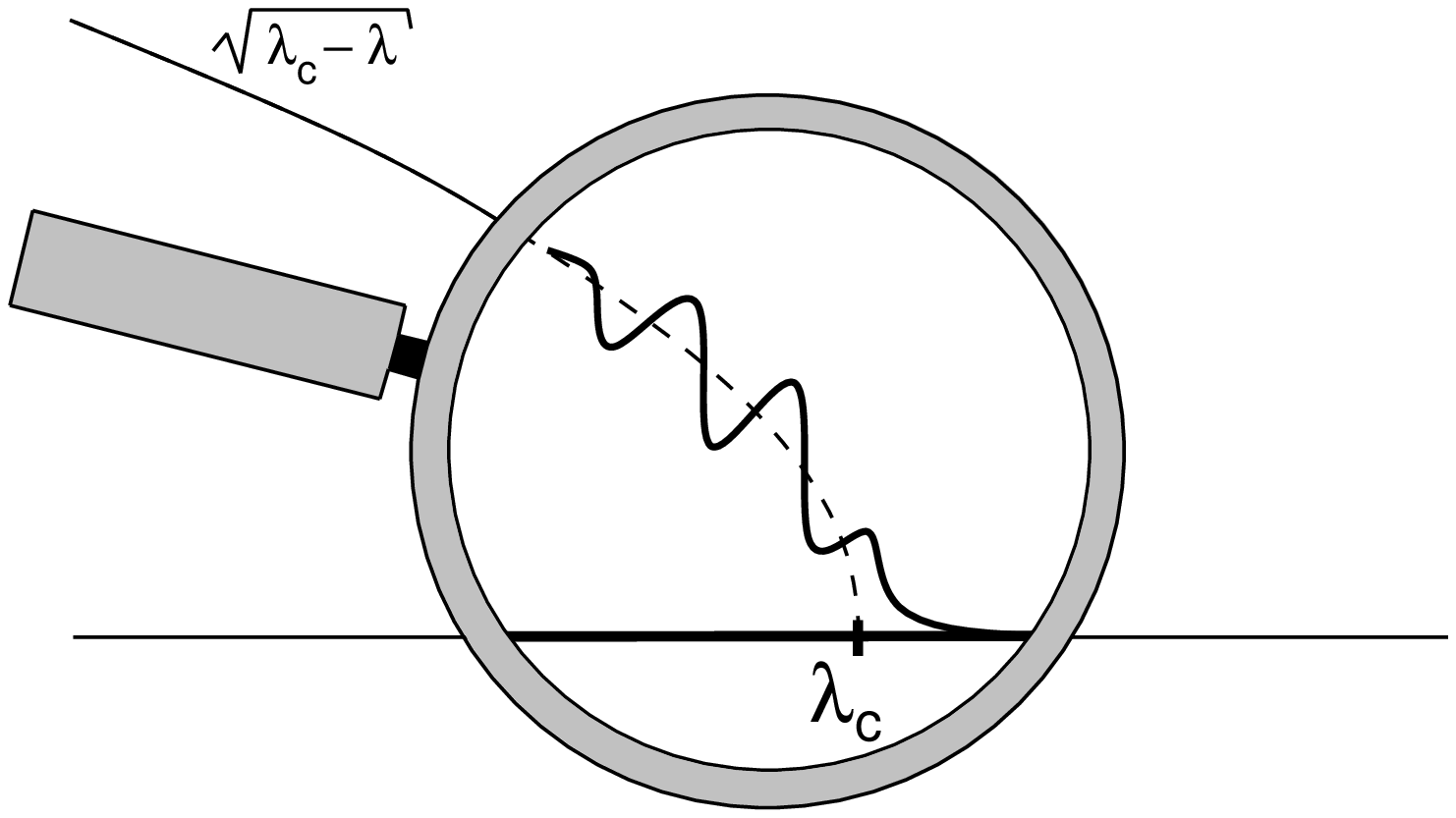}}

The edge of the eigenvalue distribution is at $\lambda=0$, and is
given by
\eqn\dscledev{
\rho(\lambda)=({\rm Ai}')^2-\lambda \kappa^{-2/3} ({\rm Ai})^2\ .}
{}From the asymptotics of Airy functions, we obtain the
picture of the eigenvalue distribution depicted in \fmag.
This completes our example of the Gaussian potential.\foot{We could have
deduced the connection to Airy functions more directly using the
WKB analysis of wavefunctions in a harmonic oscillator potential, but
that argument does not generalize to other orthogonal polynomials.}

\exercise{}

Using the asymptotics of the Airy function,
show that the double-scaled eigenvalue density behaves like:
\eqn\evasymps{
\eqalign{
\rho(\lambda)&\sim {\pi\over 4} \lambda^{1/4}\,
\ee{-{2\over 3 \kappa}\lambda^{3/2}}\qquad \lambda\to + \infty\cr
&\sim \pi (-\lambda)^{1/2} \qquad \lambda\to -\infty\cr}
}

\endexercise

Returning to the general case, we can use
\limpsi\  to find the behavior of the fermion two-point
function in the region of interest, namely, the
edge of the eigenvalue distribution. It is just
\eqn\kcont{
K(\lambda_c + a^{2/m} \lambda_1,\ \lambda_c + a^{2/m} \lambda_2)\to
a^{-2/m} K_{\rm cont}(\lambda_1,\lambda_2)\ ,}
where
\eqn\kconti{
K_{\rm cont}=\int_\mu^\infty \d z\, \psi(z,\lambda_1)\,\psi(z,\lambda_2)\ .}
(To prove this, note that the continuum limit of the
Darboux--Christoffel formula is
\eqn\kcontii{
K_{\rm cont}={\psi(\mu,\lambda_1)'\, \psi(\mu,\lambda_2)-
\psi(\mu,\lambda_2)'\, \psi(\mu,\lambda_1)
\over \lambda_1-\lambda_2}\ .
}
Taking a derivative with respect to $\mu$, we find
$\p_\mu K_{\rm cont}=-\psi(\mu,\lambda_1)\,\psi(\mu,\lambda_2)$, and
integrating gives \kconti.) From these remarks we derive
the main statement of double-scaled Fermi theory:

The nonanalytic dependence
on coupling constants in \mfreq\ comes from
the contributions in the integrals over
$\lambda$ from the
edge of the eigenvalue distribution. These
contributions  in turn may be
studied by using the double-scaled fermion field theory, i.e.,
the theory of free fermions with expansion
\eqn\dsff{\hat\psi(\lambda)=\int \d z\, a(z)\, \psi(z,\lambda)\ ,
}
where $\psi(z,\lambda)$ is the orthonormalized eigenfunction of the
Lax operator $Q$ which is exponentially decaying for $\lambda\to +\infty$
and oscillatory for $\lambda\to -\infty$.
The free oscillators satisfy
\eqn\dsoscc{\eqalign{a(z)|\mu\rangle=0 \quad  (z<\mu)\qquad\qquad
&a^\dagger(z)|\mu\rangle=0 \quad  (z>\mu)\cr
\{ a(z), a^\dagger(z')\}&= \delta(z-z')\ .\cr}}

In particular, continuum loop amplitudes are obtained from
the double-scaled operator creating macroscopic loops
\eqn\conloop{W(\ell)
=\int \d \lambda\,\ee{\ell \lambda}\,\hat\psi^\dagger \hat\psi(\lambda)\ .}
Although we integrate
$\lambda$ over the entire real axis, in fact the Laplace transform
converges. A detailed study of the asymptotics of the Baker-Akhiezer
functions shows that $\psi$ decreases exponentially fast
(as $\exp\bigl(-\lambda^{m+\ha}/\kappa\bigr)$)
off the region of perturbative support of the eigenvalue density,
and oscillates
with an algebraically decaying envelope within the region of support.
On that region, the eigenvalue
density grows algebraically and is Laplace transformable.

\newsec{Loops and States in Matrix Model Quantum Gravity} 
\seclab\slsmmqg

\subsec{Computation of Macroscopic Loops}
\subseclab\scoml

We now use the fermion formalism to calculate macroscopic
loop amplitudes in the one-matrix model. Beginning with
the one-loop amplitude we insert \conloop\ to get one
of the beautiful results of \bdss,\foot{Notice that this is valid only when the
Lax operator has a continuous spectrum. This criterion can be used to
select boundary conditions on the physically appropriate
solutions to the string equation: for example, it selects the
solutions first isolated in \rBMP.}
\eqn\oneloop{\eqalign{\bigl< W(\ell)\bigr>
&=\figins{\vcenter{\epsfxsize4in\epsfbox{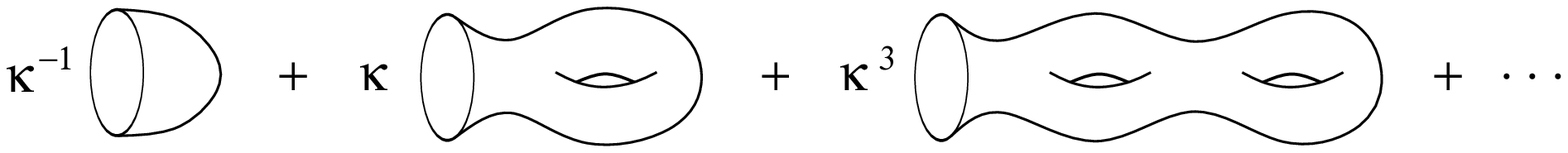}}}\cr
&=\int_{-\infty}^\infty \d\lambda\, \ee{\ell\lambda}\,
\langle \mu|\psi^\dagger\psi(\lambda)|\mu\rangle
=\int_{-\infty}^\infty \d\lambda\, \ee{\ell\lambda}\int_{\mu}^\infty \d z\,
\psi(z,\lambda)^2\cr
&=\int_{\mu}^\infty \d z\, \bigl< z|\ee{\ell Q}|z\bigr>\ .\cr}}
Similarly, the connected amplitude for two macroscopic loops is easily
shown to be \bdss
\eqn\twlp{\lbspace\eqalign{\bigl<W(\ell_1)\, W(\ell_2)\bigr>
&=\figins{\vcenter{\epsfxsize4in\epsfbox{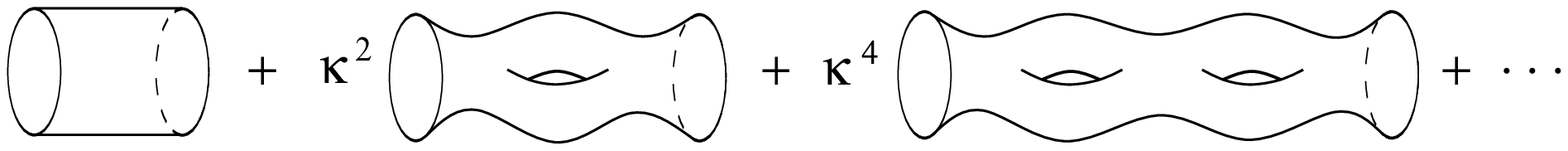}}}\cr
&=\int_{\mu}^\infty \d x\int^{\mu}_{-\infty}
\d y\, \bigl< x|\ee{\ell_1 Q}|y\bigr>\,\bigl< y|\ee{\ell_2 Q}|x\bigr>\ .\cr}}

The formulae \oneloop\ and \twlp, while elegant, do not make manifest the
physics of the models we are discussing. To address this problem, we examine
these formulae at genus zero. Since $\kappa$ counts loops, we can regard
the expectation value in \oneloop\ as a ``quantum-mechanical'' expectation
value, with $\kappa$ playing the role of $\hbar$, and obtain the genus zero
approximation to the loop formulae as follows. Using the
Campbell--Baker--Hausdorff formula to separate exponentials of $\hat p^2$ and
$u$, and then inserting a complete set of eigenfunctions
\eqn\cmplst{\langle p|x\rangle
\equiv{1\over\sqrt{2\pi \kappa}}\,\ee{ipx/\kappa}\ , }
we obtain
\eqn\gzolp{\eqalign{\bigl< W(\ell)\bigr>_{h=0}=
\int_{\mu}^\infty\d z\, \bigl< z|\ee{-\ell \hat p^2}\, \ee{-\ell u}|z\bigr>
&={1\over 2 \pi \kappa}
\int_{\mu}^\infty \d z \int_{-\infty}^\infty
\d p\, \ee{-\ell p^2}\, \ee{-\ell u}\cr
& ={1\over 2 \sqrt{\pi}\kappa\ell^{1/2}}
\int_{\mu}^\infty \d z\, \ee{-\ell u}\ .\cr}}

Let us consider this formula first for the case of pure gravity.
If we wish to calculate the expectation value of a loop with
the cosmological constant inserted,
we take a derivative with respect to $\mu$ to bring the operator
down from the action, yielding
\eqn\gzccw{\Bigl< W(\ell) \int \ee{\gamma \phi}\Bigr>
={1\over 2 \sqrt{\pi}\kappa\ell^{1/2}}\, \ee{-\ell u(\mu)}\ .}
In pure gravity, the string equation is the Painlev\'e I equation \eplv:
\eqn\paini{u^2-{\kappa^2\over 3} u''=z\ ,}
and the genus zero equation becomes simply $u(z)=z^{1/2}$.
The matrix model result for the wavefunction
of the cosmological constant is thus
\eqn\gzccwii{\eqalign{\lower10pt\hbox{$\ell$}
\figins{\vcenter{\epsfxsize30pt\epsfbox{punctan.ps}}}\lower8pt\hbox{$V$}\ \
=\Bigl< W(\ell) \int \ee{\gamma \phi}\Bigr> & \propto
\ \kappa^{-1}\mu^{1/4}\,(\sqrt{\mu}\ell)^{-1/2}\, \ee{-\ell \sqrt{\mu}}\cr
& \propto \ \kappa^{-1} \mu^{1/4} \,K_{1/2} (\ell \sqrt{\mu})\ ,\cr}}
precisely as expected from the continuum theory (e.g.\ sec.~{\it\sskpztd\/}).

\exercise{Spectrum of 2D Gravity}

Using the KPZ formula \kpzi, show that the spectrum of numbers
$\nu$ in the WdW equation \msswdw\ for the case of pure gravity is
$\nu=\nu_j=j+\half$, $j\geq 0$.

\endexercise

More generally, we may calculate the one macroscopic loop
amplitude for general perturbed $(2,2m-1)$ models coupled to
gravity along the following lines.
The genus zero limit of the string equation \eStrEqn\ can be written
\eqn\streq{\sum_{j\geq 0} t_j\, u^j=0\ .}
(Note $\mu=-t_0$. Recall that the $t_j$ describe the
coupling to the various scaling operators and if the
largest nonvanishing $t_j$ has $j=m$, we are describing
2D gravity coupled to the $(2,2m-1)$ minimal model.)
Using \streq, can explicitly evaluate the loop amplitude as
\eqn\gnzrlp{\eqalign{\bigl< W(\ell)\bigr>_{h=0}
={1\over \kappa \ell^{1/2}}\int_{-t_0}^\infty \d x\, \ee{-\ell u(x;t_i)}
&={1\over \kappa \ell^{1/2}}\int_u^\infty
\d y \bigl(\sum_{j=1}^\infty j\,t_j\, y^{j-1}\bigr)
\ee{-\ell y}\cr
&={1\over \kappa \ell}\sum_{j=1}^\infty j\,
t_j\,u^{j-1/2}\,\psi\dup_{j-1}(\tilde\ell)\ ,\cr}
}
where
\eqn\trun{\psi\dup_j(x)\equiv j!\, x^{-j-1/2}(1+x+x^2/2!+\cdots
x^j/j!)\,\ee{-x}\ ,}
and $\tilde\ell\equiv u\ell$. (We assume here that all but finitely many
$t_j$ are nonzero.)
The relation of these amplitudes to
the Bessel functions of the continuum theory is less evident
than in \gzccwii\ and will be explained in sec.~{\it\swapmm\/}.

Similarly, we can study the genus zero approximation to
the propagator as
\eqn\gzrtwlp{\eqalign{\bigl< W(\ell_1) W(\ell_2)\bigr>_{h=0}&=
{\ee{-(\ell_1+\ell_2)u}\over\sqrt{\ell_1\ell_2}\kappa^2}
\int_{-t_0}^\infty \d x\int^{-t_0}_{-\infty}\d y\,
\ee{-(x-y)^2(\ell_1+\ell_2)/(4\kappa^2\ell_1\ell_2)}\cr
&=2\sqrt{\ell_1\ell_2}\,
{\ee{-u(\mu)(\ell_1+\ell_2)}\over \ell_1+\ell_2}\ .\cr}}

\exercise{}

Prove \gzrtwlp\ by inserting \cmplst\ into \twlp. Similarly, try to prove
\eqn\zrthrlp{\bigl< W(\ell_1)\,W(\ell_2)\,W(\ell_3)\bigr>=2 {\p u\over\p t_0}
\sqrt{\ell_1\ell_2\ell_3}\,\ee{-u(\ell_1+\ell_2+\ell_3)}\ .}
(We will prove this more efficiently below.)

\endexercise

\subsec{Loops to Local Operators}
\subseclab\sltlo

By shrinking the loops we can obtain correlation functions of
the local operators. This intuition comes from the critical
string example discussed in chapt.~{\it\slascft\/} and
from the expression for the matrix model loop operator \secloop\
which is manifestly an expansion in local operators.

In eq.~\ehammj, we saw that in the matrix model there are
scaling operators $\sigma_j\propto \QT^{j-1/2}_+$ scaling like
$\ee{\alpha_j \phi}$ with $\alpha_j/\gamma=\half(m-j)$, $j=0,1,\dots\,$.
According to \exfrsj,
the macroscopic loop operator has an expansion as in \liuloco,
\eqn\loco{
W(\ell)=\sum_{j\geq 0} \ell^{j+\ha} \sigma_j\ .}

\exercise{Exponents}

Use the result \exfrsj\ to verify the expansion
\loco\ using the fact that for pure gravity we
have $Q/\gamma=5/2$ and $\alpha_j/\gamma=1-j/2$, $j=0,1,\dots$.

\endexercise

As we have already discussed in the context of semiclassical
Liouville theory,
the expansion \loco\ is not strictly true and must be treated with care. As we
see from \gnzrlp, there can be negative powers of $\ell$ in the small $\ell$
expansion of loop correlators. In sec.~{\it\sswb\/},
we showed that the $\ell\to 0$ behavior of loop amplitudes
must satisfy certain rules which imply that one can
unambiguously extract the correlators of local operators.
The rules of sec.~{\it\sswb\/} are confirmed by
explicit matrix model computations. For example,
notice that \gnzrlp\ can be written as
\eqn\exloop{
\bigl< W(\ell)\bigr>_{h=0}=\sum_{j=0}^m j!\, t_j {1\over \ell^{j+\half}}
+\sum_{n\geq 0} {\ell^{n+\half}\over
\Gamma(n+{3\over 2})}\langle \sigma_n\rangle\ .}
The divergent terms in $\ell$ are indeed analytic in the coupling constants.
Similarly, notice that \eqns{\gzrtwlp{,\ }\zrthrlp}\ are smooth as any
looplength goes to zero. In general, with the rules (1) and (2) from the end of
sec.~{\it\sswb\/} in mind, we can extract correlation functions of local
operators by shrinking macroscopic loops.

\exercise{}

Using \zrthrlp, calculate
$\langle \sigma_{n_1}\sigma_{n_2}\sigma_{n_3}\rangle$.

\endexercise

\exercise{The general amplitude}

Using rules (1) and (2) and the genus zero KdV flow equations,
we will prove that
\eqn\enlp{\Bigl<\prod_{i=1}^n W(\ell_i)\Bigr>=\prod \ell_i^{1/2}
\biggl({\p\over\p t_0}\biggr)^{n-3}\,\ee{-u \cdot \sum \ell_i}\quad.}
For example, to prove the three macroscopic loop formula proceed as follows:
\eqn\thrlpii{\eqalign{\bigl< W(\ell_1)\,W(\ell_2)\,W(\ell_3)\bigr>
&=\sum_{n=0}^{\infty}  \ell_1^{n+1/2}
\bigl<\sigma_n W(\ell_2)\,W(\ell_3)\bigr>\cr
&=\sum_{n=0}^{\infty} \ell_1^{n+1/2}
{\p\over \p t_n}\bigl< W(\ell_2)\,W(\ell_3)\bigr>\cr
&=-2 \sqrt{\ell_2\ell_3}\,\ee{-u(\ell_2+\ell_3)}
\sum_{n=0}^{\infty} (-1)^{n+1} {\ell_1^{n+1/2}\over n!}
{\p u\over \p t_0}\cr
&=2 {\p u\over\p t_0}
\sqrt{\ell_1\ell_2\ell_3}\,\ee{-u(\ell_1+\ell_2+\ell_3)}\ .\cr}
}
In the last line we may obtain the $\ell$ dependence
immediately since the
amplitude must be totally symmetric in $\ell_1,\ell_2,\ell_3$.
Give a complete proof of \enlp\ along these lines by induction.

This formula was first discovered in \ambj\
from a different point of view and then rediscovered in
\mss. The strange fact that the amplitude is essentially
only a function of the sum of the loop lengths has never
been given a simple explanation.

\endexercise

\subsec{Wavefunctions and Propagators from the Matrix Model}
\subseclab\swapmm

Let us finally match the expectations of sec.~{\it\sskpztd\/} above,
specifically the Bessel function behavior of wavefunctions,
with the results of the matrix model computations of sec.~{\it\scoml\/}.
At first the results appear to be very different but this
turns out to be a matter of working in two different bases.

One quick way to see this is to use the Gegenbauer addition formula to expand
the genus zero propagator \gzrtwlp\ in terms of Bessel functions\foot{Although
we have pulled this identity out of a hat, it is quite natural. The
Yukawa potential in {\it three\/} spatial dimensions, which is the Green's
function for the Helmholtz operator $\nabla^2-\mu$, is just
$\ee{-\mu|r_1-r_2|}/|r_1-r_2|$. This Green's function may be expanded in
partial waves and the Gegenbauer addition theorem amounts to that expansion.}
\eqn\aplgg{
\sqrt{\ell_1\ell_2}\, {\ee{-u(\ell_1+\ell_2)} \over \ell_1+\ell_2}
=\theta(\ell_2-\ell_1)\sum_{j=0}^\infty  (-1)^j(2j+1)
\,I_{j+\ha}(\ell_1 u)\, K_{j+\ha}(\ell_2 u)+[1\leftrightarrow 2]\ .}

This suggests that instead of the local operator expansion
\loco, we use a different expansion
\eqn\changebasis{W(\ell)=\sum_{j\geq 0} \ell^{j+\ha} \sigma_j=
2\sum_{j=0}^\infty \hat\sigma_j (-1)^j(2j+1)
{I_{j+\half}(\ell u)\over u^{j+1/2}}\ .}
That is, instead of expanding the loop in terms of
the functions $\ell^{1/2},\ell^{3/2},\ell^{5/2},\dots\,$,
we use the basis functions $I_{1/2},I_{3/2},\dots\,$.
Using the properties of $I$, we see that this change
of basis is upper triangular and hence the operators
$\hat \sigma_j$ are related to $\sigma_j$ by
an upper triangular transformation whose coefficients
are analytic functions of $u^2$.

Since we isolate continuum contributions from nonanalytic dependence on
couplings like $\mu$, we must not mix operators with coefficients that are
nonanalytic in $\mu$. Conversely, we are always free to make redefinitions
involving coefficients which are analytic in $\mu$. Since the critical
exponents $\alpha_j/\gamma$ are rational, there can be operator mixing, and
hence there is no unique definition of scaling operators. In order for the
change of basis \changebasis\ to satisfy this rule, $u^2$ must be an analytic
function of the couplings $t_k$. One way this can happen is by considering
perturbations around the pure gravity point where $u^2=2 \mu$.\foot{More
generally, one should look at the so-called ``conformal backgrounds,'' which,
as argued in \mss, are the precise matrix model couplings corresponding to a
tensor product with a conformal $(2,2m-1)$ model. An understanding of these
backgrounds was needed to resolve certain paradoxes about one- and two-point
functions in 2d gravity \mss.}

The wavefunctions of the operators
$\hat \sigma_j$ are given by shrinking one of the loops,
\eqn\wvfnht{
\langle \hat \sigma_j W(\ell)\rangle = u^{j+\ha} K_{j+\ha}(u \ell)\ ,
}
in complete agreement with the Euclidean on-shell wavefunctions
of sec.~{\it\sskpztd\/}! Thus we have actually done better than we had
any right to expect, the minisuperspace approximation to the
wavefunctions turned out to be exact for these boundary conditions.

In the theory of a particle, the propagator was written in terms
of on-shell and off-shell states as in \fftpr\ above. Similarly
here we may write the matrix model propagator in a way which
nicely summarizes the spectrum of the theory
\eqn\decpropx{\bigl< W(\ell_1) W(\ell_2)\bigr>\equiv\int_{0}^\infty
{\d E\over 2\pi}\, G(E)\,\psi\dup_E(\ell_1)\,\psi\dup_E(\ell_2)\ ,}
where
\eqn\cshpr{G(E)=\sum_{j=0}^\infty{(-1)^j(2j+1)\over E^2+(j+\half)^2}
={\pi\over \cosh\pi E}\ .}
It is extremely interesting to note that --- even for pure
gravity --- the third quantized universe propagator is
{\it not\/} the naive minisuperspace Wheeler--DeWitt propagator
\stanprop. In particular the ultraviolet behavior of
the propagator (in $E$) is completely different from the naive
propagator \stanprop. For example,
the $1/E^2$ behavior in the ultraviolet becomes $\ee{-\pi E}$ behavior.
Our understanding of why this is so is incomplete.
(Part of the story is explained in the next section.)

In general, we can decompose amplitudes  as
\eqn\gendecomp{
\bigl< W(\ell_1)\cdots W(\ell_n)\bigr>
=\int \prod_i \d E_i\, \psi\dup_{E_i}(\ell_i)\ A(E_1,\dots, E_n)\ ,
}
which (in the case of the four-point amplitude) we depict as
$$
\raise25pt\hbox{$\ell_1$}\llap{\raise-25pt\hbox{$\ell_4$}}
\figins{\vcenter{\epsfxsize.8in\epsfbox{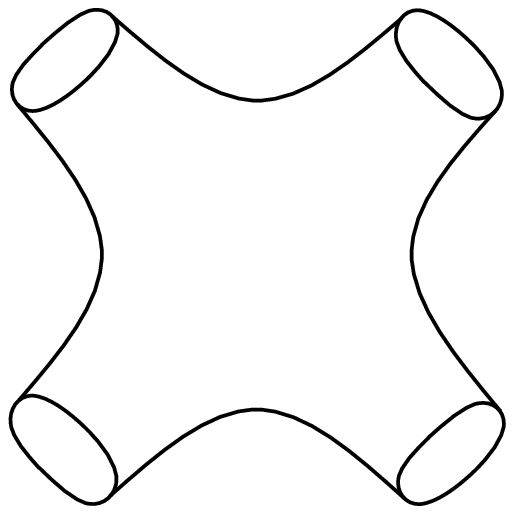}}}
\raise25pt\hbox{$\ell_2$}\llap{\raise-25pt\hbox{$\ell_3$}}
=\int \prod_i \d E_i\ \psi\dup_{E_i}(\ell_i)
\ \raise25pt\hbox{$E_1$}\llap{\raise-25pt\hbox{$E_4$}}
\figins{\vcenter{\epsfxsize.8in\epsfbox{4pt.ps}}}
\raise25pt\hbox{$E_2$}\llap{\raise-25pt\hbox{$E_3$}}\quad.
$$
By shrinking $\ell_i$ to zero we get an expansion in terms of local operators.
Alternatively, and equivalently, by doing the integral over
the $E_i$ we pick up residues corresponding to the Euclidean on-shell
states. A similar picture emerges for all the multicritical points.
The following exercise carries this out in detail
for the Ising model.

\exercise{The Ising Model}


The Ising model has a $\IZ_2$ symmetry flipping up spins for
down, which, in the matrix model formulation described in
sec.~{\it\ssIM\/} is exchange of $U\leftrightarrow V$.
Letting $W_\pm(\ell)$ denote the $\IZ_2$ odd/even loop operators
show that
\eqn\isloops{\eqalign{\bigl< W_\pm(\ell_1) W_\pm(\ell_2)\bigr>
&=\pm\sum_{j,\pm}(j+1/3)\,I_{j+1/3}
(2\sqrt{\mu}\ell_1)\,K_{j+1/3}(2\sqrt{\mu}\ell_2)\cr
&\qquad\mp\sum_{j,\mp}
(j+2/3)\,I_{j+2/3}(2\sqrt{\mu}\ell_1)\,K_{j+2/3}(2\sqrt{\mu}\ell_2)\ .\cr}}
By summing the infinite series, show that
\eqn\gforis{G(E,\pm)=2\pi(\ee{\pi E}\pm 1+\ee{-\pi E})
{\sinh \pi E\over \sinh 3\pi E}\ .}

\endexercise

{\bf Remark}: We have shown that the wavefunctions
$\bigl< \sigma W(\ell)\bigr>$ satisfy a linear WdW
equation. On the other hand, from the Schwinger--Dyson
equations \loopeq\ and their continuum analogs, we see that
$\bigl< W(\ell)\bigr>$ itself satisfies a nonlinear
equation. A precise understanding of the relation of
these has never been given (except in special cases
\tomunpb). This is an interesting problem for the future.

\ifig\ftwolc{Two loops on a continuum surface collide.}
{\epsfxsize2.5in\epsfbox{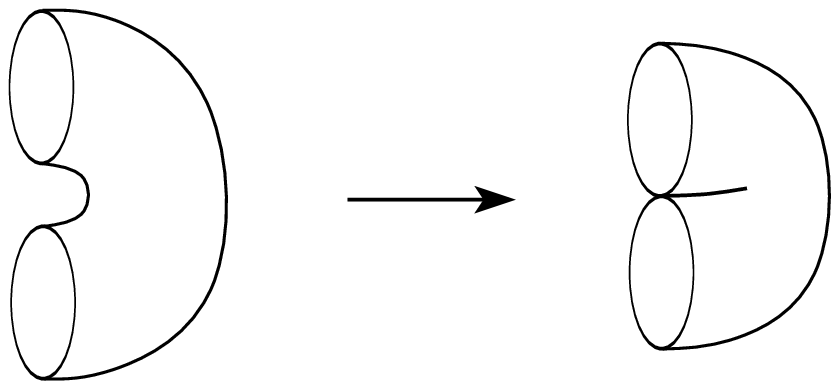}}

\subsec{Redundant operators, singular geometries and contact terms}
\subseclab\ssrpsgct

One important and not generally discussed issue is the
contribution of singular geometries to the path
integral. In the case of macroscopic loop amplitudes, there
are geometries in which loops collide to make figure-eights
as in \ftwolc, and as well more complicated geometries.
Our understanding of the contributions of these geometries is
very incomplete, but there is plenty of evidence that such
terms are responsible for several peculiarities of the matrix
model answers (e.g. the $\cosh$ propagator discovered above)
and perhaps lie at the heart of a geometrical understanding
of the Lian--Zuckerman states.
See also the discussion at the end of sec.~{\it\ssmlaco\/} below.

\newsec{Loops and States in the $c=1$ Matrix Model}
\seclab\slascomm

\subsec{Definition of the $c=1$ Matrix Model }

There are several approaches to defining a matrix model for
gravity coupled to $c=1$ matter.
The most direct method is the discretization of the
Polyakov path integral for a one-dimensional target space,
\eqn\dispol{\CZ_{\rm qg}(\kappa,g)
=\sum_\Lambda \kappa^{2h-2}\, g^{|\Lambda|}
\prod_{i=1}^V\int \d X_i \ \e{-\sum_{\langle ij\rangle} L(X_i-X_j)\ ,}}
where $|\Lambda|$ is the
number of vertices on the lattice $\Lambda$ which is summed over
Euler character $2-2h$, and the nearest neighbor
interaction $L(X_i-X_j)$ between the bosonic fields $X_{i,j}$
at vertices $i,j$ is summed over links $\langle ij\rangle$ between vertices.

The asymptotic expansion in $\kappa$ of the partition function
\dispol\ can be equivalently generated\foot{Note that this matrix model
construction works equally well for bosons $X^\mu$, $\mu=1,\ldots,D$,
to generate strings embedded in $D$ dimensions. For $D>1$, however, the
matrix model representation is no longer solvable.}
from an integral over $N\times N$ matrices,
\eqn\empf{\lbspace \CZ(N;g)=\ln\int {\rm D}\Phi\
\e{-Ng\inv\,\tr\bigl[\int\d X\,\d Y\ \half \Phi(X)\,G\inv(X-Y)\,\Phi(Y)
+\int \d X\ V\bigl(\Phi(X)\bigr)\bigr]}\ ,}
where $\Phi(X)$ is an $N\times N$ hermitian matrix field, $V$ is a polynomial
interaction of some fixed order, and the propagator $G(X)=\exp-L(X)$.
$g$ is a loop counting parameter, and therefore counts the number of
vertices in the dual graph (identified with the area of the lattice).
As in sec.~{\it\sdsmmcl\/}, the coefficient of $N^\chi$ in
an expansion of $\CZ$ in a powers of $N^2$ gives
the sum of all connected Feynman diagrams with Euler character $\chi$.
As functions of $g$, these coefficients are all singular at a critical
coupling $g\dup_c$ where the perturbation series diverges.
The continuum limit can be extracted from the leading singular behavior as
$g\to g\dup_c$, a limit which emphasizes graphs with an infinite
number of vertices.

Taking $L(X_i-X_j)=(X_i-X_j)^2$ in \dispol\ leads to the continuum limit
form $\int d^2\xi\,\sqrt g\,g^{ab}(\grad a X)(\grad b X)$, thus providing a
standard discretization of the Polyakov string \rpoly\
embedded in one dimension.
This quadratic choice corresponds to a gaussian
propagator, $G(X)\sim\exp(-X^2)$, in the matrix model \empf.  In momentum
space, the leading small momentum behavior of the gaussian form
$G\inv(P)\sim \exp P^2$ coincides with that of the Feynman form $G\inv(P)\sim
1+P^2$, which corresponds in position space to $G(X)=\exp(-|X|)$.  As
argued in \kazmig, this substitution (corresponding to $L(X_i-X_j)=|X_i-X_j|$,
with continuum form $|g^{ab}\,\del_a X\,\del_b X|^{1/2}$), should not affect
the critical properties (e.g.\ critical exponents). Due to the
ultraviolet convergence of the model, only its short distance,
i.e.\ non-universal behavior, is affected.\foot{Indeed we will see that
energies of order $\epsilon\sim 1/N$ dominate the continuum limit.}
For the same reason, continuum answers should only depend on the universality
class of the potential $V$, the necessary conditions for which will be
discussed below. For now we simply require $V(\phi)$ to go to $+\infty$
for $\phi\to \pm\infty$ in order that \empf\ is well-defined.

For the latter choice of propagators, i.e.\ the Feynman propagator,
after rescaling $\Phi$ we can write \empf\ as
\eqn\matint{
\CF_{\rm mm}(N;g,V)\equiv \lim_{T\to \infty} T^{-1} \ln\Bigl(
\int {\rm D}\Phi(X)\, \e{-N\int_0^T \d X\, \tr(\dot\Phi^2+g^{-1}
V(\sqrt{g} \Phi))}\Bigr)\ .}
Using Feynman diagrams to obtain the
large $N$ asymptotics of the function $\CF_{\rm mm}$, we write (as in
\dispol)
\eqn\aympoff{\CF(N;g,V)_{\rm mm}\sim \sum_{\Lambda} N^{2-2h}
g^{|\Lambda|} \int \d X_i \prod_{\langle ij\rangle} \ee{-|X_i-X_j|}\ ,}

The quantum mechanical model \matint\
was solved to leading order in large $N$ in \rBIPZ.
Interpreting the solution as the partition function \aympoff\
of 2d gravity on a genus zero worldsheet coupled to a single gaussian massless
field,  it was shown in \kazmig\ that the string susceptibility exponent,
defined by the leading singular behavior
$Z(g)=(g\dup_c-g)^{2-\Gamma_{\rm str}}$, satisfies $\Gamma_{\rm str}=0$, in
agreement with the continuum prediction of \rKPZ.
The emergence of such physically reasonable answers in the continuum limit
supports the assumption of universality with respect to the
choice of propagator.

We will now study the continuum limit of the integral
\matint\ to confirm and extend the above discussion.

\subsec{Matrix Quantum Mechanics}
\subseclab\smqm

{}From general principles, we see that \matint\ is simply the ground state
energy for the quantum mechanics of an $N\times N$ matrix, and was analyzed
from this point of view in \rBIPZ.

The reduction to free fermions can be established quickly using a path integral
argument given in \bkz. In the matrix quantum mechanics, we can discretize the
time coordinate $X\to X_i$ and then pass to the action
\eqn\disceact{S=N\sum_i \tr\, \Phi(X_i)\, \Phi(X_{i+1})
+ \sum \tr\, V\bigl(\Phi(X_i)\bigr)\ .}
We now analyze the model as a matrix chain model as in sec.~{\it\ssmmm\/},
diagonalizing
\eqn\diagn{\Phi(X_i) = \Omega_i\, \Lambda_i\, \Omega_i^{-1}\ ,}
where $\Lambda_i={\rm Diag}(\lambda_1(X_i),\dots \lambda_N(X_i))$.
The Vandermonde determinants all cancel except for the first and last.
Taking the time lattice spacing to zero,
we are left with a path integral for $N$ quantum
mechanical degrees of freedom $\lambda_i(t)$,
\eqn\pathfrlam{\eqalign{\CF_{\rm mm}(N;g,V)
&\equiv \lim_{T\to \infty} T^{-1} \log\Bigl(\int \prod_{i=1}^N
D \lambda_i(t)\, \Delta(\lambda_i(0))\,\Delta(\lambda_i(T))\cr
&\qquad\qquad\qquad\qquad\cdot \ee{-N\int_0^T \d x \sum_{i=1}^N\bigl(\dot
\lambda_i^2+g^{-1} V(\sqrt{g} \lambda_i)\bigr)}\Bigr)\ .\cr}}

Thus we are studying the quantum mechanics of $N$ free fermions
moving in a potential $g^{-1} V(\lambda \sqrt{g})$ with
Planck's constant equal to $\hbar=1/N$,
\eqn\fgnd{\CF_{\rm mm}(N;g,V)\equiv {1\over \hbar} E_{\rm ground}
=N\sum_{i=1}^N \epsilon_i(\hbar=1/N;\ g,V)\ ,}
where
\eqn\schrod{\Bigl(-{1\over 2N^2} {\d^2\over \d\lambda^2}+{1\over g}
V(\lambda\sqrt{g})\Bigr)\psi\dup_i=\epsilon_i \psi\dup_i\ .}

To define the double scaling limit we must:
\item{1)} Compute the asymptotic expansion as $N\to \infty$,
\eqn\expoff{\CF_{\rm mm}\sim \sum N^{2-2h} \CF_h(g,V)\ .}
\item{2)} Isolate the leading singular behavior of $\CF_h$ as $g\to g_c$.
\item{3)} Determine the scaling variable and scaling functions as
$g\to g_c$ and $N\to \infty$.

\ifig\fpotential{Generic potential with quadratic maximum at $\lambda=0$.}
{\epsfxsize3in\epsfbox{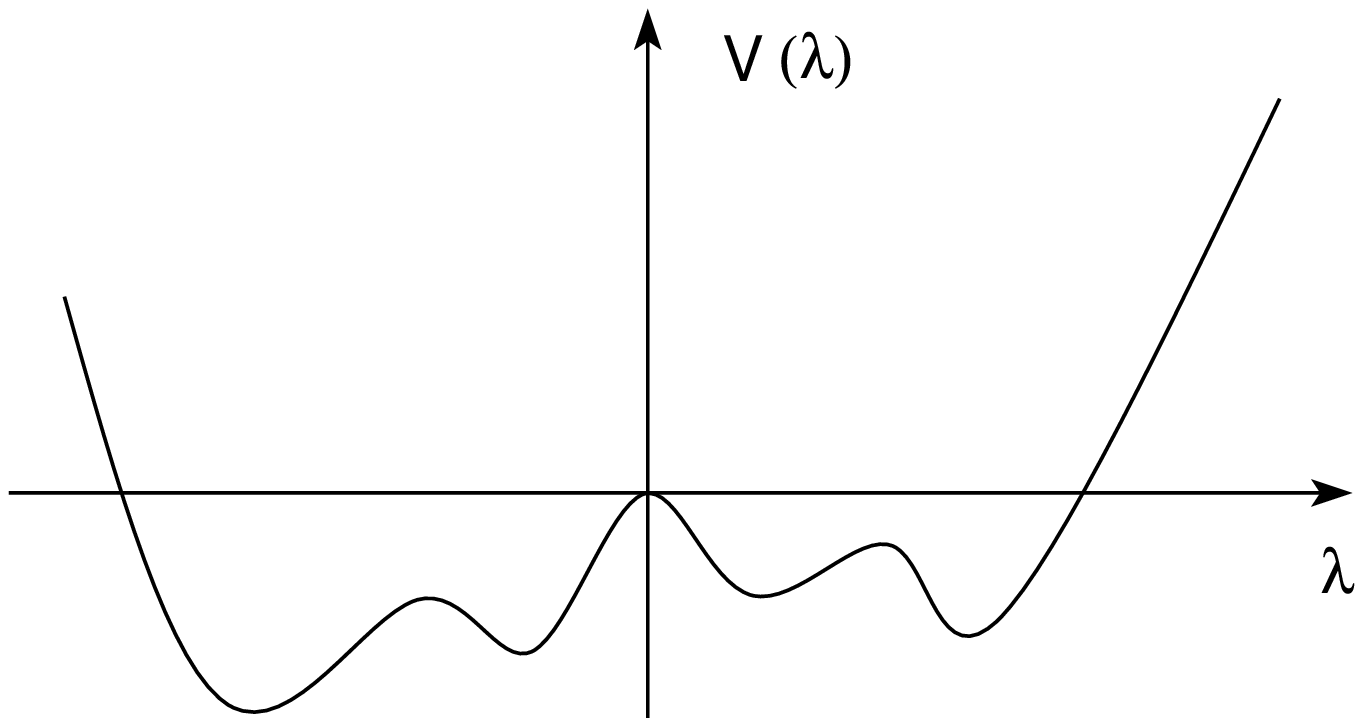}}

We begin by studying the function $\CF_0(g,V)$ corresponding to
genus zero surfaces.
The potentials of interest are polynomials that
have a quadratic maximum. We place this maximum at
$\lambda=0$ and shift $V$ so that $V(0)=0$, so that
$V$ might look as in \fpotential. To fix ideas, one can take
\eqn\vexmpl{V(\lambda)=-\ha \lambda^2 + {1\over 4}\lambda^4\ ,}
but it is important to note that the results
hold for a large class of potentials, thus providing
evidence that we are calculating true continuum results and
not lattice artifacts.

Since $\hbar=1/N$ in our problem, the function $\CF_0$ can be calculated
as the leading term in a semiclassical expansion using the WKB
approximation. Classically the energy becomes continuous
and particle states are specified by points in fermion phase space
$(\lambda,p)$. By the Pauli exclusion principle, we can put at
most one fermion in each volume element of area $2 \pi \hbar$ in
phase space. At the same time we are putting $\CO(1/\hbar)$ distinct
particles into the sea, so the sea covers a region of area $\CO(1)$.
In the classical limit, the state described
by the Fermi sea of the free fermions is thus a region in phase space.
By the Liouville theorem, the time evolution of the system
preserves the area of this region, so we may think of
the region as a fluid in phase space. We will return to
this picture in chapters \sfsdcft\ and \sscatt\ below.
The fluid has a total area determined by the Fermi level, which is
in turn fixed by the total number of fermions,
\eqn\fixnum{
N=\int {\d p\,\d\lambda\over 2 \pi \hbar}\, \theta(\epsilon_F-\epsilon)\ ,}
implying that
\eqn\effg{1=\int {\d p\,\d\lambda\over 2 \pi } \theta(\epsilon_F-\epsilon)\ ,}
where
$$\epsilon(p,\lambda)=\half p^2 +{1\over g}V(g\sqrt{\lambda})\ .$$
Thus, we require that the fluid have total area one. The total energy is
\eqn\toteng{E_{\rm ground}=\int {\d p\, \d\lambda\over 2\pi \hbar }
\,\epsilon\,\theta(\epsilon_F-\epsilon)\ ,}
which implies
\eqn\sfen{\CF_{0}(g;V)=\int
{\d p\, \d\lambda\over 2 \pi}\, \epsilon\,\theta(\epsilon_F-\epsilon)\ .}
Eq.~\effg\ determines the Fermi level $\epsilon_F$ as a
function of $g$, and then \sfen\ determines $\CF_0(g;V)$.

\ifig\fequi{Level curves in phase space. Filled Fermi levels are shaded.}
{\epsfxsize3.5in\epsfbox{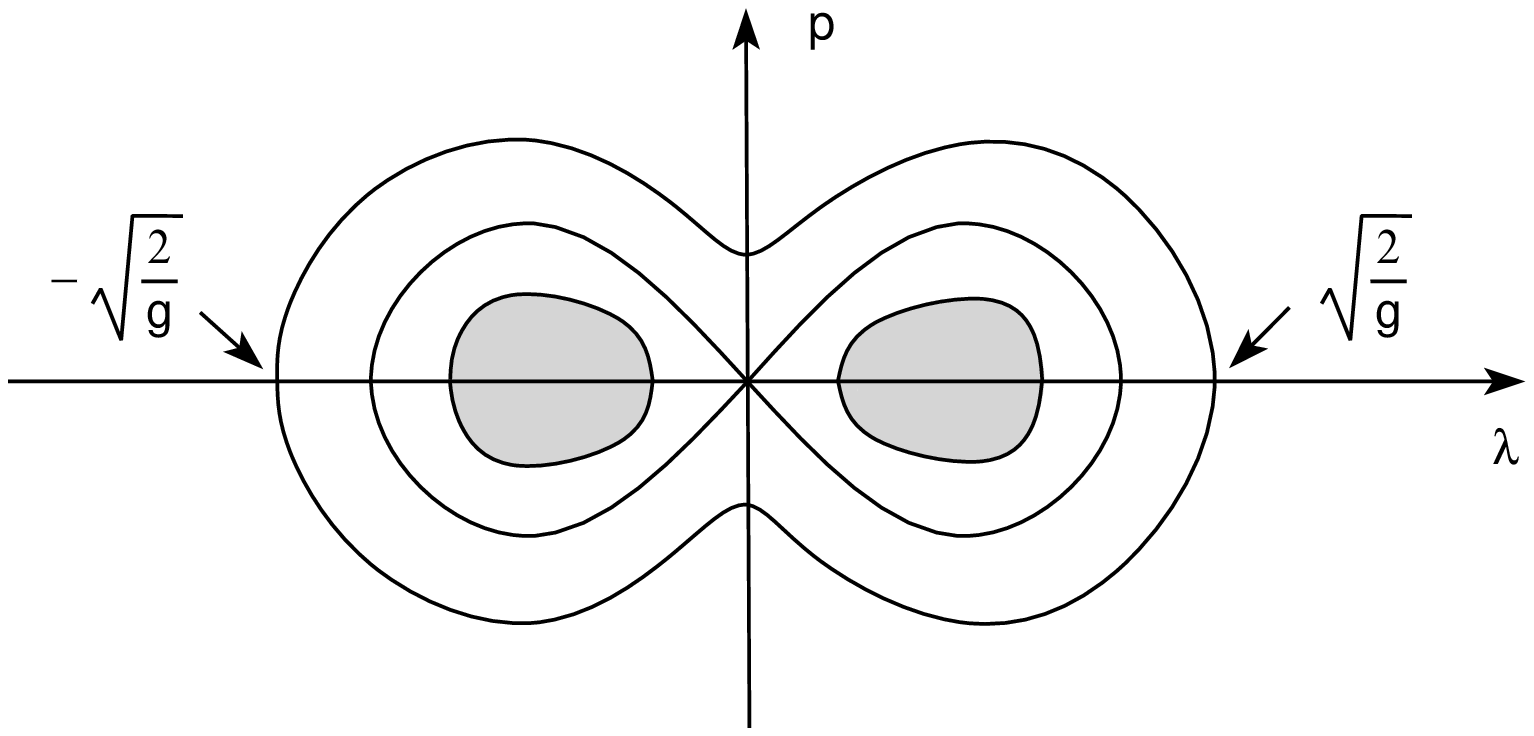}}

To see how singularities of $\CF_0$ can arise, let us
consider the specific potential \vexmpl\ for which
\eqn\epxmpl{\epsilon(p,\lambda)
=\ha p^2 -\ha \lambda^2 +{1\over 4} g \lambda^4\ .}
Level curves for $\epsilon$ are plotted in \fequi.
At small values of $g$, the lattice expansion \aympoff\
(at fixed topology) converges.
To define the continuum limit, we look for the leading singularity in $\CF_0$
due to the singularity in $g$ closest to the origin.
In general, as we vary $g$ the region of unit area defined by
\effg, and hence the weighted area \sfen, vary analytically.
As we tune $g$ from small values (as in \fequi) to large values, however,
$g$ passes through a value of order 1 where the $\epsilon=0$ line surrounds
the unit area. At this juncture the shape of the Fermi sea equipotential
changes discontinuously, resulting in nonanalytic behavior in
$\CF_0$. Thus we are interested in the limit $g\to g_c$, where $g_c$ is
defined by equating the Fermi level $\epsilon_F$ with the
top of the quadratic maximum ($=0$ here by convention).

\ifig\fblowup{Blowup near origin of \fequi.}
{\epsfxsize2.5in\epsfbox{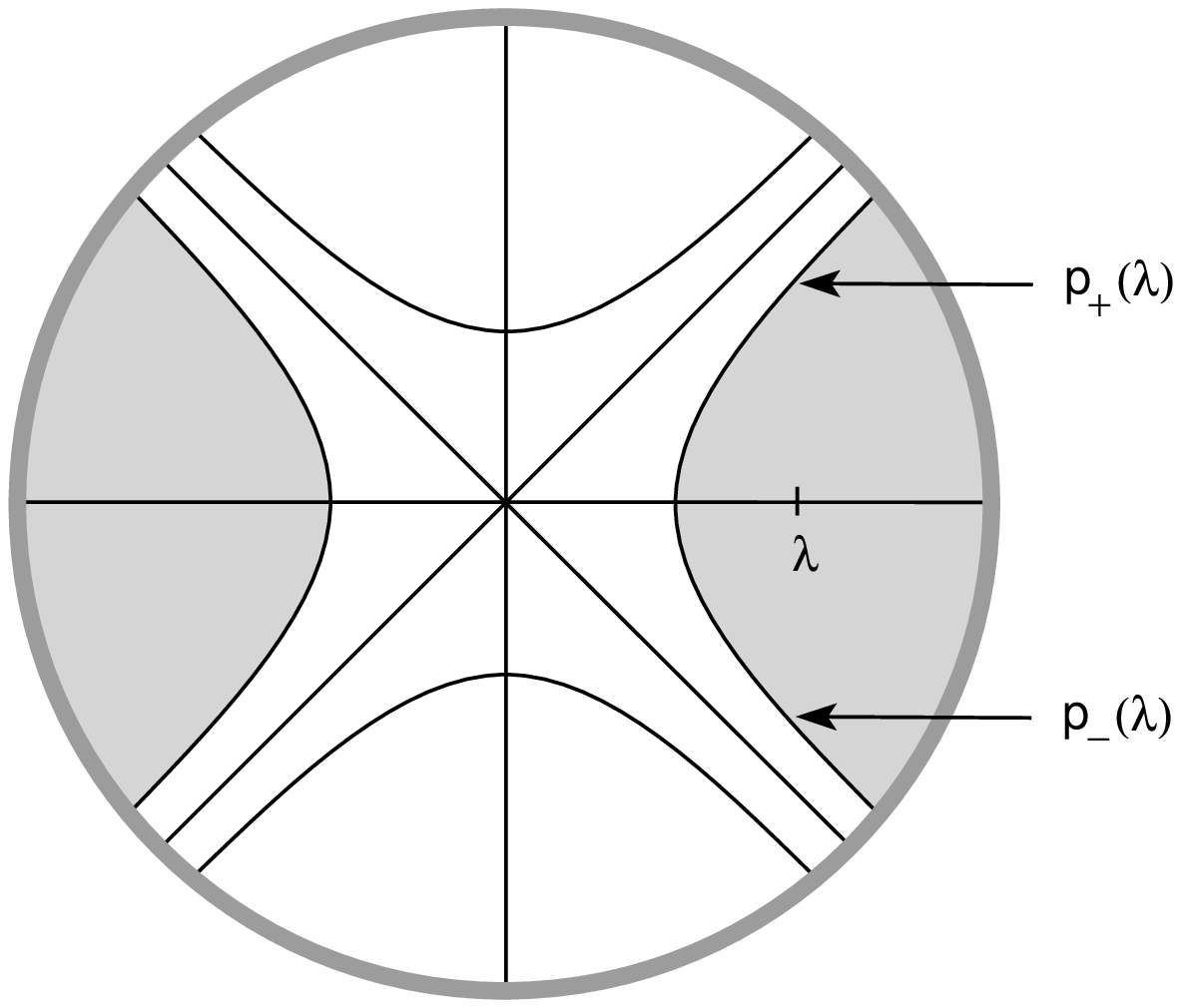}}

{}From the above discussion, it is (intuitively) clear that the
nonanalytic part of the integrals \eqns{\effg{,\ }\sfen}\ comes from
the crossing region $\lambda,p\sim 0$, and blowing up this region
gives the picture in \fblowup.
The singular dependence of $\CF_0$ can therefore be
determined by
\eqn\fnought{\eqalign{\CF_0
&=2 \int_{\lambda_1}^{\lambda_2} {\d\lambda\over 2 \pi}\,
\Bigl({1\over 6}(p_+^3-p_-^3)+(-\half \lambda^2+{g\over
4} \lambda^4)(p_+-p_-)\Bigr)\cr
&={2\over \pi}\int_{\sqrt{-2 \epsilon_F}}\d \lambda\Bigl(
{1\over 3}(\epsilon_F+\half \lambda^2)-\half \lambda^2+
{g\over 4} \lambda^4\Bigr)\sqrt{\lambda^2+2 \epsilon_F}+\cdots\cr
&={1\over \pi}\Bigl( \epsilon_F^2+{1\over
4} \epsilon_F^3 g\Bigr)\log(-\epsilon_F)+\cdots\ .\cr}}
Here $\lambda_{1,2}$ are the two turning points, $p_\pm$ define
the upper and lower branches of the Fermi surface, the
first line is exact, and
the terms omitted in the subsequent lines are analytic in $g$ and $\epsilon_F$.
Since the critical value for $g$ is of order one, we
immediately obtain the leading nonanalytic behavior as $g\to g_c$ as\foot{We
have made a constant rescaling to eliminate irrelevant numerical factors.}
\eqn\freen{\eqalign{\CF_0 &= \epsilon_F^2
\log (-\epsilon_F) + \cdots\cr
N^2 \CF_0 &= \mu^2 \log \mu + \cdots\ ,\cr}}
where
\eqn\defofmu{\mu\equiv -N \epsilon_F\ ,}
and the ellipsis in the second line of \freen\ indicates terms which are
analytic or less singular in $\mu$. (In particular, we can write
$\log\mu$ rather than $\log(-\epsilon_F)$ in \freen\
since the difference is just an analytic piece $\mu^2\log N$.)

\exercise{Doing the integrals}

For the specific example \epxmpl, all the
integrals can be done explicitly in terms of
elliptic functions. Perform these integrals and verify the statements
about nonanalytic dependence from the exact results.

\endexercise

\noindent Four important remarks:
\item{1)}From the above derivation, it is clear that the critical properties
are independent of the detailed form of $V$ and depend only on
the existence of a quadratic maximum.
\item{2)} The result \freen\ suggests that it is $\mu$
and not any power of $g\dup_c-g$ that should be taken as the scaling
variable. This is indeed the case as we will confirm in the next section.
Thus the double scaling limit is defined by varying $g$ so
that $\mu=-\epsilon_F(g) N$ is held fixed. The
free energy in this limit takes the form
\eqn\dslfg{\CF\bigl(N,g(N),V\bigr)\to \CF(\mu)\ .}
This definition of the $c=1$ double scaling limit was
given in \refs{\bkz\gzj\parisi{--}\GMil}.
\item{3)}The relation between the bare cosmological constant
$g$ and the scaling variable $-\epsilon_F N$ is subtle.
We can obtain this relation, at tree level, from the relation \effg:
\eqn\gtoep{
1=2\int_{\lambda_1}^{\lambda_2}{\d\lambda\over 2 \pi} (p_+-p_-)=
{2\over \pi}\int_{\sqrt{-2 \epsilon_F}} d
\lambda \sqrt{\epsilon_F+\half \lambda^2}+\cdots\ .}
Evaluating the singular part of the integral gives
\eqn\gtoepi{
g-g_c\sim (-\epsilon_F)\log(-\epsilon_F)+\cdots\ .}
Historically, the peculiar relation \gtoepi\ between
the ``bare'' cosmological constant and the scaling variable
caused a great deal of confusion. (For a discussion, see \kazrv.)
Unlike \ecltsc, a very complicated function of $g-g_c$ multiplies $N$
to form the scaling variable $\kappa$ that is held fixed in the double
scaling limit.  An interpretation of this result directly from the continuum
spacetime point of view has been given in \joei, and we shall reinterpret this
understanding in terms of macroscopic loop field theory
in sec.~{\it\ssmlft\/}.
\item{4)} Multicritical $c=1$ Theories.
At $c<1$ one discovers an enormous space of multicritical
points. At $c=1$ this has not been as extensively investigated, although some
results may be found in \gzj.
While the spacetime interpretation of these theories remains unclear,
there is evidence that perturbations of the conventional $c=1$ theories
by special state operators flow to these points. The reason is that the matrix
model defines flows by operators $\psi^\dagger\lambda^r\psi$  which
yield the above multicritical behavior. As discussed below, these operators
seem to be related to the special states.

\subsec{Double-Scaled Fermi Field Theory}
\subseclab\ssdsfft

Let us now turn to fermionic quantum mechanics and investigate
to all orders of perturbation theory.
Formulated in terms of a fermionic quantum field theory, the theory has
the action
\eqn\fermiact{S=N \int_{-\infty}^\infty \d X\, \d\lambda\,
 \hat\Psi^\dagger\Bigl(-i{\d\over \d X}+{\d^2\over \d\lambda^2}
+{1\over g} V\bigl(\lambda \sqrt{g}\bigr)\Bigr)\hat\Psi\ ,}
and lattice Fermi operators
\eqn\finfer{
\hat \Psi(\lambda,X)=\sum_{i=1}^\infty a_\epsilon(\varepsilon_i)
\,\psi^\epsilon(\varepsilon_i,\lambda;V)\,\ee{-i\varepsilon_i X}\ .}
The wavefunctions $\psi^\epsilon(\varepsilon_i,\lambda;V)$ are eigenfunctions
of the Schr\"odinger equation \schrod, and we may take the potential $V$ to be
symmetric since it effectively becomes so in the double-scaling limit. Thus we
also have an index $\epsilon=\pm$ which refers to the parity of the
wavefunction, and a repeated $\epsilon$ index will indicate a sum over parity
states. The $a$'s anticommute and satisfy
$\{a_\epsilon(\varepsilon_i),a_{\epsilon'}^\dagger(\varepsilon_j)\}
=\delta_{ij}\,\delta_{\epsilon,\epsilon'}$.
The Fermi sea $|N\rangle$ is defined as usual.

As we have seen from the tree-level analysis, the singular
terms in $\CF$ come from the behavior of the theory
for $\lambda\sim \CO(N^{-1/2})$ and $\epsilon\sim \CO(1/N)$.
Thus we scale equation \schrod\ by setting
$\lambda=\tilde \lambda/\sqrt{2 N}$ and $\epsilon_i=-\varepsilon/N$,
so that \schrod\ becomes
\eqn\sclschrod{\Bigl( {\d^2\over \d \tilde \lambda^2}+
{\tilde \lambda^2\over 4} +\CO(N^{-1/2})\Bigr)
\psi\dup_i=\varepsilon\, \psi\dup_i\ .}
All the details of $V$ are in the $\CO(N^{-1/2})$ piece and what remains
is the parabolic cylinder equation. (See appendix A.)
Two consequences of this are:
\par\nobreak
{\bf 1)} The density of states at the Fermi level can be calculated
to all orders of perturbation theory. The WKB matching of
the parabolic cylinder function (valid at the
tip of the potential) to the WKB functions (valid in the
other regions of $\lambda$) involves the large $\tilde \lambda$
asymptotics of the cylinder function. Matching the phases, we
find from the asymptotic formula \asyi\ for the behavior of the parabolic
cylinder function that the quantization condition
$\Phi(\varepsilon_{n+1})-\Phi(\varepsilon_{n})=\pi$ implies that
\eqn\quantcond{\rho(\varepsilon)\equiv {\p n\over \p \varepsilon}={1\over \pi}
\Phi'(\varepsilon)=-{1\over 2 \pi}\,{\rm Re}\, \psi(\half+i \varepsilon)\ ,}
where $\psi(x)={\d\over\d x}\log \Gamma(x)$ is the digamma function.
Identifying the density of states with a second derivative of the
free energy, we confirm the double scaling procedure mentioned
at the end of the last section. In particular, the
semiclassical expansion of $\rho$ is
\eqn\expans{
\rho(\varepsilon)={1\over 2 \pi}
\biggl(-\ln\varepsilon+\sum_{m=1}^\infty(2^{2m-1}-1)
{|B_{2m}|\over m}\Bigl({\hbar\over 2\varepsilon}\Bigr)^{2m}\biggr)\ ,
}
where the $B_{2m}$ are Bernoulli numbers.
This expansion shows that indeed
$N \epsilon_F$ is the correct scaling variable
and justifies the definition of the double scaling  limit in \dslfg.

{\bf 2)} We expect that, just as in the one matrix model studied in
the sec.~{\it\ssDSFT\/}, the wavefunctions themselves have
$N\to\infty$ limits
in terms of $\delta$-function normalized parabolic cylinder functions,
independent of the details of $V$:
\eqn\limswvs{
\psi^\epsilon(-\varepsilon/N,\lambda/\sqrt{2N};V)\to
\biggl({2 \pi N^{1/2}\over \log N}\biggr)^{1/2}
\psi^{\pm}(\varepsilon,\tilde \lambda)
}
where the wavefunctions $\psi^{\pm}(\varepsilon,\tilde \lambda)$
are given in \eqns{\wvfnspl{,\ }\wvfnsmi}.
The prefactor takes proper account of the normalizations of
the wavefunctions and can be verified by putting a non-universal
wall at distance $\CO(1)$ from the maximum. Thus, as in the
one-matrix model, we may take the double scaling limit of the
fermion operator by first defining an operator $\hat\Psi_N$
with a smooth $N\to\infty$ limit,
\eqn\sclfer{\hat\Psi_N(\lambda,x)\equiv {1\over (2N)^{1/4}}
\hat\Psi\bigl({\lambda\over\sqrt{2N}},N x\bigr)\ ,}
where we have substituted $x=X/N$.\foot{Note that in this chapter we have
used $X$ to denote the lattice target space coordinate and in
what follows we use
$x$ to denote the continuum target space coordinate, in minor conflict
with the notation of chapt.~\stdcst\ in which $X$ denoted the continuum
target space coordinate.}
We now rescale
\eqn\contosc{a_\epsilon(\varepsilon_i)\rightarrow {a_\epsilon(\nu)\over
\bigl({1\over \pi}\log\sqrt{2 N}\bigr)^{1/2}}\ ,}
so that in the $N\to\infty$ limit we have
\eqn\field{
\hat\Psi_N(\lambda,x)\to \hat\psi(\lambda,x)=\int \d\nu\, \ee{i\nu x}\,
a_{\epsilon}(\nu)\,\psi^\epsilon(\nu,\lambda)\ ,}
where $\psi^\epsilon(\nu,x)$ are normalized as in appendix A
(eqns.~\eqns{\wvfnspl{,\ }\wvfnsmi}).
The vacuum of the double scaled field theory is defined by
\eqn\vac{\eqalign{
a_{\epsilon}(\nu)|\mu\rangle&=0\qquad \nu<\mu\cr
a_{\epsilon}^\dagger(\nu)|\mu\rangle&=0\qquad \nu>\mu\ ,\cr}
}
where the Fermi level $\mu$ is as in \defofmu.

\subsec{Macroscopic Loops at $c=1$}
\subseclab\ssmlaco

The discussion of macroscopic loop operators given in chapters
\smmttll\ and \smmttff\
above continues to hold at $c=1$ with some minor modifications
\moore. We now wish to compute the continuum limit of the
macroscopic loop operator
\eqn\colop{W_{\rm lattice}(L,q)=
\int\d X\, \ee{i q X}\, \tr\, \ee{L \Phi(X)} \to
\int\d\lambda\,\d x\,\ee{i q x}\,
\psi^\dagger \psi(\lambda,x)\,\ee{\lambda \ell}
=W_{\rm cont}(\ell,q)\ .}
In particular, the boundary condition on the loop
is of Dirichlet type, $x(\sigma)=x$.

A subtlety that arises is that the eigenvalue density is
concentrated on both sides of the quadratic maximum, or, in
double-scaled coordinates, on $(-\infty,-2\sqrt{\mu}\,]\cup
[2 \sqrt{\mu},\infty)$. This means there are two (perturbatively)
disjoint ``worlds'' and we cannot simultaneously Laplace
transform the eigenvalue density with respect to both.
We can get around this difficulty by using a technical trick.
We compute amplitudes instead for the
Fourier transform with respect to $\lambda$,
\eqn\macropt{\eqalign{\hat\psi^\dagger \,\ee{iz\hat \lambda}\,\hat\psi
&\equiv \int_{-\infty}^{\infty}\d\lambda\, \hat\psi^\dagger(\lambda,x)
\,\ee{iz\lambda}\,\hat\psi(\lambda,x)\cr
M(z_i,x_i)&\equiv \bigl<\hat\psi^\dagger\, \ee{iz_1\hat\lambda}\hat\psi
\,\cdots\,\hat\psi^\dagger\, \ee{iz_n\hat\lambda}\hat\psi\bigr>\ .\cr}}

We will find that the answer
naturally splits into two pieces. In the first
piece we may continue $z\to i \ell$ in the upper half plane
to obtain a convergent
answer. This analytic continuation makes no sense in the second
piece, but there we can analytically continue
$z\to -i\ell$ in the lower half plane.
We interpret the two pieces as the
contributions of the two ``worlds'' defined by the two
eigenvalue cuts. Focusing on either contribution, we
can define macroscopic loop amplitudes for real loop lengths.
This rather strange reasoning can be checked in various
ways. At the level of tachyon correlation
functions, for example, the techniques of chapt.~\sscatt\ make it possible
to calculate even if we put an infinite wall at $\lambda=0$,
rendering $\psi^\dagger\, \ee{-\ell \lambda} \psi$ rigorously
well-defined. The resulting amplitudes agree to all orders.

Now we describe the calculation of the amplitudes
$M(z_i,q_i)$.
The Euclidean Green's functions of the
eigenvalue density
$\rho=\psi^\dagger\psi(\lambda,x)$, where $x$ is the
``time'' dimension of the $c=1$ matrix model, are defined by
\eqn\gfns{\eqalign{
G_{\rm Euclidean}(x_1,\lambda_1,\dots , x_n,\lambda_n)&\equiv
\langle \mu|T\Bigl( \hat\psi^\dagger\hat\psi(x_1,\lambda_1)\cdots
\hat\psi^\dagger\hat\psi(x_n,\lambda_n)\Bigr)|\mu\rangle_c\cr
G_{\rm Euclidean}(q_1,\lambda_1,\dots q_n,\lambda_n)
&\equiv \int \prod_i \d x_i\,\ee{i q_i x_i}
\,G(x_1,\lambda_1,\dots x_n,\lambda_n)\ .\cr}
}
Since the fermions are noninteracting,
these Green's functions may be written in
terms of the Euclidean fermion propagator,
\eqn\propeuc{
S_E(x_1,\lambda_1;\ x_2,\lambda_2)
=\ee{-\mu \Delta x}\int_{-\infty}^\infty {\d p\over 2\pi}\,
\ee{-ip\Delta x}\,I(p,\lambda_1,\lambda_2)\ ,
}
where $I$ is the resolvent for the upside-down oscillator
Hamiltonian $H=\half p^2-{1\over 8}\lambda^2$, or, more generally,
for a Hamiltonian $H=\half p^2 +V(\lambda)$ with the potential tuned
in the scaling region to differ from exact quadratic behavior.
In particular, for $q>0$,
\eqn\iasfg{
I(q,\lambda_1,\lambda_2)
=\bigl(I(-q,\lambda_1,\lambda_2)\bigr)^*
=\langle\lambda_1 |{1\over H-\mu-i q}|\lambda_2\rangle\ .
}

Using Wick's theorem, we evaluate \gfns\ as a sum of
ring diagrams and thereby obtain the integral representation for the
eigenvalue correlators
\eqn\nptev{\eqalign{G_{\rm Euclidean}(q_i,\lambda_i)&={1 \over n}
\int \prod_{i=1}^n {\d q_i\over 2\pi}\,
\ee{-i q_i x_i}\sum_{\sigma\in\Sigma_n}
\prod S_E(x_{\sigma(i)},\lambda_{\sigma(i)};\ x_{\sigma(i+1)},
\lambda_{\sigma(i+1)})\cr
&= {1 \over n}\, \delta(\sum q_i)
\int_{-\infty}^\infty \d q\sum_{\sigma\in\Sigma_n} \prod_{k=1}^n
I(Q^\sigma_k,\lambda_{\sigma(k)},\lambda_{\sigma(k+1)})\ ,\cr}
}
where $Q_k^\sigma\equiv q+q_{\sigma(1)}+\cdots +q_{\sigma(k)}\,$,
and the sum is over the permutation group $\Sigma_n$.

We can now obtain the formula for loop amplitudes as follows.
The resolvent of the upside down oscillator may be given the
integral representation:
\eqn\intrsv{\langle\lambda_1 |{1\over H-\zeta}|\lambda_2\rangle
=-i\int_0^{-\epsilon \infty}\d s\, \ee{-i s\zeta}
{1\over\sqrt{4\pi i \sinh s}}
\exp{i\over 4}\Bigl({\lambda_1^2+\lambda_2^2\over
\tanh s}-{2\lambda_1\lambda_2\over \sinh s}\Bigr)\ ,}
where $\epsilon={\rm sgn}({\rm Im}\,\zeta)$. Therefore, the
calculation of \macropt\ reduces to the
evaluation of a gaussian integral
with the result
\eqn\macptii{\eqalign{{\p\over\p\mu}M(z_i,q_i)
&=\ha\,i^{n+1}\delta(\sum q_i)\sum_{\sigma\in\Sigma_n}
\int_{-\infty}^{\infty} \d\xi\, {\ee{i\mu \xi}\over |\sinh \xi/2|}
\int_0^{\epsilon_1\infty} \d s_1\,\cdots\cr
\cdots \int_0^{\epsilon_{n-1}\infty} \d s_{n-1}\,
&\exp\Bigl(-\sum_{k=1}^{n-1}s_k Q^\sigma_k\Bigr)
\ \exp\Bigl({i\over 2}\coth(\xi/2)\sum z_i^2 \Bigr)\cr
&\cdot\exp\Bigl(i\sum_{1\leq i<j\leq n}
{\cosh\bigl(s_i+\cdots s_{j-1}-\xi/2\bigr)\over
\sinh(\xi/2)}z_{\sigma(i)}z_{\sigma(j)}\Bigr)\ ,\cr}}
where $\epsilon_k={\rm sgn}(Q^\sigma_k)\,$.

We now examine several special cases of the above formula.
\par\nobreak
a) One macroscopic loop:
$$M_1=\figins{\vcenter{\epsfxsize4in\epsfbox{oneloop.ps}}}$$
This is the Hartle--Hawking wavefunction. Analytically
we have
\eqn\onemacloop{
M_1={\rm Re\/}\Bigl( i \int_{0}^{\infty} {\d\xi\over \xi}\,
 \ee{i\xi+i {\mu z^2\over \xi}}\,
 {\ee{i({1\over (2 \mu)} \coth({\xi\over (2 \mu)})-{1\over \xi})\mu z^2}
 \over \sinh \xi/2 \mu}\Bigr)\ .
 }
The genus expansion is obtained by restoring the string coupling $\kappa$,
according to
\eqn\powerkap{\mu\to \mu/\kappa\ ,\quad
\ell\to \kappa^{1/2} \ell\ ,\quad
\lambda\to \kappa^{-1/2} \lambda\ .}
%
The perturbative expansion of the Hartle--Hawking wavefunction
is obtained by expanding the last factor in \onemacloop, and
continuing  $z\to i\ell$ in an integral representation for the Bessel function.

b) Two macroscopic loops:
$$M_2=\figins{\vcenter{\epsfxsize4in\epsfbox{twoloop.ps}}}$$
This gives the propagator from which we may hope to understand
the spectrum of the theory.

The integral formula for $\DM M$ becomes
\eqn\twpti{{\rm Im}\,\int_0^\infty \d\xi\, {\ee{i\mu\xi
+ \ha i(z_1^2+z_2^2)\coth(\xi/2)}
\over \sinh (\xi/2)}
\int_0^\infty \d s\, \ee{-|q|s }
\Bigl(\ee{i{\cosh(s-\xi/2)\over \sinh\xi/2}z_1z_2}-
\ee{i{\cosh(s+\xi/2)\over \sinh\xi/2}z_1z_2}\Bigr)\ .}
This formula holds for $z_i$ real. If we wish to have
real loop lengths we replace ${\rm Im}\to -\half i$ and continue
$z_i\to i \ell_i$, as discussed above.

The integral over $s$ may be written as
\eqn\expbessl{
2\pi \ee{-i\pi|q|/2}{\sinh(|q|\xi/2)\over \sin\pi|q|}J_{|q|}(2\alpha)
+\sum_{r=1}^\infty {4 i^r r\over r^2-q^2}J_r(2\alpha) \sinh(r\xi/2)
}
where $\alpha=z_1z_2/2 \sinh(\xi/2)$.
The remaining integral over $\xi$ can be done in terms of
Whittaker functions to give nonperturbative
answers. This is done in detail in \moore.

At genus zero, we have
\eqn\conepr{\eqalign{
\bigl< W(\ell_1,p) W(\ell_2,-p)\bigr>
&={\pi p\over \sin\pi p}I_{p}(2\sqrt{\mu}
\ell_1)K_{p}(2\sqrt{\mu}\ell_2)\cr
&\quad+\sum_{r=1}^\infty {2(-1)^r r^2\over r^2-p^2}\,
I_{r}(2\sqrt{\mu}\ell_1)\,K_{r}(2\sqrt{\mu}\ell_2)\ ,\cr}
}
and therefore
\eqn\decprop{\bigl< W(\ell_1,p) W(\ell_2,-p)\bigr>\equiv\int_{0}^\infty
 {\d E\over 2\pi}\, G(E,p)\,\psi\dup_E(\ell_1)\psi\dup_E(\ell_2)\ ,
}
with
\eqn\conefrm{\eqalign{
G(E,p)&={1\over E^2+p^2}{\pi E\over \sinh\pi E}\cr
&={\pi p\over \sin\pi p}{1\over E^2+p^2}+
\sum_{r=1}^\infty {2(-1)^r r^2\over r^2-p^2}{1\over E^2 +r^2}\ .\cr}}

c) Three macroscopic loops:
$$\kappa\quad\figins{\epsfysize.75in\vcenter{\epsfbox{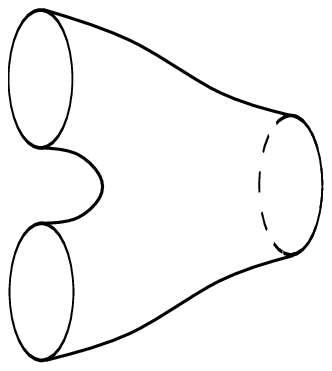}}}$$
The same techniques as above can be applied to three and four
macroscopic loop amplitudes. The final formulae are rather
complicated (see e.g.\ \msi), while at genus zero there is considerable
simplification.

The entire genus zero amplitude, together with the integer powers of
$\ell$, is summarized nicely in terms of macroscopic state
wavefunctions \msi:
\eqn\macrocr{\lbspace\Bigl<\prod_i W(\ell_i,q_i)\Bigr>
=\int_{-\infty}^{\infty}\prod_i \d E_i\,
{E_i\over E_i-iq_i}K_{iE_i}
(2\sqrt{\mu}\ell_i)\,(E_1+E_2+E_3)\coth\bigl({\pi\over 2}
(E_1+E_2+E_3)\bigr)\ .}

\exercise{macroscopic $\to$ microscopic}

Evaluate the integrals in \macrocr\ using residues by closing the
$E_i$ contours in the upper or lower half plane.
(Warning: this involves a certain amount of algebra --- the result is in
\msi.)

\endexercise

The results \decprop,\conefrm, and \macrocr\ contain a wealth of information
on the nature of contact terms and singular geometries in the path integral of
the $c=1$ theory. Note especially from \conefrm\ that the propagator agrees
with the naive Wheeler--DeWitt propagator at low values of $|E|$, but is quite
distinct, indeed exponentially decaying, at large values of $|E|$.
Correspondingly, in position space the propagator turns out to be
{\it smooth\/} at $\ell_1=\ell_2$. From a quantum gravity point of view, the
only source of violation of the naive WdW propagator in the Euclidean quantum
gravity path integral is the contribution of singular geometries such as
\ftwolc. From a target
space point of view, the smoothness of the propagator suggests the existence of
other degrees of freedom. This guess is confirmed by the pole structure of the
propagator manifested in the second line of \conefrm. These extra degrees of
freedom are clearly related to the special states. The two ideas: 1)
contributions of singular geometries, and 2) existence of new degrees of
freedom related to special states, are tied together by interpreting the new
degrees of freedom in terms of Liouville boundary operators, or, equivalently,
in terms of redundant operators in the matrix model. Recall from
sec.~{\it\ssrpsgct\/} that such operators contribute figure eights
--- the lattice version of the singular geometry of \ftwolc. In \msi\ this
interpretation was elaborated upon, and partially carried out for the data
provided by the three-point function \macrocr.

\subsec{Wavefunctions and Wheeler--DeWitt Equations}

{}From \conepr\ we can easily extract the wavefunction
of the vertex operator $V_q$ by extracting the coefficient
of $\ell_1^{|q|}$ as $\ell_1\to 0$ to get
\eqn\getwvq{
\bigl< W(\ell,-q)\, V_q\bigr>
= \mu^{|q|/2} K_q(2\sqrt{\mu}\ell)\ .}
This result is analogous to \wvfnht, and is
in complete accord with the continuum answer.

The wavefunctions to all orders of perturbation theory are not much
more complicated.
We extract the term proportional to
$z_1^{|q|}$ in \expbessl\ and perform the
remaining $\xi$ integral in terms of Whittaker functions
to find
\eqn\allordwv{
\psi\dup_q(\ell)=2 \Gamma(-|q|)
{\rm Im}\Bigl( \ee{{3 \pi i\over 4}(1+|q|)}\int_0^{|q|} \d t\,
\Gamma(\half - i \mu + t)\, \ell^{-1}\,
W_{i \mu-t+\ha |q|,\ha |q|}(i \ell^2)\Bigr)\ ,}
where $W_{\eta,\xi}$ is a Whittaker function. In particular,
the function $\chi_{\eta,\xi}(\ell)=\ell^{-1} W_{\eta,\xi}(i \ell^2)$
satisfies an equation derived from the Whittaker equation:
\eqn\whitteq{
\Bigl(-(\ell{\p\over \p \ell})^2- 4 i \eta\ell^2 + 4 \xi^2 -\ell^4\Bigr)
\chi_{\eta,\xi}=0\ .}
Therefore the all-orders Wheeler--DeWitt wavefunctions
satisfy some simple differential relations generalizing
the genus zero Wheeler--DeWitt equation.
The answer is especially simple
at $q=0$ where we find the modified Wheeler--DeWitt equation:
\eqn\wdwmdf{\biggl(-(\ell{\p\over\p\ell})^2+4\mu\ell^2-\kappa^2
\ell^4\biggr)\psi\dup_{q=0}=0\ ,}
where we have explicitly introduced the topological coupling
$\kappa$. The consequence of \whitteq\ is not so simple when
$q\not=0$.\foot{We disagree with a recent discussion of
the all orders WdW equation in \danieli.}

\subsec{Macroscopic Loop Field Theory and $c=1$ scaling}
\subseclab\ssmlft

In sec.~{\it\stdstes\/}, we discussed the tachyon field
$T(\phi,X)$. On the other hand, the
formulae of the previous section suggest the existence of
a macroscopic loop field theory in which
$W(\ell,x)$ is a field. Since $\ell$ and $\phi$ are
related, we may suspect that the two fields $T$ and $W$ are essentially
the same (recall that $X$ from sec.~{\it\stdstes\/} translates directly
to the continuum $x$ we use in this chapter).

Treated as a field, $W$ has a vacuum
expectation value given by the Hartle--Hawking wavefunction, and
correlations of fluctuations $\delta W$ are measured by higher
correlation functions. On the other hand,
we see from \getwvq\ that first order
fluctuations $\delta W$ correspond exactly
to the tachyon wavefunction, i.e.\ satisfy the Wheeler--DeWitt equation,
up to a factor of the coupling constant. This suggests the relation
\eqn\wandt{W(\ell,x)\sim \ee{-\ha Q\phi}\, T(\phi,x)\ .}
%
In the next section we shall see that tachyon $S$-matrix elements can be
extracted directly from $W$ correlators, further corroborating this result.

Replacing \wandt\ by an equality, we may transform the standard free
tachyon action
\eqn\freetach{S_0=\int_{-\infty}^\infty \d x\, \d\phi\, \ee{-Q\phi}
\Bigl((\p_\phi T)^2+\mu\, \ee{\gamma \phi}T^2 +T(H_x-{Q^2\over 4})T\Bigr)}
to the action
\eqn\freewac{S_0=\int_0^\infty{\d\ell\over\ell}\int_{-\infty}^\infty
\d x\, W\bigl(-\p_\phi^2+\mu\, \ee{\gamma\phi}+H_x\bigr)W\ .}
The interactions are complicated, but can be deduced from the formulae
of sec.~{\it\ssmlaco\/}.

Using the tachyon wavefunction, we are now prepared to adapt Polchinski's
discussion\foot{To go from the variables $\Delta$ and $\mu$ used
in \joei\ to our conventions, let $\Delta\to g-g\dup_c$, $\mu\to-\epsilon_F$.}
\joei\ to interpret the
scaling law variation \gtoepi\ and the definition \defofmu\ in
terms of the continuum theory.
As we see from \getwvq\ with $q=0$, the wavefunction for the static tachyon
background is $K_0(2\sqrt{\mu}\,\ell)$. To mimic the discreteness of matrix
models, we consider a cutoff scaling variable $\mu_B$
(analogous to $\epsilon_F$) and measure distances
$\ell$ in ``lattice units''.  Substituting in \wandt, we find that
the static tachyon configuration in the cutoff theory is given by
\eqn\ctff{T(\phi)\sim(2\sqrt{\mu_B}\,\ell)^2
\,K_0(2\sqrt{\mu_B}\,\ell)\ .}
In the $\sigma$-model approach (eq.~\pertaction), the spacetime
one-point function $\langle T \rangle$ plays the role of the cosmological
constant. With an ultraviolet cutoff $\phi\geq 0$ on the theory,
the value of the tachyon field at the cutoff is thus naturally
interpreted as the bare cosmological constant
\eqn\erelbare{g-g_c\sim T(0)\sim (2\sqrt{\mu_B})^2 K_0(2\sqrt{\mu_B})\ ,}
since working at the cutoff is equivalent in the matrix
model to multiplying the bare cosmological constant by the (unit) area of
the basic triangle. As $\mu_B\to 0$, we have
\eqn\relbare{g-g_c\sim 2 \mu_B \log \mu_B}
which is functionally equivalent to \gtoepi, giving the relation
between the bare worldsheet cosmological constant and the scaling variable.

\ifig\fkzero{The function $x^2 K_0(x)$, with a peak at $x\sim \CO(1)$.}
{\epsfxsize3in\epsfbox{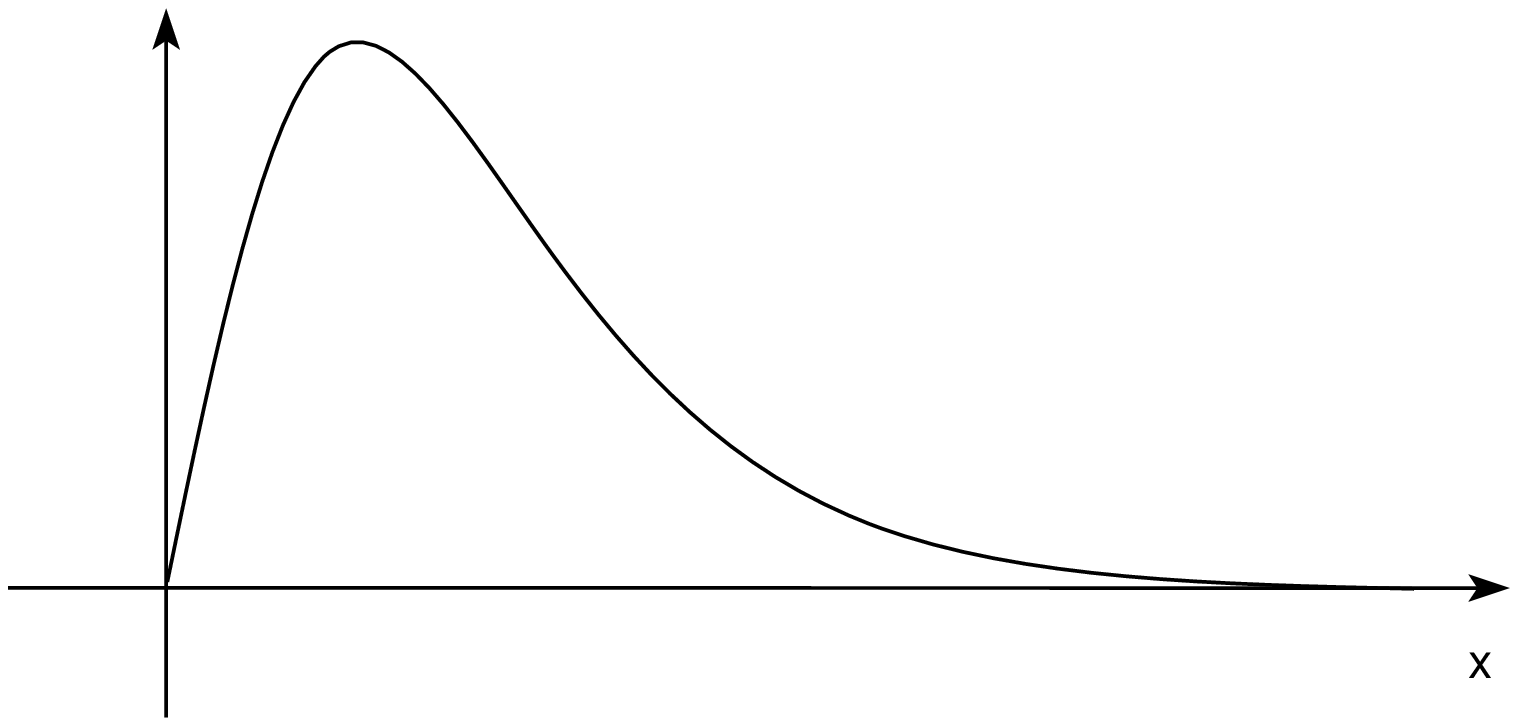}}

The function $x^2 K_0(x)$ of \ctff\ moreover behaves as in \fkzero,
with a peak at $x\sim \CO(1)$. (This is the analog of the
soliton configuration of \joei, although in our argument we use the
Wheeler--DeWitt equation for macroscopic loops rather than the properties of an
interacting tachyon theory.) As $\mu_B\to 0$, the
peak occurs at larger and larger lattice lengths
${\bar\ell}^2=\ee{\sqrt2\,\bar\phi}\sim 1/\mu_B$.
The scaling behavior at higher genus follows from perturbation theory with the
effective action \stact\ in the dilaton background
$\langle D\rangle=(Q/2)\phi=\sqrt2\,\phi$. The
effective string coupling is thus
$\kappa_{\rm eff}=\kappa\,\ee{\sqrt2\,\bar\phi}$, where we have substituted
the value at the
peak of the tachyon configuration which dominates the string scattering.
Since the bare string coupling (genus counting parameter)
in the matrix models is $\kappa=1/N$,
we see that holding fixed the effective string coupling
$\kappa_{\rm eff}=\kappa\,\ee{\sqrt2\,\bar\phi}\sim{1/(N\mu_B)}$
in the continuum limit
defines the $c=1$ double scaling limit as in the matrix models
\refs{\bkz\gzj\parisi{--}\GMil}, where the string coupling is as well related
to the bare cosmological constant via \relbare\ rather than via \ecltsc\
as in the $c<1$ models coupled to gravity.

\danger{Tachyon condensates}

In this discussion we are identifying
$\langle T\rangle \sim  \mu\ell^2 K_0(2\sqrt{\mu}\,\ell)$,
while in the $\sigma$-model discussion (see \solnii), we identified
$\langle T\rangle \sim \mu\,\ee{\gamma \phi}$. These
do not even agree in the limit $\phi\to -\infty$ ($\ell\to0$). This
confusion has plagued the subject for several years now. As mentioned
in the paragraph following \liugausii, it suggests that
we should really identify the cosmological constant operator as
$\langle T\rangle\sim\mu\phi\,\ee{\gamma\phi}$, which has the
$\mu\ell^2\ln\ell$ behavior as $\phi\to-\infty$. We comment further on the
two possible cosmological constant operators at the end of
sec.~{\it\ssposme\/}.

\subsec{Correlation functions of Vertex Operators}
\subseclab\sscfvo

As in the $c<1$ theories, vertex operators
are obtained by looking at the small
$\ell$ expansion. In particular, using the reasoning
of the exercise below \loco\ together with the formula
\eucvert, we see that
the coefficients of $\ell^{|q|}$ in the small $\ell$ expansion
of the macroscopic loop operators must be proportional to
the correlation functions of the $V_q$. We are fortunate that
for generic $q$ there is only a one-dimensional BRST cohomology
class.\foot{Historically, the first attempts at $D=1$ correlation
functions used powers of the matrix field as scaling operators
\refs{\rkbm,\rGKn}.}

String amplitudes are asymptotic expansions in the string coupling
$\kappa=1/\mu$ whose coefficients are integrals over moduli space.
These can be calculated by considering the $\ell_i\to 0$ expansion of the
matrix model loop operators:
\eqn\expan{
\langle \mu|\prod W(\ell_i,q_i)|\mu\rangle = \prod \ell_i^{|q_i|}
\CR_n(q_1,\dots q_n;\mu)\bigl(1+\CO(\ell_i^2)\bigr)+
{\rm analytic\ in}\  \ell_i\ .
}
Then, as asymptotic expansions in $\kappa=1/\mu$, we have
\eqn\correl{\CR_n(q_1,\dots q_n;\mu)
=\CA_n( \tilde{V}_{q_1}\cdots \tilde{V}_{q_n})\ ,}
where
%
\eqn\ecanr{\CA_n( \tilde{V}_{q_1}\cdots \tilde{V}_{q_n})
\sim\sum_{h\geq 0} \kappa^{-\chi}\,\CA_{h,n}
\bigl(\tilde{V}_{q_1}\cdots \tilde{V}_{q_n}\bigr)
\equiv\sum_{h\geq 0} \kappa^{-\chi}\int_{\CM_{h,n}}
\bigl<\tilde{V}_{q_1}\cdots \tilde{V}_{q_n}\bigr>}
(the CFT correlator $\bigl<\tilde{V}_{q_1}\cdots \tilde{V}_{q_n}\bigr>$
is interpreted as a differential form on moduli space $\CM_{h,n}$,
i.e.\ includes a product $\prod b\bar b$ over ghost zero modes), and
\eqn\normtach{\tilde{V}_q=\Gamma\bigl(|q|\bigr)\, c \bar{c}\,
\ee{(i q X-|q|\phi)/\sqrt{2}}\, \ee{\sqrt{2} \phi}}
is the tachyon vertex operator of \eucvert.
The normalization is fixed by comparison of computations of the
right hand side performed by Di Francesco and Kutasov \kdf,
as described in sec.~{\it\srabsm\/}. We see from \ecanr\ that the
$S$-matrix for the spacetime tachyon can be extracted from
correlation functions of the macroscopic loop operator $W$, hence
$T$ and $W$ are interpolating fields for the same asymptotic states,
as suggested in the preceding section.

\goodbreak
Remarks:\par\nobreak
\item{1)} The definition \expan\ of $\CR_n$ is ambiguous if $q_i\in \IZ$.
The functions can defined for $q_i\in \IZ$ by continuity.
\item{2)} The right hand side of \correl\ is by definition an
asymptotic expansion in the string coupling $1/\mu$.
On the other hand, the matrix model gives a nonperturbative
completion since (in contrast to the difficulties at $c<1$)
we may perform all manipulations with a potential giving
a perfectly well-defined matrix model integral.

As an example of the use of \correl\ we may immediately
extract from the small $\ell$ expansion of the all-orders
Wheeler--DeWitt wavefunction \allordwv\ the two-point function
of the tachyon:
\eqn\twpttach{\eqalign{
&{\p\over\p\mu}\langle V_q\, V_{-q} \rangle=\bigl(\Gamma(-|q|)\bigr)^2
\,{\rm Im}\,\biggl(\ee{i\pi |q|/2}
\Bigl({\Gamma(|q|+\half-i\mu)\over\Gamma(\half-i\mu)}-
{\Gamma(\half-i\mu)\over\Gamma(-|q|+\half-i\mu)}\Bigr)
\biggr)\cr
&\quad\sim (q\Gamma(-|q|))^2\mu^{|q|}
\Bigl({1\over |q|}-(|q|-1){(q^2-|q|-1)\over
24}\mu^{-2}\qquad \cr
&\qquad+\prod_{r=1}^3(|q|-r)
{(3q^4-10|q|^3-5q^2+12|q|+7)\over 5760}\mu^{-4}\cr
&\qquad-\prod_{r=1}^5(|q|-r)
{(9q^6-63|q|^5+42q^4+217|q|^3-205|q|-93)\over 2903040}
\mu^{-6}+\cdots \Bigr)\ .\cr}
}

In sec.~{\it\ssnpsm\/} below we will describe a much
better way to compute tachyon correlators which easily yields
the generalization of \twpttach\ to arbitrary tachyon
correlation functions.

\danger{Special states in the matrix model}

If one wishes to interpret the integer powers of $\ell$ in
terms of an operator expansion it is necessary to introduce the
redundant operators $\CB_{r,q}$ corresponding to moments
of $\lambda^r$. It can already be seen from \conepr\ that
to define such operators we must take
\eqn\spb{\CB_{r,q}=\lim_{\ell\to 0}
\Bigl({\p\over\p \ell}\Bigr)^r W(\ell,q)\Big|_{\ell=0} \sim
\int \d x\, \d\lambda\, \ee{iqx}\, \psi^\dagger\lambda^r\psi\ ,}
where $q>r$, otherwise the limit
$\ell\to 0$ diverges. We then analytically continue to any $q$.
The physical origin of the divergence
at $\ell\to 0$, or, at $\lambda\to \infty$, is in the
ultraviolet region of the worldsheet integral, and is
probably connected with the fact that the special state
operators are irrelevant operators.

By upper triangular transformations of the basis of operators,
analogous to the change of basis in sec.~{\it\swapmm\/} relating
$\sigma_j$ to $\hat \sigma_j$, we obtain the full
operator expansion of the macroscopic loop \msi:
\eqn\locopexp{W_{\rm in}(\ell,p)=
\tilde V_p\,\Gamma(|p|+1)\, \mu^{-|p|/2}
I_{|p|}(2\sqrt{\mu}\ell)-
\sum_{r=1}^\infty\hat\CB_{r,p}{2(-1)^r r\over r^2-p^2}
\mu^{-r/2}I_r(2\sqrt{\mu}\ell)\ .}

\newsec{Fermi Sea Dynamics and Collective Field Theory} 
\seclab\sfsdcft

\subsec{Time dependent Fermi Sea}

Another source of intuition, very different from the
macroscopic loop approach, comes from the motions of the
Fermi sea of the upside-down oscillator and the associated
collective field theory, also known as the Das--Jevicki--Sakita theory
\refs{\dj,\wadia,\gki}.
As we saw in our analysis of the tree-level free energy
(sec.~{\it\smqm\/}),
one is naturally lead to think about the fermionic
phase space. This point of view leads to a very
beautiful description of tree-level $c=1$ dynamics \joesea.

In sec.~{\it\smqm\/}, we studied the ground state from the
point of view of a fluid in phase space.
When describing dynamics, we have to perturb the system
so we are now looking at time-dependent Fermi seas
resulting from the disturbances produced by
various operators. The possible dynamical solutions of the
system can be described in the following way
\joesea. Consider the generating functional for
correlators in the theory,
\eqn\joecor{
Z[J]=\int \d\psi\,
\e{-\int \psi^\dagger(i\d/\d t + \d^2/\d \lambda^2 + \lambda^2)\psi
+J \psi^\dagger \psi}\ ,}
where we imagine $J$ has been turned on and off during a
finite time interval. (In this section, the $c=1$ coordinate
$t$ will always be taken to be a Minkowskian time coordinate.).
The source $J$ acts as an external force on the fermions.
After it has been turned off, the state evolves as some
time-dependent solution of the system. It is clear that
the points simply move along trajectories in phase space
appropriate to the upside-down oscillator, that is, they
move along lines of constant $p^2-\lambda^2$.

\ifig\fgenfermi{A generic initial configuration of the Fermi sea.}
{\epsfysize1.5in\epsfxsize1.4in\epsfbox{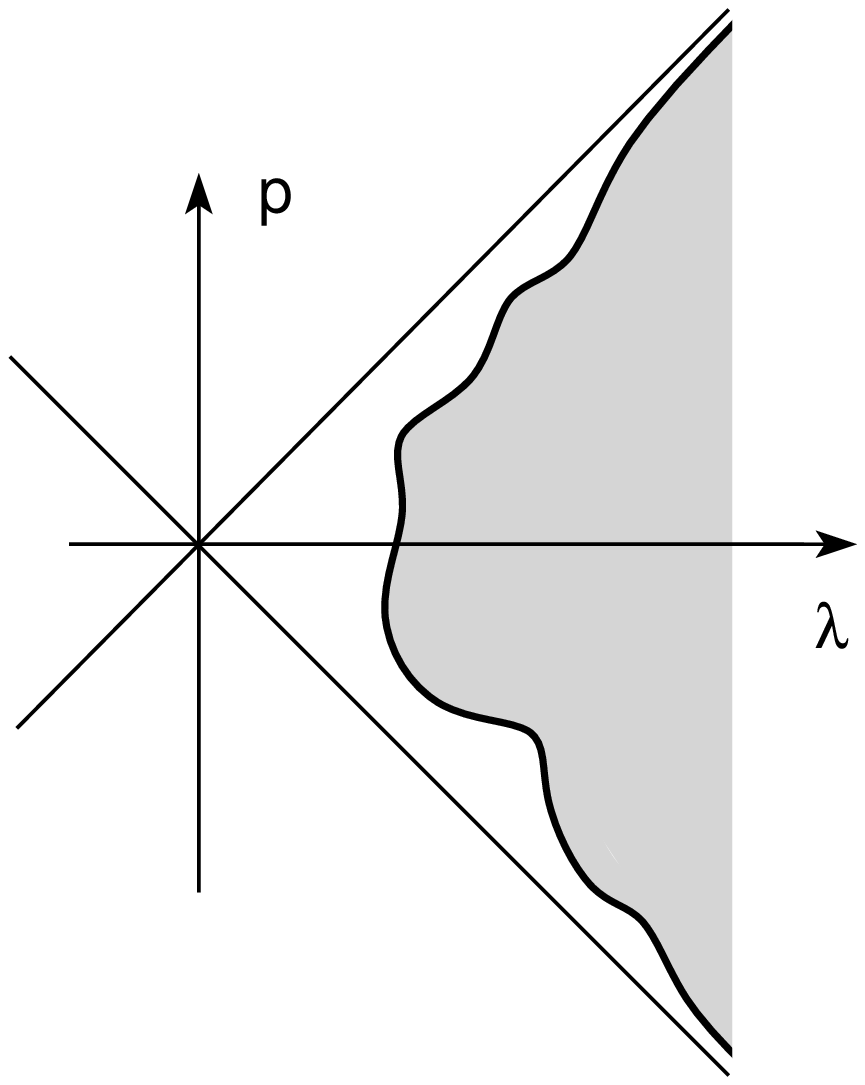}}

Thus to write down the general time-dependent motion of the system we imagine
at time zero a generic Fermi sea as in \fgenfermi,
which we may describe as a parametrized curve
%
$$\lambda=(1+a(\sigma))\,\cosh(\sigma)\,,\qquad
p=(1+a(\sigma))\,\sinh(\sigma)\ ,$$
where $a(\sigma)$ is a smooth function subject to the constraint that
the initial Fermi surface $(\lambda,p)$ be physically reasonable.
Hamilton's equations
\eqn\evolve{\eqalign{\p_t p &= \{H,p\} \cr
&=\lambda - p\, \p_\lambda p\ .\cr}}
then give the general solution for time evolution:
\eqna\frms
$$\eqalignno{\lambda &=(1+a(\sigma))\,\cosh(\sigma-t)&\frms a\cr
p&=(1+a(\sigma))\,\sinh(\sigma-t)\ .& \frms b\cr}$$

\subsec{Collective Field Theory}

We are now in a position to derive the collective
field theory of $c=1$. Consider the case in which
the Fermi sea only has two branches $p_\pm$.
The functions $p_\pm(\lambda,t)$ may be thought of as on-shell
fields related by a boundary condition
$p_+(\lambda_*,t)=p_-(\lambda_*,t)$
where $\lambda_*$ is the leftmost point of the sea. As in
sec.~{\it\smqm\/}, the energy, or Hamiltonian, is
given by
\eqn\hamil{\eqalign{
H&=\int {\d p\,\d\lambda\over 2 \pi}\, \epsilon\, \theta(\epsilon_F-\epsilon)
+{\mu\over 2} N\cr
&=\int d\lambda \Bigl(\bigl({p_+^3\over 6}-\lambda^2 {p_+\over 2}\bigr)
-\bigl({p_-^3\over 6}-\lambda^2 {p_-\over 2}\bigr)\Bigr)
+{\mu\over 2}\int \d \lambda\,(p_+-p_-)\ .\cr}}

To interpret \hamil\ as a field theory of the eigenvalue density, define
\eqn\djfld{
p_\pm = -\kappa^2 \Pi_{\chi} \pm \pi \p_{\lambda}\chi\ .
}
In terms of $\zeta$, the eigenvalue density is given by
\eqn\evdens{\rho(\lambda)=p_+-p_- = 2 \pi\, \p_{\lambda}\chi\ .}
After rescaling, the Hamiltonian in these variables may be written as
\eqn\djham{
H=\int \d\lambda\, \Bigl( {\kappa^2\over 2}\chi'
\pi_\chi^2  + {\pi^2\over 6 \kappa^2}
(\chi')^3+{v(\lambda)\over \kappa^2}\chi'\Bigr)
+{\mu\over 2 \kappa^2}\int \d \lambda\ \chi'\ ,}
where $v(\lambda)$ is the double-scaled matrix model potential
$-\half \lambda^2$.
This Hamiltonian appeared from very different points of
view in \refs{\dj,\wadia,\gki}\
as the field theory of an eigenvalue density field $\rho(\lambda,t)$.
The present derivation overcomes some of the difficulties with
understanding the Jacobian for the change
$\prod \d \lambda_i\to \d \rho(\lambda,t)$.

\exercise{}

Derive the Lagrangian corresponding to the Hamiltonian \djham.

\endexercise

\ifig\ffoldfermi{A configuration of the Fermi sea with folds.
$p$ is a multivalued function of $\lambda$.}
{\epsfysize1.5in\epsfxsize1.4in\epsfbox{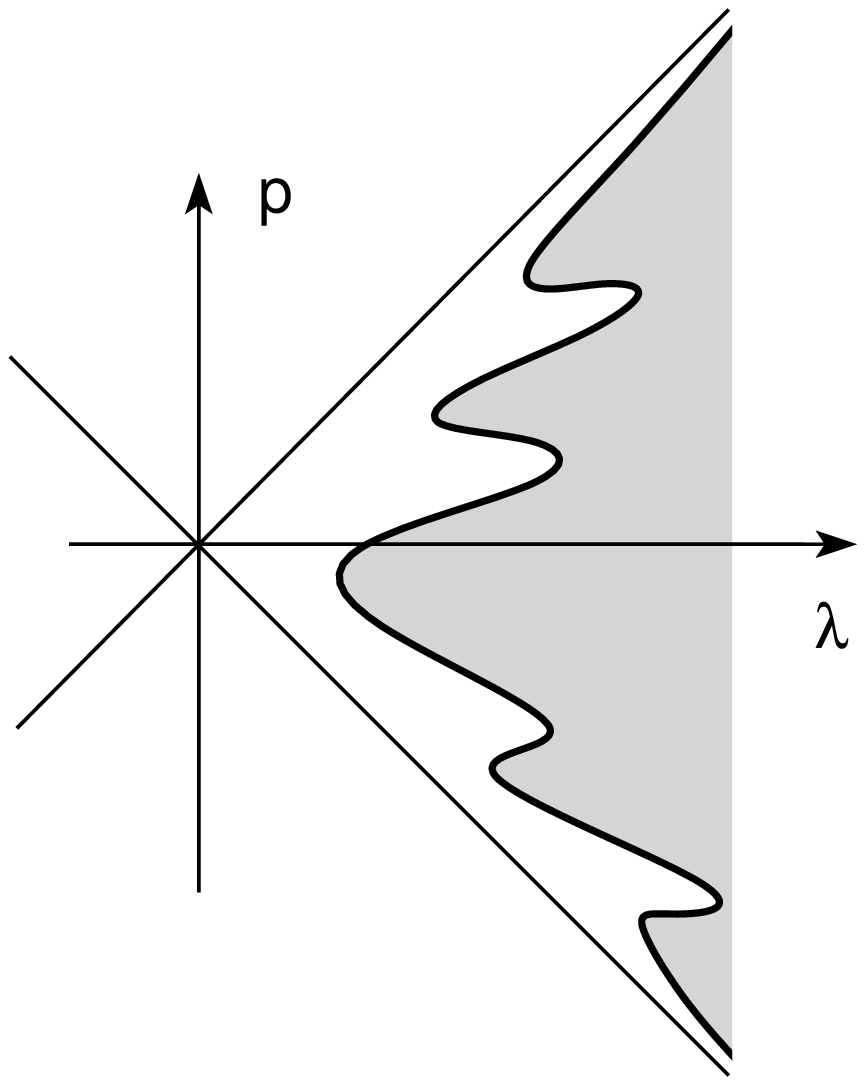}}

\danger{Folds}
As pointed out in \joesea, there can be solutions to \frms{a,b}\
which have four or
more branches $p(\lambda,t)$ for a given $\lambda$ (e.g.\ \ffoldfermi). These
solutions are perfectly sensible from the free fermion point of view but are
quite strange from the collective field theory point of view. Curiously, the
number of folds is {\it not\/} conserved in time. A surface with two branches
can very well evolve into one with four or more branches, and vice versa. These
fold-solutions are extremely interesting in the context of collective field
theory as a model of string field theory, for they show that the ``obvious''
string field might be a bad description of the string degrees of freedom for
some perfectly sensible backgrounds. It has been suggested that if the $c=1$
model is equivalent to a model of $1+1$-dimensional black holes, then
folding solutions will be an important piece of the puzzle.

\subsec{Relation to 1+1 dimensional relativistic field theory}

It is natural to rewrite the collective field theory as a relativistic theory.
Let us consider solutions such that there are only two branches $p_\pm$ of
the Fermi sea.
Consider the classical equations of motion for a particle in
an upside-down oscillator:
\eqn\clss{\ddot \lambda = \lambda\ ,}
solved by $\lambda(t)=A \cosh(t+B)$. Thus if we
change our spatial coordinates to $\lambda=2\sqrt{\mu}\cosh \tau$,
motion in $\tau,t$ space is relativistic: $\tau(t)=t+B$.

Let us now explain this point within the collective field theory.
We are interested in the fluctuations $\delta p_\pm$
of the Fermi sea. Since there is a nonzero background field configuration
corresponding to the genus zero eigenvalue density,
\eqn\coneev{p_\pm=\pm \sqrt{\lambda^2-\mu}
=\pm(\lambda-{\mu\over 2 \lambda})+\CO(\lambda^{-2})\ ,}
we make a field redefinition
\eqn\fldred{
p_\pm(\lambda,t)=\pm \lambda \mp {\mu\over 2 \lambda}\chi_\pm \ ,
}
so that, up to a constant shift independent of
$\chi_\pm$, the Hamiltonian \hamil\ becomes
\eqn\redham{\eqalign{H&=
+{\mu^2\over 8}\int \d \tau \Bigl((\chi_+^2+\chi_-^2)
{\sqrt{(1-e^{-6 \tau})(1-e^{-2 \tau})}\over (1+e^{-2 \tau})^2}
\cr
&\qquad\qquad\qquad-{1\over 6}\bigl(  \chi_+^3+
  \chi_-^3 \bigr)e^{-2 \tau}{1-e^{-2 \tau}\over (1+e^{-2 \tau})^3}\Bigr)\cr
&={\mu^2\over 8}\int \d \tau \Bigl((\chi_+^2+\chi_-^2)
-{e^{-2 \tau}\over 6}\bigl(  \chi_+^3+
  \chi_-^3 \bigr)\Bigr)\bigl(1+\CO(e^{-2 \tau})\bigr)\ .\cr}}
%
Far from the edge of the eigenvalue density, we may define
$\chi\dup_\pm = \pm \pi_S - \p_\tau S\,$, where $S$ is a free massless
scalar field.

Dirichlet boundary conditions imply the free propagator for the fermions
$\chi\dup_\pm$:
\eqn\djpr{\int_0^\infty \d E\, {\cos E\tau_1 \cos E\tau_2\over E^2+p^2}\ .}
The change of variables to \redham\ can be pursued more carefully and
perturbation theory calculations can be performed in this formalism.
See \dj\ and \klebrev\ for details.

{\bf Remark}: The one-loop free energy in this 1+1 dimensional relativistic
field theory can be calculated \dj\ and goes as
$\log\mu$, which is hence interpreted as the volume of $\phi$-space.
Further attempts to interpret this result may be found in \kazrv\
(see also sec.~{\it\ssmlft\/}).

\subsec{$\tau$-space and $\phi$-space}
\subseclab\sstsaps

Collective field theory is a theory of a massless boson that represents
fluctuations in the eigenvalue density. On the other hand,
the massless boson of
string theory is a scalar field $T(\phi,t)$. In this section we discuss the
nontrivial relation between these two bosonic fields \msi.\foot{Many
authors continue to identify $\tau$ with
the Liouville coordinate $\phi$. While both spaces share many
qualitative features, they cannot be the same. The matrix
model coordinate $\lambda$ has no (obvious) geometric
meaning in the discrete worldsheet sum.
Rather, it is the loop operator
$W(L)$ that has a geometric meaning and is related to
the worldsheet metric, and therefore to the Liouville field $\phi$.}

As we have seen, the macroscopic loop and
tachyon field are essentially the same. In turn, $W$ and $\rho$ are related by
a Laplace-like transform,
\eqn\lpchi{W(\ell,x)
=\int_{2\sqrt{\mu}}^\infty \d\lambda\, \ee{-\ell\lambda}\rho
={2\sqrt{\mu}\over\ell}K_1(2\sqrt{\mu}\ell)+\int_0^\infty \d\tau\,
\ee{-2\ell\sqrt{\mu}\cosh\tau}\,\p_\tau \zeta\ ,}
where in the second equation we have shifted the field $\rho$ by its
genus zero one-point function $\rho=\sqrt{\lambda^2-4\mu}+\p_\lambda\chi$,
and changed variables to $\lambda=2\sqrt{\mu}\cosh\tau$.

Using this transformation on fields, we can understand the
relation between the tree level propagators of the $W$-theory,
\eqns{\decprop{,\ }\conefrm}\ and
those of the collective field theory \djpr.
The key relation is provided by the kernel $\ee{-2\ell\sqrt{\mu}\cosh\tau}$,
which satisfies the differential equation:
\eqn\krdfeq{H_\phi\, \ee{-2\ell\sqrt{\mu}\cosh\tau}
\equiv \Bigl(-{\p^2\over \p \phi^2}+4 \mu
\ell^2\Bigr)\ee{-2\ell\sqrt{\mu}\cosh\tau}=-{\p^2\over \p \tau^2}
\,\ee{-2\ell\sqrt{\mu}\cosh\tau}\ .}
It follows that if we define a nonlocal transformation of functions:
\eqn\formaltmn{
\eqalign{
\hat B(\phi)&=\int_0^\infty \d\tau\, \ee{-\ell\cosh\tau}B(\tau)\cr
\check{B}(\tau)&=\int_0^\infty {\d\ell\over \ell}\,\ee{-\ell\cosh\tau}
B(\phi)\ ,\cr}}
then we have
$$f(H_\phi)\, \hat B(\phi)=\int_0^\infty \d\tau\, \ee{-\ell\cosh\tau}
f(-{\p^2\over \p \tau^2}) B(\tau)\ ,$$
if $B$ is such that integration by parts is valid. In this way
we may establish the classical function identities:
\eqn\jvtm{\eqalign{K_{iE}(2\sqrt{\mu}\ell)&=\int_0^\infty \d\tau\,
\ee{-2\sqrt{\mu}\ell \cosh\tau} \cos E\tau\cr
{\cos E\tau\over E \sinh \pi E}&=\int_0^\infty{\d\ell\over\ell}\,
\ee{-2\sqrt{\mu}\ell \cosh\tau}K_{iE}(2\sqrt{\mu}\ell)\ ,\cr}}
relating eigenfunctions of the Bessel and Laplace operators.
Comparing, we now see that \lpchi\ indeed maps \eqns{\decprop{,\ }\conefrm}\
to \djpr. As a further check, the tree level 3-point function
of the eigenvalue density has been calculated in \jevscatii\ to be
\eqn\eigcr{\eqalign{\bigl< \rho(\lambda_1,q_1)\,
\rho(\lambda_2,q_2)\,\rho(\lambda_3,q_3)\bigr>_c
&=\delta(q_1+q_2+q_3)\,{1\over 8\mu^{3/2}
\sinh\tau_1 \sinh\tau_2 \sinh\tau_3}\cr
\cdot\int_{-\infty}^{\infty} \prod_i \d E_i\, \Bigl({E_i\over
 E_i-iq_i}\cos E_i\tau_i\Bigr)&(E_1+E_2+E_3)\coth\bigl({\pi\over 2}
(E_1+E_2+E_3)\bigr)\ ,\cr}}
which is related to \macrocr\ by \lpchi.

The transform \lpchi\ is very subtle. While it is nonlocal, it
can be shown to map exactly the quadratic $\tau$-space action \redham\
to the $\phi$-space WdW action \freewac.
On the other hand, the interaction terms will not be locally related.
The nonlocality of the Lagrangian for $T(\phi,t)$ is not
a surprise. It is present in the covariant formulations of
closed string field theory \bz\ and has also been found within
the context of 2D string theory by  Di Francesco and Kutasov
using continuum methods (see below).  (The detailed comparison
of the $W$-field theory with the above two formulations has
not been carried out.)

As a second application, the origin of
the Wheeler--DeWitt equation from the point of view of the
eigenvalue dynamics can be understood as follows:

\exercise{Variations on WdW}

a) Derive the WdW equation for tachyon wavefunctions
using the dynamical Fermi sea picture as follows. Write
$$W(\ell,t)
=\int \d\lambda\, \ee{\ell \lambda}
\bigl(p_+(\lambda,t)-p_-(\lambda,t)\bigr)\ .$$
Using the flow equations, show that
$$\p_t W= \half \ell\int \d\lambda\, \ee{\ell
\lambda} (p_+(\lambda,t)^2-p_-(\lambda,t)^2)\ .$$
Then take another derivative to obtain
$$\eqalign{&\Bigl(\p_t^2 -(\ell{\p\over \p \ell})^2\Bigr)W
= 2 \ell^2 \int \d\lambda\, \ee{\ell \lambda}\, \CH(\ell)\cr
&\CH(\lambda)=\Bigl({1\over 6}p_+(\lambda,t)^3-\lambda^2 p_+\Bigr)-
\Bigl({1\over 6}p_-(\lambda,t)^3-\lambda^2 p_-\Bigr)\ .\cr}
$$
Take the variation of the loop to get the wavefunction of
the tachyon and from this recover the WdW equation:
\eqn\wddw{
\Bigl(\p_t^2- {\p^2\over \p \ell^2}\Bigr)\delta W= 2 \mu\ell^2\, \delta W\ .}

b)  Generalize part a) to arbitrary Hamiltonians of the
form $H=p^2 + \half V(\lambda)$, where $V$ is a polynomial,
to obtain
\eqn\genwdw{
\Bigl(\p_t^2+\ell^2 V({\p\over \p \ell})+
\half \ell V'({\p\over \p \ell})\Bigr)\delta W= 2 \mu\ell^2\, \delta W\ ,
}
for fluctuations in $W$ along the Fermi surface, $H(p,\lambda)=\mu$.
This equation was derived differently in \refs{\danieli,\danielii}.

It should be emphasized that \eqns{\wddw{,\ }\genwdw}\ are only
valid at genus zero.

\endexercise

{\bf Remark}:\foot{Based on conversations with A.B. Zamolodchikov.}
A further interesting property of the transform \lpchi\ not
directly related to 2D gravity is that it relates massive and massless field
theories in 2 spacetime dimensions. To see this, consider the Lagrangian of a
massive Klein--Gordon field in 2 Euclidean dimensions:
\eqn\kgfld{S_{KG}
=\int \d^2 w\, \Bigl(\p_w \Psi \p_{\bar{w}} \Psi+ 4 \mu^2 \Psi^2\Bigr)\ .}
Making a change of variables $w=\ee{z}$, $z=\half(\phi+i X)$,
the action becomes
\eqn\kgfldi{ \eqalign{
S_{KG}&=\int \d^2 z\,
\Bigl(\p_z \Psi \p_{\bar{z}} \Psi+ 4 \mu^2 |w|^2\Psi^2\Bigr)\cr
&=\int \d \phi\, \d X \Bigl(\p_\phi \Psi \p_\phi \Psi
+ 4 \mu^2 \ee\phi \Psi^2+\p_X \Psi \p_X \Psi\Bigr)\ .\cr}}
It is precisely this action which is mapped to a massless field on the
half-space $\tau\geq 0$ by \lpchi.

\subsec{The $w_{\infty}$ Symmetry of the Harmonic Oscillator}

Collective field theory reduces genus zero matrix-model dynamics to the
dynamics of a phase-space fluid under the influence of an upside-down harmonic
oscillator. This system has a very interesting symmetry algebra, following from
the existence of an infinite-dimensional symmetry of the harmonic
oscillator.\foot{This symmetry of the harmonic oscillator appears to have been
noticed first by matrix-model theorists in 1991! Although it was well-known
that one could construct a phase space realization of the wedge subalgebra of
$w_\infty$, the important point that this is a dynamical symmetry of the
oscillator appears to have been overlooked.}

Consider the functions $a_-=\lambda-p$, $a_+=\lambda+p$ on phase-space.
Under the Hamiltonian flow defined by
$H=\half(p^2-\lambda^2)$, we have $a_\pm (t)=a(0)\,\ee{\pm t}$,
so the functions
\eqn\chrges{
\tilde{\CC}_{n,m}=(a_+)^n (a_-)^m \ee{(m-n)t}
}
are, in fact, time-independent. As functions
on (phase space)$\times\IR$, under Hamiltonian flow they satisfy
\eqn\totalder{
{\d \tilde{\CC}_{n,m}\over \d t}=
{\p \tilde{\CC}_{n,m}\over \p t} +\{H, \tilde{\CC}_{n,m}\} =0\ ,
}
and should be considered as conserved charges with
explicit time-dependence.
It is also evident that they form a closed algebra under
Poisson brackets,
\eqn\clwi{
\bigl\{ \tilde{\CC}_{n,m}\,,\,\tilde{\CC}_{n',m'}\bigr\} = 2(m' n-m n')\,
 \tilde{\CC}_{n+n'-1,m+m'-1}\qquad n,m\geq 0\ .}
As we will see below, this defines the ``wedge subalgebra'' of
$w_{1+\infty}$. Notice that  $\tilde{\CC}_{1,1}$ is itself the
Hamiltonian. Upon quantization, we obtain a quantum
$W_{\infty}$-type algebra which is in fact a spectrum-generating
algebra.\foot{Note that the ordinary harmonic oscillator action is minus the
Euclidean action of an inverted oscillator. Thus the above
results apply to the ordinary harmonic oscillator. The formulae
differ in some factors of $i$ arising from the analytic continuation
of Euclidean to Minkowskian time.}

\exercise{Classical $w_{1+\infty}$ and its subalgebras}

There is a bewildering choice of bases and algebras in the
literature all related to $w_{1+\infty}$ but differing
in slight, yet important, ways. In this exercise we survey
some of them.

Classical $w_{1+\infty}$ \bakas\
is the algebra generated by
basis vectors $W_{s,n}$, $s=0,1,2,\dots$, $n\in \IZ$, subject
to the relations
\eqn\winfrel{
[W_{s,n},W_{s',n'}]=(s' n-s n') W_{s+s'-1,n+n'}\ .
}
The basis generators are parametrized by a semilattice of
points $(s,n)$ in $\IR^2$. Equivalent bases in the literature
are obtained by applying affine transformations to this semilattice.
For example, one could instead take generators $V_{s,n}$,
$s=1,2,\dots$, $n\in \IZ$, related by $V_{s,n}=W_{s-1,n}$ to
give
\eqn\winfreli{
[V_{s,n},V_{s',n'}]=((s'-1) n-(s-1) n')\, V_{s+s'-2,n+n'}\ .}
Several subalgebras are notable:

$\bullet$ $w_{\infty}$: generated by $V_{s,n}$ but with
$s\geq 2$. In the study of extended chiral algebras of
rational conformal field theories, one encounters these
algebras where $V_{s,n}$ are the modes of spin $s$
currents generating the algebra. It is therefore hardly
surprising to find the next subalgebra:

$\bullet$ Witt algebra = Virasoro $(c=0)$: the algebra generated by
the elements with $s=2$:
\eqn\wittalge{
[V_{2,n},V_{2,n'}]=(n-n')\, V_{2,n+n'}\ .}

$\bullet$ $\vee w$: the Wedge subalgebra, is generated
by $Q_{j,m}$ with $j=0,\ha,1,\dots$, $m\in\{-j,-j+1,\dots,j-1,j\}$
with relations
\eqn\winfrelii{
[Q_{j,m},Q_{j',m'}]=(j' m-j m')\, Q_{j+j'-1,m+m'}\ .}

$\bullet$ $\vee^2 w$: the double-wedge subalgebra is the subalgebra
of the wedge algebra generated by $Q_{j,m}$ with
$j=1,{3\over2},2,\dots$, $|m|\leq j-1$. Clearly we can continue the
process and form a filtration of wedge algebras $\vee^n w$,
defined by restricting $|m|< j-n+1$.

$\bullet$ ${\it Vir\/}^+$: the Borel subalgebra of the Virasoro algebra.
${\it Vir\/}^+$ may be embedded in the $\vee w$ algebra in many ways:
\eqn\borel{\eqalign{
L_{2s} &= Q_{s+1,s} \qquad s=0,\ha,1,\dots\cr
L_{2s} &= Q_{s+1,-s} \qquad s=0,\ha,1,\dots\cr
L_{s} &= V_{s,2-s} \qquad s=1,2, \dots\ .\cr}}

$\bullet$ $w^+$: Borel subalgebra of $w$.

This is generated by $V_{s,n}$ where $s=2,3,\dots$ and
$n\geq -s+1$, and is the analog the Borel of Virasoro.
It plays a role in the $W$-constraints of the
$c<1$ models. For $w^+_{1+\infty}$ include $s=1$.


a) Show that \winfrelii\ is a subalgebra of \winfrel.

b)  Show that $\vee^2 w/\vee^3 w$ contains both of the ${\it Vir\/}^+$ algebras
defined in the first two lines of \borel.

\endexercise

The $w_\infty$ symmetry of the inverted oscillator was nicely
reformulated in terms of symplectic geometry in
\grndrng. The action of the oscillator, in first order
form, is $S=\int \d \alpha$, where $\alpha=p\,\d q- H \d t$ is a
$1$-form on (phase space)$\times \IR$.
A transformation on this space that takes $\alpha\to \alpha+\d\beta$
is a symmetry. Symmetries are thus transformations preserving the 2-form
$\omega=\d\alpha$.
For the inverted oscillator, we may write
\eqn\sympform{\eqalign{\omega
=\d \alpha &= \d p\, \d q -(p\, \d p - q\, \d q)\d t= \d p'\, \d q'\cr
p'&= \cosh t p - \sinh t q\cr
q'&=-\sinh t p + \cosh t q\ .\cr}
}
The symmetries are thus generated by the Hamiltonian vector fields
\eqn\symmvect{
V_g = {\p g(p',q')\over \p q'} {\p\over \p p'}-{\p g(p',q')\over \p p'}
{\p\over \p q'}
}
associated to the charges $g$,
where $g$ is a polynomial in $p',q'$. By standard symplectic
geometry:
\eqn\homomorph{
[V_{g_1},V_{g_2}]=V_{\{g_1,g_2\}}\ ,}
so we may invariantly characterize the wedge algebra as the
algebra of area-preserving polynomial vector fields on $\IR^2$.

\exercise{Realizations of $w$-algebras}

Verify that:

a) The wedge
algebra may be realized as a Poisson algebra by
\eqn\poissi{
Q_{j,m}=2 a_+^{j+m} a_-^{j-m}\ .
}

b) The Borel algebra
 realization occurs naturally in phase space via
\eqn\poissii{
V_{s,n}=p^{n+s-1} \lambda^{s-1}\ .
}

c) Using \poissii, show that ${\it Vir\/}^+$
corresponds under Poisson action to the algebra
of analytic coordinate changes in $\lambda$.

\endexercise

\subsec{The $w_{\infty}$ Symmetry of Free Field Theory}

{\it Classical Theory\/}.
Finally let us note that what is true of harmonic oscillators is
necessarily true of free field theory: any free field
theory contains an infinite set of $w_{\infty}$ algebras.
Spacetime locality considerably limits the set of interesting
algebras. For example, if $\phi(x,t)$ is a free massless field
in $1+1$ dimensions, we can consider the spin $s$ currents:
\eqn\spiness{
V_{s}(x,t)={1\over s} (\p \phi)^s \qquad s=1,2,\dots\ ,
}
whose moments form a classical $w_{1+\infty}$ algebra.

\exercise{Poisson brackets}

Use the Poisson brackets $\{ \p \phi(x), \p \phi(y)\}= 2 \pi \delta'(x-y)$
to show that the modes of $V_s$
\eqn\wmodes{
V_s(x)=\sum_{n\in \IZ} V_{s,n}\, \ee{i n x} \qquad 0\leq x<2 \pi
}
obey a classical $w_{1+\infty}$ algebra:
$$\{ V_{s,n},V_{s',n'} \}=i\bigl((s'-1) n-(s-1) n'\bigr)
V_{s+s'-2,n+n'}\ .$$

\endexercise

{\it Quantum Theory\/}.
There is a large literature on quantum extensions of $w_\infty$.
One of particular interest to us is $W_{1+\infty}$ which may be realized
as the algebra of modes of the Fermion bilinears
$\colon\p^k \psib(z)\, \p^l\psi(z)\colon$ where $\psib,\psi$ comprise a
Weyl fermion in 2 dimensions. By bosonization this may
be related to the algebra generated by the modes of the
currents
$V_s={1\over s} \colon\ee{- \phi(z)}\, \p^s \ee{\phi(z)}\colon$.
The structure constants are very complicated and can be found in
\refs{\pope,\kawaii}.  

{\bf Remark}: It should be clear from the above discussion that
$w_{1+\infty}$ symmetry is generic, and occurs whenever there is
a massless scalar field in the problem. This symmetry is so robust
that its seeming presence in completely wrong or meaningless formulae
has deceived many an author.

\subsec{$w_{\infty}$ symmetry of Classical Collective Field Theory}
\subseclab\sswsccft

Let us apply the results of the previous section to
collective field theory. Both in $\phi$-space and in
$\tau$-space, we have asymptotic conformal field theories
(=massless scalars) in {\it spacetime}. Thus, we expect on
{\it a priori} grounds to find a spacetime $w_\infty$
symmetry of the $S$-matrix. (See sec.~{\it\sTrr\/} below.)

One approach, pursued by Avan and Jevicki \wviii,
is to form the  charges
\eqn\avjev{
Q_{j,m}=\int \d\lambda\, \int_{p_-}^{p_+} \d p\,
(p+\lambda)^{j+m+1}\,(p-\lambda)^{j-m+1}\ ,
}
interpreting $p_\pm$ in terms of the
collective field as in \djfld.
The integrals don't converge so the expression is somewhat formal,
but, working formally, one can use the Poisson bracket
structure to show that the charges satisfy the correct
algebra. Although collective field theory is not a free
theory Avan and Jevicki show that it has a spectrum
generating algebra given by these charges. They go on
to interpret the collective field action in terms of
coadjoint orbit quantization for a group of area-preserving
diffeomorphisms \avjvwinft.
Similar work has been undertaken in a series of
papers by Wadia and collaborators \wix.

The $w_\infty$ symmetry may also be seen in the Fermi fluid picture
\refs{\wvii,\msi},
where the charges have exactly the realizations  in terms of phase
space coordinates described in the previous section. In the Fermi sea picture,
the wedge algebras $\vee w$, $\vee^2 w$ have pretty geometrical
interpretations discussed in \refs{\grndrng,\kutmarsei}. The phase space
charges have associated Hamiltonian vector fields inducing  diffeomorphisms
of the $(\lambda,p)$ plane.  The double-wedge algebra $\vee^2 w$ is the
algebra of area-preserving diffeomorphisms that preserves the hyperbola $a_+
a_-=0$. Therefore, by conjugating with an appropriate diffeomorphism, we can
turn it into the  algebra preserving the collective field ground state at
$\mu>0$. Notice that these diffeomorphisms fix the Fermi level as a set, but
not pointwise. Similarly the triple wedge subalgebra $\vee^3 w$ preserves
the Fermi sea pointwise. The quotient $\vee^2 w/\vee^3 w$ contains two
copies of the Virasoro Borel, ${\it Vir\/}^+$, corresponding  to
diffeomorphisms of the upper and lower branches of the Fermi sea.

It was first proposed in \wittbh\
that the $w_\infty$ symmetry of  the matrix model is
related to the extra complexity of the   BRST cohomology found in the Liouville
approach and discussed in sec.~{\it\ssstdgmb\/} above.
The best evidence for the connection is:

1) {\it Algebraic structures\/}.
As we will discuss in sec.~{\it\ssastdcc\/}
below, in the continuum approach (at least, at $\mu=0$) one
discovers very similar algebraic structures, in particular, a
realization of the $\vee^2 w$ algebra associated with
the charges $\CA_{j,m}$ discussed in sec.~{\it\ssstdgmb\/}.
However, in view of the generic nature of such symmetries, and
the nontrivial relation between the matrix model coordinate
$\tau$ and the Liouville field $\phi$, we should be cautious
about such identifications.

2) {\it Quantum numbers\/}.
{}From the local operator expansion \locopexp\ we see
that the operators $\hat \CB$, which are simply related
to the moments $\CB_r$ of $\lambda^r$, have wavefunction
\eqn\strwvf{
\langle\hat\CB_{r,q}\,W(\ell,-q)\rangle
=-r \mu^{r/2}\,K_r(2\sqrt{\mu}\ell)\ ,}
as one would expect for the Wheeler--DeWitt wavefunctions of the
ghost number $G=2$ special operators of sec.~{\it\ssstdgmb\/}. In particular,
after the transform to $\phi$-space these operators have the
correct Liouville quantum numbers.

3) {\it Behavior of Redundant Operators\/}.
The transformations
\eqn\dellam{\delta^{s,q} \lambda=(\lambda+i\dot\lambda)^{(s+q)/2-1}
(\lambda-i\dot\lambda)^{(s-q)/2-1}(q\lambda -is\dot\lambda)\,\ee{iqt}\ ,}
where $s=1,2,\dots$ and $q\in \IR$ form a closed algebra if we
interpret the fractional powers by expanding in $ \lambda/ \dot \lambda$
and dropping nonpolynomial terms. The algebra of such
transformations can be shown to be \msi:
\eqn\genalg{
[\delta^{s_1,q_1},\delta^{s_2,q_2}]=(q_1 s_2-q_2 s_1)
\,\delta^{s_1+s_2-2,q_1+q_2}\ .
}
For $q\notin \{-s,-s+2,\dots s\}$,
these transformations are not
symmetries of the harmonic oscillator action
\eqn\smin{S={1\over 4} \int \d t\,(\dot\lambda^2 - \lambda^2 )\ ,}
but rather induce the variation
\eqn\varacti{
\delta^{s,q}S=-{1\over (s-1)!}\prod_{r=0}^s\bigl( q-(s-2r)\bigr)
\int\d t\,\lambda^s\,\ee{i qx}\ .}
In other words, the operators $\CB_s(q)$ are redundant operators, with only
contact term interactions if $q\notin \{-s,-s+2,\dots s\}$.
For $q\in \{-s,-s+2,\dots s\}$, they are not redundant and hence are bulk
operators. The failure of these operators to be redundant in the latter case
is a signal of
the appearance of an extra cohomology class, as is indeed predicted by the
continuum formalism.

One weakness of the matrix-model approach to understanding
the special states is that one cannot tell
which of the four cohomology classes at discrete values of
$(p_\phi,p_X)$ is represented by the matrix model operators.

One can try to use the transformations
\genalg\ to obtain Ward identities for insertions of
special state operators. This works nicely for
$s=1$ \msi. However, as shown by the results of the
next chapter, for $s\geq 2$ the measure and ordering problems
present serious obstacles to this approach.

\newsec{String scattering in two spacetime dimensions} 
\seclab\sscatt

\subsec{Definitions of the $S$-Matrix}

We are finally ready to calculate the scattering of strings in
two spacetime dimensions described physically in sec.~{\it\sstdgatds\/}
(i.e.\ \fwall).  Recall that scattering takes place in Minkowski
space. In this chapter we study the theory of sec.~{\it\stdstms\/}.A:
the Liouville coordinate $\phi$ is regarded as space,
the time coordinate $t$ is a negative signature $c=1$ field
obtained by analytically continuing $X$.
The tachyon background
\eqn\tcond{ \bigl< T(\phi,t)\bigr> = \mu\, \ee{\sqrt{2} \phi}
}
acts as a repulsive wall for incoming bosons and the
dilaton background leads to a spatially-varying coupling
\eqn\sptcp{
\kappa\dup_{\rm eff}(\phi)=\kappa\dup_0\,\ee{\ha Q \phi}\ .}

Because the $S$-matrix of massless bosons in two-dimensions is
a subtle object, we begin with some precise mathematical
definitions of what we are talking about.
We begin with the string definition. As explained in sec.~{\it\stdstms\/},
the vertex operators are $V_\omega^\pm$ given by \minkvrtx.
Using \ecanr, we write

\noindent
{\bf Def 1}: The connected string scattering matrix
elements are asymptotic expansions in $\kappa$ given by
\eqn\stringess{
\CS_c^{ST}\Bigl(\sum_{i=1}^k \omega_i\to \sum_{i=1}^l \omega_i'\Bigr)
=\CA_n(V_{\omega_1}^-,\dots V_{\omega_k}^-,V_{\omega_1'}^+,\dots
V_{\omega_l'}^+)\ .}

Mathematically it is easier to use a Euclidean
signature boson $X$ via the analytic continuation
$|q|\to -i \omega$:
\eqn\continue{\eqalign{&V_{\omega}^+\to V_q \qquad q>0\cr
&V_{\omega}^-\to V_q \qquad q<0\ .\cr}}
We'll refer to the $S$-matrix elements calculated with
$V_q$ as the ``Euclidean $S$-matrix.''

According to the matrix model hypothesis, these amplitudes
may be calculated via the $c=1$ matrix model according to
the discussion of sec.~{\it\sscfvo\/}. If one is interested in
the $S$-matrix and not in the macroscopic loop amplitudes
(which contain much more information), then it is most
efficient to calculate the {\it collective field\/} $S$-matrix
which we describe next.\foot{Indeed,
defining the $S$-matrix directly via asymptotics in $\tau$-space
\mpr, as presented below, was an important technical advance over the original
method \moore\ of calculating loop amplitudes and then shrinking the loops.}

In collective field theory we define the $S$-matrix according
to the coordinate-space version of the LSZ prescription,
that is, we isolate the piece of the large spacetime asymptotics
of time-ordered Green's functions which is proportional to
the product of on-shell incoming and outgoing wavefunctions.

An incoming or outgoing boson of energy $\omega>0$ has wavefunction
$\psi_\omega^L(t,\tau)=\ee{-i\omega(t+\tau)}$,
$\psi_\omega^R(t,\tau)=\ee{-i \omega(t-\tau)}$,
respectively. Therefore we define the $S$-matrix according
to

\noindent
{\bf Def 2}: Consider the asymptotic behavior of the time-ordered, connected,
Minkowskian collective field Green's function:
\eqn\timord{G(t_1,\tau_1,\dots t_n,\tau_n)\equiv
\langle 0|T\prod \p_\tau \chi(t_i,\tau_i)|0\rangle\ ,
}
as $\tau_i\to +\infty$, $t_i\to -\infty$ $(1\leq i\leq k)$,
$t_i\to +\infty$ $(k+1\leq i\leq n=k+l)$.
Then we define the connected $S$-matrix element for
the process $|\omega_1,\dots \omega_k\rangle \to
|\omega'_1,\dots \omega'_l\rangle$
to be the function
$S_c^{\rm CF}$ in the asymptotic formula:
\eqn\dfessi{\eqalign{
&G\sim \int_0^\infty\prod_{i=1}^k \d \omega_i\, \prod_{i=1}^l \d \omega'_i\,
\delta(\sum \omega_i - \sum \omega'_i)
\prod_{i=1}^k (\psi^L_{\omega_i})^* \prod_{i=1}^l \psi^R_{\omega'_i}\cr
&\qquad\qquad\cdot\sqrt{\,k!\, l!\,}\, S_c^{\rm CF}
\Bigl(\sum_{i=1}^k \omega_i\to \sum_{i=1}^l \omega_i'\Bigr)
+\hbox{off-shell terms}\ .\cr}}
The plane wave states are normalized such that
$\langle \omega|\omega'\rangle=\omega\,\delta(\omega-\omega')$.
An equivalent definition has been used in \refs{\rColl,\jevscatii}\
to compute the $S$-matrix from standard Feynman perturbation
theory applied to collective field theory.

While this definition is physically satisfying, it is not
the best mathematical definition. An equivalent definition
is obtained by continuing the Minkowskian Green's functions
to Euclidean space $\Delta t \to -i \Delta X$. Fourier
transforming the Euclidean Green's functions with respect to
$X_i$, we obtain mixed Green's functions
\eqn\mixedgreen{
G_E(q_1,\tau_1,\dots q_n,\tau_n)\equiv
\int \prod_i \d X_i\,\ee{i q_i X_i}
G_{\rm Euclidean}(X_1,\tau_1,\dots X_n,\tau_n)\ ,
}
in terms of which we may define the $S$-matrix via:

\noindent
{\bf Def2$'$}: The large $\tau_i$ asymptotics
\eqn\defessii{
G_E(q_1,\tau_1,\dots q_n,\tau_n)\sim \delta(\sum q_i)
\prod_i \ee{-|q_i| \tau_i} \CR(q_1,\dots q_n)
\bigl(1+\CO(\ee{-\tau_i})\bigr)
}
defines a function $\CR_n(q_1,\dots q_n)$ from which we may obtain the
connected $S^{\rm CF}$-matrix elements via analytic continuation
$|q|\to -i \omega$, where $q<0$ corresponds to the incomers,
and $q>0$ corresponds to the outgoers. Specifically,
$S_c\to -{i^{k+l+1}\over \sqrt{k!\, l!}} \CR$.

{\bf Remark}: The non-obvious property that the function
$\CR_n$ is independent of the order in which the
$\tau_i$ are taken to $\infty$ was demonstrated in \mpr.

The equivalence of the collective field theory $S$-matrix and the correlators
defined by shrinking macroscopic loops is demonstrated using the relation
between $\tau$-space and $\phi$-space explained in sec.~{\it\sstsaps\/} above.
In particular, transforming asymptotic wavefunctions according to \lpchi, we
relate $\ell\to 0$ and $\tau\to \infty$ asymptotics via the integral
\eqn\intgl{\int^\infty \d\tau\,
\ee{-\ell 2\sqrt{\mu}\cosh\tau}\,\ee{-|q|\tau}\sim
(\ell\sqrt{\mu})^{|q|}\Gamma(-|q|)\ ,}
plus terms regular in $\ell$. Notice that
the two prescriptions only make complete sense
when $q$ is nonintegral. Otherwise
we must use the full identity
\eqn\carinte{\eqalign{
\int_A^\infty \d\tau\, \ee{-2\ell\sqrt{\mu}\cosh\tau}\,\ee{-|q|\tau}
&=-{\pi\over \sin\pi|q|}I_{|q|}(2\sqrt{\mu}\ell)\cr
-\sum_{n\geq 0}
(-1)^n &(\ell\sqrt{\mu})^n\sum_{m=0}^n{1\over m!(n-m)!}
{\ee{A(m-n-|q|)}\over m-n-|q|}\ .\cr}}
The pole in the $\Gamma$--function in \intgl\
is a warning that we cannot unambiguously separate
the two terms in \carinte\ via nonanalyticity in $\ell$.

\danger{Leg Factors}
According to the arguments of chapt.~\slascomm, we expect that the Euclidean
$S$-matrices $\CS^{ST}$ and $\CS^{CF}$ should agree up to an overall
normalization $f(q)$ of the vertex operators $V_q$. This is because, for
$q\notin \IZ$, the BRST cohomology with the relevant quantum numbers is
one-dimensional. Indeed, comparison with vertex operator calculations in
Liouville theory, which will be described in sec.~{\it\srabsm\/} below, shows
that
\eqn\compamps{
\CA_{0,n}(V_{q_1},\dots V_{q_n})=(-i)^{n+1}
\prod_{i=1}^n {\Gamma(-|q_i|)\over \Gamma(|q_i|)}
\CR_n(q_1,\dots q_n)\ .}
The factors
\eqn\wvfnfctr{f(q) = {\Gamma(-|q|)\over \Gamma(|q|)}}
are called ``leg factors.'' Notice that for the Minkowskian $S$-matrix they are
pure phases, but the phases for incomers and outgoers are {\it not\/} complex
conjugated. Comparison with the first quantized wavefunction for the spacetime
boson, described by the Wheeler--DeWitt equation \onshmn,
indicates that neither
normalization in \compamps\ is the correct physical normalization since
standard first-quantized scattering theory predicts that the genus zero $1\to
1$ $S$-matrix is ${\Gamma(iE)\over \Gamma(-i E)}$.
This suggests that the correct
normalization of the vertex operators is obtained by taking the squareroot of
\wvfnfctr. All this needs to be clarified!

\subsec{On the Violation of Folklore}

The $c=1$ $S$-matrix violates several standard
aspects of $S$-matrix folklore. It is commonly
said, for example, that one cannot define an
$S$-matrix for massless bosons. For example,
the standard LSZ prescription appears to be
problematic because if we make a field redefinition
\eqn\fldredef{
\Phi\to \chi + a_2\, \chi^2 + a_3\, \chi^3+\cdots\ ,
}
in the massless case there is no gap between the one-particle and two-particle
thresholds, so $S$-matrix elements appear to depend on the choice of
interpolating field.\foot{For a discussion of the independence of the
$S$-matrix from a choice of interpolating field, see \refs{\sidney,\benlee}.}
More physically, we cannot expect to tell the difference between (say) a
rightmoving boson of energy $E$ and two rightmoving bosons of energy $E/2$. A
related mathematical point is that the momentum-space Green's functions should
have cuts, not poles, so we can't isolate an $S$-matrix element by
extracting the residues at poles.

In the present case we find that there are no cuts, but there are instead
kinematic regions, and the momentum space Green's functions are
continuous, but not differentiable, across regions. The $S$-matrix will have a
large symmetry group related to $W_\infty$, which is nonlinear in the momentum
and allows us to distinguish a rightmoving boson of energy $E$ and two
rightmoving bosons of energy $E/2$.

Another objection to massless $S$-matrices is that by a simple conformal
transformation one can fill the vacuum with particles.\foot{Indeed,
calculations of Hawking radiation in the CGHS theory \rCGHS\
are based on this phenomenon.}
In our case, the ``defect'' or wall at $\tau=0$ breaks conformal
invariance enough to forbid such freedom.

We have also violated folklore in another way. The exactly solvable $S$-matrix
presented below has particle production, yet at the same time has a large
$W_\infty$ symmetry. Typically, exactly solvable $S$-matrices in field theories
with infinite numbers of conservation laws \zamosq\ do not have particle
production.

There are several related issues, connected with the interpretation of the
wavefunction factors $f(q)$. The resolution of these issues will probably
require careful specification of how $S$-matrix elements are to be measured.

Finally, we remark that the $c=1$ $S$-matrix bears a great similarity to a
number of other physical problems which have been of interest in recent years.
These include the Kondo effect, the Callan--Rubakov effect, Hawking radiation
and particle scattering off a black hole (especially in the CGHS model
\rCGHS) and $1+1$ linear dilaton electrodynamics. Massless $S$-matrices have
played a role in the theory of exactly solvable field theories, for example
they have appeared in past discussions of the XXX and XYZ models \xyzmdl\ and
more recently have begun to play a more central role in the massless flows
between conformal field theories \zamii.

\subsec{Classical scattering in collective field theory}

We now consider the classical scattering problem for the collective field using
the picture of the time-dependent Fermi sea. Suppose the solution is given by
\frms{}\ and represents an incoming wavepacket which is dispersed as it travels
in phase space. We will derive a functional relation between the incoming and
outgoing wavepackets \joesea.

Let us return to the general solution \frms{a,b}, and assume there are no
folds.\foot{The conditions for this are given in \mrpl.}
We may solve the first equation to obtain
$\sigma_\pm(\lambda,t)$. If there were no dispersion
of the wavepacket, we would find $\bar\sigma_\pm(\lambda,t)=t\pm \tau$.
Denote the difference by
$\sigma_\pm=\bar \sigma_\pm + \delta \sigma_\pm$.
{}From the spacetime asymptotics of the solution \frms{}\ above, we find
\eqn\intmedi{\delta \sigma_\pm(t\pm\tau)=
\mp \log\Bigl(1+a\bigl(t\pm \tau+\delta \sigma_\pm(t\pm \tau)\bigr)\Bigr)\ ,
}
and, in particular, $\delta \sigma_\pm$ becomes a function
of one variable. The asymptotic behavior defining
in- and out-waves is
\eqn\limbeh{
p_\pm(\lambda,t)\rightarrow \pm \lambda \mp {1\over 2\lambda}\bigl(1+
\psi\dup_\mp(t\mp \tau)\bigr)+\CO(1/\lambda^2)\ ,
}
where $\lambda\to +\infty$ holding $t\mp \tau$ fixed.
Plugging this into the expression $(p_\pm \mp \lambda)\lambda$ and
comparing with the general solution, we find that the waves can be expressed
in terms of the function $a$ as:
\eqn\psitoa{
1+\psi\dup_\pm
=\Bigl(1+a\bigl(t\pm \tau+\delta \sigma(t\pm \tau)\bigr)\Bigr)^2\ .}
Thus we can calculate the time-delay, namely the relation
between $t$ and $t'$ such that
$\psi\dup_+(x')=\psi\dup_+(t'+\tau)=\psi\dup_-(t-\tau)=\psi\dup_-(x)$:
\eqn\timdel{x'+\delta \sigma_+(x')=x + \delta
\sigma_-(x)\quad\Longrightarrow\quad
x'=x+2 \delta \sigma_-(x)\ ,}
since from \intmedi\ we see that $\delta\sigma_+(x')=-\delta \sigma_-(x)$.
It follows from \intmedi\ and \psitoa\ that
that we have the functional relation between in- and out-
waves:
\eqn\scateq{\eqalign{\psi\dup_-(x)&=\psi\dup_+(x')\cr
&=\psi\dup_+\Bigl(x+\log\bigl(1+\psi\dup_-(x)\bigr)\Bigr)\ .\cr}}
{}From the derivation we see the essential physics:
different parts of the wavepacket suffer different time delays.

We now solve the equation \scateq, thus solving the
classical field scattering and, in principle, the
tree level $S$-matrix of the theory.
The solution of \scateq\ was given in \mrpl\
and is derived as follows.
Suppose  $\Psi_\pm$ constitute a solution of
the classical scattering equations \scateq , and suppose
further that
$\Psi_\pm + \gamma_\pm$ is a nearby solution,
where $\gamma_\pm$ are small. To first order in
the variations, \scateq\ becomes
\eqn\perti{
\gamma_+(\tilde x)\,\d\tilde x=\gamma_-(x)\, \d x\ ,
}
where $\tilde x=x+\log(1+\Psi_-(x))$.
Taking a Fourier transform of this equation, with
\eqn\dffour{
\gamma_\pm(x)\equiv \int_{-\infty}^\infty\d\xi\  \gamma_\pm(\xi)\,
\ee{i\xi x}\ ,}
leads to
\eqn\solfour{\gamma_+(\xi)={1\over 2\pi}
\int\d x\, \ee{-i \xi x}\gamma_-(x)\bigl(1+\Psi_-(x)\bigr)^{-i \xi} \ .}
This may be regarded as a first-order differential equation in
function space. Integrating this equation with the
boundary condition $\psi\dup_+=0 \Rightarrow \psi\dup_-=0$,
we obtain the general solution of Polchinski's scattering equations:
\eqn\gensol{2\pi \psi\dup_\pm(\xi)={1\over 1\mp i\xi}
\int_{-\infty}^\infty \d x\,\ee{-i\xi x}
\Bigl(\bigl(1+ \psi\dup_\mp(x)\bigr)^{1\mp i\xi}-1\Bigr)\ .}
In position space this takes the form
\eqn\gensoli{
\psi\dup_{\pm}(x)=-\sum_{p\geq 1}{\Gamma(\pm\p_x+p-1)\over \Gamma(\pm\p_x)}
{(-\psi\dup_\mp(x))^p\over p!}\ .}
(The ratio of $\Gamma$-functions is interpreted as a polynomial in
derivatives.)

This completely solves the classical scattering problem.

\subsec{Tree-Level Collective Field Theory $S$-Matrix}
\subseclab\sstlcft

{}From the classical scattering matrix, we may derive the
tree-level quantum $S$-matrix by interpreting the
left- and right-moving fields as incoming and outgoing quantum fields:
\eqn\defflds{\psi\dup_\pm\rightarrow -\sqrt{ \pi} {1\over \mu}
(\p_t\ \pm  \p_\tau) \chi_\pm\qquad
\eqalign{&\chi\dup_+=i\int_{-\infty}^\infty {\d\xi\over \sqrt{4 \pi} \xi}\,
\alpha_+(\xi)\,\ee{i \xi (t+\tau)} \cr
&\chi\dup_-=i\int_{-\infty}^\infty {\d\xi\over \sqrt{4 \pi} \xi}\,
\alpha_-(\xi)\,\ee{i \xi (t-\tau)} \cr
&[\alpha_\pm(\xi),\,\alpha_\pm(\xi')]= - \xi\, \delta(\xi+\xi')\ .\cr}}

Now following Polchinski, we interpret the relation \gensoli\
as a relation between incoming and outgoing Fourier modes:
\eqn\inout{\alpha_\pm(\eta)= \sum_{p\geq 1} ({1\over \mu})^{p-1}
{\Gamma(1\mp i\eta)\over \Gamma(2\mp i\eta-p)}
{1\over p!}\int_{-\infty}^\infty \d^p\xi\
\delta(\eta-\sum \xi_i) \,\colon \alpha_\mp(\xi_1)\cdots
 \alpha_\mp(\xi_p)\colon\ .}
Quantum mechanically, the Fourier modes
in \defflds\ are creation and annihilation operators
for left- and right-moving particles.
Let us consider the $S$-matrix element for one incoming
left-mover of energy $\omega$
to decay to $m$ outgoing particles of
energies $\omega=\sum \omega_i$:
\eqn\selmnt{S_c( \omega\to \sum_{i=1}^m \omega_i) =
\langle 0|  \alpha_-(-\omega)
\prod_{j=1}^m \alpha_+(\omega_j) |0\rangle_c\ ,
}
where
the vacuum is defined by $\alpha_+(-\omega)|0\rangle=0$
for $\omega>0$. From \inout\ we may read off without
further calculation the result:
\eqn\onetom{S_c^{CF}(\omega\to \sum_{i=1}^m \omega_i)
 = -i ({1\over \mu})^{m-1}  \omega\prod_{k=1}^m \omega_k
{\Gamma(-i\omega)  \over \Gamma(2-m-i \omega)}\ .}
The corresponding Euclidean $S$-matrix is
\eqn\onetomi{\mu^{|q|}\CR_{m+1}(q_1,\dots q_m,q)=({1\over \mu})^{m-1}
i^m|q| \prod |q_i| ({\p\over \p \mu})^{m-2} \mu^{|q|-1}\ ,}
a formula we will obtain in the next chapter via continuum methods.

Other $S$-matrix amplitudes can be derived analogously \mrpl.
The $S$-matrix is {\it not} analytic in the energies $\omega_i$ and
does not satisfy crossing symmetry.
In general we must divide momentum space into {\it kinematic regions}.
These are defined as follows. (It is convenient to work in
Euclidean space here).
For any set $S$ of momenta we let
$H(S)=\{\vec q\in \IR^k|\sum_{q\in S} q=0\}$. Then we take connected
components of the region
\eqn\edivreg{\Bigl\{\sum q_i=0\Bigr\}\cap \Bigl[\IR^k-\cup_{S} H(S)\Bigr]
=\coprod_\alpha \CC_\alpha\ ,}
where $\cup_{S}$ is over proper subsets $S$ of momenta, and
the $\CC_\alpha$ are the disjoint kinematic regions.

{}From the above formulae one can show that
$S$ is continuous on $\{\sum q_i=0\}\cap \IR^k$, and indeed in
each region $\CC_\alpha$, $S$-matrix elements are polynomials
in the $q_i$. The polynomials change from region to region, however,
so the $S$-matrix elements are not differentiable across regions.

\noindent 
{\bf Example}: Four-point function.\par\nobreak
Both cases $S_c(\omega_1\to \omega_2+\omega_3+\omega_4)$
and  $S_c(\omega_1 +\omega_2 \to \omega_3+\omega_4)$
are covered by the formula
\eqn\twotwo{
i ({1\over \mu})^2 \prod_{i=1}^4 \omega_i
\bigl(1+i\,{\rm max}\{\omega_i\}\bigr)\ ,}
where analyticity is lost due to the appearance of the
maximal value ${\rm max}\{\omega_i\}$.

\subsec{Nonperturbative $S$-matrices}
\subseclab\ssnpsm

The tree level $S$-matrix can be extended to all
orders of perturbation theory, and can even be given an
unambiguous  nonperturbative definition by returning to
the eigenvalue/macroscopic loop correlators of chapt.~\slascomm.
According to {\bf Def2} and \intgl\ above we
must isolate the large $\lambda$ asymptotics of \nptev.
These in turn follow from the large
$\lambda$ asymptotics of the Euclidean fermion
propagator \iasfg, which we now describe.

The function $I$ can be
written in terms of parabolic cylinder functions,
whose asymptotics are well-known. In this way
we find the asymptotics for $\lambda_i\to+\infty$ to be:
\eqn\asympi{\eqalign{
I(q,\lambda_1,\lambda_2)\sim {-i\over\sqrt{\lambda_1\lambda_2}}
\Bigl(\ee{-q|\tau_1-\tau_2|}\ee{i\mu|G(\tau_1)-G(\tau_2)|}&
\cr
+R_q \ee{i\mu(G(\tau_1)+G(\tau_2))} \ee{-q(\tau_1+\tau_2)} &\Bigr)
\bigl(1+\CO(\ee{-\tau_i})\bigr)\quad q>0\cr
I(q,\lambda_1,\lambda_2)\sim {i\over\sqrt{\lambda_1\lambda_2}}
\Bigl(\ee{q|\tau_1-\tau_2|}\,\ee{-i\mu|G(\tau_1)-G(\tau_2)|}&
\cr
+(R_q)^* \ee{-i\mu(G(\tau_1)+G(\tau_2))}\, \ee{q(\tau_1+\tau_2)}&\Bigr)
\bigl(1+\CO(\ee{-\tau_i})\bigr)\quad q<0\ ,\cr}
}
where $G(\tau)$ is the WKB wavefunction factor.
The two terms in \asympi\ may be understood intuitively
as those
corresponding to direct and reflected propagation of
the fermions in the presence of a wall.
The function $R_q$ is a Euclidean continuation of the
fermion reflection factor $R(E)$ for potential
scattering with $V(\lambda)\sim - \lambda^2$.
In particular, for scattering on a half-line $\lambda\in [0,\infty)$,
we have
\eqn\rqx{\eqalign{R(E) &=i \mu^{i E}\sqrt{1+i \ee{-\pi E }\over
1-i \ee{-\pi E}}
\,\sqrt{\Gamma(\half-i E)\over\Gamma(\half+i E)}\cr
&= \mu^{i E}\sqrt{2\over\pi}\ee{3 i\pi/4}
\cos\bigl({\pi\over 2}(\half+i E)\bigr)\,\Gamma(\half-i E)\ .\cr}}
The corresponding Euclidean ``bounce factor'' is given
by $R_q=R(\mu+i |q|)$. Using the rule $|q|\to -i\omega$, we can pass easily
back and forth from the Euclidean to the Minkowskian picture (keeping in mind
that $q<0$ corresponds to incomers and $q>0$ to outgoers).

In order to obtain the $S$-matrix from \nptev\ we must substitute \asympi\ into
\nptev\ and isolate only the terms corresponding to the coefficients of the
on-shell wavefunctions. In particular, we are only interested in the terms
where (1) the factors of $\ee{i \mu G(\tau)}$ cancel, and (2) the overall
$\tau$-dependence is proportional to $\prod \ee{-|q_i| \tau_i}$. The
decomposition of $I$ in terms of direct and reflected propagation is easily
encapsulated in a diagrammatic formalism whose
detailed derivation is given in \mpr. The final result is
sufficiently intuitive that the reader should be satisfied with our
presentation here without proof.

\ifig\foto{$1\to 1$ scattering}
{\epsfxsize3in\epsfbox{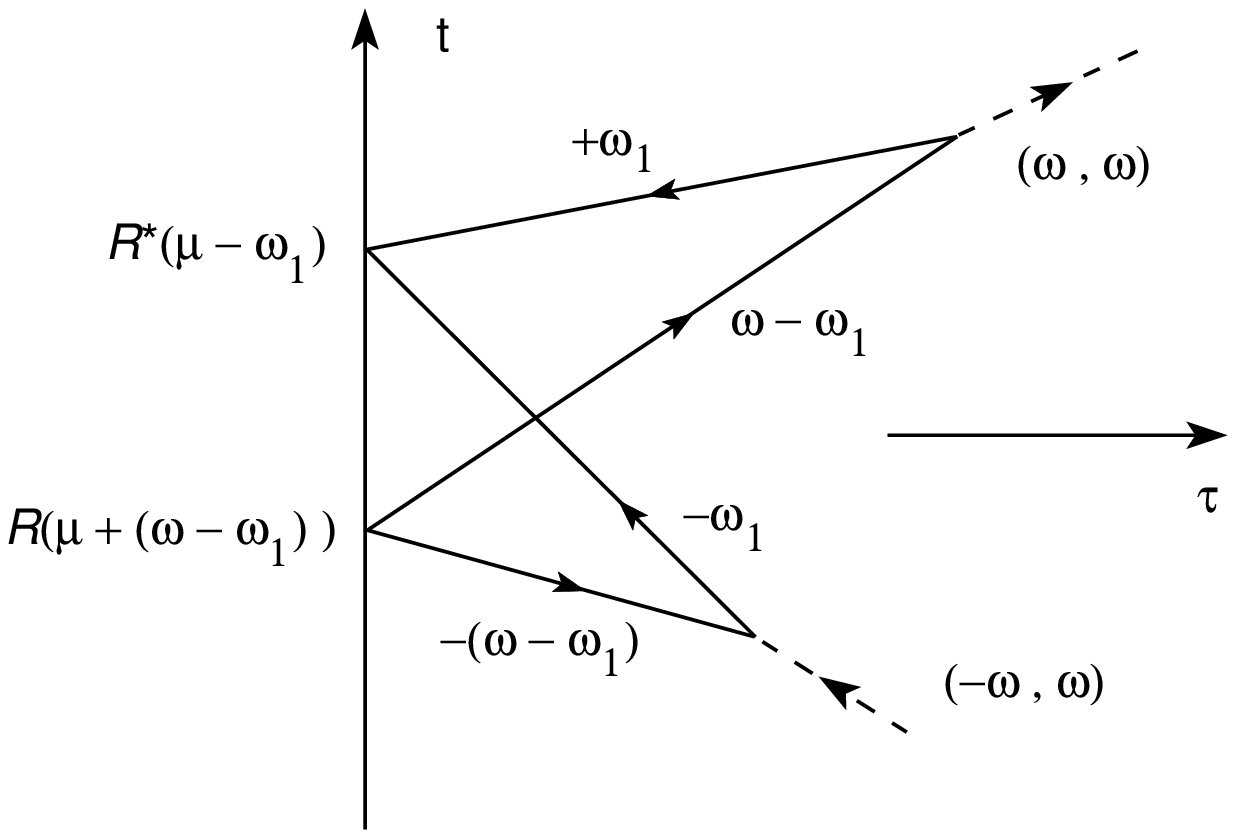}}

Consider the case of $1\to 1$ scattering, illustrated by \foto.
We have depicted an incoming relativistic boson, which may be fermionized to a
particle--hole pair. The particle and hole undergo potential scattering, and
reflect back from the wall. They may then be rebosonized. The amplitude for
this process is simply an integral over possible particle--hole energies
weighted by the reflection factor for the particle and hole, that is, we have
the $1\to 1$ $S$-matrix element:
\eqn\oneonesc{S(\omega\to \omega)= \int_0^\omega\d\omega_1\,
R^*(\mu-\omega_1)\, R\bigl(\mu+(\omega-\omega_1)\bigr)\ .}

\ifig\fipnm{a) A pictorial version of the integral $I$ for
positive momentum. b) A pictorial version of the integral $I$ for
negative momentum}{\epsfxsize4in\epsfbox{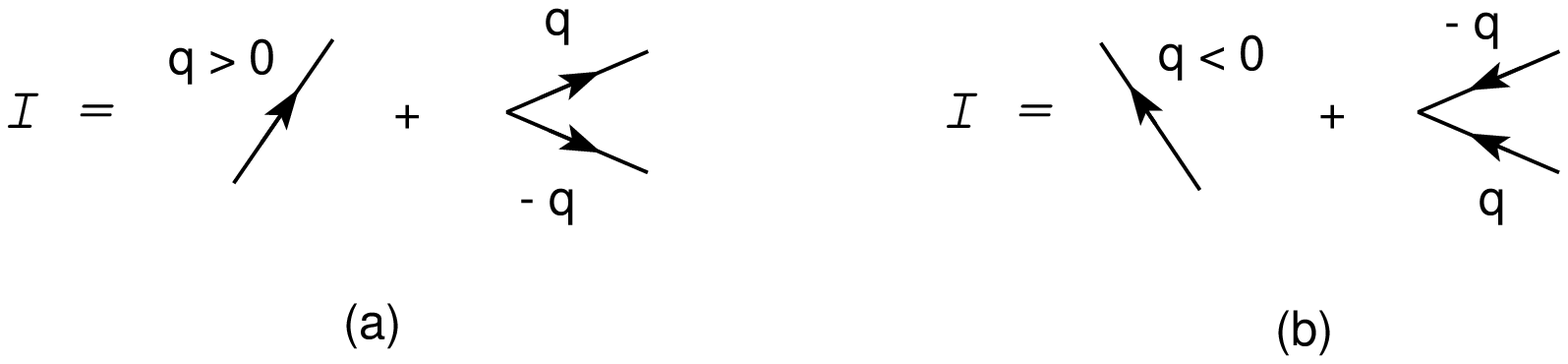}}

\ifig\fiov{Incoming and outgoing vertices.
The dotted line carrying negative (positive) momentum $q_i$ should be
thought of as an incoming (outgoing) boson with energy $|q_i|$.
Momentum carried by lines is always conserved as time flows upwards.}
{\epsfxsize2.25in\epsfbox{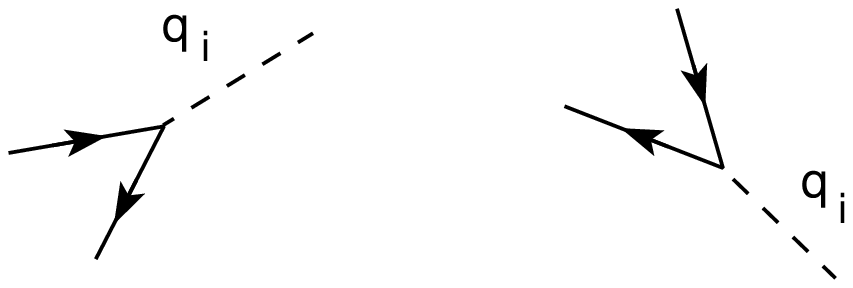}}

This intuitive description may be formalized by the following set of general
rules:\foot{The following is paraphrased directly from \mpr.}
To each incoming and outgoing boson associate a vertex in the $(t,\tau)$
half-space. Connect points via line segments to form a one-loop graph. Since
the expression for $I$ in \asympi\ has two terms, we have both direct and
reflected propagators as in \fipnm. Each line segment carries a momentum and an
arrow. Note that the reflected propagator in \fipnm, which we call simply a
``bounce,'' is composed of two segments with opposite arrows and momenta. These
line segments are joined according to the following rules:

\smallskip

{\parindent=30pt
\item{RH1.} Lines with positive (negative) momenta slope upwards to the
right (left).

\item{RH2.} At any vertex arrows are conserved and momentum
is conserved as time flows upwards. In particular momentum
$q_i$ is inserted at the vertex as in \fiov.

\item{RH3.} Outgoing vertices at $(t_{\rm out},\tau_{\rm out})$ all have
later times than incoming vertices
$(t_{\rm in},\tau_{\rm in})$: $t_{\rm out}>t_{\rm in}$.
\par}

\smallskip

\ifig\frules{Bounce factors for reflected propagators. The Minkowskian
factors are shown at the left, and their Euclidean analogs are shown at the
right.} {\epsfxsize3.25in\epsfbox{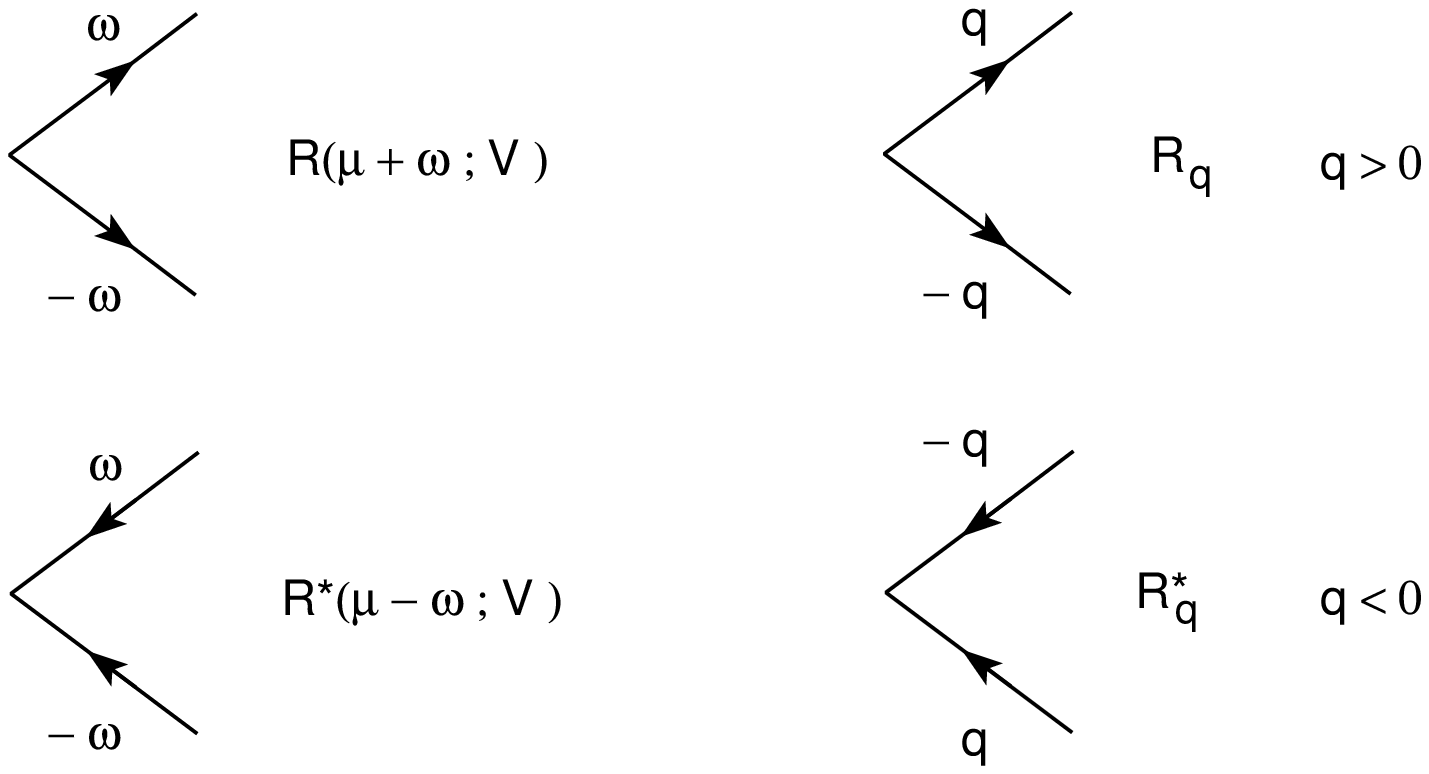}}

To each graph we associate an amplitude, with bounce factors $R$
for reflected propagators as in \frules.
and $\pm 1$ for upwards (downwards) sloping direct propagators.
Finally, we sum over graphs and integrate over kinematically allowed momenta,
thus getting a formula for the Euclidean amplitudes $\CR_n$ which reads
schematically:
\eqn\radmis{\CR=i^n\sum_{\rm graphs}\pm \int \d q \prod_{\rm bounces}
R_Q\,(-R_Q)^*\ .}
See \mpr\ for more details.

\ifig\fott{$1\to 2$ scattering}
{\epsfxsize2.25in\epsfbox{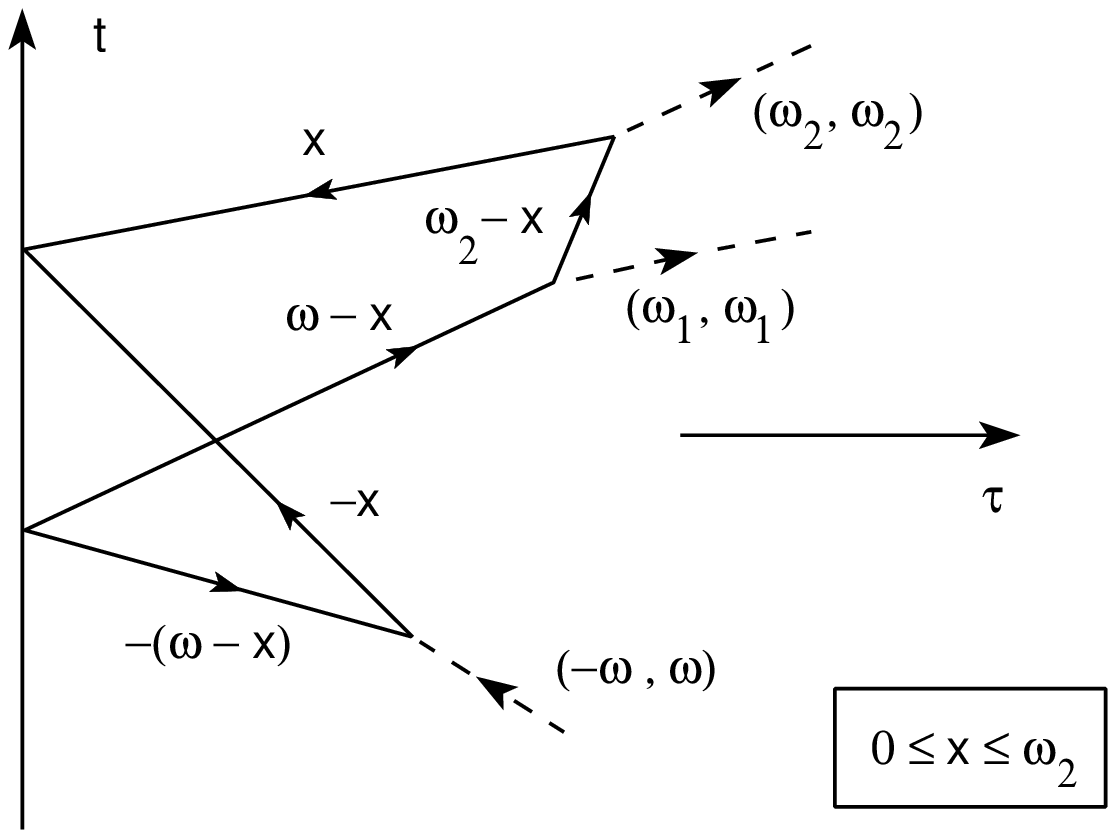}\qquad\epsfxsize2.25in\epsfbox{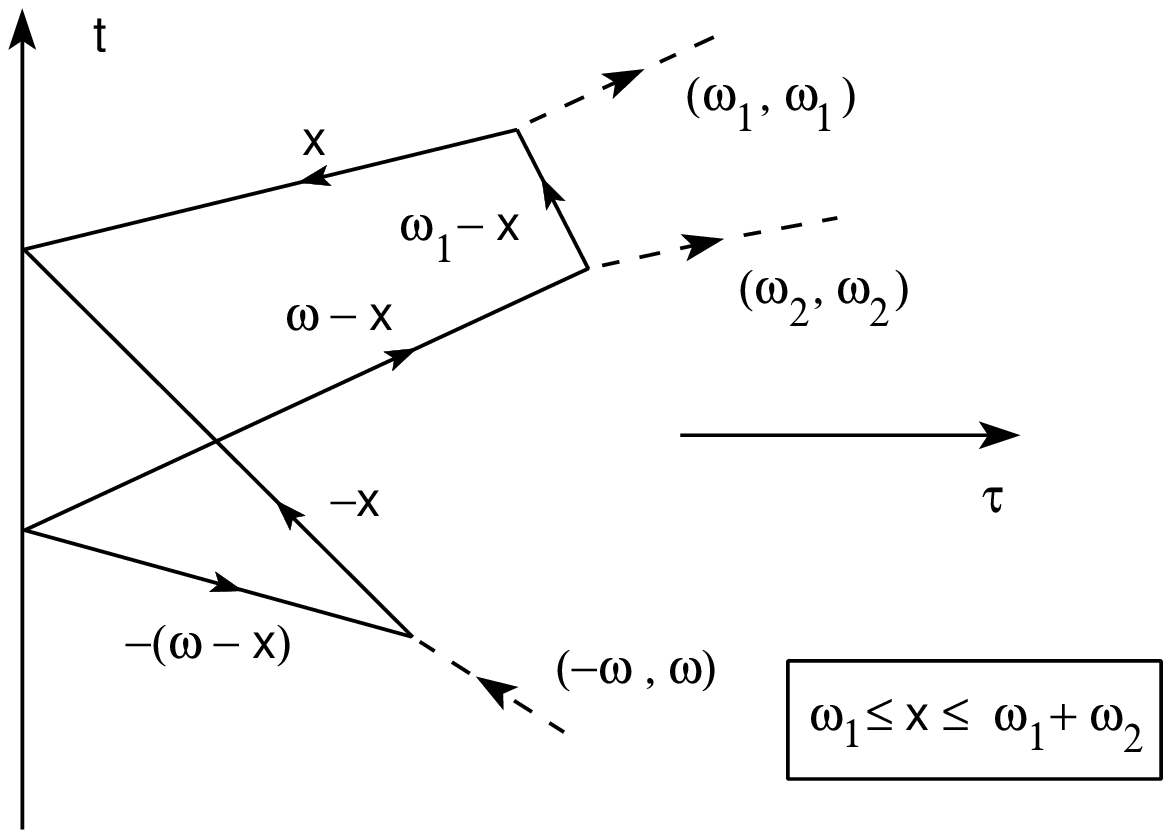}}

\exercise{}

Returning to Minkowski space, derive the $1\to 2$ scattering matrix by showing
that the two diagrams in \fott\ correspond to
\eqn\onetotwo{\sqrt{2} S(\omega\to \omega_1+\omega_2)=
\int_0^{\omega_2} \d x\,R(\mu+\omega-x)\,R^*(\mu- x) -
\int_{\omega_1}^{\omega}\d x\, R(\mu+\omega-x)\,R^*(\mu- x)\ . }

\endexercise

Finally, we must relate these nonperturbative $S$-matrices
to string perturbation theory.
By double-scaling, the string perturbation series can be extracted by
restoring the string coupling $\mu\to \mu/\kappa$ and taking
$\kappa\to 0$ asymptotics. This is the same as taking
$\mu\to\infty$ asymptotics holding $p_i$ or $\omega_i$
fixed. Thus we need the asymptotic behavior of the bounce factors.
To all orders of perturbation theory, we can replace the
expression \rqx\ by the simpler expression for the Euclidean
bounce factor at $p>0$:
\eqn\newbounce{R_p=(-i \mu)^{-p}
{\Gamma(\half-i\mu +p)\over \Gamma(\half-i \mu)}
\sim 1+\sum_{k=1}^\infty {Q_k(p)\over \mu^k}\ .}
Here the $Q_k$ are polynomials in $p$.

\subsec{Properties of $S$-Matrix Elements}
\subseclab\ssposme

{}From the above algorithm one can calculate any
$S$-matrix element. Some general properties of the $S$-matrix
elements following from the above construction are the following.

First let us define some notation. By KPZ scaling,
the Euclidean $S$-matrix elements
\eqn\imprtfct{
\Bigl< \prod_{i=1}^k T_{q_i}^+\Bigr>_h =
\mu^{2h+k-1-\half \sum |q_i|} F_h(q_1,\dots q_k)
}
define certain functions $F_h(q_1,\dots q_k)$
associated to the moduli spaces $\CM_{h,k}$
of curves with $h$ handles and $k$ punctures.
Defining different kinematic regions $\CC_\alpha$
as in eq.~\edivreg, one can then show:

\noindent
{\bf Some Properties of perturbative amplitudes}:
\par\nobreak
\item{i)} $F_h$ is parity-invariant: $F_h(q_i)=F_h(-q_i)$.
\item{ii)} $F_h$ is continuous on $\{\sum q_i=0\}\cap \IR^k$
\item{iii)} In $\CC_\alpha$, $F_h$ is a polynomial in the momenta with rational
coefficients. In general the polynomial is {\it different\/} in different
regions. That is, the expressions are continuous but not continuously
differentiable.
\item{iv)} The degree of the polynomial is $2k+4h-3$.
\item{v)} As any momentum goes to zero, we have
\eqn\loweng{
\Bigl< \prod_{i=1}^k T_{q_i}^+\Bigr>
\ {\buildrel q_i\to 0\over \sim} \
|q_i| {\p\over \p \mu}\Bigl< \prod_{j\not= i} T_{q_i}^+\Bigr>\ .}
\par\nobreak
\item{vi)} If $q_i\in \IZ$, then $F_h=0$ for sufficiently large genus,
specifically, for $2h-2+k > \sum |q_i|$.

Property (i) follows from the integral representations
of macroscopic loops \moore.
Properties (ii),(iii) and (v) are proved in \mpr.
(Property (iii) was first noted in \refs{\moore,\joesea,\kdf}.)
Properties (iv) and (vi) are proved in \mrsg.

Properties (ii--v) have interesting physical interpretations:
Properties (ii) and (iii) result from having
derivatively coupled massless bosons. Usually, massless
particles lead to cuts in the $S$-matrix. In our case,
the cuts become simple discontinuities of the derivatives
with respect to energy.
Property (iv) essentially says that at large
spacetime energies the string coupling becomes effectively
energy dependent, $\kappa\dup_{\rm eff}(\omega)\sim \omega^2/\mu$. This
effective energy-dependence of the string coupling
has been discussed from the continuum Liouville theory
point of view in \seibshen.

\exercise{Energy-Dependent Effective String Coupling}

Derive the rule $\kappa\dup_{\rm eff}(\omega)\sim \omega^2$
from the Liouville theory as
follows \seibshen: From the formula for Liouville energy, compute the turning
point in $\phi$. Plug into the formula for the spatially-dependent string
coupling to find $\kappa\dup_{\rm eff}(\omega)
=\kappa\dup_0\,\ee{\ha Q \phi}=\kappa\dup_0\,\omega^2/\mu$.

\endexercise

A related phenomenon is the inapplicability of the string perturbation
expansion for high energy scattering \moore. The asymptotic expansion for
string perturbation amplitudes is an expansion at fixed $\omega_i$ for $\mu\to
+\infty$. Ordinarily in physics we measure physical values of the coupling
constants (e.g. $\alpha={1\over 137}$) and we probe physical laws by building
ever larger and more expensive accelerators, i.e., by increasing the energies
$\omega_i$.\foot{Ignore renormalization group flow of couplings, for the sake
of this argument.} In the $c=1$ model we would find, at fixed $\mu$ and
sufficiently high energies, that the string perturbation series ceases even to
be an asymptotic expansion. At such energies new physics must emerge and the
string approximation --- which is now seen to be only a {\it low energy}
approximation --- breaks down. In the present context the ``underlying
physics'' which we would discover would be the spacetime matrix model fermions.
It remains to be seen if this situation is typical of nonperturbative string
theory.

\danger{Decoupling of the cosmological constant}
The low-energy theorem, property (v), is probably related to the decoupling of
one of the two cosmological constant operators, and plays a key role in the
analysis of \seibshen. Taking a naive $q\to 0$ limit of the tachyon vertex
operator, we obtain
$V_q\to \ee{\sqrt{2} \phi} (1+ (i q X -|q| \phi)/\sqrt{2}+\CO(q^2))$.
One may therefore try to interpret property (v) in
terms of the decoupling of the cosmological constant operator $\ee{\sqrt{2}
\phi}$ and as well the ``operator'' $X \ee{\sqrt{2} \phi}$, since the leading
term in amplitude goes as $|q|$ which
multiplies the ``operator'' $\phi\,\ee{\sqrt{2} \phi}$. This fits in well with
the Seiberg bound \seibound. By a limiting process, we may interpret
$\ee{\sqrt{2} \phi}$ and $\phi\,\ee{\sqrt{2} \phi}$ as the two KPZ dressings of
the unit operator. We choose the root of the KPZ equation \ealph\
so that the exponential grows at
$\phi\to - \infty$, this being the root we expect to correspond to a local
operator. In the present case we must choose the root
$\phi\,\ee{\sqrt{2}\phi}$, as anticipated
in the paragraph following \liugausii, and in accord with the argument given
at the end of sec.~{\it\ssmlft\/}.

\exercise{Spacetime interpretation of the bounce factor}

Apply the low energy theorem, property (v),
to the two-point function to show that
the ``bounce factor'' is the one-point function
of the tachyon zeromode \dmrpl:
\eqn\vacx{\langle T_0\rangle =i\log R(\mu;V)\ .}

\endexercise

We regard property (vi) as intriguing: it
strongly hints at a topological field theory
interpretation of $c=1$.

\subsec{Unitarity of the $S$-Matrix}

One immediate application of the algorithm of
sec.~{\it\ssnpsm\/} is that
we can give a very simple and conceptual discussion
of the unitarity of the $S$-matrix \mpr.

\ifig\fcomptm{Composition of three maps: fermionization,
free-fermion potential scattering, and rebosonization.}
{\epsfxsize3in\epsfbox{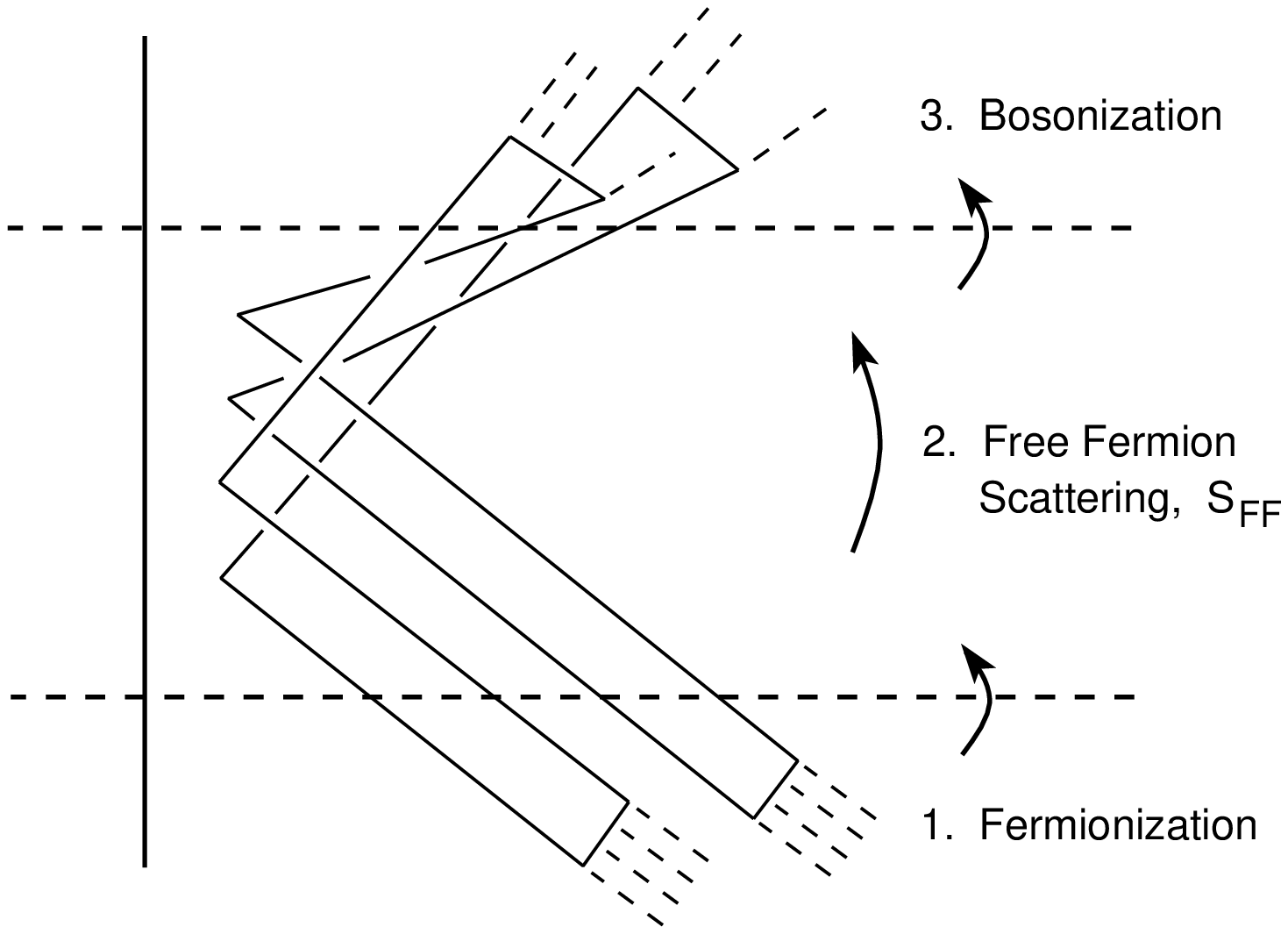}}

The key observation is that the combinatorics of
connecting lines according to the diagrammatic rules
of the previous section is identical to the combinatorics
of bosonization. We can then describe the algorithm
as a three-step process: fermionization, then
free-fermion potential scattering, then rebosonization,
as shown in \fcomptm.

To be more precise, we describe in/out bosonic Fock spaces
$\CF^{\rm in/out}$ made
from the Heisenberg algebra of in/out massless bosons:
$\alpha(\eta)^{\rm in/out}$
where $\eta\in\IR$, $[\alpha(\eta),\alpha(\eta')]=\eta' \delta(\eta+\eta')$ and
the in/out vacua are defined by $\alpha(\eta)|0\rangle=0$ for $\eta<0$. Now,
the Hilbert space of the theory may also be described in terms of the Fermionic
Fock space $H_{\rm FF}$ defined by the oscillators $a(E)$ of
sec.~{\it\ssdsfft\/} (see \eqns{\field{,\ }\vac}).\foot{We are describing only
one world so we drop the $\epsilon$ label.} As is well-known, the
fermionization map
\eqn\fermionztion{\iota_{b\to f}\ : \ \alpha(\eta)\to
\int_{-\infty}^\infty \d \xi\, a(\mu+\xi)\,a^\dagger(\mu-(\eta-\xi))}
defines an isometry $H_{\rm FF}^0\cong \CF^{\rm in/out}$,
where the superscript $0$ indicates restriction to the sector with
the difference ${\rm\# particles} - {\rm\# holes} = 0$.
Thus, the prescription of sec.~{\it\ssnpsm\/}
may be summarized by writing the collective field
$\CS^{CF}$ as a composition of three maps:
\eqn\compthree{\CS_{\rm CF}
=\iota_{f\to b}\circ S_{\rm FF}\circ \iota_{b\to f}\ ,}
where $\iota_{b\to f}$ is the fermionization map,
$\iota_{f\to b}$ is the  inverse bosonization map,
and $S_{\rm FF}$ is the free-fermion potential scattering
$S$-matrix defined by \rqx.
Although standard bosonization is definitely not exact for the
nonrelativistic fermion systems, the asymptotic bosonization is exact
for fermions in a potential approaching $V(\lambda)\sim - \lambda^2$ at
infinity, and this suffices for computation of the $S$-matrix.

{}From \compthree, we immediately deduce that {\it the
$S$-matrix is nonperturbatively unitary if and
only if $S_{\rm FF}$ is unitary\/}. There are two
immediate consequences of this remark.

1) In theories with no infinite wall where the reflection factors have absolute
value smaller than one, the theory will
fail to be nonperturbatively unitary. This
is not because bosons can tunnel, but because a single fermion in a
particle--hole pair can tunnel,
thus leaving a nontrivial soliton sector on either side of
the world. Put another way, if we insist that the (left and right) Hilbert
space of the theory be $H_{\rm FF}^0$ (again the sector with
${\rm\# particles} - {\rm\# holes} = 0$)
then the model will be non-unitary.
If we allow nonzero \# particle $-$ \# hole number,
i.e., nonzero soliton sectors, then
nonperturbative unitarity will be restored. A target space string
interpretation of the solitons would be quite interesting.

2) By making small perturbations of the matrix model potential \fpotential,
we can produce
infinitely many nonperturbatively unitary completions of the string $S$-matrix
\mpr.
In other words, the requirement of nonperturbative unitarity is a very weak
constraint on nonperturbative formulations of string theory. Strangely, the
situation is opposite to that of unitary $c<1$ models
coupled to gravity, where no satisfactory
nonperturbative definitions exist. In either case, we see that matrix models
have been somewhat disappointing as a source of nonperturbative physics.

\subsec{Generating functional for $S$-matrix elements}
\subseclab\ssgffsm

The key formula \compthree\ leads to a concise generating functional for all
$S$-matrix elements \dmrpl. A very intriguing aspect of this formula is that it
involves the asymptotic conformal field theory in {\it spacetime\/}
in a natural way.

We have mentioned above that the collective field theory, or equivalently the
spacetime tachyon theory $T(\phi,t)$, is asymptotically a conformal field
theory. In fact there are {\it two\/} asymptotic conformal field theories
corresponding to the two different null infinities $\CI^\pm$ in the past and
the future. According to \compthree, the entire content of broken conformal
invariance in the interior is summarized by the potential scattering of
fermions:
\eqn\potscat{\eqalign{a(E)_{\rm out} &= R(E) a(E)_{\rm in}
= S^{-1}\,a(E)_{\rm in}\, S\cr
S&\equiv \exp\Bigl( \int_{-\infty}^\infty
\d E\, \log \bigl(R(E)\bigr)
\bigl(a^\dagger(E)\, a(E)\bigr)_{\rm in}\Bigr)\ .\cr}}
As we have noted, unitarity of the $S$-matrix is equivalent to
the identity $R(E) R(E)^*=1$ on the reflection factors.

We may use \potscat\ to summarize the entire $S$-matrix as follows.
Define vertex operators with normalization
\eqn\newnorml{
\tilde V_{\omega}^\pm =
{\Gamma(-i \omega)\over \Gamma(i \omega)}\mu^{1+i \omega/2}V_{\omega}^\pm
}
relative to the normalization of \minkvrtx, and define the generating
functional
\eqn\mnkgenfun{
\mu^2 \CF\bigl[t(\omega),\bar{t}(\omega)\bigr]\equiv
\Bigl<\Bigl< \ee{\int_0^\infty \d\omega\, t(\omega) \tilde{V}_{\omega}^+}
\,\ee{\int_0^\infty \d \omega\, \bar{t}(\omega) \tilde{V}_{\omega}^-}
\Bigr>\Bigr>_c\ ,}
where $\langle\langle\ldots\rangle\rangle$ indicates a sum over genus and
integral over moduli space,
$\sum_{h\geq 0} \kappa^{-\chi}\int_{\CM_{h,n}}$ (as in \ecanr), and the
subscript $c$ indicates the connected part.
The genus expansion of \mnkgenfun\ is given by
$\CF=\CF_0 + {1\over \mu^2} \CF_1+\cdots$, and thus
by KPZ scaling, combining \compthree\ and \potscat\
 we have the formula \dmrpl:
\eqn\mnkgnfuni{\mu^2 \CF\bigl[t(\omega),\bar{t}(\omega)\bigr]=
\langle 0| \ee{\mu \int_0^\infty \d \omega\, t(\omega) \alpha(-\omega)}
\cdot S \cdot \ee{\mu \int_0^\infty \d \omega\,
\bar{t}(\omega)\alpha(\omega)}|0 \rangle_c\ .}

The expression \mnkgnfuni\ has a simple compactified Euclidean space analog.
If we take the Euclidean coordinate $X$ to have finite radius $\beta$,
then from \refs{\rdOneGK,\kleblow}, 
we see that the only modification in \nptev\ is
that bosonic momenta instead lie on a lattice
$q\in {1\over \beta}\IZ$ and the fermions, now interpreted
as being at finite temperature ${1\over \beta}$, have
Matsubara frequencies ${1\over \beta}(\IZ+\half)$.

\ifig\fdmmw{The Euclidean spacetime of the matrix model in
natural coordinates. Note that the asymptotic conformal
field theory on spacetime is concentrated in the
``ultraviolet'' region at the center of the disk.}
{\epsfxsize3in\epsfbox{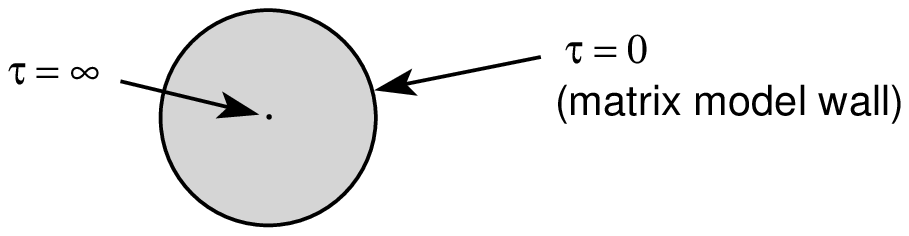}}

The analytic continuation of the asymptotically
conformal collective field is given by the standard
$c=1$ scalar field
\eqn\scalar{\p\phi^{\rm in/out}(z)
= \sum_n \alpha_n^{\rm in/out}\, z^{-n-1}\ ,}
where $z=\ee{-\tau+i X}$, so that the Euclidean spacetime
in the $z$-plane looks as in \fdmmw.

In particular, the bosonization becomes the standard one
with Weyl fermions in the Neveu-Schwarz sector:
\eqn\expwyl{\eqalign{&\psi(z)=\sum_{m\in\IZ}
\psi\dup_{m+\ha}\, z^{-m-1}\qquad
\psib(z)=\sum_{m\in\IZ} \psib_{m+\ha}\, z^{-m-1}\cr
&\qquad\{\psi\dup_{r},\psib_s\}=\delta_{r+s,0}\qquad\qquad
\p \phi = \psi(z)\bar\psi(z)\ .\cr}}
The Euclidean analog of \potscat\ is
\eqn\trans{\eqalign{
\psi^{\rm in}_{-(m+\ha)}&=R(\mu+i {p_m})\,\psi^{\rm out}_{-(m+\ha)}\cr
\bar\psi^{\rm in}_{-(m+\ha)}&=R(\mu-i{p_m})^*\,
\bar\psi^{\rm out}_{-(m+\ha)}\ .\cr}}
where $p_m\equiv (m+\half)/\beta$.

Thus defining Euclidean equivalents
$\tilde{V}_q=\mu^{1-|q|/2}{\Gamma(|q|\over\Gamma(-|q|)}V_q$
of \newnorml,  the Euclidean
analog of \mnkgenfun\ becomes
\eqn\genfun{\eqalign{\mu^2 \CF &\equiv
\Bigl<\Bigl< \ee{\sum_{n\geq 1} t_n \tilde{V}_{n/\beta} +
\sum_{n\geq 1} \bar t_n
\tilde{V}_{-n/\beta}}\Bigr>\Bigr>_c\cr
&=-{1 \over \beta}\langle 0|\ee{i\mu \sum_{n\geq 1} t_n \alpha_{n} } S\,
\ee{i\mu \sum_{n\geq 1} \bar t_n \alpha_{-n} }|0\rangle_c\ .\cr}}
where $|0\rangle$ is the standard $SL(2,\IC)$ invariant vacuum
and the scattering operator is now given by
\eqn\unitran{
S =\ \colon\exp\Bigl(\sum_{m\in\IZ} \log
R_{p_m}\psi^{\rm out}_{-(m+\ha)}\psib^{\rm out}_{m+\ha}\Bigr)\colon\ .}

The formulae \eqns{\mnkgenfun{,\ }\genfun}\ are enormous
simplifications over previous
expressions for $c=1$ amplitudes. They also clarify
several mathematical properties of the $c=1$ $S$-matrix, in
particular, its connections to integrable systems.\foot{M. Green and
T. Eguchi have pointed out some intriguing similarities
between the present discussion of the $c=1$ $S$-matrix and the
topological--antitopological fusion of \rcecvaf. Indeed, a picture of in- and
out- disks joined along the $\tau=0$ boundary of \fdmmw\ defines exactly the
same geometrical setup.}

\subsec{Tachyon recursion relations}
\subseclab\sTrr

{}From the previous formulae we can obtain some
interesting relations between tachyon amplitudes.
We will restrict attention to genus zero with
$X$ uncompactified in this section.

We may interpret the solution \gensol\ or \inout\ to the classical scattering
problem in terms of operators in the coherent state representation acting on
the generating functional $Z$ of all amplitudes. This leads immediately to the
$w_{1+\infty}$ flow equations for genus zero amplitudes. The equations are most
elegantly stated at the self-dual radius, or by working at infinite radius but
restricting to integer momenta. In either case we have:
\eqn\gzrwif{
\mu^{-2}{\p\over \p t(n)} Z=\oint \d w\,
{1\over n+1}\ \colon\Bigl(\bar\p  \phi(w)\Bigr)^{n+1}\colon\ Z\ ,}
where we have the coherent state representation:
\eqn\chrphi{
\bar\p  \phi = w^{-1}+ \sum_{k=1}^\infty k\, \bar t(k)\, w^{k-1}
+{1\over \mu^2}\sum_{k=1}^\infty w^{-k-1}{\p\over \p \bar t(k)}\ ,
}
and in \gzrwif\ we only keep terms to leading order in
the $1/\mu^2$ expansion for any given correlator.

In terms of explicit constraints on amplitudes, these flow equations lead to
the following relations between tachyon amplitudes \mrpl.
The identities are most simply written in terms of
\eqn\eTq{\CT_q = {\Gamma (|q|) \over \Gamma (1-|q|)} V_q\ .}
Consider first the insertion of a ``special tachyon,''
with $q\in \IZ_+$.
If we continue $\omega\to i\,n$ with $n\in \IZ_+$ then
the series \inout\ truncates after $n+1$ terms. These
terms have a ``universal'' effect in correlation functions.
Specifically, an insertion of $\CT_n$ is given by
\eqn\ward{\eqalign{&\langle \CT_q \prod_{i=1}^m \CT_{q_i} \rangle
= \sum_{k=2}^{m}{\Gamma(n) \over \Gamma(2+q-k)}
\sum_{l=1}^{\min (m_-,k-1)} \sum_{|T|=l} \theta\bigl(-q(T)-q\bigr)\cr
&\qquad\qquad\cdot\sum_{S_1, \ldots S_{k-l}}
\prod_{j=1}^{k-l} \biggl( \theta\bigl(q(S_j)\bigr)\, q(S_j)\,
\Bigl<\CT_{-q(S_j)} \prod_{S_j} \CT_{q_i} \Bigr> \biggr)\ ,\cr}}
where $q>0$.\foot{In \mrpl\ these were written, with no loss of
generality, for positive integer $q$.
The case $m_-=m$ is  exceptional (\ward\ vanishes while the correlator does
not) but the amplitude is known from \onetom.
This ungainly feature is not shared by \gzrwif.}
The notation is as follows: Let $S = \{q_1 \ldots q_m\}$,
and let $S^-$ denote the subset of $S$ of negative momenta.
Denote $m_-=|S_-|$.
The sum on $T$ is over subsets of $S^-$ of order $l$.
The subsequent sum is over distinct disjoint decompositions
$S_1 \amalg \ldots  \amalg S_{k-l}= S\setminus T$. $q(T)$
denotes the sum of momenta in the set $T$.
The momenta $q_i$ are taken to be generic so
that the step functions are unambiguous. This entails no
loss of generality since
the amplitudes are continuous (but not differentiable)
across kinematic boundaries \moore.

The first two examples of \ward\ are:
\eqn\bndryop{n=1:\qquad
\Bigl<  \CT_{1}\prod_{i=1}^n\CT_{q_i}\Bigr>=
\sum_{q_i<-1} |q_i+1| \Bigl< \CT_{q_i+1}\prod_{j\not= i}\CT_{q_j}\Bigr>}
\eqn\wardtwo{\lbspace\eqalign{&n=2:\qquad
\Bigl< \CT_2 \prod_{i=1}^m \CT_{q_i} \Bigr>
=  \sum_{q_i < -2} |q_i + 2|
\Bigl< \CT_{q_i+2} \prod_{j\ne i} \CT_{q_j} \Bigr> \cr
&\qquad\qquad+\sum_{{\scriptstyle q_i+q_j < -2}\atop{\scriptstyle q_i,q_j<0}}
|q_i+q_j+2|
\Bigl< \CT_{q_i+q_j+2} \prod_{k\ne i,j} \CT_{q_k} \Bigr>\cr
&+ \sum_{q_i<-2}\ \,\sum_{S_1\amalg S_2=S\setminus \{ q_i\} }\!\!
\theta\bigl(q(S_1)\bigr)\,q(S_1)\,\theta\bigl(q(S_2)\bigr)\,q(S_2)\,
\Bigl< \CT_{-q(S_1)} \prod_{S_1} \CT_{q_j} \Bigr>
\Bigl< \CT_{-q(S_2)} \prod_{S_2} \CT_{q_j} \Bigr> \ .\cr}
}
Note that in \wardtwo\ there is a change in tachyon number
by one in the second line,
and the product of two correlators in the third line. The pattern
continues for higher $n$: there are terms with $|T|=l=k-1$ removing
$l$ incoming tachyons, which are linear in the correlators, and
terms with a product of $k-l$ correlators.
Using the representation \mnkgenfun\ one can write the analog of \gzrwif,
which is valid to all orders of $1/\mu$ perturbation theory. Essentially,
the $w_{1+\infty}$ algebra is replaced by the $W_{1+\infty}$ algebra \dmrpl.

\subsec{The many faces of $c=1$}

In recent years many authors have tried to relate other interesting physical
systems to the $c=1$ matrix model. These include:
\smallskip
\noindent 1) Two-dimensional black holes.\nl
It was originally proposed by
Witten \wittbh\ that the $SL(2,\IR)/U(1)$ model of black holes would be, in
some sense, equivalent to the $c=1$ model. This fascinating conjecture has
inspired an enormous literature, but, despite all the work, the situation
remains confused. Space does not allow a proper review here. A small sampling
of the vast literature includes the following proposals:

\item{a)} The models are equivalent after a non-local integral transform on the
field variables. In \martshat, a transform from the Liouville equations of
motion to the $SL(2,\IR)/SO(2)$ equations of motion is proposed. See also
\brshkutbh. In \rtatastuff, this transform was composed with the $\tau$-space
to $\phi$-space transform described in sec.~{\it\sstsaps\/}. The results so far
have been limited to transforms of the tachyon equations of motion and, when
treated nonperturbatively, have some difficulties with singularities at the
horizon and/or the singularity. There have been many variants of these
proposals in the literature. See, for example, \russo.

\item{b)} The models have different operators turned on corresponding to
non-normalizable modes. Consequently the (Euclidean) black hole and the $c=1$
model are in different ``superselection sectors'' \seibshen.

\item{c)} The 2D black hole and the $c=1$ model are equivalent; the $c=1$
$S$-matrix includes black hole formation and evaporation as an intermediate
process, but the black hole physics is difficult to recognize because of the
exact solubility of the model. Specifically the $w_\infty$ symmetry of the
theory makes black holes difficult to recognize in the $c=1$ $S$-matrix. This
has been advocated in \wittconf.

\item{d)} The 2D black hole and the $c=1$ model are related,
but are different cosets of $SL(2,\IR)$ current algebra \dvvbh.

\item{e)} The $c=1$ model is equivalent to a twisted $N=2$
supersymmetric $SL(2,\IR)/U(1)$ model \mukhivafa.

\item{f)} The models are {\it not}
equivalent and we will learn nothing about black holes from
the $c=1$ model. Several physical arguments may be advanced
in favor of this viewpoint.

\smallskip
\noindent 2) Topological Field Theory.\nl
In \refs{\wittnmatrix,\mukhivafa}, a
relation between the $c=1$ model with $X$ compactified on a self-dual radius
and a certain topological field theory has been proposed. This is potentially
significant because the $c=1$ model has, as we have seen, local physics and a
nontrivial $S$-matrix. This topological field theory is a twisted
$SL(2,\IR)/U(1)$ Kazama--Suzuki model at level $k=-3$ coupled to topological
gravity. Among other things, this interpretation would equate the tachyon
$S$-matrix for $T_{k_i}$ with the Euler character of the vector bundle
$\CV\to\CM_{g,n}$ whose fiber at a Riemann surface $\Sigma$ is
\eqn\fiber{
\CV |_{\Sigma}=H^0\bigl(\Sigma;\, K^2\otimes_{i=1}^n \CO(z_i)^{1-k_i}\bigr)\ ,
}
where $K$ is the canonical bundle of $\Sigma$. This
conjecture has been checked for the free energy \distlervafa\
and for the four-point function \refs{\wittnmatrix,\mukhivafa}.
Checking this in other cases appears to be quite
nontrivial.

\smallskip
\noindent 3) 2D QCD.\nl
Very recently  \mpdoug, a connection with two-dimensional QCD
has been advocated.

\newsec{Vertex Operator Calculations and Continuum Methods}
\seclab\svoccm

Matrix model reasoning is extremely indirect. It is therefore
important to verify matrix model results directly
via vertex operator calculations. Aside from logical
consistency, it is useful to see how matrix model results
are explained by standard string-theoretic ideas
(for example in terms of operator product expansions, etc.).
Moreover, vertex operator calculations are the only known
approach to the supersymmetric models.

\subsec{Review of the Shapiro-Virasoro Amplitude}
\subseclab\srsva

Many of the important ideas of string perturbation
theory are nicely summarized in one of its oldest results: the
Shapiro-Virasoro amplitude for 4-point scattering
of string tachyons. Although this material is
completely standard, it is good to review it before
plunging into the bizarre world of 2D string theory.

The relevant density on moduli space for the scattering
of four on-shell closed string tachyons is
\eqn\svsai{
\Omega(V_{p_1},V_{p_2},V_{p_3},V_{p_4})
=\d z\wedge \d\zb\, |z|^{2 p_1p_3}|z-1|^{2p_2p_3}\ ,
}
where $z={z_{13} z_{24}/ z_{12} z_{34}}$ and $\ha p_i^2=1$.

In this case, the integral over moduli space
$\CM_{0,4}$ can be done using the formula
\eqn\oldsva{
\int_{\IC} \d^2 z\, |z|^{2 a}|z-1|^{2b}=
\pi\, \Delta(1+a)\,\Delta(1+b)\,\Delta(-a-b-1)\ ,
}
and hence
\eqn\oldsvai{
\CA_{0,4}(V_{p_1},V_{p_2},V_{p_3},V_{p_4})
=\pi {\Gamma(1+p_1\cdot p_3)\over \Gamma(-p_1\cdot p_3)}
{\Gamma(1+p_2\cdot p_3)\over \Gamma(-p_2\cdot p_3)}
{\Gamma(1+p_3\cdot p_4)\over \Gamma(-p_3\cdot p_4)}\ .}

The left hand side of \oldsva\ converges when the
arguments of all the $\Gamma$-functions are positive.
Amplitudes in other kinematic regimes,
obtained by {\it analytic continuation\/}
in the external momenta,
have an infinite set of poles at the values:
\eqn\plpstns{\eqalign{
\half (p_1+p_3)^2&=1,0,-1,\dots \cr
\half (p_2+p_3)^2&=1,0,-1,\dots \cr
\half (p_3+p_4)^2&=1,0,-1,\dots \cr}
}
These poles have both  spacetime and worldsheet interpretations:

{\bf Spacetime interpretation}.
The poles signal the existence of new particles
in the theory. At the above values of $t,u,s$, there
is an on-shell particle in the respective channel.
This is the first signal of the infinite
tower of string states of arbitrarily large target space spin.

{\bf Worldsheet interpretation}.
The poles arise from the terms in the operator product
expansion. The poles in the $t$ channel, for example,
are best understood by considering the operator
product expansion of operators $V_{p_1}$ with $V_{p_3}$.
Then we have:
\eqn\splope{\eqalign{
&c\bar{c}\, \ee{i p_3\cdot X}(z,\zb)\, c\bar{c}\, \ee{i p_1\cdot X}(0)
\sim \sum \Phi^s(0)
\, \bigl< \Phi_s(\infty)\, c\bar{c}\, \ee{i p_3\cdot X}(z,\zb)
\,c\bar{c} |p_1\bigr> \cr
}
}
where $\Phi^s$ has ghost number 4.
Thus we can interpret the expansion of $\Omega$
in powers of $z$ as a statement about the factorization
properties of the correlator:
\eqn\svafct{
\langle VVVV\rangle \to \sum_s \langle VV  \Phi^s\rangle
\langle \Phi_s VV \rangle\ .}

\exercise{Gaussian OPE}

Express the operators in  \splope\
in terms of Schur polynomials of $\p^k c, \p^k X$.

\endexercise

The expansion is only convergent for $|z|<1$, so we must
separate the integral over moduli space into two parts:
$|z|<\rho$ and $|z|>\rho$, where $\rho<1$.
In the first integral we may
use the OPE and integrate term by term to get:
\eqn\plesum{
\CA_{0,4} =
2 \pi \sum_{n\geq 1} { \rho^{2n+2+2 p_1\cdot p_3}\over 2n+2+2 p_1\cdot p_3}
\langle V_1 V_3  \Phi^s\rangle\, \langle \Phi_s V_2 V_4 \rangle
+\int_{|z|\geq \rho} \Omega\ .
}
Thus we see that the poles in the $t$-channel come from
the contribution of operators in the $V_1,V_3$ OPE that
satisfy $\half (p_1+p_3)^2 =1,0,-1,\dots\,$. Furthermore, the Fock space
for the Gaussian conformal field theory (tensored with
ghosts) may be decomposed into states which are
BRST cohomology representatives, unphysical states, and
trivial states, in a manner invariant under
the conjugation $\Phi_s\to \Phi^s$. That is, the
factorization behaves schematically like:
\eqn\schemfact{\rm
\sum |phys\rangle \langle phys|
+\sum |unphys\rangle \langle trivial|
+\sum |trivial \rangle \langle unphys|\ .}
Since the $V_i$ are BRST invariant, only the BRST invariant operators can
contribute to the sum in \plesum. Thus we finally conclude that the infinite
sum of poles in the scattering amplitude stem from the BRST cohomology classes
in the operator product expansion. Similarly, the poles in the $s,u$ channels
arise from the other two boundaries of moduli space.

Note, in particular, that it would be inconsistent with unitarity to truncate
the string spectrum to lowest lying states.

\exercise{BRST puzzle}

When $p_1+p_3$ is not on-shell, every term in \splope\ is a BRST commutator so
$V_{p_1} V_{p_3}$ is BRST trivial. Explain why this does {\it not\/} imply that
the four-point function is zero.

\endexercise

\subsec{Resonant Amplitudes and the ``Bulk $S$-Matrix''}
\subseclab\srabsm

Unfortunately the Liouville theory is incalculable: we
cannot even write the density on moduli space in general,
much less integrate it. We can of course
calculate in the free theory at $\mu=0$. This has led to
a large literature on the ``Bulk scattering matrix'' to
be contrasted with the ``Wall scattering matrix''
or $W$-matrix discussed in the previous chapter.

Bulk scattering is scattering in the $\mu=0$ theory with
the condition $s=0$, where $s$ is the KPZ exponent \kpzexp, is
imposed as a kinematical condition. This makes best
physical sense if we rotate $\phi\to i t$ (as we may
when $\mu=0$), and regard $X$ as a spatial variable.
We are therefore discussing theory B of sec.~{\it\stdstms\/}.
As explained there, the vertex operators are given by
\bsvrtx\ and we have energy and momentum conservation
laws for the amplitude
$\langle \prod T^+_{k_i}\prod T^-_{p_i}\rangle$ given
by
\eqn\bscons{\eqalign{
&\qquad\sum k_i +\sum p_i =0\cr
&s =2- \sum(1+\half k_i) - \sum (1-\half p_i)=0\ .\cr}
}
Standard vertex operator calculations now give
\eqn\virshap{
\Bigl< \prod T^+_{k_i}\prod T^-_{p_i}\Bigr>
=\int \prod_{i=4}^N \d^2 z_i\, \prod_{i<j} |z_{ij}|^{-2 s_{ij}}
\ ,}
where we take the three points at $0,1,\infty$ as usual,
$s_{ij}=\beta_i \beta_j-\half k_i k_j$,
$\beta=\sqrt{2}+k/\sqrt{2}$ for $T^+_k$, and
$\beta=\sqrt{2}-p/\sqrt{2}$ for $T^-_p$.
 The amplitude \virshap\ is
known as the ``shifted Virasoro--Shapiro amplitude''
and is quite similar to the familiar expressions for
strings in Minkowski spacetime. Let us examine these
amplitudes more closely.

Consider first the case where all vertex operators but one have the same
chirality, without loss of generality we take say $(+,-^{N})$. If the
``effective masses'' $m_i=\half(\beta_i^2-\half p_i^2)=1-p_i>0$, and
$p_i+p_j>1$, then the integral \virshap\ is {\it convergent\/} and
well-defined, and results in
\eqn\svsa{\langle T^+_{k}\prod_{i=1}^N T^-_{p_i}\rangle
=\prod_{i=1}^N \Delta(m_i) {\pi^{N-2}\over (N-2)!}\ .}
This has been shown in \refs{\polss,\kdf}\ by analytic arguments
and in \klebpasq\ by an elegant algebraic technique.
Note in particular that:
\item{1)} $p_i,k_i\in \IR$. We have put $\mu=0$ so there is no longer any
rationale to impose the Seiberg bound \seibound.
\item{2)} As in 26 dimensions, we can continue to other momenta for which the
integral representation does not converge. Then there are poles, but in this
case they occur for $p_i=1,2,\dots\,$. These are known as the ``leg poles.''
\item{3)} We already see a remarkable difference between $D=2$ and
$D>2$ strings since in general there is no simple
closed formula for \virshap\ for $N>4$.

Let us now consider other combinations of chiralities. We find a new surprise.
Because of kinematic ``coincidences'', one cannot define the integrals, even by
analytic continuation, since one is always sitting on top of a
$\Gamma$-function pole or zero. Indeed it has been argued in
\refs{\kdf,\polss} that these amplitudes are zero, at least for generic
external momenta.

{\bf Example}. Let us consider the most general four-point function. In string
theory, we apply the fundamental identity \oldsva. Usually we use this
expression in conjunction with analytic continuation in the momenta to define
the scattering matrix in all kinematic regimes. Let us apply this to the
amplitude $\bigl< T^+_{k_1} T^+_{k_2}T^-_{p_1} T^-_{p_2}\bigr>$. The
kinematic constraints \bscons\ force $p_1+p_2=-(k_1+k_2)=2$. Thus the third
factor of \oldsva\ becomes $\Delta(2)=0$, while the first two factors remain
nonsingular for generic momenta.

\ifig\fbsp{Several nonvanishing bulk processes. One can also
take the parity conjugate of each the above processes.}
{\epsfxsize3in\epsfbox{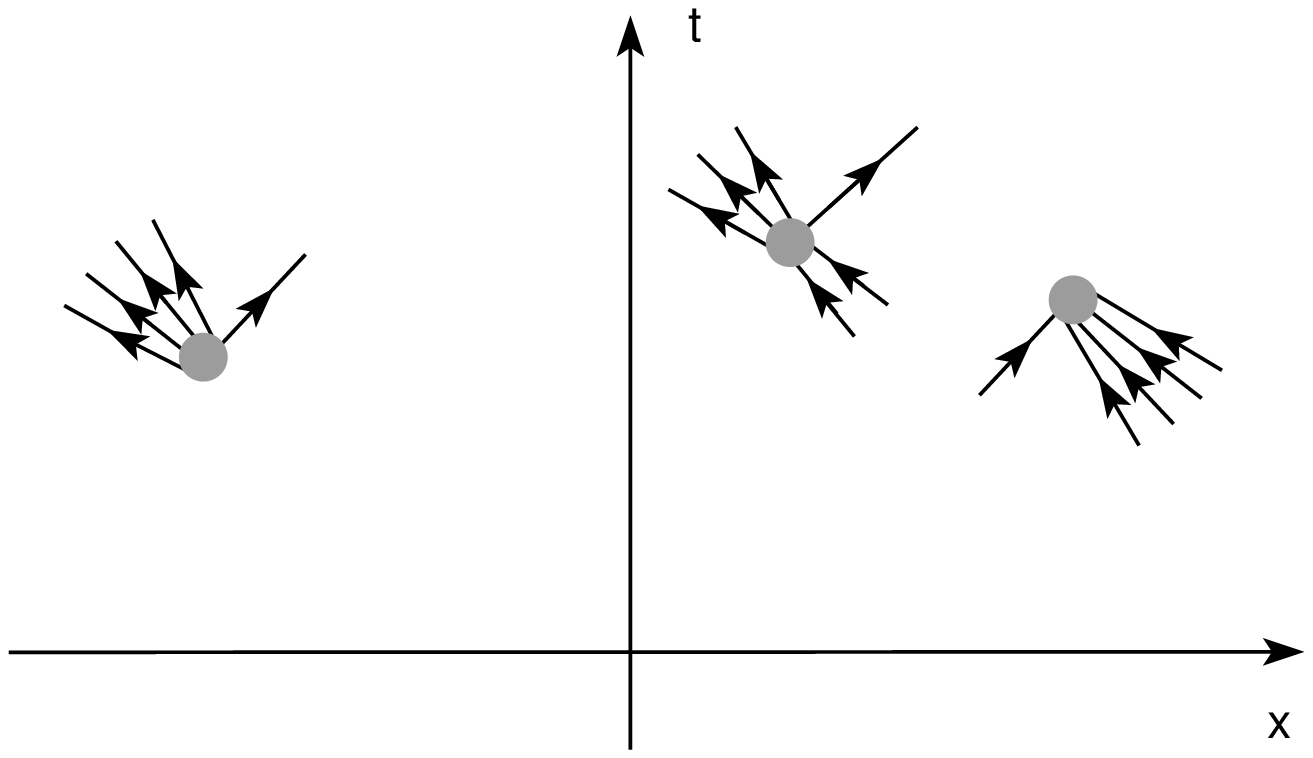}}

The mixed chirality amplitudes are put to zero by some authors
\refs{\polss,\kdf,\tanii}\ and argued (on the basis of unitarity equations) to
be proportional to $\delta$-functions in momenta by others
\refs{\minicyang,\lowe}. Taken together these amplitudes define the ``Bulk
$S$-matrix,'' or $B$-matrix for short. Bulk scattering is quite peculiar, some
examples of processes are drawn in \fbsp. The existence of particle
creation/annihilation in some processes is not surprising given the
time-variation of the background, and in particular of the coupling constant.
The existence of $\delta$-function singularities in the other $S$-matrix
elements suggests that the spacetime background with $\mu=0$ is highly
unstable.

\danger{Factorization on discrete states}

It should be emphasized that the simplicity
of the formula for
$N$-tachyon scattering amplitudes \svsa\ is extremely
remarkable. The analogous singularity structure for
the 26-dimensional string would be vastly
more complicated.
Physically this arises because in $2D$ there is only
one propagating degree of freedom.  Nevertheless,
since the amplitudes \svsa\ were calculated using
free-field operator products, the standard discussion
of sec.~{\it\srsva\/} applies here as well, with some
small modifications implied by the kinematic
laws \bscons. This has been carried out in detail in
\refs{\minicyang,\tanii}. Using the free field OPE as in
sec.~{\it\srsva\/}, in \tanii\ it is shown that the ``leg poles''
in \svsa\ may be interpreted in terms of on-shell intermediate
discrete states. The vanishing of the mixed chirality
amplitudes is important in their discussion.
While this makes sense from the worldsheet point of view,
the existence in spacetime of (normalizable!) modes which
are only physical at discrete momenta is quite peculiar
and has not been adequately interpreted.

\subsec{Wall vs.\ Bulk Scattering}
\subseclab\swvbs

We finally discuss the relation of the $B$-matrix to the $W$-matrix,
that is, we
compare amplitudes at $\mu=0$ with amplitudes at $\mu>0$. Since we cannot
expand in $\mu$ we have no right to expect a simple relation. Moreover, the
perturbative $W$-matrix does not have a good $\mu\to 0$ limit. Nevertheless,
there is an interesting series of conjectures explored in
\refs{\rfftech,\kdf,\polss} on the relation between these $S$-matrices. We
describe these here.

To compare, we must continue back to Euclidean
space and impose the Seiberg bound \seibound. Thus we
only consider processes with $T_k^+, k>0$, and $T_k^-$, $k<0$.
The chirality rule thus becomes the rule that amplitudes
are generically zero unless all but one of the momenta $k_i$
have the same sign. Without loss of generality, we take
$k_1,\dots k_N<0$, hence $s=0$ implies that $k_{N+1}=N-1$.
We now try to relate the $\mu=0$ and $\mu>0$ theories
by integrating over the Liouville zero mode as in
sec.~{\it\slcfac\/}, splitting $\phi=\phi_0+\tilde{\phi}$. This gives:
\eqn\glee{
\langle V\cdots V\rangle=\mu^s\, \Gamma(-s)\,
\langle V\cdots V\rangle_{\tilde{\phi},\mu=0}\ .
}
Since $s=0$, the RHS is ill-defined, but the pole of the
$\Gamma$-function has a nice physical interpretation. Returning
to fixed area correlators we see that it arises from an
ultraviolet $A\to 0$ divergence. That is, in spacetime
terms, a divergence from an integration over the volume
of the  $\phi$-coordinate. We may therefore regulate
the theory and consider $\log \mu$ the regularized volume of the world,
\eqn\regvol{
\mu^s\Gamma(-s)|_{s=0}\rightarrow \int_\epsilon^\infty {\d A\over A} \,
\ee{-\mu A}\rightarrow \log(\mu\, \epsilon)\ .
}
To extract the residue of the
$s=0$ pole, we divide by the volume of $\phi$-space.
As mentioned in sec.~{\it\slcfac\/}, the Liouville interaction is effectively
zero in ``most'' of $\phi$-space so we
should be able to treat $\phi$ as a free field in this
regime and calculate the residue of the
$s=0$ pole with free-field techniques. But this calculation
just leads to the $B$-matrix. Using the free-field result
\svsa\ gives ($k_i<0$, $i=1,N$):
\eqn\kdff{\eqalign{
\langle V_{k_1}\cdots V_{k_{N+1}}\rangle &=
\mu^s \Gamma(-s)|_{s=0}\langle V\cdots
V\rangle_{\tilde{\phi},\mu=0}\cr
&=\mu^s \Gamma(-s)|_{s=0}\prod_{i=1}^N \Delta(m_i)
{\pi^{N-2}\over (N-2)!}\cr
&=\pi^{N-2}\prod_{i=1}^N \Delta(m_i) {\Gamma(1-(N-1)+\epsilon)\over
\Gamma(N-1)}(N-2)!\cr
&=\prod_{i=1}^{N+1}\Delta(m_i)\bigl({\p\over \p
\mu}\bigr)^{N-2}\mu^{s+N-2}|_{s=0}\cr}
}
where now $m_i=1-|k_i|$ and $|k_{N+1}|=N-1$ and we
regulate by taking $s=\epsilon\to 0$.
Di Francesco and Kutasov \kdf\ have generalized this result to positive
integer values for $s$ by carefully taking limits $k_i\to 0$, and find
\eqn\kdffi{
\langle V\cdots V\rangle\propto \prod_{i=1}^{N+1}\Delta(m_i)
\bigl({\p\over \p \mu}\bigr)^{N-2} \mu^{s+N-2}\ .}
%
The equation \kdffi\ has an obvious ``continuation'' to $s\notin\IZ_+$
with $s=1-N+|k_{N+1}|$, where the RHS becomes
well-defined and finite. Remarkably, comparison with the
matrix model result \onetom\ shows that we obtain an
identical amplitude \eqns{\onetom{,\ }\onetomi}\
differing only by ``wavefunction renormalization factors''
$f(q)$ \wvfnfctr, and the continuation $|p|\to -i \omega$
appropriate to the $W$-matrix.
In order for this story to be consistent, we should understand from the
$W$-matrix
why the mixed chirality amplitudes vanish. The reason is that in these
kinematic regimes, the KPZ exponent is typically fractional. For example, for
$2\to n$ scattering with $p_1+p_2=k_1+\cdots k_n$ we have $s=2-n+p_1+p_2$, even
when leg factors blow up $\mu^s$ is fractional, there is no $\log \mu$
dependence and hence no ``bulk'' piece proportional to the volume of the world.

Di Francesco and Kutasov \kdf\ have argued that is is nevertheless possible to
use the
data of the $B$-matrix to obtain the remaining $W$-matrix elements at
$\mu\not=0$. The crucial point is that one must use spacetime reasoning \kdf.
First, note that \kdffi\ depends (up to wavefunction factors) on the momenta
only through a polynomial. Therefore, we can construct a local spacetime field
theory of the tachyon field analogous to the macroscopic loop field theory of
sec.~{\it\ssmlft\/}.\foot{``Local'' means we have an finite set of local finite
derivative interactions for each interaction involving $n$ fields. In total,
the Lagrangian involves an infinite set of interactions.} In \kdf, it is shown
that \kdffi\ uniquely fixes all the interactions in the Lagrangian and that one
can proceed to calculate the amplitudes in other kinematic regimes at
$\mu\not=0$ using this field theory. In all cases where the procedure has been
checked (five- and six-point functions), the amplitudes obtained from this
procedure agree with the matrix model amplitudes. Thus, the $B$-matrix element
\svsa\ at $\mu=0$ completely determines the $\mu\not=0$ $W$-matrix. These
arguments have not been extended to higher genus and the equality of
$S$-matrices thus remains conjectural (although physically plausible).

\noindent{\bf Remarks}:\nl
1) As we discussed in the previous section the leg factors of the $B$-matrix
give
poles corresponding to on-shell intermediate discrete tachyons. On the other
hand, for the Euclidean $W$-matrix the analogous factors are of the form
\wvfnfctr, and have poles at $|q|\in \IZ_+$. In the physical regime of the
$W$-matrix, these correspond to phase factors
$${\Gamma(iE)\over \Gamma(-i E)}$$
which do not have poles in the physical regime. This makes perfectly good
sense. As we have discussed, the poles of the leg-factors correspond to
non-normalizable states with imaginary momentum, they cannot appear in
intermediate channels for physical scattering. Thus, we see that at $\mu>0$ the
different nature of the Liouville OPE essentially changes the physics and
alters the standard discussion of sec.~{\it\srsva\/}. In particular, the
$W$-matrix has the peculiar property, unique among string theories, that the
tachyon $S$-matrix is a unitary scattering matrix in the absence of all other
string states.

2) These calculations have been extended to the open string in \berkut, in
which case the
amplitudes have a pole structure much more complicated than the closed
string bulk amplitudes above. Explaining these amplitudes remains an important
challenge for the matrix model approach.

\subsec{Algebraic Structures of the 2D String: Chiral Cohomology}
\subseclab\ssastdcc

We have seen that, at least at $\mu=0$, the 2D string has a rich spectrum of
cohomology. As mentioned in sec.~{\it\sshrrss\/}, this may be taken as an
indication that the $D=2$ string background is a much more symmetric background
for string theory. By contrast, the Minkowski background of standard critical
strings would seem to be a very {\it asymmetric} background, not at all a good
place to look for underlying symmetries and principles of string theory. With
these motivations in mind, several groups have intensively investigated the
algebraic structures defined by the BRST cohomology of $D=2$ string theory
\refs{\witzwie,\vermster,\grndrng,\kutmarsei,\klebpasq,\klebpol,\klebward}.

Quite generally, the operator product algebra of the chiral operators in a
conformal field theory defines an example of a mathematical object known as a
{\it vertex operator algebra} \flm. Indeed much of the work on conformal field
theory (especially RCFT) has been an investigation of these algebraic
structures \refs{\msrcft,\bouwlect}. In string theory, where there is a BRST
operator $Q$, additional structures arise. This is nicely illustrated in the
example of the operator product algebra of the 2D string.

First let us consider the absolute chiral cohomology at the self-dual radius.
As we have described in chapt.~\stdcst, this is spanned by operators at ghost
numbers $G=0,1,2$ for the $(+)$-states:
\eqn\cohorev{\eqalign{
G=0 \qquad & \qquad \CO_{j,,m}\cr
G=1 \qquad & \qquad a\CO_{j,,m}\qquad Y^+_{j,m}\cr
G=2 \qquad & \qquad a Y^+_{j,m}\ ,\cr}
}
together with the $(-)$-states at ghost numbers
3,2,1, which are dual via the tilde-conjugation.

The operator product of the ground ring operators $\CO\in\CG$
can be restricted to the BRST cohomology:
\eqn\chrlgr{\CO_1(x)\,\CO_2(y)\sim \CO_3(y)\quad {\rm mod}\{Q,* \}\ ,
}
since the operator product is nonsingular and and ghost number is additive.
Thus, the ground ring operators form a ring.
One can show that the BRST reduction of the operator product algebra is
\grndrng:
\eqn\chrlgri{
\CO_{j_1,m_1}(x)\, \CO_{j_2,m_2}(y)=\CO_{j_1+j_2,m_1+m_2}(y)\quad
{\rm mod}\{Q,* \}\ .
}
This result almost follows simply from consideration
of ghost and momentum quantum numbers. The fact that
the structure constant is unity requires more
detailed analysis \grndrng. With the identifications
$x\equiv \CO_{1/2,1/2}$, $y \equiv \CO_{1/2,-1/2}$, we
identify the chiral ground ring with the
algebra of polynomials in $x$ and $y$, denoted $\IC[x,y]$.

Of course, the existence of a ring in the OPA of the BRST cohomology does not
require us to restrict to ghost number $G=0$. To describe the full operator
product algebra, we first introduced some geometry.

\noindent
{\bf Geometry of the BRST operator product algebra}\nl An old observation
\rwoso\ is that the BRST cohomology of string theory resembles cohomological
structures of manifolds. The operator product algebra of the 26-dimensional
string has proven too complicated to pursue this line of thought very far, but
the 2D string example has provided some very interesting realizations of that
idea \refs{\grndrng,\witzwie}. From \chrlgri, we see that the ground ring is
the ring of polynomial functions on the $x,y$ plane. Witten and Zwiebach
\witzwie\ show that the remaining cohomology can be identified with polynomial
vectors and bi-vectors via the introduction of an area-form $\omega=\d
x\wedge\d y$. Indeed we have the correspondence:
\eqn\cohcorr{\eqalign{
\CO_{j,m} \qquad \leftrightarrow &
\qquad f_{j,m}\equiv x^{j+m} y^{j-m}\cr
Y^+_{j,m} \qquad \leftrightarrow &
\qquad V_{j,m}\equiv
{\p f_{j,m}\over \p y}{\p\over \p x}
-{\p f_{j,m}\over \p x}{\p\over \p y} \cr
a \CO_{j,m} \qquad \leftrightarrow & \qquad
X_{j,m}= x^{j+m}y^{j-m}
\biggl(x{\p\over \p x}+
y{\p\over\p y}\biggr) \cr
a Y^+_{j,m} \qquad \leftrightarrow & \qquad
f_{j,m}(x,y) {\p\over \p x}\wedge {\p\over \p y}\ .\cr}
}
In the third line, the vector field $X$ is an
area non-preserving diffeomorphism and satisfies
$\CL_{X_{j,m}} \omega= f_{j,m} \omega$, or,
$\p_i X^i = f_{j,m}$.
With these identifications,
we can elegantly summarize the operator product
algebra as the ring structure on $\Lambda^*T=\oplus_{i=0}^2 T$,
where $T$ is the polynomial tangent bundle on the $x,y$ plane
\refs{\witzwie,\lziii}.

Since we are working at $\mu=0$ we must also consider the
$(-)$ states. These may be nicely incorporated into the
theory. The full structure has been elucidated by Lian and
Zuckerman \lziii\ in terms of an algebraic structure they
call a Gerstenhaber algebra. Related algebraic structures
have also figured prominently in several recent works on
string field theory and topological string theory. See
\refs{\getzler,\graeme}.

\exercise{Explicit Ring Structure}

Show that the ring structure in the natural basis is
\eqn\explrng{\eqalign{
\CO_{j_1,m_1}\cdot \CO_{j_2,m_2} &=\CO_{j_1+j_2,m_1+m_2}\cr
\CO_{j_1,m_1}\cdot Y^+_{j_2,m_2} &=\alpha\, Y^+_{j_1+j_2,m_1+m_2}+
\beta a\,\CO_{j_1+j_2+1,m_1+m_2}\cr
Y^+_{j_1,m_1}\cdot Y^+_{j_2,m_2}&
= a\, Y^+_{j_1+j_2-1,m_1+m_2}\ .\cr}
}

\endexercise

\noindent{\bf Remarks}:\nl
\item{1)} There is a dual interpretation replacing polyvectors
by differential forms. In this formulation, $b_0$
essentially plays the role of an exterior derivative.
See \witzwie.
\item{2)} It is natural to ask for the analog of the ground
ring at $c<1$. This has been discussed in
\refs{\kutmarsei,\sarmadi}. The operator product ring is
$\IC((w))\otimes \IC[x,y]$
with relations $x^{p-1}\sim y^{q-1}\sim 1$
(where $\IC((w))$ designates
the ring of Laurent series with finite order poles).

\noindent{\bf Lie Algebra of Derivations}\nl
Let us investigate more closely some consequences of the above assertions. The
operator product algebra of the ghost number $G=1$ operators is the Lie algebra
of vector fields. When restricted to the area-preserving vector fields
$Y^+_{j,m}$, this may be identified with the $\vee w$ Lie algebra as follows.
The operator $Y$ has the structure $Y_{j,m}=c W_{j,m}$,where $W$ is a
ghost-free operator of dimension one, so applying the descent equations to the
BRST invariant zero-form $\Omega^{(0)}=Y_{j,n}$ gives a dimension one operator
$\Omega^{(1)}=W$. The associated Lie algebra can be deduced by direct
calculations of the operator products to be \klebpol\
\eqn\wopealg{
W_{j_1,m_1}(z)W_{j_2,m_2}(0)\sim {2(j_1 m_2-j_2 m_1)\over z}
W_{j_1+j_2-1,m_1+m_2}(0)\ .
}
Again, much of this formula is fixed simply by considering
the quantum numbers. The expression \wopealg\ is
in agreement with the commutator of polynomial vector fields.

Associated with the Lie algebra of currents are the
charges
\def \CQ{{\cal Q}}
$\CQ(Y^+_{j,m})=\oint W_{j,m}$. These act on the
ground ring as derivations. To prove this,
let $ \CO_1(P)$, $\CO_2(Q)$ be two ground ring operators,
and let $\CC$ be a contour
surrounding points $P,Q$, and $\CC_1$, $\CC_2$ surround only
$P$ and $Q$, respectively. We have:
\eqn\surrndtwo{
\oint_{\CC} W_{j,m} \biggl(\CO_1(P)\CO_2(Q)\biggr)
=\biggl( \oint_{\CC_1} W_{j,m} \CO_1(P)\biggr)\CO_2(Q)
+\CO_1(P)\Bigl( \oint_{\CC_2} W_{j,m} \CO_2(Q)\Bigr)\ .
}
Since the BRST invariant contribution to the operator
product is
independent of the difference $z(P)-z(Q)$, the action of
the charges descends to a derivation on the ground ring.
In the geometrical interpretation this is just the
action of polynomial vector fields on polynomial functions.

\exercise{Two viewpoints}

Show that the second description of the operator
algebra of ghost number $G=0$ and $G=1$ states
is equivalent to the ring structure on $\Lambda^* T$.
Use the fact
 that if $\CL_W \omega=0$ is area preserving and
$\CL_V \omega=f \omega$, then
$\CL_{[V,W]} \omega =W(f) \omega$.

\endexercise

\noindent
{\bf Tachyon Modules}: Away from the self-dual radius, there are
new BRST cohomology classes $V_q=c\, \ee{i q X/\sqrt{2}}\,
\ee{\sqrt{2}(1-\half |q|)\phi}$, with $q\notin\IZ$. The ring of BRST
operators acts on these new cohomology classes
via operator products. Since the position-dependence
of the operators is a BRST commutator, the tachyon
operators form a module representing the ring
$\Lambda^* T$ \kutmarsei.

First let us determine the action of the ground ring.
An easy free-field calculation, using
the explicit formulae for $x,y$ given in chapt.~\stdcst,
shows
\eqn\tachmodle{\eqalign{
\CO_{1/2,1/2}\cdot V_q &=q\, V_{q+1} \qquad q>0\cr
\CO_{1/2,-1/2}\cdot V_q &=0\qquad q>0\cr
\CO_{1/2,1/2}\cdot V_q &=0 \qquad q>0\cr
\CO_{1/2,-1/2}\cdot V_q &=q\, V_{q-1}\qquad q>0\ .\cr}
}
So irreducible representations are classified by
${\rm Sign}(q)$ and $q\ \mod\,1$.
The remaining ring action is somewhat complicated,
but can be largely obtained by
considering the
$X,\phi$ quantum numbers. For example, $Y^+_{j,m}\cdot V_p$
is a state with $p\dup_X=(p+2m)/\sqrt{2}$ and
$-i p\dup_\phi/\sqrt{2}=|p|+2j-2$. Thus the resulting
state can only lie on the tachyon dispersion line
if $|p+2m|=|p|+2j-2$.
Therefore, for example, we can immediately
conclude that
\eqn\explact{Y^+_{j,m}\cdot V_p=0\ ,}
for $p\notin \half \IZ_+$ if $p+2m<0$, $p>0$, or if $p+2m>0$, $p<0$.
If $p+2m$ and $p$ have the same sign, then we still require
$|m|=j-1$ for a nonzero product. In the latter case, the
nonvanishing product is most simply described as
\eqn\conviract{
\oint W_{j,j-1} V_p = {(-1)^{2j-1}\over (2j-1)!}(p)_{2j-1} V_{p+2(j-1)}
}
for $p>0$, with a similar formula for $W_{j,1-j}$ for $p<0$
(and $(p)_m=\Gamma(p+m)/\Gamma(p)$ is the Pochammer symbol).
Thus, when the $\vee w$ algebra generated by
the currents $W_{j,m}$ acts on the tachyon module,
only the ${\it Vir\/}^+$ subalgebra $\vee^2 w/\vee^3 w$ acts
nontrivially on $V_p$.

\subsec{Algebraic Structures of the 2D String:
Closed String Cohomology}
\subseclab\ssascsc

The algebraic structures for the closed string case are quite similar. The only
subtlety occurs in combining left and right-moving structures.

Consider first the ground ring for the self-dual compactification. The ghost
number $G=0$ cohomology classes are spanned by
$\CR_{j,m,m'}=\CO_{j,m}\bar{\CO}_{j,m'}$. We must use the same spin $j$ even at
the self-dual radius, since left- and right-moving Liouville momenta must
match. The geometrical interpretation of this ring emerges when one writes
ground ring elements as
$x^n y^m \bar{x}^{\overline n}\bar{y}^{\overline m}$. Equating
left and right Liouville-momenta we have $n+m = \overline{n}+\overline{m}$.
The ground
ring is therefore always generated by polynomials in the expressions $a_1=x
\bar{x}$, $a_2=y\bar{y}$, $a_3=x\bar{y}$, $a_4=y\bar{x}$. Note that the $a$'s
obey the relation $a_1 a_2=a_3 a_4$, defining a three-dimensional quadric cone
$Q$. At infinite radius we only have ground ring generators $\CR_{j,m,m}$ and
the ground ring again becomes the ring of polynomial functions on the $x,y$
plane.

In a manner analogous to the previous section, one can consider the other
algebraic structures and their geometrical interpretations in terms of the cone
$Q$. For example, the symmetries associated to the ghost number $G=1$
cohomology are the volume preserving diffeomorphisms of $Q$. Further results
may be found in \witzwie.

As in the chiral case, the ground ring and discrete charges act on the tachyon
operators $V_p$. Indeed, recall from sec.~{\it\ssstdgmb\/} that we may apply
the descent equations to the ghost number one BRST classes
$\CJ_{j,m}=Y^+_{j,m}\bar{\CO}_{j-1,m}$ and its holomorphic conjugate. The first
step in the descent equations gives a current
\eqn\nnhlcrrnt{\Omega^{(1)}_{j,m}=W^+_{j,m}\bar{\CO}_{j-1,m}\, \d z -
cW^+_{j,m}\bar{X}\, \d\bar z\ ,
}
where $|\bar X\rangle = \bar b_{-1}|\bar\CO_{j-1,m}\rangle$.
This is an unusual current: although it has dimension
$(1,0)$, it is not purely holomorphic. Moreover, its charge
is only conserved up to BRST exact states.
Nevertheless, we can let these discrete currents
act on tachyons. The story is very similar to the
chiral case. In BRST cohomology, the only nonzero
actions occur for $p>0$, $\CJ_{j,j-1}$ or $p<0$, $\CJ_{j,1-j}$.
In this case we have
$\CA_{j,j-1}={(-1)^{2j}\over (2j+1)!} L_{2j}$
we find that, for $p>0$:
\eqn\schrgact{
[L_n, \tilde V_p]=p\, \tilde V_{p+n}\ .
}
So, again ${\it Vir\/}^+$ (see \borel) acts.


String theory Ward identities as applied to 2D string theory have been
described in \refs{\witzwie,\vermster,\kutmarsei,\klebward,\klebpasq}.

Further extensions of this formalism and likely directions for future
progress, including applications in physical contexts, are deferred
to \rbook.

\newsec{Achievements, Disappointments, Future Prospects}
\seclab\sadfp

Quantum gravity has been a theoretical challenge for 70 years.
String theory has been evolving for 25 years. In the
past 3--4 years, some new ideas have been applied to these
old problems. It is time to assess the harvest of this recent effort.

\exercise{Missing lessons}

Determine which of the lessons below are covered quite elegantly in portions of
text that have been omitted from these lecture notes [0] but will be restored
for the book version \rbook.

\endexercise

\subsec{Lessons}

\noindent From the quantum gravity point of view, the main
lessons we have learned from the matrix model are:
\item{$\bullet$} Euclidean Quantum Gravity makes sense, at least in
two dimensions.
\item{$\bullet$}The nature of quantum states in Euclidean
quantum gravity, and their interpretation within the
quantum mechanical framework is surprising, and requires the
introduction of non-normalizable wavefunctions as well as
normalizable wavefunctions.
\item{$\bullet$} The  Wheeler--DeWitt constraint is violated in
topology-changing processes.
\item{$\bullet$} The contributions of {\it singular\/} geometries
to the path integral of quantum gravity are important.
\item{$\bullet$} There is a phase of topological gravity which
can be connected to phases of nontopological gravity.

\medskip

\noindent From the string theory point of view, the main lessons
we have learned from the matrix model are:
\item {$\bullet$} Nonperturbative definitions of string physics, at least
in some target spaces, exist.
\item{$\bullet$} There are backgrounds with large unbroken symmetries,
e.g., $w_{1+\infty}$ and volume preserving diffeomorphism algebras.
\item{$\bullet$} The large order behavior of perturbation theory at order $g$
has the typically ``stringy'' $(2g)!$ growth.
\item{$\bullet$} In solvable string theories, there is a beautiful
mathematical framework (KP flow, $W$-constraints, etc.) that
relates string physics in different backgrounds.
\item{$\bullet$}With current understanding, it is fundamentally
impossible to achieve complete background independence: There
is always dependence on boundary and initial conditions associated
with non-normalizable states.
\item{$\bullet$}There is a phase of string theory which is
topological, and can be connected to nontopological phases with
local physics (such as string scattering in two dimensions).

\subsec{Disappointments}

\noindent From the quantum gravity point of view,
our main disappointments thus far are:

\item{$\bullet$} It is not yet obvious how to apply our new insights into
quantum gravity in two dimensions to treat the case of quantum gravity in
four dimensions.
\item{$\bullet$} Even in two dimensions, the matrix model results have
not yet provided solutions to fundamental problems of quantum
gravity, such as the ultimate nature of singularities, whether Hawking
radiation violates fundamental principles of quantum mechanics,
and related paradoxes.
\item{$\bullet$} Some nonperturbative aspects of gravity have been
investigated, but no clear lessons have been drawn and there remain
many important open problems.
\medskip
\noindent From the string theory point of view,
our main disappointments thus far are:
\item{$\bullet$} The spacetime physics for $c<1$ conformal matter coupled
to quantum gravity, while not fully elucidated, seems rather uneventful due
to the lack of a time dimension, i.e.\ due to the lack of fully developed
spacetime field theory.
\item{$\bullet$} Spacetime physics of the $c=1$ matter coupled to quantum
gravity is essentially that of a free boson.
We have as yet no understanding of the infinite tower of
string states or of backreaction. It may be that strings propagating in
two target space directions, i.e.\ with no transverse dimensions,
is not representative of strings propagating in higher dimensions. Even for
strings in two target space dimensions, we have not progressed so far beyond
the $\sigma$-model point of view to a conceptually new formulation.
\item{$\bullet$} The biggest disappointments have been from the standpoint
of nonperturbative physics:
 \itemitem{$\bullet$} There are stable non-perturbative solutions for
the minimal (2,5) model (Yang--Lee edge singularity), and higher non-unitary
models coupled to quantum gravity,
but again the dynamics is limited due to the lack of time coordinate and
consequent lack of spacetime interpretation.
 \itemitem{$\bullet$} For the $c<1$ {\it unitary\/}
models coupled to quantum gravity, there is no nonperturbative theory.
 \itemitem{$\bullet$} For $c=1$ matter coupled to quantum gravity, we have the
opposite problem: there are infinitely many nonperturbative
completions of the $c=1$ $S$-matrix, i.e., there are infinitely many
``$\theta$-parameters.''
\item{$\bullet$} Our lessons on background dependence are sobering:
there are infinitely many superselection sectors.

\subsec{Future prospects and Open Problems}

\noindent{\it\ Singularity is almost invariably a clue.\/} --- Sherlock Holmes
\smallskip\nobreak
Each paragraph in the text marked with the ``dangerous bend sign''
\raise4pt\hbox{\manual\char127} represents an opportunity.

\item{$\bullet$} The quantum Liouville theory remains unsolved, and is
still needed to calculate answers to many physics questions,
so major surprises remain possible.

\item{$\bullet$} We need a better understanding of backgrounds. At present,
we seem to have an infinite dimensional manifold
of solutions to string theory, and an infinitely large
class of superselection sectors.
Are all these solutions related by some symmetry?

\item{$\bullet$} Can we use these backgrounds to understand anything
about time-dependence in string theory?

\item{$\bullet$} Natural nonperturbative definitions of 2D string theory and
2D gravity are still lacking!
One might have hoped that imposing some physical criterion
such as unitarity would strongly constrain the possible nonperturbative
definitions of the theory, but this does not occur in the case of the $c=1$
model coupled to gravity. There we found infinitely many nonperturbative
completions all of which seem perfectly natural from the matrix model point of
view, and we thus obtain little guidance in this regard.

\item{$\bullet$}  Can the comprehensive picture
of the $c<1$ backgrounds, unified via the KP formalism, be generalized
to the case of 2D string backgrounds?
Is there e.g.\ a multiparameter space of theories
which encompasses both the black hole and $c=1$ spacetimes?
Finding a unified picture of all $2D$ or $c\leq 1$ backgrounds
remains an interesting open problem.

\item{$\bullet$} We need to find
new ways of cancelling the tachyonic divergences of
string theory --- i.e., of making sense of the integrals
over moduli spaces. This is essentially the problem of
going beyond the ``$c=1$ barrier.''

\item{$\bullet$} Does the $c=1$ model teach us how to understand better the
covariant closed string field theory of \bz ?

\item{$\bullet$} One of the great open puzzles in the subject
is the absence of
backreaction on the metric and other ``special state'' degrees of
freedom, and in particular, the role of 2D black holes in
$c=1$ string theory.

\item{$\bullet$} Are there interesting supersymmetric extensions of
the theories we consider here (i.e.\ with potentially interesting
spacetime properties such as the construction of \kutseibss)?

\bigskip
\centerline{\bf Acknowledgements}

We thank especially N. Seiberg for a long series of collaborative efforts on
the subject of 2d gravity. For commentary on various portions of the manuscript
we would like to thank N. Seiberg and M. Staudacher.
We also would like to thank many people for discussions and for teaching us
much of the above material.
In particular we thank T. Banks, R. Dijkgraaf, M. Douglas, J. Horne,
C. Itzykson, I. Klebanov, D. Kutasov, B. Lian, E. Martinec, R. Plesser,
S. Ramgoolam, H. Saleur, G. Segal, N. Seiberg, R. Shankar, S. Shatashvili,
S. Shenker, M. Staudacher, A.B. Zamolodchikov, G. Zuckerman, and B. Zwiebach.
GM is supported by DOE grant DE-AC02-76ER03075
and by a Presidential Young Investigator Award, and PG by
DOE contract W-7405-ENG-36.

\appendix{A}{Special functions}

\subsec{Parabolic cylinder functions}

Unfortunately, there are four notations commonly used for parabolic cylinder
functions \refs{\gradsh,\abram}. Our wavefunctions $\psi^\pm(a,x)$ are the
$\delta$-function normalized even and odd solutions of
$({d^2\over dx^2}+{x^2\over 4})\psi=a\psi$.
In terms of degenerate hypergeometric $_1 F_1(\alpha,\beta;x)$
and Whittaker functions $M_{\mu,\nu}(x)$, $D_a(x)$, we have even and odd parity
wavefunctions:
\eqn\wvfnspl{\eqalign{&\qquad\psi^+(a,x)
={1\over \sqrt{4 \pi (1+\ee{2\pi a})^{1/2}}}\,(W(a,x)+W(a,-x))\cr
&={1\over \sqrt{4 \pi (1+\ee{2\pi a})^{1/2}}}2^{1/4}\biggl|{\Gamma(1/4+i a/2)
\over \Gamma(3/4+i a/2)}\biggr|^{1/2}\ee{-i x^2/4}
\,{}_1F_1(1/4-ia/2;1/2;i x^2/2)\cr
&={\ee{-i \pi/8}\over 2\pi} \ee{-a \pi/4}|\Gamma(1/4+i a/2)|{1\over \sqrt{|x|}}
M_{ia/2,-1/4}(i x^2/2)\ ,\cr}}
\eqn\wvfnsmi{\eqalign{&\qquad\psi^-(a,x)
={1\over \sqrt{4 \pi (1+\ee{2\pi a})^{1/2}}}\,(W(a,x)-W(a,-x))\cr
&={1\over \sqrt{4 \pi (1+\ee{2\pi a})^{1/2}}}2^{3/4}\biggl|{\Gamma(3/4+i a/2)
\over \Gamma(1/4+i a/2)}\biggr|^{1/2} x \ee{-i x^2/4}
\,{}_1F_1(3/4-ia/2;3/2;i x^2/2)\cr
&={\ee{-3 i \pi/8}\over \pi} \ee{-a \pi/4}|\Gamma(3/4+i a/2)|
{x\over |x|^{3/2}}
M_{ia/2,1/4}(i x^2/2)\ .\cr}}

\subsec{Asymptotics}

Define
\eqn\asydefs{\eqalign{\Phi(\mu)
&\equiv{\pi\over 4}+\half {\rm arg}\,\Gamma(\half+i\mu)\cr
k(\mu)&=\sqrt{1+\ee{2\pi\mu}}-\ee{\pi\mu}=\CO(\ee{-\pi\mu})\cr
k(\mu)\inv&=\sqrt{1+\ee{2\pi\mu}}+\ee{\pi\mu}
=2 \ee{\pi\mu}+\CO(\ee{-\pi\mu})\ .\cr}}
The asymptotic properties of the wavefunctions \abram\ are:\hfil\break
1) $\mu\gg\lambda^2$
\eqn\pcfi{\eqalign{
\psi^+(\mu,\lambda)&\sim {\ee{-\pi\mu/2}\over (2\pi)^{1/2}\mu^{1/4}}
\cosh\bigl(\sqrt{\mu}\lambda\bigr)\cr
\psi^-(\mu,\lambda)&\sim {\ee{-\pi\mu/2}\over (2\pi)^{1/2}\mu^{1/4}}
\sinh\bigl(\sqrt{\mu}\lambda\bigr)\ .\cr}}
2) $-\mu\gg\lambda^2$
\eqn\pcfii{\eqalign{
\psi^+(\mu,\lambda)&\sim {1\over (4\pi)^{1/2}|\mu|^{1/4}}
\cos(\sqrt{-\mu}\lambda)\cr
\psi^-(\mu,\lambda)&\sim {1\over (4\pi)^{1/2}|\mu|^{1/4}}
\sin(\sqrt{-\mu}\lambda)\ .\cr}}
3) $\lambda\gg |\mu|$
\eqn\asyi{\eqalign{\psi^{\pm}(\mu,\lambda)&\sim  {1\over
(2\pi\lambda\sqrt{1+\ee{2\pi\mu}})^{1/2}}
\Bigl(\sqrt{k(\mu)}\cos\bigl(\lambda^2/4-\mu\log\lambda+\Phi(\mu)
\bigr)\cr
&\quad\pm 1/\sqrt{k(\mu)}\sin\bigl(\lambda^2/4-\mu\log\lambda+\Phi(\mu)\bigr)
\Bigr)\ .\cr}}
4) $X\equiv\sqrt{\lambda^2-4\mu}\gg 1$
\eqn\asyii{\eqalign{\psi^{\pm}(\mu,\lambda)
\sim {1\over(2\pi X \sqrt{1+\ee{2\pi\mu}})^{1/2}}
&\Bigl(\sqrt{k(\mu)}\cos\Bigl({1\over 4}\lambda X-\mu \tau(\lambda,\mu)
+{\pi\over 4}\Bigr)\cr
&\pm 1/\sqrt{k(\mu)}\sin\biggl({1\over 4}\lambda X-\mu \tau(\lambda,\mu)
+{\pi\over 4}
\Bigr)\biggr)\ .\cr}
}

\listrefs\bye